\documentclass[]{aastex62}
\usepackage{subfigure,graphicx}
\usepackage{rotating}
\usepackage{hhline}
\usepackage{amsmath}
\usepackage[normalem]{ulem}
\usepackage{soul}
\usepackage{tabularx}

\begin{document}

\title{Analysis of GRB Closure Relationship in Multi-wavelengths}

\author[0000-0002-0786-7307]{M. G. Dainotti}
\affiliation{National Astronomical Observatory of Japan, 2 Chome-21-1 Osawa, Mitaka, Tokyo 181-8588, Japan\\}
\affiliation{Department of Astronomical Sciences, The Graduate University for Advanced Studies, SOKENDAI, Shonankokusaimura, Hayama, Miura District, Kanagawa 240-0193, Japan \\}
\affiliation{Space Science Institute, 4765 Walnut Street Ste B, Boulder, CO 80301, USA}

\author[0000-0003-4709-0915]{S. Bhardwaj}
\affiliation{National Astronomical Observatory of Japan, 2 Chome-21-1 Osawa, Mitaka, Tokyo 181-8588, Japan\\}
\affiliation{Department of Astronomical Sciences, The Graduate University for Advanced Studies, SOKENDAI, Shonankokusaimura, Hayama, Miura District, Kanagawa 240-0193, Japan \\}

\author[0000-0001-9935-8106]{E. Bissaldi}
\affiliation{Dipartimento Interateneo di Fisica “M. Merlin”, Politecnico di Bari, Via E. Orabona 4, 70125 Bari, Italy}
\affiliation{Istituto Nazionale di Fisica Nucleare -- Sezione di Bari, Via E. Orabona 4, 70125 Bari, Italy}

\author[0000-0002-0173-6453]{N. Fraija}
\affiliation{Instituto de Astronomía, Universidad Nacional Autónoma de México Circuito Exterior, C.U., A. Postal 70-264, 04510 México D.F., México}

\author[0000-0002-0169-4003]{S. Sourav}
\affiliation{Department of Physics, Washington University in St. Louis, MO 63130, USA}

\author{A. Galvan-Gamez}
\affiliation{Instituto de Astronomía, Universidad Nacional Autónoma de México Circuito Exterior, C.U., A. Postal 70-264, 04510 México D.F., México}

%\author{R. Bhattacharjee}
%\affiliation{Department of Physics, St. Xavier's College (Autonomous), Kolkata 700016, West Bengal, India}



\begin{abstract}

Gamma-ray bursts (GRBs) are intense pulses of high-energy emission associated with massive stars’ death or compact objects’ coalescence. Their multi-wavelength observations help verify the reliability of the standard fireball model.  We analyze 14 GRBs observed contemporaneously in gamma-rays by the \textit{Fermi} Large Area Telescope (LAT), in X-rays by the \textit{Swift} Telescope, and in the optical bands by \textit{Swift} and many ground-based telescopes. 
We study the correlation between the spectral and temporal indices using closure relations according to the synchrotron forward-shock model in the stratified medium ($n \propto r^{-k}$) with $k$ ranging from 0 to 2.5.
We find that the model without energy injection is preferred over the one with energy injection in all the investigated wavelengths. In gamma-rays, we only explored the $\nu > $ max\{$\nu_c,\nu_m$\} (SC/FC) cooling condition (where $\nu_c$ and $\nu_m$ are the cooling and characteristic frequencies, namely the frequencies at the spectral break). In the X-ray and optical bands, we explored all the cooling conditions, including $\nu_m < \nu < \nu_c$ (SC), $\nu_c < \nu < \nu_m$ (FC), and SC/FC, and found a clear preference for SC for X-rays and SC/FC for optical. 
Within these cooling conditions, X-rays exhibit the highest rate of occurrence for the density profile with $k = 0$, while the optical band has the highest occurrence for $k$ = 2.5 when considering no energy injection. Although we can pinpoint a definite environment for some GRBs, we find degeneracies in other GRBs.

\end{abstract}

%\keywords{Classical Novae (251) --- Ultraviolet astronomy (1736) --- History of astronomy (1868) --- Interdisciplinary astronomy (804)}


\section{Introduction} \label{sec:intro}
Gamma-ray bursts (GRBs) are immensely energetic transient events emitting
radiation spanning the electromagnetic spectrum, from gamma-rays down to X-rays, optical and radio, and up to TeV energies.
%in multi-wavelengths, 
%gamma-rays, X-rays, optical, and radio, 
%although their energy can extend up to the TeV range. 
%primarily detected in the MeV-GeV energy range, although their energy can extend up to the TeV range. 
They are conventionally categorized as short (SGRBs) and long (LGRBs), according to $T_{90}$ duration.
$T_{90}$ is the time in which the 90\% of the counts, including the subtracted background, is emitted between the 5\% and 95\% of the total prompt emission as measured in the 50--300 keV band.
LGRBs have $T_{90}>2$s, while SGRBs have $T_{90}< 2s$ \citep{kouveliotou93}.

%the initial emission burst: < 2s or > 2s, respectively. 
The prompt emission is typically observed at high energies, from gamma-rays to hard and soft X-rays and sometimes also optical \citep[see][for a review]{2014IJMPD..2330002Z, zhang_2018}. 
%The spectrum of GRBs is commonly characterized by a phenomenological function known as the Band function \citep{1993ApJ...413..281B}, or it can include multiple functions, including a thermal component \citep{2005ApJ...625L..95R, 2015ApJ...813..127P}. 
After the prompt emission, the long-lasting emission known as afterglow can be typically observed in soft X-rays, optical, and radio bands \citep[see][for a review]{2015PhR...561....1K, zhang_2018}. 
The afterglow emission is sometimes detected at high energies, too, in the 
MeV-GeV-TeV  
energy range.
GRB 090510 is one of the earliest examples where such behavior was observed \citep{2010ApJ...716.1178A}.

The \emph{Swift} X-ray light curves (LCs) have demonstrated that these LCs display complex characteristics beyond a simple power-law. 
These characteristics have been thoroughly investigated by \citet{Tagliaferri2005,Nousek2006,OBrien2006,Zhang2006,2007ApJ...666.1002Z, 2007ChJAA...7....1Z, zhang07a, zhang07b,sakamoto07, Zhang2019, 2019ApJ...885...29F}. 
A notable characteristic identified in LCs is the presence of a plateau, which refers to a period of relatively constant luminosity that follows the prompt emission of GRBs and comes before the subsequent decay of the afterglow. 
The occurrence of the plateau has been observed in different wavelengths, such as X-ray \citep{OBrien2006,Zhang2006,Nousek2006, sakamoto07,Evans2009}, optical \citep{Dainotti2020ApJ, Dainotti2022}, and radio \citep{2022ApJ...925...15L}. 
These plateaus typically last between $10^2$ and $10^5$ seconds and tend to be attributed to a central engine supplying energy for a long time \citep{dai98,rees98,sari2000,zhang2001,Zhang2006,2007ChJAA...7....1Z,2007ApJ...670..565L, 2008ApJ...675..528L,zhang11}. 
This injection of energy can occur through mechanisms such as the fall-back of accreting matter onto a black hole \citep{Kumar2008,Cannizzo2009,cannizzo2011,Beniamini2017,Li2018,Metzger2018} or the spin-down luminosity from a newborn magnetar \citep{zhang2001,troja07,Toma2007,Rowlinson2010,dallosso2011,gompertz2013, rowlinson2013,rowlinson14,lu2014,gompertz2014,gompertz2015,lu2015,rea15,BeniaminiandMochkovitch2017,Li2018,Metzger2018,Stratta2018,Fraija2020}. Another significant feature often observed in GRB afterglow LCs is the presence of an achromatic break, commonly referred to as jet break, which indicates that GRB jets are highly collimated. A subsequent steep decline is observed in the LC following this jet break. This decline happens because the jet decelerates as it interacts with the surrounding ambient medium, and its emission becomes less beamed. This steepening of the LC is driven by two primary factors: the relativistic beaming of the emission and the lateral expansion of the jet as it spreads out and widens \citep{Rhoads99,SPH99,kumar15}.

The standard fireball model has become widely accepted as one of the most reliable models for describing both the prompt emission and the long-lasting afterglow of GRBs \citep{1996ApJ...473..204S, 1995ApJ...455L.143S, 1999A&AS..138..537S,  2000ApJ...532..286K, 2002ApJ...568..820G,zhang2004,Zhang2006}. 
The long-lasting afterglow emission can be explained by this model %\textcolor{blue}{attributing it to the interplay between the relativistic outflow and the ambient external medium}
 through its association with the interplay between the relativistic outflow and the external medium in vicinity \citep{1999A&AS..138..537S, 2000ApJ...532..286K}. 
An essential component of the conventional fireball model involves the interaction of the shells of the expanding plasma with the external medium. This interaction generates the so-called external forward shock (FS). A quick test of the standard fireball model can be verified using closure relations (CRs) \citep[see][for a review]{Gao2013}. 
CRs are described by the equations that establish the relationships between the spectral index, denoted as $\beta$, and the temporal index, denoted as $\alpha$, of a given segment of the LC. The $\alpha$ and $\beta$ parameters are related to the flux according to the convention $F_v \propto t^{-\alpha}v^{-\beta}$.
In the external FS, electrons primarily undergo acceleration and cooling through synchrotron radiation. 
The formulation of the CR equations relies on specific assumptions concerning a hypothetical astrophysical environment.
These assumptions encompass scenarios such as a uniform density interstellar medium (ISM), stellar wind environment, as well as the stratified medium due to plasma instabilities that adhere to a power-law relationship, denoted as $n(r) \propto r^{-k}$, where $k$ falls within the range of 0 to 2.9 \citep{2000ApJ...532..286K,2007ARA&A..45..177C,2012ApJ...746..122D,2012ApJ...751...57D,Gao2013,2013ApJ...776..120Y,2020ApJ...895...94Y,2020ApJ...896...25F, Dainotti2023Galax..11...25D}.

Most studies about CRs have focused on investigating the environments that correspond to ISM and stellar wind, characterized by values of $k=0$ and $k=2$, respectively
\citep{Chevalier2000, Panaitescu2000, Ramirez-Ruiz2001, Ramirez-Ruiz2005,Dainotti2023Galax..11...25D}. 
The inclusion of the wind medium is typically associated with the stellar wind the massive star emits before its collapse.
The presence of a stratified medium is crucial in understanding the evolution of the relativistic blast wave generated by GRBs.  As the relativistic ejecta from the GRB interacts with the surrounding medium, the blast wave undergoes a transition between different phases of this evolution, from being relativistic to non-relativistic \citep{2012ApJ...746..122D}. The stratified medium serves as a key factor in this transition between the relativistic phase, governed by the Blandford–McKee self-similar solution, and the subsequent Newtonian phase described by the Sedov–Taylor solution \citep{2012ApJ...746..122D}.
%}

%\textcolor{red}{There is generally believed that SGRBs typically arise by the merger of binary systems that involve compact objects, such as a neutron star and a black hole (NH-BH), or two neutron stars (NS-NS) \citep{1976PhDT.........3L, 1992ApJ...395L..83N, 1992ApJ...392L...9D, 1992Natur.357..472U, 1994MNRAS.270..480T, 2007PhR...442..166N, 2017ApJ...846L...5G, 2017ApJ...848L..14G, 2017ApJ...848L..12A}, surrounded by a circumstellar medium constant density  ISM \citep{Sari+98}. In contrast, the possibility of a stellar wind environment is frequently contemplated in the context of LGRBs, which may arise from the gravitational collapse of massive stars \citep{1994MNRAS.269L..41M,DaigneMochkovitch98,2000ApJ...536..195C, Panaitescu+00}. The predominant emphasis in the majority of studies pertaining to CRs linked to GRBs has been on investigating the environments that correspond to ISM and stellar wind, characterized by the values of $k=0$ and $k=2$ respectively. Nevertheless, recent research have found certain cases in which environments demonstrate density profiles that adhere to a power-law relationship, denoted as $n(r) \propto r^{-k}$, where $k$ falls within the range of 0 to 2.9 \citep{2012ApJ...746..122D,2013ApJ...776..120Y,2020ApJ...895...94Y,2020ApJ...896...25F}.}
Previous research has been conducted in separated wavelengths in high-energy gamma-rays, X-rays, and optical.

\subsection{Previous studies on the CRs in gamma-rays}\label{CRs in gamma-rays}
%
%Apart from the FS, there are other physical phenomena like the reverse shock  \citep[where the material propagates back into the GRB shell;][]{1995ApJ...455L.143S} or association with supernovae (SNe), and both of these phenomena can also emit light in the optical range, adding complexity for testing the CRs on optical GRB afterglows. 
%CRs are tools used to test the consistency of observed data with the predictions of the standard fireball model, particularly focusing on the afterglow resulting from the FS. The challenge lies in discerning whether the optical emissions are primarily due to the FS or if these other physical processes influence them.
%Despite these challenges, for the optical GRB afterglows in our multi-wavelength GRB sample, we investigate them using a set of CRs described by the standard fireball model.}

The temporally prolonged emission at high energies, typically lasting for hundreds to thousands of seconds and occurs at energies $\geq$ 100 MeV, is often explained using the synchrotron FS model. 
Consequently, this high-energy emission is expected to conform to the CRs associated with the synchrotron FS model \citep{2009MNRAS.400L..75K, Kumar2010MNRAS}. 
The Large Area Telescope (LAT) instrument onboard the \emph{Fermi}
Gamma-ray Space Telescope \citep[\emph{Fermi}-LAT,][]{Atwood:09} plays a crucial role in detecting and measuring these high-energy GRBs ranging from 20 MeV to over 300 GeV.
The work of \cite{2019ApJ...883..134T}, \citet{2021ApJS..255...13D}, and \cite{Dainotti2023Galax..11...25D} have delved into CRs within the realm of gamma-rays.
\cite{2019ApJ...883..134T} analysed 59 GRBs taken from \emph{Fermi}-LAT.  
They selected these GRBs based on stringent criteria, requiring an uncertainty on their temporal indices and spectral indices to be less than 1/2 and 1/3, respectively.
Their analysis revealed that while the standard synchrotron FS emission model effectively characterizes the spectral and temporal indices for most cases, a substantial fraction of GRBs could not be adequately characterized within this framework. 
Among their key findings, they also discovered that GRBs that fail to conform to any CRs have a temporal decay index $\alpha_{\rm LAT}<1$, indicative of a relatively gradual decay. 
% The results of their analysis revealed that while the standard synchrotron FS emission model effectively characterizes the spectral and temporal indices for most cases, a substantial fraction of GRBs could not be adequately characterized within this framework. 
% Among their key findings, they also discovered that GRBs that fail to conform to any CRs have a temporal decay index $\alpha_{\rm LAT}<1$, indicative of a relatively shallow decay. 
There are few cases in which the plateau emission has been discovered also at high energies \citep{2019ApJ...878...52A}.
\cite{2021ApJS..255...13D} examined CRs for three specific GRBs (090510A, 090902B, and 160509A). The study determined that these set of GRBs conformed to a slow-cooling (SC) environment \footnote{where the timescale of cooling for shocked electrons is of the same order or longer than the GRB jet's dynamic timescale.} ($\nu_{m}<\nu_{\rm LAT}<\nu_{c}$), rather than a fast-cooling (FC) environment ($\nu_{c}<\nu_{\rm LAT}<\nu_{m}$), where $\nu_{m}$ and $\nu_{c}$ are the characteristic and cooling frequencies at the spectral break, regardless of whether they were situated in a constant-density ISM or a stellar wind medium \citep[see][]{Sari+98}. 
On the other hand, the analysis performed by \cite{Dainotti2023Galax..11...25D} differs from that done by \cite{2019ApJ...883..134T}, as they used a bigger sample  \citep[86 vs 59 in][]{2019ApJ...883..134T} and focused on CRs with both broken power law (BPL) and simple power law (PL) fitting, not only PL fitting, as done by \cite{2019ApJ...883..134T}. 
Furthermore, \cite{Dainotti2023Galax..11...25D} employed a frequentist approach to classify the fulfillment of CRs rather than Bayesian probability performed by \cite{2019ApJ...883..134T}.
However, they also found similar results, with most of the GRBs in their sample fulfilling the CRs in the SC regime either in a constant-density ISM or stellar wind environment.
\cite{Dainotti2023Galax..11...25D} found that out of 86 GRBs in their sample, taken from the \emph{Fermi}-LAT Second Gamma-ray Burst Catalog \citep[2FLGC, ][]{2019ApJ...878...52A}, 74 
of them comply with at least one CR, indicating that many of the features observed in high-energy GRBs can be explained by the external FS model. %GRBs 
%\textcolor{blue}{conformed to at least one CR, indicating that the external FS model can elucidate numerous characteristics observed in high-energy GRBs}. 
Thus, 12 GRBs from their sample do not fulfill any CRs, which is interesting for the current study. 
After summing the contribution of the fulfillment rates in each case, they observed a preference for CRs without energy injection over those assuming energy injection. 
They found that for CRs without energy injection, 35 GRBs satisfy at least one CR. 
They also considered a subsample of the 21 GRBs fitted with a BPL and found that 8 GRBs (090926A, 091003, 110731A, 130504C, 160509A, 160816A, 171010A, 171120A) have $\alpha$ and $\beta$ parameters in alignment with the $\nu > $ max\{$\nu_c,\nu_m$\} regime for all values of $k$. 
However, in the case of CRs with energy injection, they found 15 GRBs that failed to adhere to any of the CRs. They also performed MCMC simulations, which indeed support these conclusions.

\cite{2024MNRAS.527.1884F} derived the CRs in a stratified medium with variations of microphysical parameters of the synchrotron and synchrotron self Compton (SSC) FS model (for the introduction of the thermal component, refer to \cite{Warren2022ApJ...924...40W}). 
In the realm of the analytical investigation in some works \citep{Fukushima2017ApJ...844...92F}, it is shown that even if the emission mechanism is switching from synchrotron to SSC, the gamma-ray LCs can be a smooth PL, which agrees with the observed LC at the GeV emission.

Furthermore, \cite{2023ApJ...958..126F} estimated the CRs in an off-axis FS scenario to investigate the spectral and temporal index evolution of the bursts reported in 2FLGC. Finally, \cite{2024MNRAS.527.1674F} introduced the SSC reverse-shock (RS) scenario in a stratified environment for the thick- and thin-shell regimes and showed that this emission can reproduce the early LCs exhibited in some bursts reported in 2FLGC. 

Since a significant fraction of the afterglow phase in GeV emission in the \emph{Fermi}-LAT data cannot be explained by CRs of the standard fireball synchrotron FS model, \cite{2020ApJ...905..112F} proposed the possibility of a significant contribution from the SSC process \citep{2014ApJ...787..168V, Warren2022ApJ...924...40W}. In this regard, \cite{2022ApJ...934..188F} studied the CRs for \emph{Fermi}-LAT GRBs in the SSC afterglow model context using the 2FLGC, and they could explain a considerable portion of bursts with a constant or stellar wind medium. A recent study by \cite{2023MNRAS.525.1630F} examined the CRs for the \emph{Fermi}-LAT GRBs in the framework of the SSC afterglow model also accounting for the intermediate density profile ($\propto r^{-k}$) with $0 \leq k \leq 2.5$, which considers several scenarios, including the adiabatic/radiative regime and the presence or absence of energy injection. They investigated these aspects for all possible values of the electron spectral index, $p$. The study's findings revealed that the afterglow SSC model with an intermediate density profile successfully explains a significant subset of GRBs that do not follow the stellar wind or constant medium.

% \begin{equation*}
% \Gamma = \frac{3}{\left(\frac{m_p c^2}{2\pi k}\right)^{3/4} (2\pi k)^{1/4}} \frac{1}{(1+z)^{3/4}} A_k^{1/4} E^{1/4} \left(\frac{t}{1+z}\right)^{-1/4}
% \end{equation*}

\subsection{Previous studies on the CRs in X-rays}
In the regime of X-rays, \citet{2009ApJ...698...43R}, \citet{Srinivasaragavan2020ApJ}, and \citet{2021PASJ...73..970D} have performed studies for CRs. 
Their analysis showed that most GRBs within their sample can be ascribed to the external FS model as predicted by the standard fireball model.
These models assume not a structured jet as \citet{Ryan2020ApJ...896..166R}. The most preferred scenario is the SC regime, regardless of a constant or a stellar medium environment. 
\citet{2009ApJ...698...43R} explored the CRs for both the ISM and stellar wind environments, considering cases with and without energy injection. 
\citet{Srinivasaragavan2020ApJ} fitted 455 X-ray LCs that exhibit the plateau phase.
They investigated whether these LCs follow the CRs in two different astrophysical environments and cooling regimes within the external FS model. 
They found that the most favored environments in the SC regime are a wind medium or a constant-density ISM. 
They also confirmed the existence of the 3D fundamental plane relation (also known as the Dainotti relation) between the rest-frame time and luminosity at the end of the plateau emission and the peak prompt luminosity with a much larger sample compared to previous studies \citep[see][for 3D fundamental plane relation]{2016ApJ...825L..20D, 2020ApJ...904...97D, Dainotti2022}. 
They further segregated the sample of GRBs following the Dainotti relation within groups corresponding to distinct astrophysical environments found by studying the CRs. This plane emerges as a model discriminator for these environments. The metrics used to determine if a given environment and energy emission mechanism can be promoted as possible standard candles are obtained by looking at the best-fit parameters and their dispersions. The smaller the dispersion of the fundamental plane, the better the sample is considered a standard candle. They found that the sample of GRBs that have peculiar CRs exhibit an intrinsic scatter $\sigma$ that is compatible within a $1\sigma$ range of the ``Gold" GRBs, a subset of LGRBs featuring relatively flat plateaus. Although this analysis has not led to a smaller dispersion for these samples, it is still a pathway to lead to standard candle samples comparable to the golden sample dispersion.
\citet{2021PASJ...73..970D} also analyzed 455 X-ray LCs to examine these GRBs' emission mechanisms and astrophysical environments by studying CRs within the time interval of the plateau emission.
They found that most recurrent environments for the electron spectral distribution, $p > 2$, are wind SC and ISM SC for cases where the parameter $q$, indicating the flatness of the plateau emission and incorporating energy injection, is 0 and 0.5, respectively. They also find that for SGRBs, all ISM environments with $q$ = 0 have the smallest $\sigma = 0.04 \pm 0.15$ in terms of the fundamental plane relation. They have shown that most GRBs presenting plateau emission fulfill the CRs, including the energy injection, with a particular preference for the wind SC environment. Again, similarly to the study of the post-plateau phase, in this case, GRBs that fulfill the given relations can be used as possible standard candles.
 Moreover, these findings offer insights into possible strategies for diminishing the intrinsic scatter observed in these investigated relationships \citep{Dainotti2017ApJ,2022MNRAS.514.1828D}.

%\textcolor{magenta}{[ADD THE Liang ET AL> 2007, 2008, Gao et al.]}
\subsection{Previous studies on the CRs in the optical domain}
%[ADD THE references Li optical different component]
 In the optical domain, a previous study by \citet{oates2012} examined 48 GRBs observed by \emph{Swift}. 
Their analysis indicated that almost half of the GRBs in their sample conformed to the standard fireball model determined by the CRs.
They evaluated CRs in three different density profiles: one for a wind-like density environment, one for a constant-density ISM environment, and one independent of the density profile of the external medium, i.e., $\nu_{\rm opt} > \nu_{c}$.  In the first scenario, 6 GRBs follow a constant medium; in the second scenario, 7 GRBs follow the wind medium, and 8 GRBs follow the third scenario.
 \cite{2022A&A...662A.126J} investigated the optical afterglow of LGRB 190919B and concluded that it follows the CR for an SC regime with constant ISM.
Recently, \citet{2022ApJ...940..169D} studied CRs in the optical band using a sample of 82 GRBs.
Their study found that the most favored regime is $\nu > $ max\{$\nu_c,\nu_m$\} for both ISM and stellar wind medium. 
Similarly to what has been done in X-rays, they tested the 2D Dainotti correlation between the rest-frame end time of the plateau and the luminosity at that time \citep{Dainotti2008, dainotti2010, Dainotti11b, Dainotti2013b, Dainotti2015b, delvecchio16, Dainotti2017a, Dainotti2020ApJ, Dainotti2022} for GRBs that satisfy the favored CRs, to understand if these samples are better suited for cosmological analysis and have a physical grounding in the framework of the standard fireball model.
They found that the scatter within this sample in the 2D Dainotti relation is compatible with the previous values in the X-ray \citep{Dainotti2013b, 2021PASJ...73..970D, Srinivasaragavan2020ApJ}, optical \citep{Dainotti2020ApJ, Dainotti2022}, and radio \citep{2022ApJ...925...15L}, within the 1$\sigma$ range, both before and after correcting for selection biases. 
Like X-rays, this method identifies subsets of GRBs underlying a physical emission mechanism or a peculiar environment that could pave the way for using GRBs as standard candles.

\subsection{Previous study on the CRs in multi-wavelengths}
\begin{figure*}
\gridline{\fig{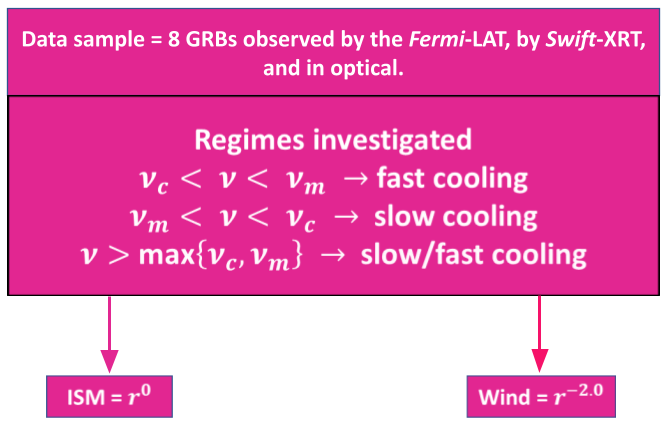}{0.51\textwidth}{}
          }
\gridline{\fig{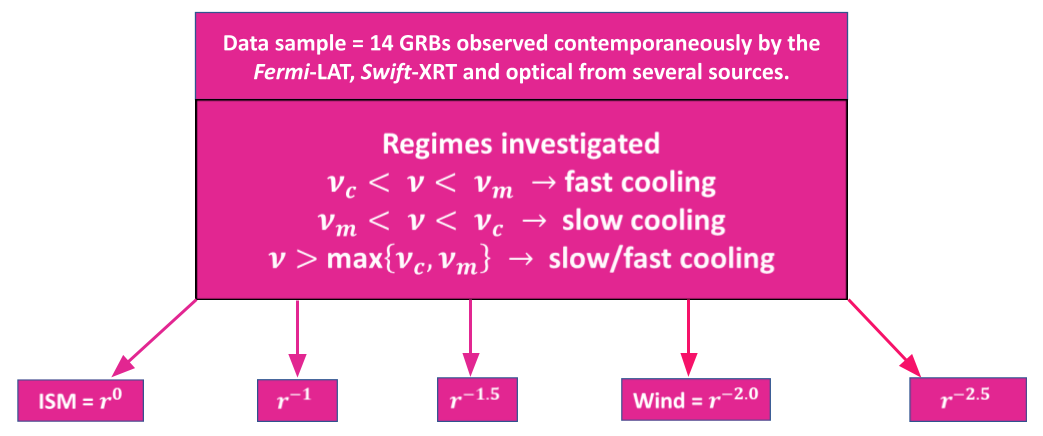}{0.81\textwidth}{}
         }
\caption{Top: Illustration depicting the GRB sample and regimes explored in prior multi-wavelength studies. Bottom: Schematic of the GRB sample and regimes investigated in this study in multi-wavelength.} 
\label{fig:current-studies}
\end{figure*}

These studies investigated what fraction of GRBs satisfied the CRs within the standard fireball model to determine what environment and energy mechanism they favored. 
The top panel in Figure \ref{fig:current-studies} summarizes the previous studies conducted in the multi-wavelength domain. 

\cite{2011A&A...526A.154A} analyzed GRB 050502B to understand the possible correlation between its X-ray and optical data. They found that GRB 050502B follows the CR in the fireball synchrotron forward shock model for both constant-density ISM and stellar-wind medium. 
%\textcolor{blue}{\citet{Zhang+11} studied the physical nature of GRBs across the gamma, X-ray, and optical wavelengths.}
\citet{2013ApJ...763...71A} and \citet{2015ApJS..219....9W} have examined CRs in both X-rays and optical wavelengths.
%Comparable outcomes emerged from their investigation, as they both observed a similar preference for the SC regime, whether situated within a stellar wind or a constant-density ISM environment, respectively. 
\citet{2013ApJ...763...71A} analyzed CRs involving X-ray and optical data for GRB\,110731A, which had also been observed by \emph{Fermi}-LAT. They found that GRB 110732A favored the SC regime in a stellar wind environment.
%While 
\citet{2015ApJS..219....9W} conducted a comprehensive study involving 85 GRBs observed by \emph{Swift}. 
Their analysis also focused on testing the CRs in X-ray and optical, aiming to test their consistency with the synchrotron FS model. Within their sample, they identified 45 out of 85 GRBs that exhibited an achromatic break, and these GRBs were found to align with the standard CRs of the external FS model across all segments of their afterglows. 
Furthermore, their study identified an additional 37 GRBs that did not entirely satisfy the CRs in one or more segments of their afterglows but did display an achromatic break. 
This suggests that the synchrotron FS model could partially describe a substantial portion of their sample, even though there were deviations in some regions of their afterglow LCs. In the study of \citet{Fukushima2017ApJ...844...92F}, numerical models were employed to simulate X-ray and optical LCs within an ISM environment, accounting for both FC and SC regimes. 
They applied these simulation models specifically to the case of GRB 130427A and found that a more complex and refined model is required to describe the behavior of this burst.
\cite{gompertz2018environments} analyzed a sample of 56 LGRBs detected by \textit{Fermi} (LAT and GBM) in gamma-rays and by the \textit{Swift} X-ray Telescope (XRT) and by optical and NIR telescopes by performing a fitting of the temporal and spectral indices of these GRBs using the synchrotron CRs.  We note that between our study and this one, we have 7 GRBs in common, but our study includes 230812, so it includes new data for more than 8 years. The difference is that we restrict ourselves to the \textit{Fermi}-LAT data. 
Another key difference is that they focused on two density profiles (wind medium and ISM), while we also explore the stratified medium ($r^{-1}$, $r^{-1.5}$, and $r^{-2.5} $) along with the wind and ISM environment, providing a comprehensive analysis of the CRs to test the external synchrotron FS model.

Recently, a study conducted by \citet{2021ApJ...911...14K} entails a comparative analysis of the observed patterns in radio LCs and their counterparts in X-ray and optical domains. They used a sample of 21 GRBs with radio LCs in their afterglow phase. Their findings indicated a substantial incompatibility between the radio LCs and the observed patterns exhibited by X-ray and optical LCs. They concluded that the radio LCs are inconsistent with the standard fireball model.
Similarly, \citet{2021MNRAS.504.5685M} investigated the afterglow of GRB 190114C across several wavelengths, encompassing X-ray, optical, and radio observations. Their analysis revealed that the X-ray and radio LCs exhibited behaviors inconsistent with predictions based on the standard fireball model for GRB afterglows.

In a recent study by \citet{2022NatCo..13.5611D}, a detailed study was conducted on a sample comprising 13 GRBs exhibiting well-defined X-ray plateaus along with optical counterparts. The study identified three distinct phases within the LC of these GRBs: firstly, the plateau phase; second, the transition from the plateau to the decaying LC (marking the first break as the end of the plateau); and third, a second, late-time break, called jet-break, which typically occurs when the relativistic jet begins to decelerate significantly. They explored two cooling regimes, namely fast ($\nu_{m} > \nu_{c}$) and slow ($\nu_{c} > \nu_{m}$), each of them divided into three regions. For fast cooling, the regions are defined as A ($\nu < \nu_{c} < \nu_{m}$), B ($\nu_{c} < \nu < \nu_{m}$), and C ($\nu_{c} < \nu_{m} < \nu$). For slow cooling, the regions are D ($\nu < \nu_{m} < \nu_{c}$), E ($\nu_{m} < \nu < \nu_{c}$), and F ($\nu_{m} < \nu_{c} < \nu$). Their study primarily focused on regions C, E, and F, demonstrating that the CRs are independently satisfied within each wavelength band and during each phase of the LC.

%\textcolor{red}{Nevertheless, the 
In this work, we present the first comprehensive analysis and evaluation of CRs using multi-wavelength observations in the context of the synchrotron FS model % has not yet been undertaken 
for a combined sample observed in gamma-rays, X-rays, and optical wavelengths. 
%\textcolor{red}{Therefore, it is imperative to conduct a more comprehensive analysis into the multi-wavelength aspect 
%We aim to gain a thorough understanding of the validity of the standard fireball model.
We aim to understand the validity of the standard fireball model.
%EB: The next sentence is a repetition! I suggest to remove it. In this context, our research primarily \textcolor{orange}{focuses} on the examination of CRs across multiple wavelengths, specifically gamma-rays, X-rays, and optical \textcolor{orange}{wavelengths}. 
%EB: I suggest not to use the future tense.
%We will 
%\begin{figure}[t!]
%    \centering
%    \includegraphics[width=0.9\textwidth]{flow-chart-present-studies.png}
%    \caption{Schematic of the GRB sample and regimes investigated in this study in multi-wavelength.}
%    \label{fig:current-studies}
%\end{figure}
The main idea is to conduct a comprehensive analysis by comparing individual GRBs, employing a novel approach.
%This will enable us to ascertain the presence or absence of a jet break. 
A deeper understanding of GRB's fundamental characteristics can be achieved by examining CRs across multiple wavelengths. This could advance our ability to categorize and standardize GRBs for cosmological studies in forthcoming studies.
%The novelty here is that the one-to-one GRB comparison in multi-wavelength (gamma-rays, X-rays, and optical) has not been done to date. 
%This will allow us to check if a jet break is present or not.
%In this paper, we investigate the CRs for GRBs observed across gamma-rays, X-rays, and optical wavelengths, and for which the LCs can be fitted with a PL or a BPL or different combinations thereof.
%\textcolor{blue}{EB: this sentence is again a repetition of something already said few lines before!}

The bottom panel in Figure \ref{fig:current-studies} shows the summarized overview of our study. % presented in this paper. 
The paper is structured as follows: Section \ref{sec:datasample} provides an overview of the data sample used in our analysis. 
In Section \ref{sec:methodology}, we show the initial data analysis, which requires fitting the LCs, a necessary step for evaluating the CRs. In that section, we also present the methodology used to examine the theoretical CRs associated with constant ISM, stellar wind, and stratified environment.
%\textcolor{red}{In Section \ref{sec:methodology}, we examine the theoretical CRs associated with both constant ISM, \textcolor{magenta}{stellar wind, and stratified ISM} environment.}
Finally, Section \ref{sec:results} provides a detailed summary of our results and conclusions.

\section{Data Sample}\label{sec:datasample}
%

%Following \cite{Dainotti2023Galax..11...25D}, we selected the sample of 86 GRBs from the 2FLGC observed from 2088 to 2018, showing temporally extended emission. It contains 21 GRBs fitted with BPL and 65 GRBs fitted with PL.
The sample used in this study comprises 14 GRBs contemporaneously observed in gamma-rays, X-rays, and optical wavelengths. To obtain our final sample of 14 GRBs for our analysis, we have initially taken 33 GRBs from \textit{Fermi}-LAT, observed from August 18th, 2008 to August 12th, 2023, each with measured redshift. This group includes the BOAT GRB (221009A), and since the \textit{Fermi}-LAT data for this GRB is not yet public, it was excluded from our analysis. Thus, we are left with 32 GRBs from \textit{Fermi}-LAT, each with measured redshift, temporal indices ($\alpha$), and spectral indices ($\beta$). We further excluded three GRBs due to their high relative errors ($\delta\alpha_{\rm{LAT}} / \alpha_{\rm{LAT}} > 1$ and $\delta\beta_{\rm{LAT}} / \beta_{\rm{LAT}} > 1$), resulting in a sample of 29 GRBs. The \emph{Fermi}-LAT data has been taken from the official \emph{Fermi}-LAT analysis presented in the 2FLGC\footnote{\url{https://www-glast.stanford.edu/pub_data/953/2FLGC/}} \citep{2019ApJ...878...52A}.

\begin{figure}
    \centering
    \includegraphics[scale=0.95]%,\height=0.60]
    {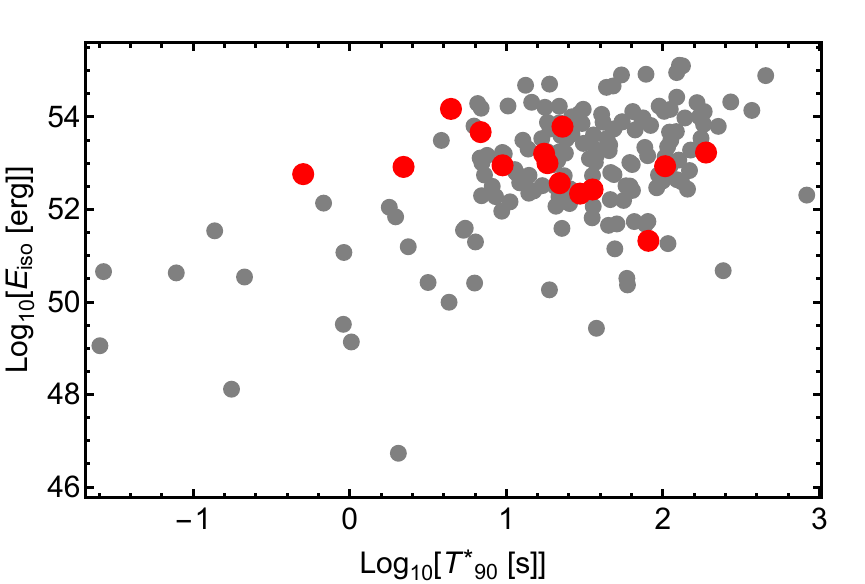}
       \caption{The distribution of $\log(E_{iso})$ vs. $\log(T^{*}_{90})$ for the bursts in our sample (large red dots), together with all other GRBs with redshift reported in the Fermi-GBM catalog (gray dots). Quantities marked with * refer to rest-frame-calculated values. }
       \label{logEisovsT90rest}
    
\end{figure}

In the same period as the \textit{Fermi}-LAT observations, 349 GRBs with measured redshifts were documented in the optical catalog \citep[see][]{DainottiOpticalCatalog}. Of these, 27 GRBs overlapped with our \textit{Fermi}-LAT sample. Within this subset, 17 GRBs had documented $\beta_{\rm{opt}}$ from the literature and satisfied the criterion $\delta\beta_{\rm{opt}} / \beta_{\rm{opt}} < 1$. From these, we selected 16 GRBs with redshift information obtained from spectroscopy rather than photometry. One GRB was excluded because it only had two data points from \textit{Swift}-XRT, resulting in a sample of 15 GRBs. We then conducted color evolution analysis on the optical data of these 15 GRBs (see Section \ref{sec:methodology}) and excluded another GRB that had only two data points in the optical spectrum after the color evolution analysis. This process yielded a final data sample of 14 GRBs.
%with observed redshift, and all the spectral indices and temporal indices are well constrained in all wavelengths, fulfilling the condition $\delta_{\alpha}/\alpha<1$, $\delta_{\beta}/\beta<1$. Another condition that must be fulfilled is the absence of temporal evolution for the optical LCs to guarantee that we can rescale them safely by adding multiple data points. For details about this analysis, refer to \cite{DainottiOpticalCatalog}. 
%The \emph{Fermi}-LAT data has been taken from the official \emph{Fermi}-LAT analysis presented in the 2FLGC\footnote{\url{https://www-glast.stanford.edu/pub_data/953/2FLGC/}} \citep{2019ApJ...878...52A}. 
The XRT data has been taken from the \emph{Swift}-BAT+XRT repository\footnote{\url{https://www.swift.ac.uk/xrt_live_cat/}} \citep{Evans2009}, while the optical data for all the 14 GRBs has been taken from \cite{DainottiOpticalCatalog}.
%\textcolor{blue}{This first sentence is difficult to understand, please rephrase. I suggest to simply say "The sample used in this study comprises 15 GRBs contemporaneously observed in Gamma-rays, X-rays and in the optical. Fermi-LAT data was taken from...for 14 cases, etc etc etc.}%These newly added GRBs were also observed in gamma-rays, X-rays, and optical.
More precisely, the optical data is gathered from \emph{Swift}-UVOT \citep{2005SSRv..120...95R} and 416 ground-based telescopes, as outlined in \cite{Dainotti2022}.
The complete list of our sample of 14 GRBs is given in Tables \ref{table:datasample1}, \ref{table:datasample2}, and \ref{table:datasample3}.  All the temporal and spectral fits are tabulated with their corresponding uncertainties. When a model does not provide a reliable fit, it is replaced by a ``-" in all Tables. 
In Table \ref{table:datasample1}, we also provide the $T_{90}$ and $E_{iso}$ for each GRB in our sample 
Figure  \ref{logEisovsT90rest} shows the distribution of $\log E_{iso}$ vs. log  $\log T^{*}_{90}$  for the bursts in our sample (large red dots), together with all other GRBs with redshift reported in the \textit{Fermi}-GBM catalog (gray dots). 
Quantities marked with $*$ refer to rest-frame-calculated values. 
In Fig. \ref{logEisovsT90rest}, there is a positive correlation with a Pearson coefficient of 0.52, which is aligned with the previous analysis in the literature. \cite{Dainotti2015b} showed a positive correlation between the energy released during the prompt episode vs. the time of the pulses, which on average roughly corresponds to the $T^{*}_{90}$.
There is no particular clustering of the data in any region of the $\log E_{iso}$ vs. $\log T^{*}_{90}$, but the data tend to be on the higher end of the $E_{iso}$ distribution.

We here stress that the optical spectral indices have been taken from various sources in the literature, which are quoted in the seventh column of Table \ref{table:datasample3}, and that we consider only the GRBs for which we have not found color evolution. For a more detailed analysis of how we determine the color evolution, refer to Section \ref{sec:methodology}.

.%\citep{DainottiOpticalCatalog}.

%EB: In Table 1 the parameter alpha 1 is listed twice?
%shows the sample of 15 GRBs utilized in this study.

%% gamma-ray data table
\begin{longrotatetable}
\begin{deluxetable*}{lccccccccccccc}
%\tablenum{1}
\tablecaption{%The table represents the set of 15 GRBs observed in gamma-rays. The first column indicate the name, the second the model to the fit the temporal slope, the third and forth column shows the $\alpha$ parameters table also shows the best-fitting models in gamma-rays and their corresponding temporal and spectral indices, denoted by $\alpha_{\rm{\gamma}}$ and $\beta_{\rm{\gamma}}$, respectively. \label{table:datasample1}
Best-fit parameters of the temporal and spectral indexes from the gamma-ray analysis. Columns 1 and 2 list the GRB names and the best-fit models, respectively. Columns 3 and 4 list the $T_{90}$ (in the 50-300 keV energy range$^{a}$) and $E_{iso}$$^{b}$ for each GRB, respectively. Columns 5 and 6 list the first temporal slope of the BPL fit ($\alpha_{\rm{\gamma_1}}$) and the second temporal slope of the BPL fit ($\alpha_{\rm{\gamma_2}}$), respectively. When the BPL is not a viable model for the paucity of data points, we adopt the PL model. %\textcolor{blue}{EB: HERE WE NEED TO ADD A SENTENCE ABOUT THE FOUR CASES FIT BY THE SIMPLE PL. }
%\textcolor{magenta}{Is it now clear?}
Column 7 lists the spectral parameter, $\beta_{\rm{\gamma}}$, which is the same in the two segments of the LC. Columns 8 and 9 lists the start and the end times of the LC, respectively. Column 10 lists the break time between the two segments of BPL. Columns 10 and 11 lists the start and end times of the energy injection. Column 12 lists the duration of the plateau. %EB: I AM NOT SURE ABOUT THIS TABLE. WHICH ONE IS THE TEMPORAL AND WHICH ONE THE SPECTRAL PARAMETERS? PLEASE USE THE SAME TYPE OF SHORT DESCRIPTION ALSO FOR TABLES 2 AND 3.}
\label{table:datasample1}}

%\tablewidth{0pt}
\tabletypesize{\scriptsize}
\tablehead{
%\colhead{GRB Name} & \colhead{gamma-ray LC best-fit model} & \colhead{$\alpha_{\rm{\gamma_1}} \pm \delta_{\alpha_{\rm{\gamma_1}}}$} & \colhead{$\alpha_{\rm{\gamma_2}} \pm \delta_{\alpha_{\rm{\gamma_2}}}$} & \colhead{$\beta_{\rm{\gamma}} \pm \delta_{\beta_{\rm{\gamma_{err}}}}$} \\
\colhead{GRB} \vspace{-0.2cm} & \colhead{gamma-ray} & \colhead{$T_{90}$} & \colhead{$E_{iso}$} & \colhead{$\mathbf{\alpha_{\rm{\gamma_1}} \pm \delta_{\alpha_{\rm{\gamma_1}}}}$} & \colhead{$\mathbf{\alpha_{\rm{\gamma_2}} \pm \delta_{\alpha_{\rm{\gamma_2}}}}$} & \colhead{$\mathbf{\beta_{\rm{\gamma}} \pm \delta_{\beta_{\rm{\gamma_{err}}}}}$} & \colhead{$\log(t_{start})$} & \colhead{$\log(t_{end})$} & \colhead{$\log(t_{break})$} & \colhead{$\log(t_{start,EI})$} & \colhead{$\log(t_{end,EI})$} & \colhead{$log (t_{plateau})$} \\
%\colhead{Name} & \colhead{LC best-fit} & \colhead{} & \colhead{(100 MeV--10 GeV)} & \colhead{ } & \colhead{ } & \colhead{ } & \colhead{} & \colhead{} & \colhead{} & \colhead{} & \colhead{} & \colhead{} \\
\colhead{Name} & \colhead{LC best-fit model} & \colhead{(s)} & \colhead{($\times 10^{52}$ erg)} & \colhead{ } & \colhead{ } & \colhead{ } & \colhead{(s)} & \colhead{(s)} & \colhead{(s)} & \colhead{(s)} & \colhead{(s)} & \colhead{(s)}
%\multicolumn2c{Distance} & \colhead{} & \colhead{V} \\
%\colhead{Number} & \colhead{Number} & \nocolhead{Name} & \colhead{Type} &
%\multicolumn2c{(kpc)} & \colhead{Constellation} & \colhead{(mag)}
}
%\decimalcolnumbers
\startdata
090328A & BPL & $61.7 \pm 1.8$ & $2.7 \pm 0.4$   & $0.73 \pm 0.43$  & $1.06 \pm 0.14$ & $1.20 \pm 0.13$ &    1.32 & 5.00 & 2.40 &  1.32 & 2.40 & $2.36$ \\%3.838 \\ 
090510A & BPL & $0.96 \pm 0.14$ & $5.8 \pm 0.5$  & $2.32 \pm 0.18$  & $1.34 \pm 0.18$ & $1.05 \pm 0.06$ &   -0.17 & 4.96 & 0.60 & -0.17 & 0.60 & $3.3$ \\%0.055 \\
090902B & BPL & $19.3 \pm 0.3$ & $47 \pm 2$    & $1.87 \pm 0.17$  & $1.24 \pm 0.23$ & $0.94 \pm 0.04$ &    0.35 & 4.99 & 2.20 &  0.35 & 2.20 & $2.20$\\%2.604 \\
090926A & BPL & $13.8 \pm 0.3$ & $149 \pm 8$   & $1.77 \pm 0.17$  & $1.10 \pm 0.17$ & $1.14 \pm 0.05$ &    0.60 & 4.98 & 2.00 &  0.60 & 2.00 & 1.98\\%1.599 \\
% 100414A & BPL & $1.80 \pm 0.22$ & $0.3   \pm 0.6$   & $0.80 \pm 0.12$ \\
% 110731A & BPL & $1.77 \pm 0.11$ & $0.4   \pm 1.1$   & $1.29 \pm 0.16$ \\
120711A & PL & $44.0 \pm 0.7$ & $10 \pm 2$     & $-$               & $1.63 \pm 0.24$ & $1.06 \pm 0.17$ &   2.65 & 4.96 & $-$  &  2.65 & 4.96 & 4.96\\ %1526.621 \\
130427A & BPL & $138 \pm 3$ & $8.6 \pm 0.4$      & $0.79 \pm 0.16$ & $1.42 \pm 0.10$ & $0.99 \pm 0.04$ &   0.26 & 4.96 & 2.70 &  0.26 & 2.70 & 2.70\\%8.323 \\
141028A & PL & $31.49 \pm 2.4$ & $9 \pm 2$     & $-$               & $0.97 \pm 0.03$ & $1.44 \pm 0.23$ &   1.15 & 4.98 & $-$  &  1.15 & 4.98 & 4.98 \\%1595.089 \\
%160509A & BPL & $0.88 \pm 0.27$ & $1.3   \pm 0.3$   & $1.38 \pm 0.12$ \\
160625B & PL & $453.4 \pm 0.6$ & $17 \pm 1$    & $-$                & $2.24 \pm 0.28$ & $1.35 \pm 0.07$ &  1.93 & 4.99 & $-$  &  1.93 & 4.99 & 4.99 \\%1616.108 \\
170405A & PL & $78.6 \pm 0.6$ & $16 \pm 7$     & $-$                & $1.27 \pm 0.01$ & $1.79 \pm 0.35$ &  1.54 & 4.98 & $-$  &  1.54 & 4.98 & 4.98 \\%1602.112 \\
171010A & BPL & $107.3 \pm 0.8$ & $0.21 \pm 0.03$ & $2.24  \pm 0.73$  & $0.97 \pm 0.29$ & $1.04 \pm 0.13$ &  2.58 & 4.99 & 2.90 &  2.58 & 2.90 & 2.61 \\%6.873 \\
180720B & BPL & $48.9 \pm 0.4$ & $2.2 \pm 0.2$   & $1.46 \pm 0.19$  & $3.20 \pm 0.56$ & $1.23 \pm 0.10$ &  1.22 & 4.99 & 2.37 &  1.22 & 2.37 & 2.34 \\%3.632 \\
210822A & PL & $180 \pm 40^{c}$ & $95 \pm 8^{d}$    & $-$          & $0.57 \pm 0.18$ & $1.11 \pm 0.36$ &  2.99 & 4.97 & $-$  &  2.99 & 4.97 & 4.98 \\%1539.137 \\
220101A & PL & $128 \pm 16$ & $364 \pm 23^{e}$ & $-$                & $1.10 \pm 0.53$ & $1.54 \pm 0.25$ &  1.62 & 4.99 & $-$  &  1.62 & 4.99 & 4.99 \\%1609.392 \\
230812B & PL & $3.3 \pm 0.1$ & $ 12 \pm 1^{f}$ & $-$                & $1.14 \pm 0.07$ & $1.16 \pm 0.14$ & -0.74 & 4.54 & $-$  & -0.74 & 4.54 & 4.54 \\ %581.897 \\
\enddata
\end{deluxetable*}
\tablenotetext{a}{From \href{https://heasarc.gsfc.nasa.gov/W3Browse/all/fermilgrb.html}{https://heasarc.gsfc.nasa.gov/W3Browse/fermi/fermigbrst.html}}
\tablenotetext{b}{From \citet{Ajello2019apj}}
\tablenotetext{c}{From \citet{2021GCN.30689....1L} in the $15-350$ keV range}
\tablenotetext{d}{From \citet{2021GCN.30694....1F} in the 20 keV--20 MeV range}
\tablenotetext{e}{From \citet{2022GCN.31433....1T} in the 20 keV--20 MeV range}
\tablenotetext{f}{From \citet{2024MNRAS.530....1H} in the 10 keV--1 MeV range}
\end{longrotatetable}

%% X-ray data table
%\begin{sideways}{lcccccccccc}
%\begin{landscape}
\begin{longrotatetable}
\begin{deluxetable*}{lccccccccccl}
%\begin{deluxetable*}{lcccccccccccl}
%\begin{deluxetable*}{lllrrrrrrll}
%\tablenum{2}
\tablecaption{Best-fit parameters of the temporal and spectral indices from the X-ray analysis. Columns 1 and 2 list the GRB names and the best-fit model, respectively. Columns 3 and 7 lists the first temporal slope of the PL+BPL+PL fit, $\alpha_{\rm{X_1}}$, and the corresponding spectral parameter, $\beta_{\rm{X_1}}$, respectively. Columns 4 and 8 lists the second temporal slope of the PL+BPL+PL fit, $\alpha_{\rm{X_2}}$, and the corresponding spectral parameter, $\beta_{\rm{X_2}}$, respectively. Columns 5 and 9 lists the third temporal slope of the PL+BPL+PL fit, $\alpha_{\rm{X_3}}$, and the corresponding spectral parameter, $\beta_{\rm{X_3}}$, respectively. Columns 6  and 10 list the fourth temporal slope of the PL+BPL+PL fit, $\alpha_{\rm{X_4}}$, and the corresponding spectral parameter, $\beta_{\rm{X_4}}$, respectively.
%When PL+BPL is present, the first temporal slope is absent and is replaced by a "-", while in the presence of BPL, both the first and second temporal slopes are absent and are marked as "-". When PL is present, the first, second, and third temporal slopes are absent and are replaced by a "-". %Column 5th lists the spectral parameters, $\beta$. 
%All the temporal and spectral fits are tabulated with their corresponding uncertainties. 
\label{table:datasample2}}
%\tablewidth{700pt}
\tabletypesize{\scriptsize}
\tablehead{
\colhead{GRB} \vspace{-0.2cm} & \colhead{X-ray LC} & \colhead{$\alpha_{\rm{X_1}} \pm \delta_{\alpha_{\rm{X_1}}}$} & \colhead{$\alpha_{\rm{X_2}} \pm \delta_{\alpha_{\rm{X_2}}}$} & \colhead{$\alpha_{\rm{X_3}} \pm \delta_{\alpha_{\rm{X_3}}}$} & \colhead{$\alpha_{\rm{X_4}} \pm \delta_{\alpha_{\rm{X_4}}}$} & \colhead{$\alpha_{\rm{X_5}} \pm \delta_{\alpha_{\rm{X_5}}}$} & \colhead{$\beta_{\rm{X_1}} \pm \delta_{\beta_{\rm{X_{1}}}}$} & \colhead{$\beta_{\rm{X_2}} \pm \delta_{\beta_{\rm{X_{2}}}}$} & \colhead{$\beta_{\rm{X_3}} \pm \delta_{\beta_{\rm{X_{3}}}}$} & \colhead{$\beta_{\rm{X_4}} \pm \delta_{\beta_{\rm{X_{4}}}}$} &
\colhead{$\beta_{\rm{X_5}} \pm \delta_{\beta_{\rm{X_{5}}}}$} \\
\colhead{Name} & \colhead{best-fit} & \colhead{ } & \colhead{ } & \colhead{ } & \colhead{ } & \colhead{ } & \colhead{ } & \colhead{ } & \colhead{ } \\
\colhead{ } & \colhead{model} & \colhead{ } & \colhead{ } & \colhead{ } & \colhead{ } & \colhead{ } & \colhead{ } & \colhead{ } & \colhead{ } 
%\multicolumn2c{Distance} & \colhead{} & \colhead{V} \\
%\colhead{Number} & \colhead{Number} & \nocolhead{Name} & \colhead{Type} &
%\multicolumn2c{(kpc)} & \colhead{Constellation} & \colhead{(mag)}
}
%\decimalcolnumbers
\startdata
%080916C & BPL  & $-$ & $-$ & $2.1 \pm 0.4$   & $1.01 \pm 0.16$ & $-$ & $-$ & $0.8 \pm 0.4$   & $0.7 \pm 0.3$\\
090328A & PL & $-$ & $-$ & $-$ & $-$ & $1.64 \pm 0.07$ & $-$ & $-$ & $-$ & $-$ & $0.97 \pm 0.24$\\ 
090510A & BPL & $-$ & $-$ & $-$ & $0.63 \pm 0.04$ & $2.06 \pm 0.08$ & $-$ & $-$ & $-$ & $0.64 \pm 0.12$ & $0.86 \pm 0.18$\\
090902B & PL & $-$  & $-$ & $-$ & $-$ & $1.33 \pm 0.03$ & $-$ & $-$ & $-$ & $-$ & $0.76 \pm 0.12$\\
090926A & PL & $-$ & $-$ & $-$ & $-$ & $1.43 \pm 0.03$ & $-$ & $-$ & $-$ & $-$ & $1.03 \pm 0.12$\\
%100414A & PL   & $-$ & $-$ & $-$               & $1.70 \pm 0.07$ & $-$ & $-$ & $-$               & $0.7 \pm 0.6$\\
%110731A & PL+BPL & $-$ & $1.268 \pm 0.037$ & $1.110 \pm 0.024$ & $1.316 \pm 0.036$ & $-$ & $0.97 \pm 0.06$ & $0.78 \pm 0.08$ & $0.91 \pm 0.21$\\
120711A & PL & $-$ & $-$ & $-$ & $-$ & $1.60 \pm 0.01$ & $-$ & $-$ & $-$ & $-$ & $0.82 \pm 0.08$\\
130427A & PL+BPL+PL & $-$ & $2.099 \pm 0.002$ & $0.94 \pm 0.01$ & $1.45 \pm 0.08$ & $1.28 \pm 0.01$ & $-$ & $0.61 \pm 0.02$ & $0.50 \pm 0.02$ & $0.85 \pm 0.17$& $0.70 \pm	0.04$\\
141028A & PL & $-$ & $-$ & $-$ & $-$ & $0.92 \pm 0.26$ & $-$ & $-$ & $-$ & $-$ & $1.00 \pm 0.45$ \\
%160509A & PL+BPL    & $-$ & $0.62 \pm 0.27$ & $1.26 \pm 0.11$ & $1.99 \pm 0.23$ & $-$ & $0.9 \pm 0.1$ & $1.03 \pm 0.10$ & $0.76 \pm 0.27$\\
160625B & BPL & $-$ & $-$ & $-$ & $1.27 \pm 0.12$ & $2.13 \pm 0.89$ & $-$ & $-$ & $-$ & $0.66 \pm 0.52$ & $0.80 \pm 0.35$\\
170405A & PL+BPL & $-$ & $-$ & $1.98 \pm 0.04$ & $0.95 \pm 0.14$ & $1.83 \pm 0.17$ & $-$ & $-$ & $0.62 \pm 0.16$ & $0.85 \pm 0.09$ & $1.20 \pm 0.50$ \\
171010A & BPL & $-$ & $-$ & $-$ & $1.34 \pm 0.04$ & $1.94 \pm 0.17$ & $-$ & $-$ & $-$ & $0.93 \pm 0.15$ & $0.61 \pm 0.29$\\
180720B & PL+BPL+BPL & $0.980 \pm 0.009$ & $0.53\pm 0.02$ & $1.16\pm 0.09$ & $2.16 \pm 2.48$ & $1.51 \pm 2.46$  & $0.70 \pm 0.02$ & $0.70\pm 0.02$ & $0.76 \pm 0.04$ & $-$ & $0.74 \pm 0.06$\\
210822A & PL+BPL & $-$ & $-$ & $1.03 \pm 0.09$ & $1.13 \pm 0.04$ & $1.82 \pm 0.05$ & $-$ & $-$ & $0.74 \pm 0.02$ & $0.60 \pm 0.12$ & $0.78 \pm 0.11$ \\
220101A & PL+BPL+PL & $-$ & $0.04 \pm 0.07$  & $1.084 \pm 0.004$ & $1.18 \pm 0.05$ & $1.70 \pm 1.50$ & $-$ &$-0.07 \pm 0.03$ & $0.68 \pm 0.19$ & $0.63 \pm 0.04$ & $0.86 \pm 0.10$\\
230812B & PL & $-$ & $-$ & $-$ & $-$ & $1.28 \pm 0.04$ & $-$ & $-$ & $-$ & $-$ & $0.74 \pm 0.15$
% 221009A & BPL & $-$ & $-$ & $1.51 \pm 0.06$ & $1.643 \pm 0.026$ & $-$ & $-$ & $0.751 \pm 0.011$ & $0.78 \pm 0.05$
\enddata
\end{deluxetable*}
%\end{landscape}
%\end{sideways}
\end{longrotatetable}

%% X-ray data table
%\begin{sideways}{lcccccccccc}
%\begin{landscape}
\begin{longrotatetable}
\begin{deluxetable*}{lcccccccccccccl}
%\begin{deluxetable*}{lcccccccccccl}
%\begin{deluxetable*}{lllrrrrrrll}
%\tablenum{2}
\tablecaption{Best-fit models from the X-ray analysis. Columns 1 and 2 list the GRB names and the best-fit model, respectively. Columns 3 and 4 lists the start and the end times of the LC, respectively. Column 5 lists the time at the end of PL and the start of BPL for the case of PL+BPL, PL+BPL+PL, and PL+BPL+BPL. Column 6 lists the break time between the two segments of BPL in the case of BPL, PL+BPL, PL+BPL+PL, and PL+BPL+BPL. Column 7 lists the time at the end of BPL and the start of PL and BPL in the case of PL+BPLP+PL and PL+BPL+BPL, respectively. Column 8 lists the break time between the two segments of the second BPL in the case of PL+BPL+BPL. Columns 9 and 10 lists the start and end times of the energy injection. Column 11 lists the duration of the plateau.}
%When PL+BPL is present, the first temporal slope is absent and is replaced by a "-", while in the presence of BPL, both the first and second temporal slopes are absent and are marked as "-". When PL is present, the first, second, and third temporal slopes are absent and are replaced by a "-". %Column 5th lists the spectral parameters, $\beta$. 
%All the temporal and spectral fits are tabulated with their corresponding uncertainties. 
\label{table:datasample2-XRAY}
%\tablewidth{700pt}
\tabletypesize{\scriptsize}
\tablehead{
\colhead{GRB} \vspace{-0.2cm} & \colhead{X-ray LC} & \colhead{$\log(t_{start})$} & \colhead{$\log(t_{end})$} & \colhead{$\log(t_{break_{1}})$} & \colhead{$\log(t_{break_{2}})$} & \colhead{$\log(t_{break_{3}})$} & \colhead{$\log(t_{break_{4}})$} & \colhead{$\log(t_{start,EI})$} & \colhead{$\log(t_{end,EI})$} & \colhead{$t_{plateau}$} \\
\colhead{Name} & \colhead{best-fit} & \colhead{} & \colhead{} & \colhead{} & \colhead{} & \colhead{} & \colhead{} \\
\colhead{} & \colhead{model} & \colhead{(s)} & \colhead{(s)} & \colhead{(s)} & \colhead{(s)} & \colhead{(s)} & \colhead{(s)}& \colhead{(s)} & \colhead{(s)} & \colhead{(s)} \\
%\colhead{ } & \colhead{\bf model} & \colhead{(s)} & \colhead{(s)} & \colhead{(s)} & \colhead{(s)} & \colhead{(s)} & \colhead{(s)} 
%\multicolumn2c{Distance} & \colhead{} & \colhead{V} \\
%\colhead{Number} & \colhead{Number} & \nocolhead{Name} & \colhead{Type} &
%\multicolumn2c{(kpc)} & \colhead{Constellation} & \colhead{(mag)}
}
%\decimalcolnumbers
\startdata
%080916C & BPL  & $-$ & $-$ & $2.1 \pm 0.4$   & $1.01 \pm 0.16$ & $-$ & $-$ & $0.8 \pm 0.4$   & $0.7 \pm 0.3$\\
090328A & PL & 4.76 & 5.90 & $-$ & $-$ & $-$ & $-$ & 4.76 & 5.90 & 5.87 \\ %13831.273 \\ 
090510A & BPL & 2.00 & 4.80 & $-$ & $3.13 \pm 0.05$ & $-$ & $-$ & 2.00 & $3.13 \pm 0.05$ & 3.10 \\ %20.868 \\
090902B & PL & 4.67 & 6.20 & $-$ & $-$ & $-$ & $-$ & 4.67 & 6.2 & 6.19 \\%23672.704 \\
090926A & PL & 4.67 & 6.29 & $-$ & $-$ & $-$ & $-$ & 4.67 & 6.29 & 6.28 \\%28656.198 \\
%100414A & PL   & $-$ & $-$ & $-$               & $1.70 \pm 0.07$ & $-$ & $-$ & $-$               & $0.7 \pm 0.6$\\
%110731A & PL+BPL & $-$ & $1.268 \pm 0.037$ & $1.110 \pm 0.024$ & $1.316 \pm 0.036$ & $-$ & $0.97 \pm 0.06$ & $0.78 \pm 0.08$ & $0.91 \pm 0.21$\\
120711A & PL & 3.89 & 5.39 & $-$ & $-$ & $-$ & $-$ & 3.9 & 5.39 & 5.38 \\%3841.058 \\
130427A & PL+BPL+PL & 2.20 & 7.19 & $2.782 \pm 0.006$ & $3.77 \pm 0.18$ & $4.90 \pm 0.40$ & $-$ & $2.782 \pm 0.06$ & $3.77 \pm 0.18$  & 3.72 \\%88.959 \\
141028A & PL & 4.50 & 5.08 & $-$ & $-$ & $-$ & $-$  & 4.50 & 5.08 & 4.95 \\%1421.108 \\
%160509A & PL+BPL    & $-$ & $0.62 \pm 0.27$ & $1.26 \pm 0.11$ & $1.99 \pm 0.23$ & $-$ & $0.9 \pm 0.1$ & $1.03 \pm 0.10$ & $0.76 \pm 0.27$\\
160625B & BPL & 3.99 & 6.59 & $-$ & $6.26 \pm 1.14$ & $-$ & $-$ & 3.99 & $6.26 \pm 1.14$ & 6.26 \\
170405A & PL+BPL & 2.30 & 5.04 & $2.84 \pm 0.04$ & $3.76 \pm 0.19$ & $-$ & $-$ & 2.84 & $3.76 \pm 0.19$ & 3.70 \\%84.845 \\
171010A & BPL & 4.39 & 6.21 & $-$ & $5.54 \pm 0.14$ & $-$ & $-$ & 4.39 & $5.54 \pm 0.14$ & 5.51 \\
180720B & PL+BPL+BPL & 2.30 & 6.46 & $3.020 \pm 0.009$ & $3.32 \pm 0.11$ & $4.71 \pm 0.94$ & $4.72 \pm 0.94$ & $3.020 \pm 0.009$ & $3.32 \pm 0.11$ & 3.02 \\%17.409 \\
210822A & PL+BPL & 1.80 & 5.70 & $2.75 \pm 0.38$ & $4.06 \pm 0.07$ & $-$ & $-$ & $2.75 \pm 0.38$  & $4.06 \pm 0.07$ & 4.04 \\
220101A & PL+BPL+PL & 1.87 & 6.04 & $2.141 \pm 0.009$ &  $3.59 \pm 0.28$ & $4.89 \pm 0.04$ & $-$ & $3.59 \pm 0.28$ & $4.89 \pm 0.04$ & 4.87 \\%1.065 \\
230812B & PL & 4.39 & 6.19 & $-$ & $-$ & $-$ & $-$ & 4.39 & 6.19 & 6.18 \\%23281.008
% 221009A & BPL & $-$ & $-$ & $1.51 \pm 0.06$ & $1.643 \pm 0.026$ & $-$ & $-$ & $0.751 \pm 0.011$ & $0.78 \pm 0.05$
\enddata
\end{deluxetable*}
%\end{landscape}
%\end{sideways}
\end{longrotatetable}

%% optical data table
%\begin{deluxetable*}{cchlDlc}
\begin{longrotatetable}
%\begin{footnotesize}
%\begin{adjustbox}{height=0.95\textheight,center}
%\scalebox{0.9}{
\begin{deluxetable*}{lcccccccccccl}
%\tablenum{3}
\tablecaption{Best-fit parameters of the optical temporal and spectral analysis. Columns 1 and 2 list the GRB name and the best-fit model, respectively. Columns 3, 4, and 5 lists the first temporal slope of the PL+BPL fit ($\alpha_{\rm{opt_1}}$), the second temporal slope of the PL+BPL fit ($\alpha_{\rm{opt_2}}$), and the third temporal slope of the PL+BPL fit ( $\alpha_{\rm{opt_3}}$)
%, and the fourth temporal slope of the PL+BPL+PL fit ($\alpha_{\rm{opt_4}}$)
, respectively. %When PL+BPL is present, the first temporal slope is absent and is replaced by a "-", while in the presence of BPL, both the first and second temporal slopes are absent and are marked as "-". When PL is present, the first, second, and third temporal slopes are absent and are replaced by a "-". 
Column 6 lists the spectral parameter, $\beta_{\rm{opt}}$, which is considered to be the same for all segments of the LCs. Column 7 provides the references from where the $\beta_{\rm{opt}}$ has been extracted. Columns 8 and 9 lists the start and the end times of the LC, respectively. Column 10 lists the time at the end of PL and the start of the BPL in the case of PL+BPL ($\log t_{break_1}$) and its uncertainty. Column 11 lists the break time between the two segments of BPL in the case of both BPL and PL+BPL $\log t_{break_2}$ and its uncertainty. 
%All the temporal and spectral fits are tabulated with their corresponding errors.
\label{table:datasample3}}
\tablewidth{0pt}
\tabletypesize{\scriptsize}
\tablehead{
\colhead{GRB} \vspace{-0.2cm} & \colhead{Optical LC} & \colhead{$\alpha_{\rm{opt_1}} \pm \delta_{\rm{\alpha_{opt_1}}}$} & \colhead{$\alpha_{\rm{opt_2}} \pm \delta_{\rm{\alpha_{opt_2}}}$} & \colhead{$\alpha_{\rm{opt_3}} \pm \delta_{\rm{\alpha_{opt_3}}}$} &  \colhead{$\beta_{\rm{opt}} \pm \delta_{\beta_{\rm{opt}}}$} & \colhead{$\beta_{\rm{opt}}$ Source} & \colhead{$\log(t_{start})$} & \colhead{$\log(t_{end})$} & \colhead{$\log(t_{break_{1}})$} & \colhead{$\log(t_{break_{2}})$} \\
\colhead{Name} & \colhead{best-fit} & \colhead{ } & \colhead{ } & \colhead{ } & \colhead{ } & \colhead{} & \colhead{} & \colhead{} & \colhead{} & \colhead{} \\
\colhead{} & \colhead{model} & \colhead{ } & \colhead{ } & \colhead{ } & \colhead{ } & \colhead{ } & \colhead{(s)} & \colhead{(s)} & \colhead{(s)} & \colhead{(s)}
%\multicolumn2c{Distance} & \colhead{} & \colhead{V} \\
%\colhead{Number} & \colhead{Number} & \nocolhead{Name} & \colhead{Type} &
%\multicolumn2c{(kpc)} & \colhead{Constellation} & \colhead{(mag)}
}
%\decimalcolnumbers
\startdata
%080916C & PL        & $-$ & $-$ & $-$               & $1.6 \pm 0.6$     &  $0.38 \pm 0.10$ & \citet{2010ApJ...720.1513K} \\
090328A & PL & $-$ & $-$ & $0.95 \pm 0.09$  & $1.19 \pm 0.21$  & {\citet{2012ApJ...758...27L}} & 4.76 & 6.03& $-$ & $-$ \\ 
090510A & BPL & $-$ & $6.96 \pm 1.56$ & $2.57 \pm  0.69$ & $0.85 \pm 0.05$ & \citet{2018ApJS..234...26L} & 4.35& 4.55 & $-$ & $4.38 \pm 0.04$ \\
090902B & BPL & - & $2.16 \pm 1.64$ & $0.77 \pm 0.07$ & $0.68 \pm 0.11$ & \citet{2012ApJ...758...27L} & 3.72 & 6.01 & $-$ & $4.41 \pm 0.30$ \\
090926A & BPL & -&  $0.003 \pm 0.213$ & $1.34 \pm 0.16$ & $0.72 \pm 0.17$ & \citet{2012ApJ...758...27L} & 4.84 & 5.98 & $-$ & $4.96 \pm 0.09$ \\
%100414A & PL        & $-$ & $-$ & $-$               & $1.2 \pm 0.6$   & $1.20 \pm 0.20$ & \citet{2012ApJ...748L...4U} \\
%110731A & PL        & $-$ & $-$ & $-$               & $1.01 \pm 0.12$ & $0.66 \pm 0.03$ & \citet{2013ApJ...763...71A} \\
120711A & PL+BPL & $1.75 \pm 0.02$ & $0.37 \pm 0.04$ & $1.48 \pm 0.08$ & $0.53 \pm 0.02$ & \citet{2018ApJS..234...26L} & 2.11 & 5.57 & $3.19 \pm 0.04$ & $4.78 \pm 0.05$ \\
130427A & PL+BPL    & $1.19 \pm 0.04$ & $1.000 \pm 0.002$ & $0.33 \pm 0.04$ & $0.92 \pm 0.10$ & \citet{2018ApJS..234...26L} & 2.64& 8.0 & $3.07 \pm 0.04$ & $6.00 \pm 0.02$ \\
141028A & PL & $-$ & $-$ & $1.01 \pm 0.08$ & $1.29 \pm 0.07$ & \citet{Burgess2016} & 4.57 & 5.18 & $-$ & $-$ \\
%160509A & PL        & $-$ & $-$ & $-$ & $0.65 \pm 0.05$ & $0.63 \pm 0.19$ & \citet{2020ApJ...894...43K} \\
160625B & BPL & $-$ & $0.949 \pm 0.007$ & $1.54 \pm 0.06$ & $0.68 \pm 0.07$ & \citet{2020ApJ...894...43K} & 4.51 & 6.67 & $-$ & $6.13 \pm 0.033$ \\
170405A & PL & $-$ & $-$ & $1.38 \pm 0.12$ & $0.80 \pm 0.09$ & \citet{Dainotti2022} & 2.31 & 3.77 & $-$ & $-$ \\
171010A & BPL & $-$ & $6.68 \pm  0.28$ & $0.10 \pm 0.01$& $1.33 \pm 0.19$ & \citet{2019MNRAS.490.5366M} & 5.11 & 7.01 & $-$ & $5.56 \pm 0.05$ \\
180720B & PL & $-$ & $-$ & $0.87 \pm 0.04$ & $0.80 \pm 0.04$ & \citet{2019ApJ...885...29F} & 4.01 & 4.60 & $-$ & $-$ \\
210822A & PL & $-$ & $-$ & $1.34 \pm 0.02$ & $0.77 \pm 0.03$ & \citet{2024MNRAS.527.8140A} & 2.25 & 2.97 & $-$ & $-$ \\
220101A & BPL & $-$ & $2.39 \pm 0.18$ & $0.47 \pm 0.14$ & $0.70 \pm 0.05$ & \citet{2023arXiv230102407J} & 2.2 & 4.8 & $-$ & $3.28  \pm 0.12$ \\
230812B & PL+BPL & $1.44 \pm 0.02$ & $0.016 \pm 0.011$ & $3.45 \pm 0.52$ & $0.74 \pm 0.02$ & \citet{2024ApJ...960L..18S} & 4.48& 6.52 & $5.46 \pm 0.01$ & $6.29 \pm 0.01$ \\
% 221009A & BPL       & $-$ & $-$ & $0.555 \pm 0.006$ & $1.412 \pm 0.003$ & $0.750 \pm 0.016$ & \citet{2023ApJ...948L..12K}
\enddata
%\tablecomments{This table ``hides'' the third column in the \latex\ when compiled. The Distance is also centered on the decimals. Note that when using decimal alignment you need to include the {\tt\string\decimals} command before {\tt\string\startdata} and all of the values in that column have to have a space before the next ampersand.}
% \label{table:datasample3}
\end{deluxetable*}
%\end{adjustbox}
%}
%\end{footnotesize}
\end{longrotatetable}

\begin{deluxetable*}{lcccccc}
\tablecaption{Best-fit models from the optical analysis. Columns 1, 2, 3, and 4 list the GRB name, the redshift, the source for the redshift, and the best-fit model, respectively. Columns 5 and 6 list the energy injection's start and end times, respectively. Column 7 lists the duration of the plateau in the log scale.} \label{table:datasample3-OPT}
\tablewidth{0pt}
\tabletypesize{\scriptsize}
\tablehead{
\colhead{GRB Name} & \colhead{$z$} & \colhead{Source} & \colhead{Optical LC} & \colhead{$\log(t_{\mathrm{start,EI}})$ (s)} & \colhead{$\log(t_{\mathrm{end,EI}})$ (s)} & \colhead{$\log(t_{\mathrm{plateau})}$ (s)}
}
\startdata
090328A & 0.736 & JG$^{a}$ & PL & 4.76 & 6.03 & 6.00 \\
090510A & 0.903 & VLT emission$^{b}$  & BPL & $4.38 \pm 0.04$ & 4.55 & 4.06       \\
090902B & 1.822 & JG$^{a}$ & BPL    & $4.42 \pm 0.30$  & 6.01  & 6.00       \\
090926A & 2.106 & JG$^{a}$ & BPL    & $4.96 \pm 0.09$ & 5.98 & 5.94   \\
120711A & 1.405 & JG$^{a}$ & PL+BPL & $3.19 \pm 0.035$ & $4.78 \pm 0.05$ & 4.77   \\
130427A & 0.339 & Gemini North absorption$^{b}$  & PL+BPL & $3.07 \pm 0.04$   & $6.00 \pm 0.02$   & 6.00      \\
141028A & 2.330 & JG$^{a}$   & PL     & $-$  & $-$   & $-$   \\
160625B & 1.406 & JG$^{a}$  & BPL    & 4.51  & $6.13 \pm 0.03$   & 6.12       \\
170405A & 3.510 & GTC absorption$^{b}$                    & PL     & 2.31 & 3.77 & 3.75    \\
171010A & 0.329 & JG$^{a}$  & BPL    & $-$  & $-$   & $-$       \\
180720B & 0.654 & VLT absorption$^{b}$                    & PL      & $-$  & $-$   & $-$ \\
210822A & 1.736 & NOT absorption$^{b}$                    & PL     & $-$  & $-$   & $-$    \\
220101A & 4.618 & NOT absorption$^{b}$                    & BPL    & $3.28 \pm 0.12$  & $4.08$   & $4.01$       \\
230812B & 0.360 & JG$^{a}$ & PL+BPL & $5.46 \pm 0.01$ & $6.29 \pm 0.01$ & 6.22 \\
\enddata
\tablenotetext{a}{From \url{https://www.mpe.mpg.de/~jcg/grbgen.html}}
\tablenotetext{b}{From \url{https://swift.gsfc.nasa.gov/archive/grb_table/}}
\end{deluxetable*}

\section{Methodology}\label{sec:methodology}
\subsection{Data Analysis}\label{sec:dataanalysis}
We performed the fitting of the GRB LCs with the simple PL and the BPL model or combinations thereof.
The PL function describing the temporal evolution of the GRB flux is defined as follows:
\begin{equation}
    F(t) = F_{0}\biggl(\frac{t}{T_{0}}\biggl)^{-\alpha}
\label{eq:PL}
\end{equation}
where $F_{0}$ represents the normalization flux, $T_{0}$ represents the GRB trigger time, $t$ represents the time of the observation relative to $T_{0}$, and $\alpha$ indicates the temporal decay index. 
The BPL function, on the other hand, is expressed as follows:
%\begin{eqnarray*}
%   && F(T) =
%    \begin{cases}
%        F_{b}\biggl(\frac{t}{T_{b}}\biggl)^{-\alpha_{\rm 1}} t < T_{b},  F_{b}\biggl(\frac{t}{T_{b}}\biggl)^{-\alpha_{\rm 2}} t \geq T_{b}
%    \end{cases}
%\label{eq:BPL}
%\end{eqnarray*}
\begin{equation}
    F(T) = F_{b} 
    \begin{cases}
    \begin{aligned}
        \biggl(\frac{t}{T_{b}}\biggl)^{-\alpha_{\rm 1}} t < T_{b}, \\
        \biggl(\frac{t}{T_{b}}\biggl)^{-\alpha_{\rm 2}} t \geq T_{b}
    \end{aligned}
    \end{cases}
\label{eq:BPL}
\end{equation}
where $T_{b}$ represents the break time, $F_{b}$ represents the flux at $T_{b}$, and $\alpha_{\rm 1}$ and $\alpha_{\rm 2}$ denote the temporal decay indices before and after $T_{b}$, respectively. 
For \emph{Fermi}-LAT, we used the values for $\alpha_{\rm{\gamma}}$ and $\beta_{\rm{\gamma}}$ given in \cite{2019ApJ...878...52A}, which are based on either a PL model (7 GRBs) or a BPL model (7 GRBs). % as \textcolor{red}{reported} in 2FLGC \citep{2019ApJ...878...52A}.
%In our sample, we have 12 GRBs fitted with BPL and 4 GRBs fitted with PL in the
In the case of X-rays, we employ the \texttt{grbLC} package, which is an automated LC fitting tool \citep{DainottiOpticalCatalog}
to determine the best-fit model among the different models (PL, BPL, or their combinations). We have performed different fittings and decided which is the best-fitting model based on the comparison among the models. This comparison has been performed in several steps: first, we find the minimum of the Akaike Information Criterion \citep[AIC;][]{1100705} among the considered models, denoted by AIC$_{\rm min}$. 
Then, we compute the $B_{\rm{i}}$ = $e^{(\rm AIC_{\rm min} - \rm AIC_{i}) / 2}$ for each model, where $B_{\rm{i}}$ is the Akaike model weight, and AIC$_{\rm{i}}$ refers to the AIC of the respective model. 
Finally, for every model, we determine the relative likelihood, denoted as $p_{\rm{i}}$ = $B_{\rm{i}}$ / $\sum_{\rm{i}}$ ($B_{\rm{i}}$), where the sum is taken over all models under consideration. 
The model with $p > 0.95$ is chosen as the best-fit LC model. For GRBs where none of the fitting models achieves a $p$-value greater than 0.95, we manually select the best-fit model by examining the contour plots. This manual inspection allows us to determine the most suitable model visually. 
Based on the best-fit model, we extracted the value of $\alpha_{\rm{X}}$. 
Subsequently, we obtained the value of the photon index gamma ($\Gamma_{\rm{X}}$) from the \emph{Swift} \footnote{\url{https://www.swift.ac.uk/burst_analyser/}} repository by using the time-sliced spectra, which corresponds to the time range of the several segments of the LC. 
Finally, we calculated $\beta_{\rm{X}}$ as $\Gamma_{\rm{X}}$ = $\beta_{\rm{X}}$ - 1.
For the optical data, as it has been taken from \cite{DainottiOpticalCatalog}, the LC data has already been stacked together to allow better coverage of the LCs. %\st{before proceeding to the LC fitting} \citep[in prep]{DainottiOpticalCatalog}
To stack the data, we followed several steps following \cite{DainottiOpticalCatalog}: first, we flagged the outliers 
%(the data points that are 5-sigma away from the LC trend) as 
``yes" in the initial data of magnitudes. We classify outliers into three categories: bad photometry points, where the photometry is deemed unreliable according to their sources (e.g., some GCNs); data points with magnitudes that deviate by at least 5$-\sigma$ from other data points at the same epoch; non-simultaneous outliers, for which the coincident epoch criterion (as defined in Eq. \ref{Eq:outlier} taken from \cite{DainottiOpticalCatalog}) cannot be applied, and their magnitudes deviate by more than 5$-\sigma$ from the closest data points in time \citep{DainottiOpticalCatalog}.

The coincident criterion is defined as follows:
\begin{equation}
    \frac{|{t_{f} - t_{g}}|}{t_{f}} \leq 0.025 \longrightarrow \mbox{$t_{f}$ and $t_{g}$ are coincident,}
    \label{Eq:outlier}
\end{equation}

where, $t_{f}$ and $t_{g}$ represent the midpoints of the observations in the given filter band, denoted with $f$, and $g$ denote a generic band from which we start the calculation, respectively, measured in seconds after the satellite trigger. In our case, we are rescaling in the R band.
Second, we remove these outliers data points; third, we convert the magnitudes into the AB system; fourth, we correct for the Galactic extinction; fifth, we apply the \emph{k}-correction; and sixth, the color evolution analysis \citep{DainottiOpticalCatalog}.

Here, we define color evolution as the changes in color over time. More specifically, the color is the difference in magnitude between two bands, one of which serves as a reference at a specific time. To investigate whether a GRB displays color evolution, we first compute the rescaling factors using Eq. \ref{Eq:rescalingfactor},
taken from \cite{DainottiOpticalCatalog}:

\begin{equation}
    rf_{mn,f} = a * \log_{10}(t_{f}) + b
    \label{Eq:rescalingfactor}
\end{equation}

where, $f$ denotes the $f$ band, $mn$ represents the most numerous filter, $a$ is the slope, and $b$ is the normalization. This Eq. is computed for each data point, allowing us to analyze the behavior across different filters.
The rescaling factor is fitted with slope $a=0$ and the normalization $b$. We then calculate the probability $P$ that the fitting is drawn by chance, the reduced $\chi^{2}$, and the Bayesian Information Criteria (BIC) value. If the $P\geq0.05$, we categorize the GRB as exhibiting no color evolution. Conversely, if $P<0.05$, the GRB is considered to exhibit color evolution.
We then stacked the LCs in different bands only when no color evolution was detected among the bands. For the filters that exhibit color evolution, we removed those filters.
%For the GRBs exhibiting color evolution, the rescaling of color is taken into account. 
%Given the availability of optical data in multiple filters (e.g., U, B, V, R, I, u, g, r, i, z, etc.), our approach involved selecting the filter with numerous data points for conducting the LC fitting and obtain the value of $\alpha_{\rm opt}$. 
Finally, we converted the magnitudes into flux for proceeding with the LC fitting to obtain the $\alpha_{\rm opt}$ parameter. 
We followed the same approach for fitting the optical LC as used for the X-ray LC fitting. 
The $\beta_{\rm{opt}}$ values were obtained from \cite{DainottiOpticalCatalog}.
%In Table \ref{table:datasample1}, we present the GRB sample used in this study, along with their best-fit model for each wavelength.
%The table also provides the corresponding $\alpha$ and $\beta$ values for the respective wavelengths.
%Tables \ref{table:datasample1} and \ref{table:datasample2} display the GRB sample used in this study and their best-fit models for the gamma-ray and X-ray LCs, respectively.
Tables \ref{table:datasample1}, \ref{table:datasample2}, and \ref{table:datasample3} display the GRB sample used in this study and their best-fit models for the gamma-ray, X-ray, and optical LCs, respectively.
The table also presents each wavelength's corresponding $\alpha$ and $\beta$ values. In Table \ref{table:datasample1} and \ref{table:datasample3}, we also provide the start and end times of the LC and the break times for gamma-ray and optical LCs, respectively. Table \ref{table:datasample1} also details the energy injection time and the duration of the plateau for gamma-ray LCs. Table \ref{table:datasample2-XRAY} details the start and end times of the X-ray LC along with the break times, the energy injection time, and the duration of the plateau. Table \ref{table:datasample3-OPT} presents the energy injection time and the duration of the plateau for optical LCs.
% The table also presents the corresponding $\alpha$ and $\beta$ values for each wavelength.
%Table \ref{table:datasample3} presents the same sample of GRBs, showcasing their best-fit model for the optical LC and their corresponding $\alpha$ and $\beta$ values.

\subsection{Derivation of the bulk Lorentz factor}
We here derive the equation for the bulk Lorentz factor in a stratified density profile (=$A_{k}\times r^{-k}$) with for $0 \leq k < 3$.

During the deceleration phase, the bulk Lorentz factor of the relativistic outflow  becomes 
$E_k=\frac{4\pi}{3} m_pc^2 n(r) r^{3} \Gamma^2$ \citep[Blandford-McKee solution;][]{bm76} with the radial distance given by 
\begin{equation}
r\simeq \frac{2 c}{1+z}\Gamma^2 t
\label{radius}
\end{equation}
This equation considers the approximation of the volume of the expanding plasma with a sphere shape.  We then substitute $n(r)$ and $r$ in the Blanford-Mckee solution, we obtain

\begin{equation}
E_k = \frac{4 \pi}{3} (2c)^{3-k} m_{p}c^{2} (1+z)^{3-k} A_k \Gamma^{8-2k} t^{3-k}
4\end{equation}

Now we solve for $\Gamma$:
%\begin{equation}
%    \Gamma^{8-2k} t^{3-k} = \biggl(\frac{3}{4 \pi m_p c^2 (2c)^{3-k}} \biggl) (1+z)^{3-k} A_k^{-1} E_k
%\end{equation}
\begin{equation}
    \Gamma = \biggl(\frac{3}{4 \pi m_p c^2 (2c)^{3-k}} \biggl)^{\frac{1}{2(4-k)}} (1+z)^{\frac{3-k}{2(4-k)}} A_k^{-\frac{1}{2(4-k)}} E_k^{\frac{1}{2(4-k)}} t^{-\frac{(3-k)}{2(4-k)}}
    \label{Eq:bulklorentz}
\end{equation}
From this equation, we can compute the bulk Lorentz factor considering the time we are in the emission and then substitute this value in Eq. \ref{radius}.
%\begin{equation*}
%\Gamma = \frac{3}{\left(\frac{m_p c^2}{2\pi k}\right)^{3/4} (2\pi k)^{1/4}} \frac{1}{(1+z)^{3/4}} A_k^{1/4} E^{1/4} \left(\frac{t}{1+z}\right)^{-1/4}
%\end{equation*}

%\textcolor{orange}{
%\begin{equation}
%    P(X > \mu) = \frac{2^{-\mu/2}}{\Gamma(\mu/2)} \int_{\mu}^{+\infty} e^{-x/2}x^{-1+(\mu/2)} dx
%    \label{Eq:probability}
%\end{equation}
%}
%\textcolor{orange}{where, $\mu$ represents the product between the reduced $\chi^{2}$ value and the number of data points used in the fitting process.}

%\section{Methodology}\label{sec:methodology}
\subsection{Testing the Closure Relations}\label{sec:Testing CR}

We followed a similar approach to \cite{Dainotti2023Galax..11...25D} for comprehensively examining CRs between $\alpha$ and $\beta$. 
These CRs are associated with different astrophysical environments, including the density profile of the surrounding medium, the $p$ index, and the electron cooling regime. 
Furthermore, we have examined three sets of CRs, one with energy injection, one with no energy injection, and the other for the jet break. 
We have also taken into account different density profiles encompassing a constant-density ISM characterized by $n \propto r^{0}$, a stellar wind medium, $n \propto r^{-2}$, and a stratified density profile, $n \propto r^{-k}$, with $k$ values of 1, 1.5, and 2.5. 

\begin{figure*}
    \centering
    \begin{minipage}[t]{0.32\textwidth}
        \centering
        \includegraphics[width=1\textwidth,height=0.18\textheight]{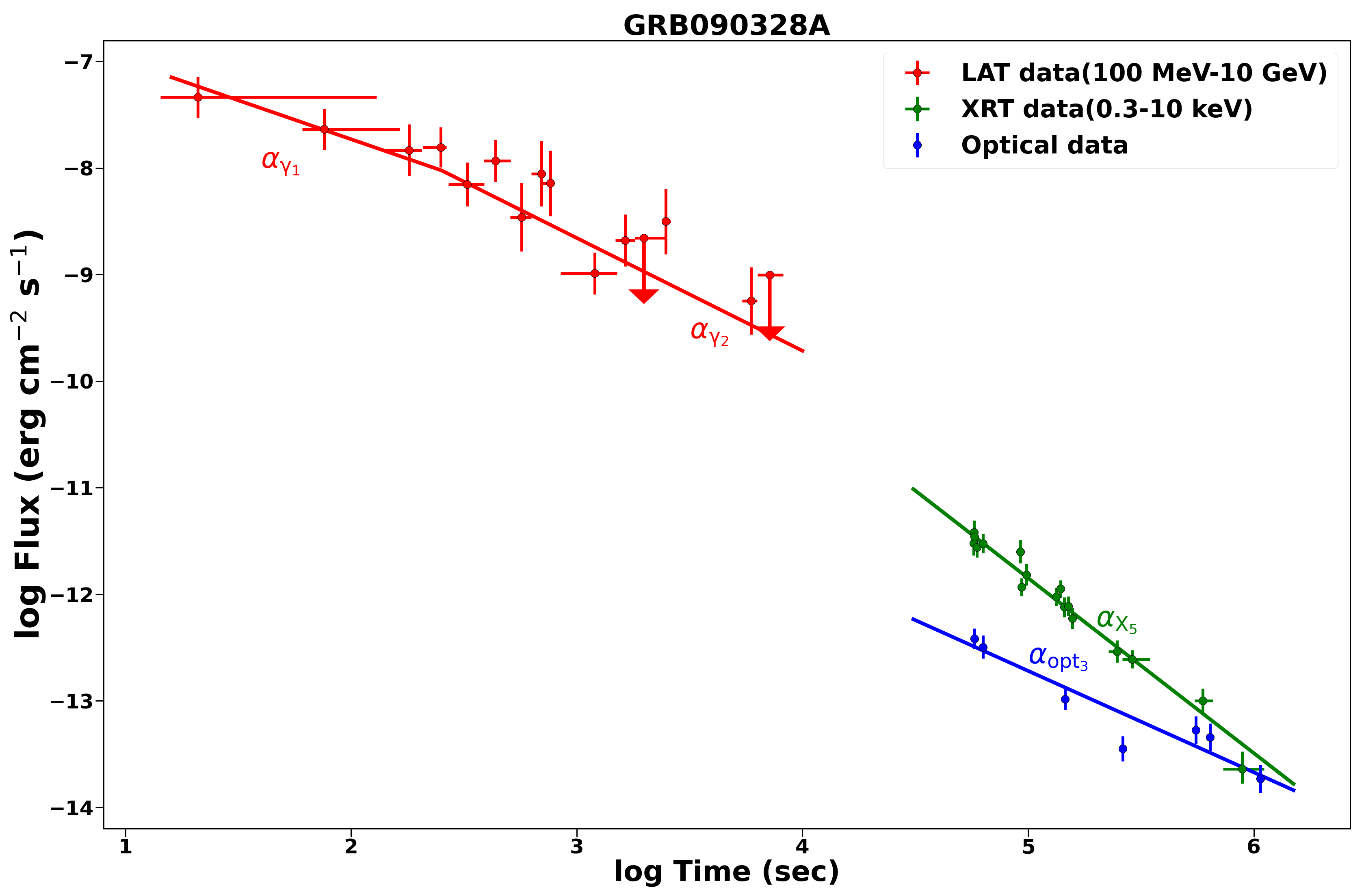}
    \end{minipage}
    \begin{minipage}[t]{0.32\textwidth}
        \centering
        \includegraphics[width=1\textwidth,height=0.18\textheight]{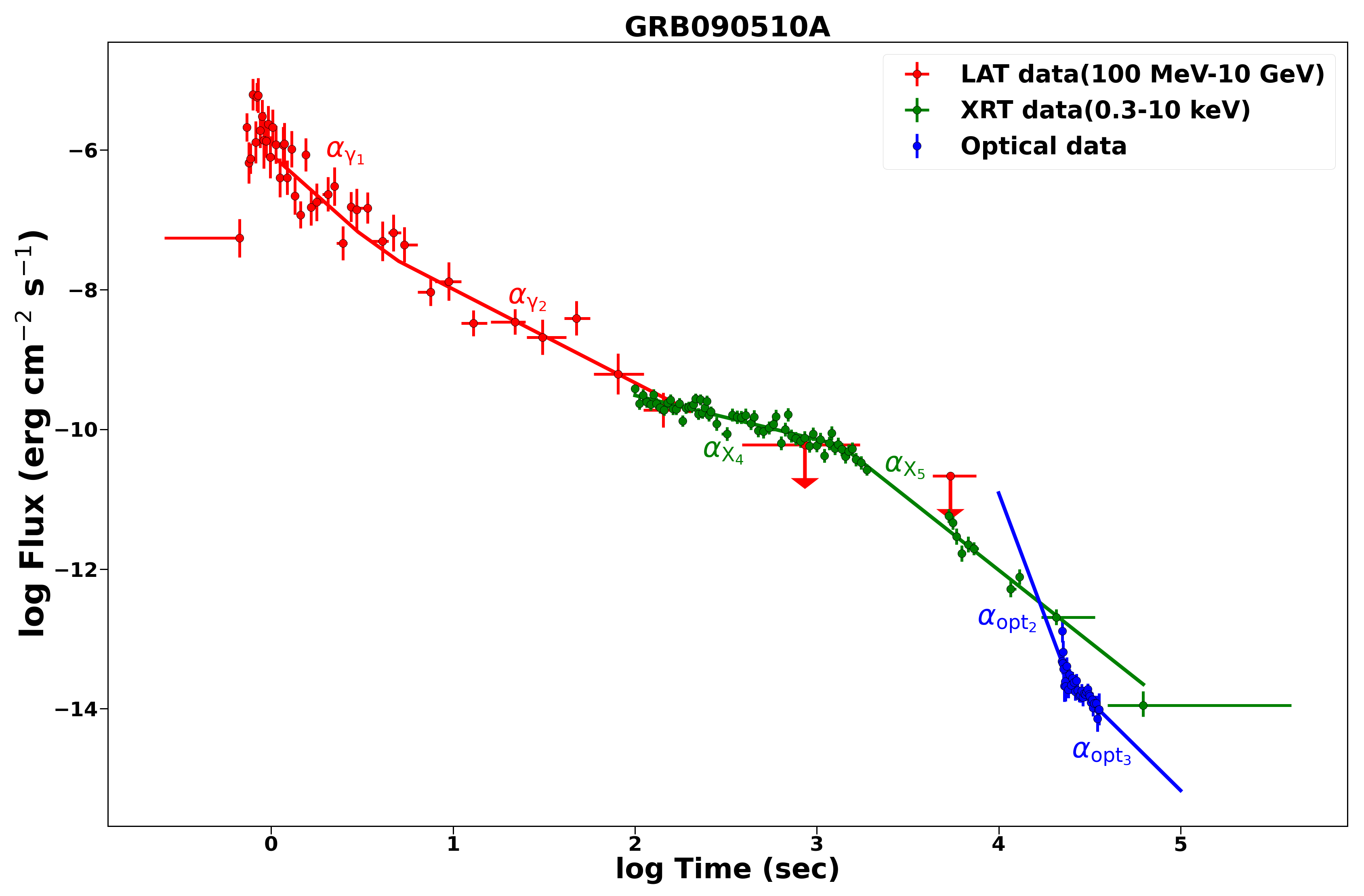}
    \end{minipage}
    \begin{minipage}[t]{0.32\textwidth}
        \centering
        \includegraphics[width=1\textwidth,height=0.18\textheight]{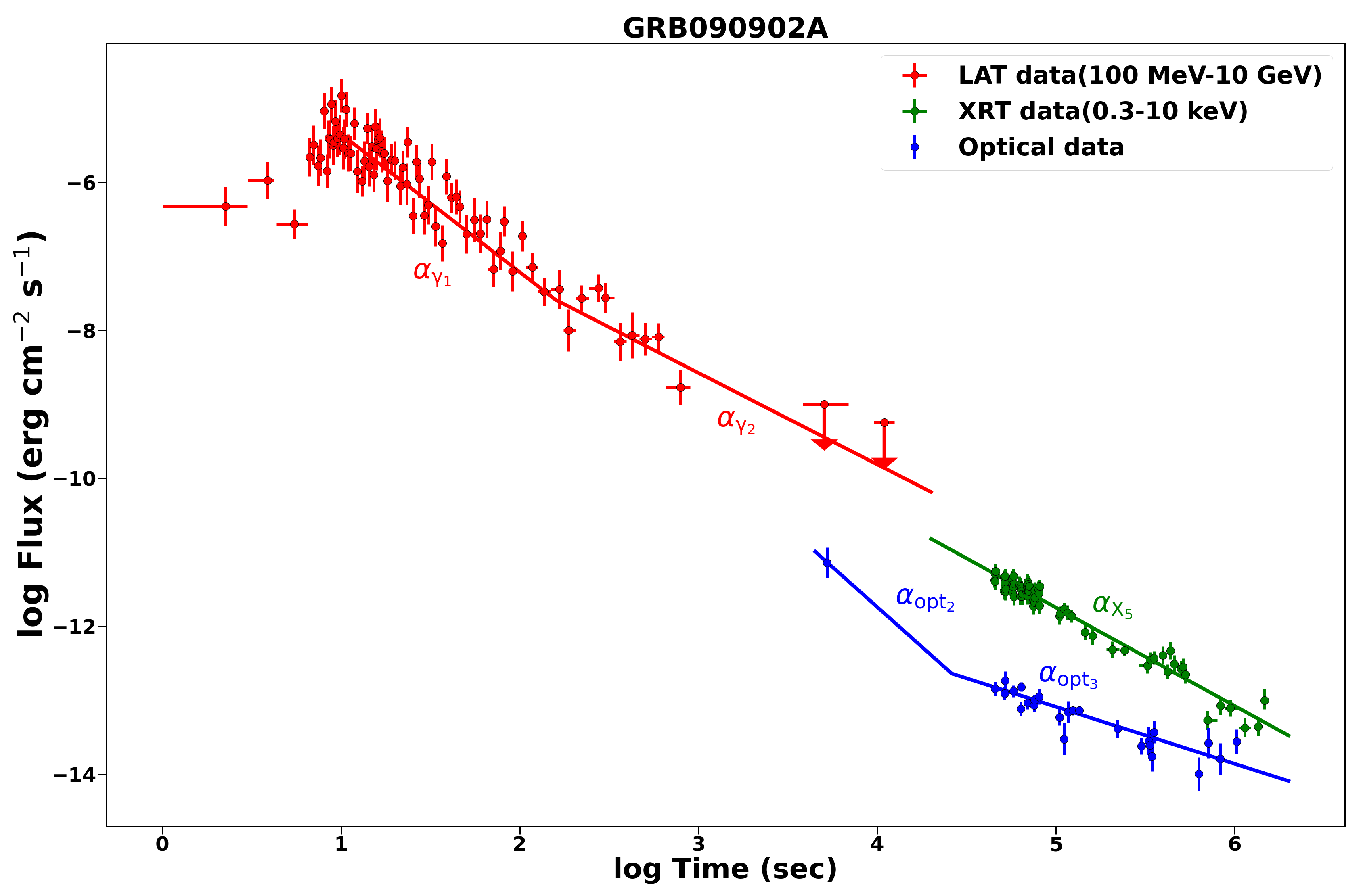}
    \end{minipage}
    
    \vspace{0.25cm} % Adjust space between rows

    \begin{minipage}[t]{0.32\textwidth}
        \centering
        \includegraphics[width=1\textwidth,height=0.18\textheight]{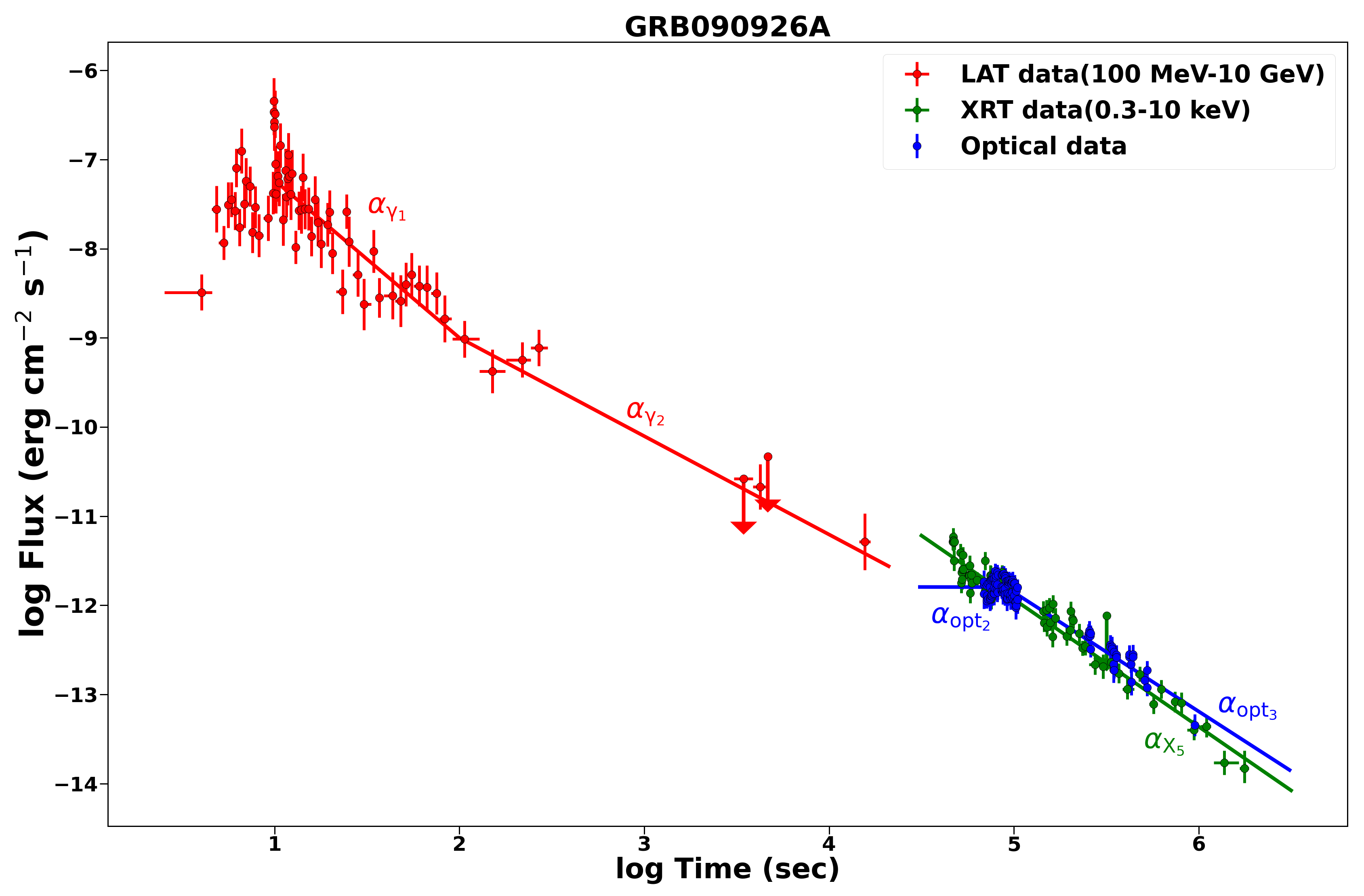}
    \end{minipage}
    \begin{minipage}[t]{0.32\textwidth}
        \centering
        \includegraphics[width=1\textwidth,height=0.18\textheight]{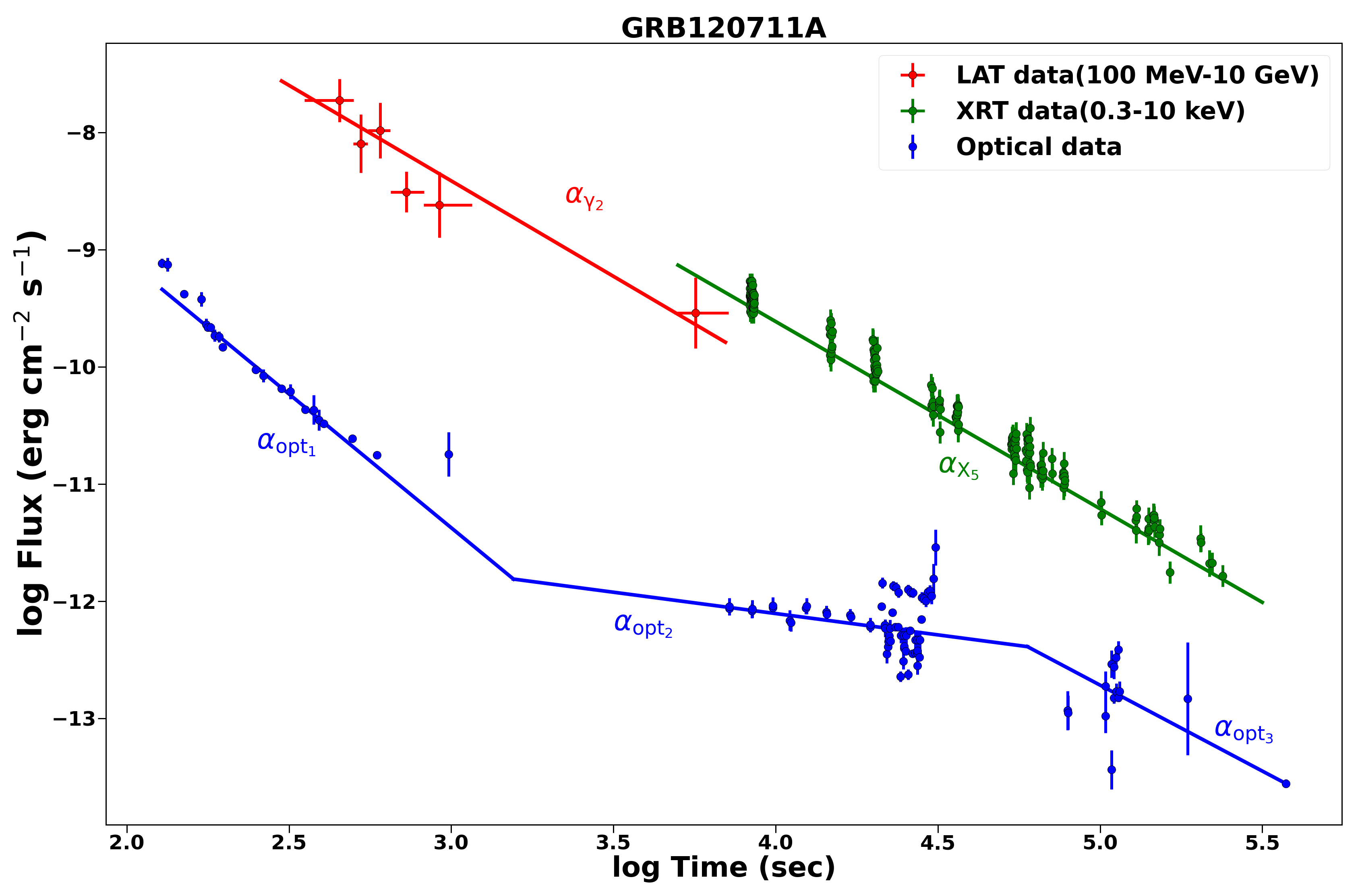}
    \end{minipage}
    \begin{minipage}[t]{0.32\textwidth}
        \centering
        \includegraphics[width=1\textwidth,height=0.18\textheight]{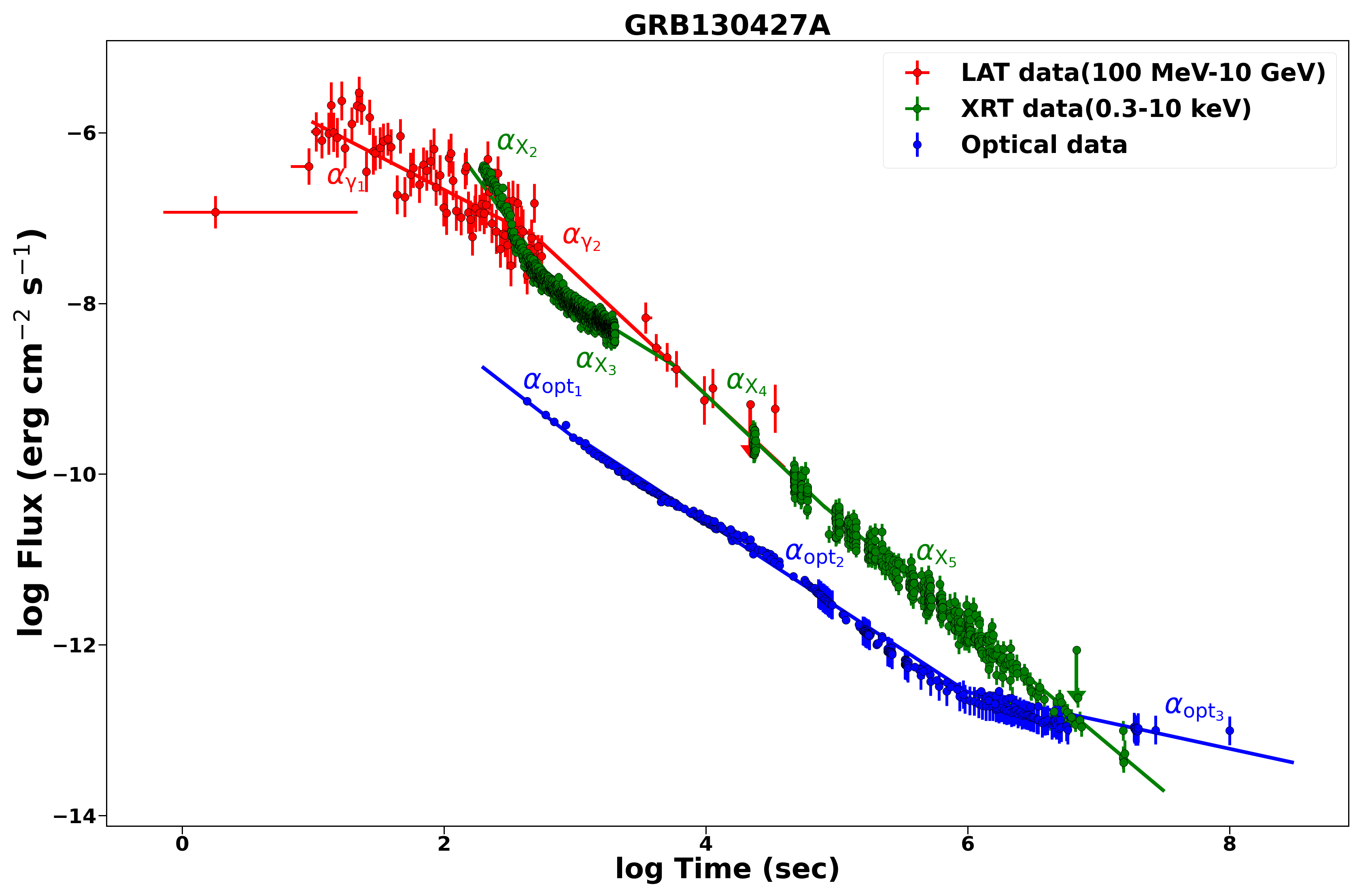}
    \end{minipage}

    \vspace{0.25cm} % Adjust space between rows

    \begin{minipage}[t]{0.32\textwidth}
        \centering
        \includegraphics[width=1\textwidth,height=0.18\textheight]{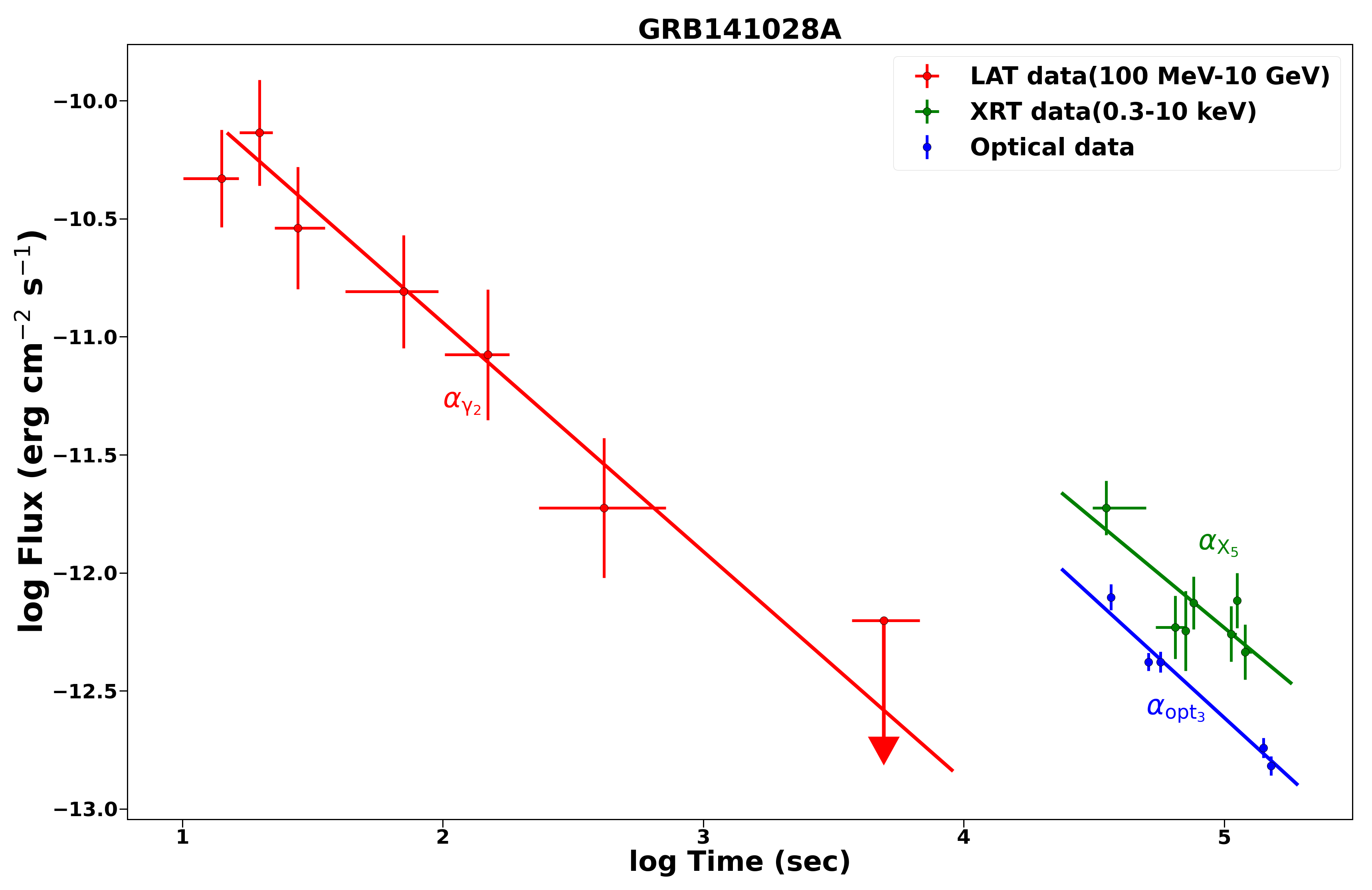}
    \end{minipage}
    \begin{minipage}[t]{0.32\textwidth}
        \centering
        \includegraphics[width=1\textwidth,height=0.18\textheight]{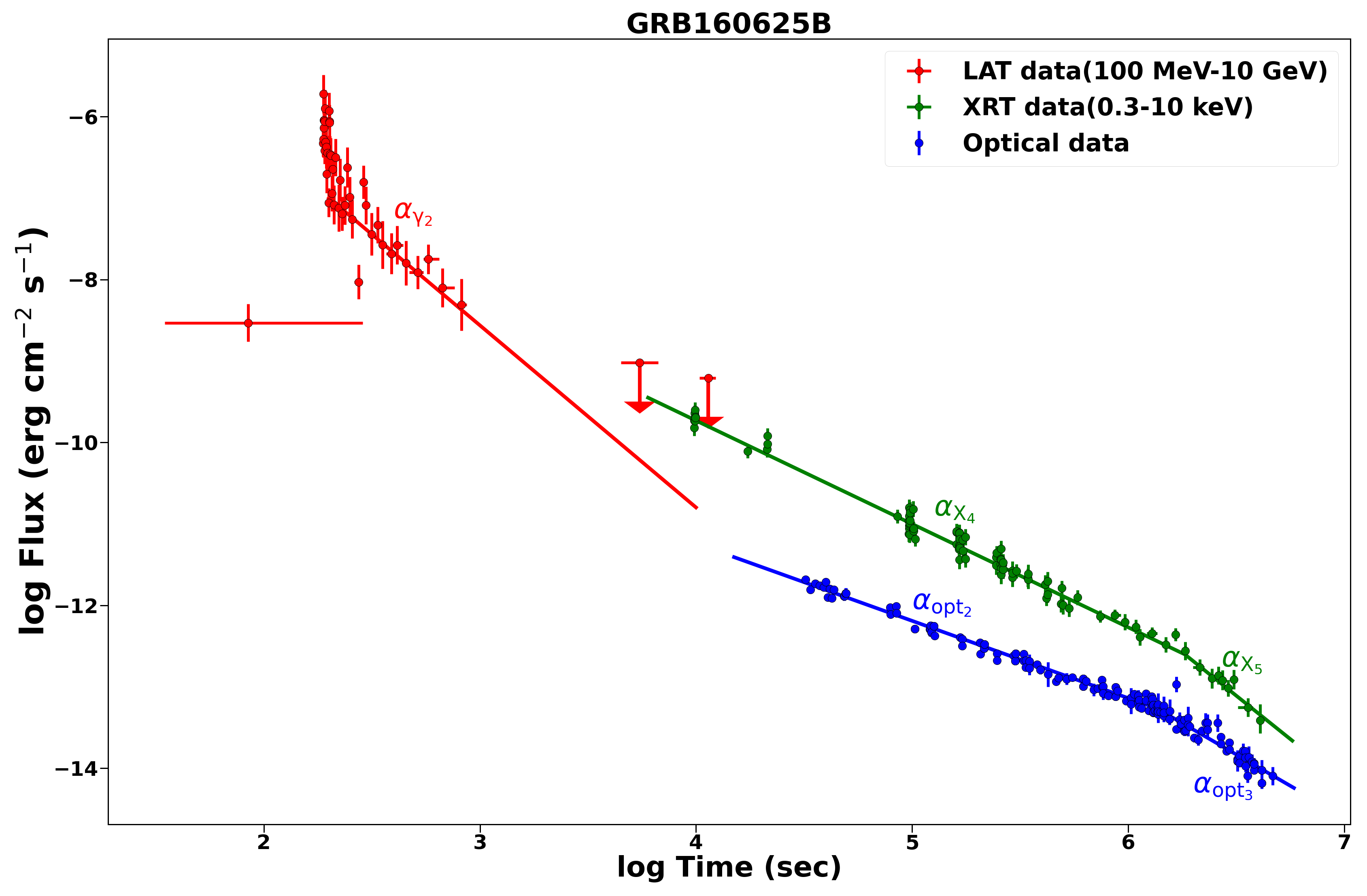}
    \end{minipage}
    \begin{minipage}[t]{0.32\textwidth}
        \centering
        \includegraphics[width=1\textwidth,height=0.18\textheight]{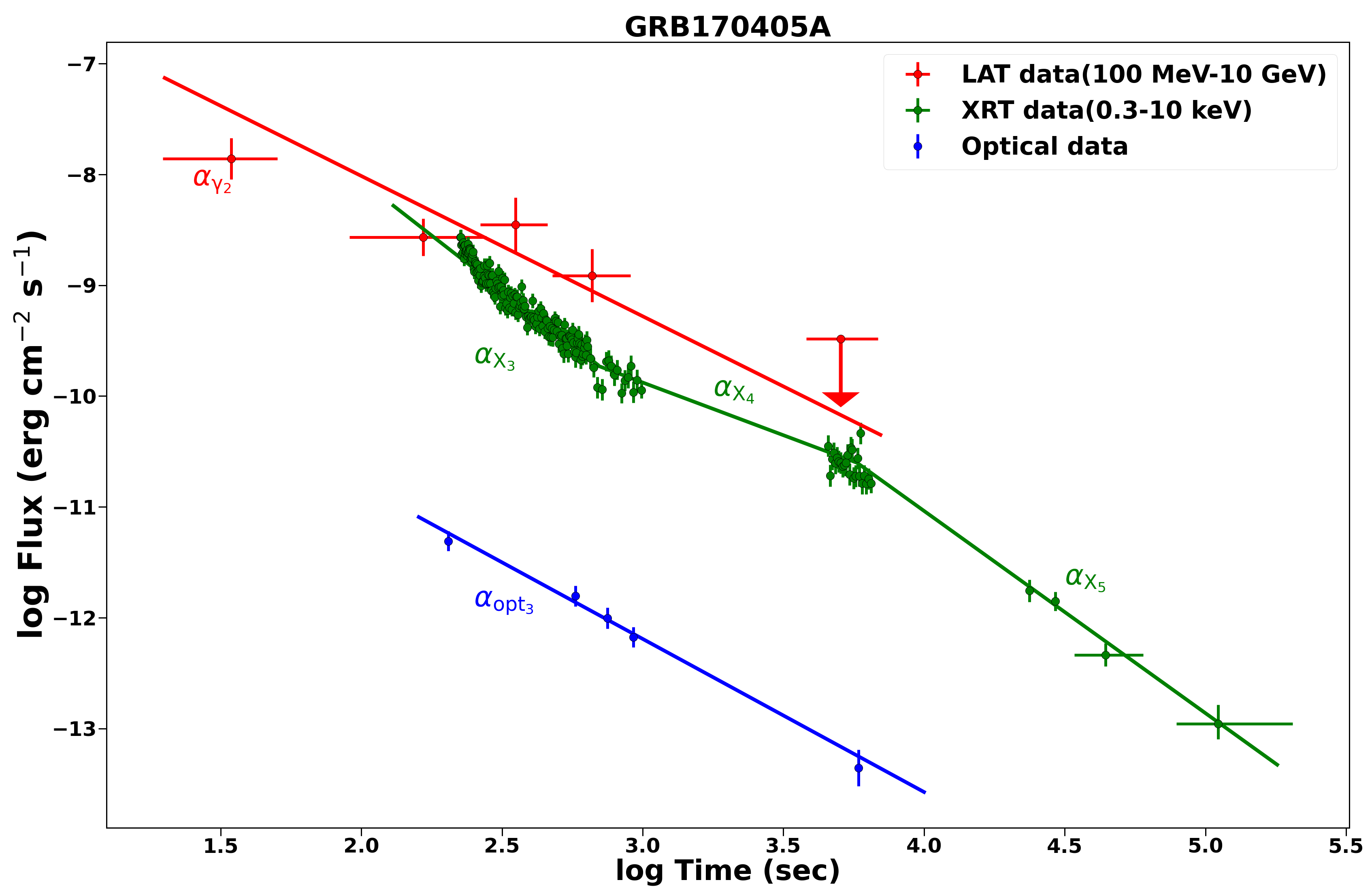}
    \end{minipage}

    \vspace{0.25cm} % Adjust space between rows

    \begin{minipage}[t]{0.32\textwidth}
        \centering
        \includegraphics[width=1\textwidth,height=0.18\textheight]{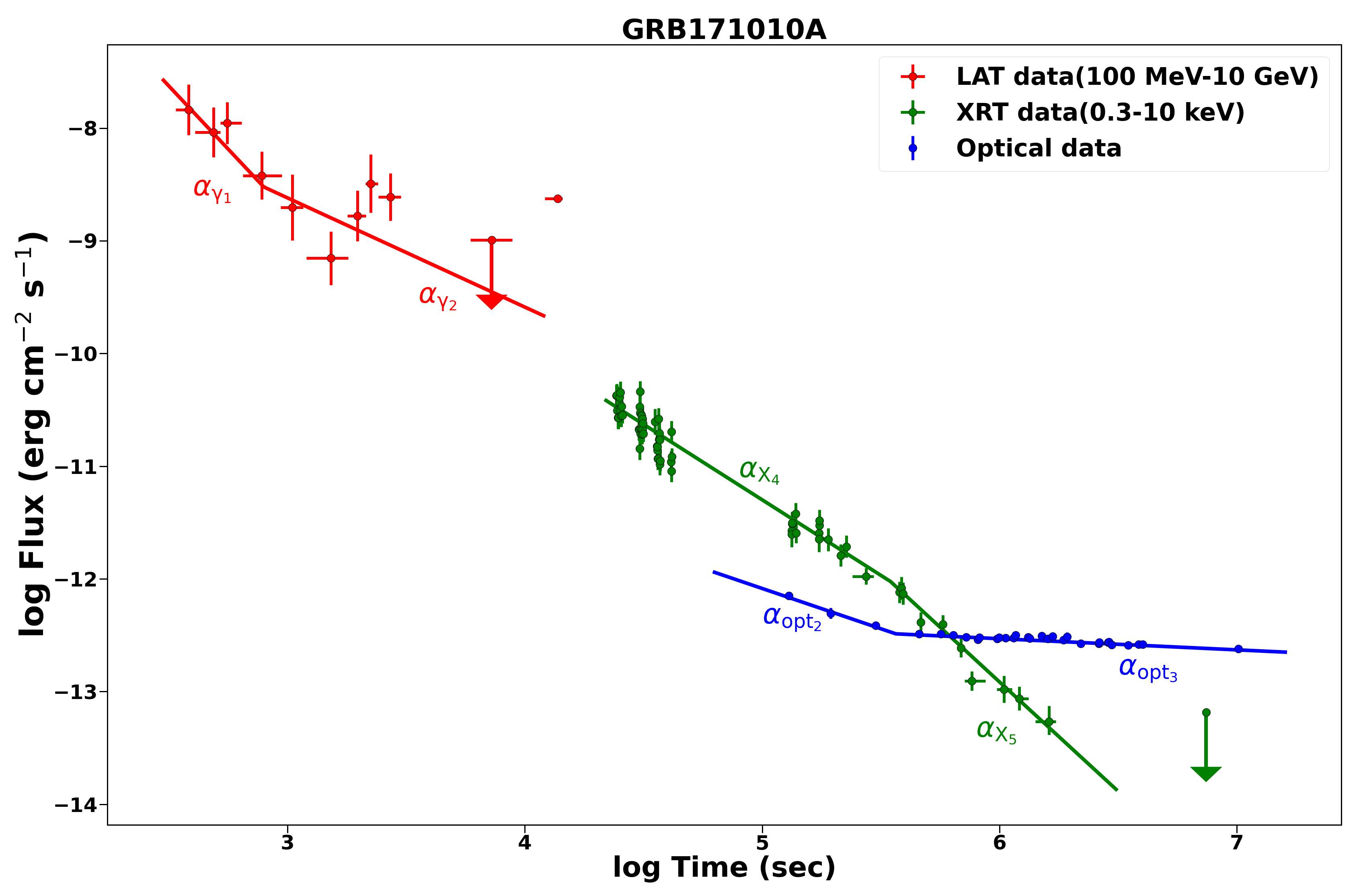}
    \end{minipage}
    \begin{minipage}[t]{0.32\textwidth}
        \centering
        \includegraphics[width=1\textwidth,height=0.18\textheight]{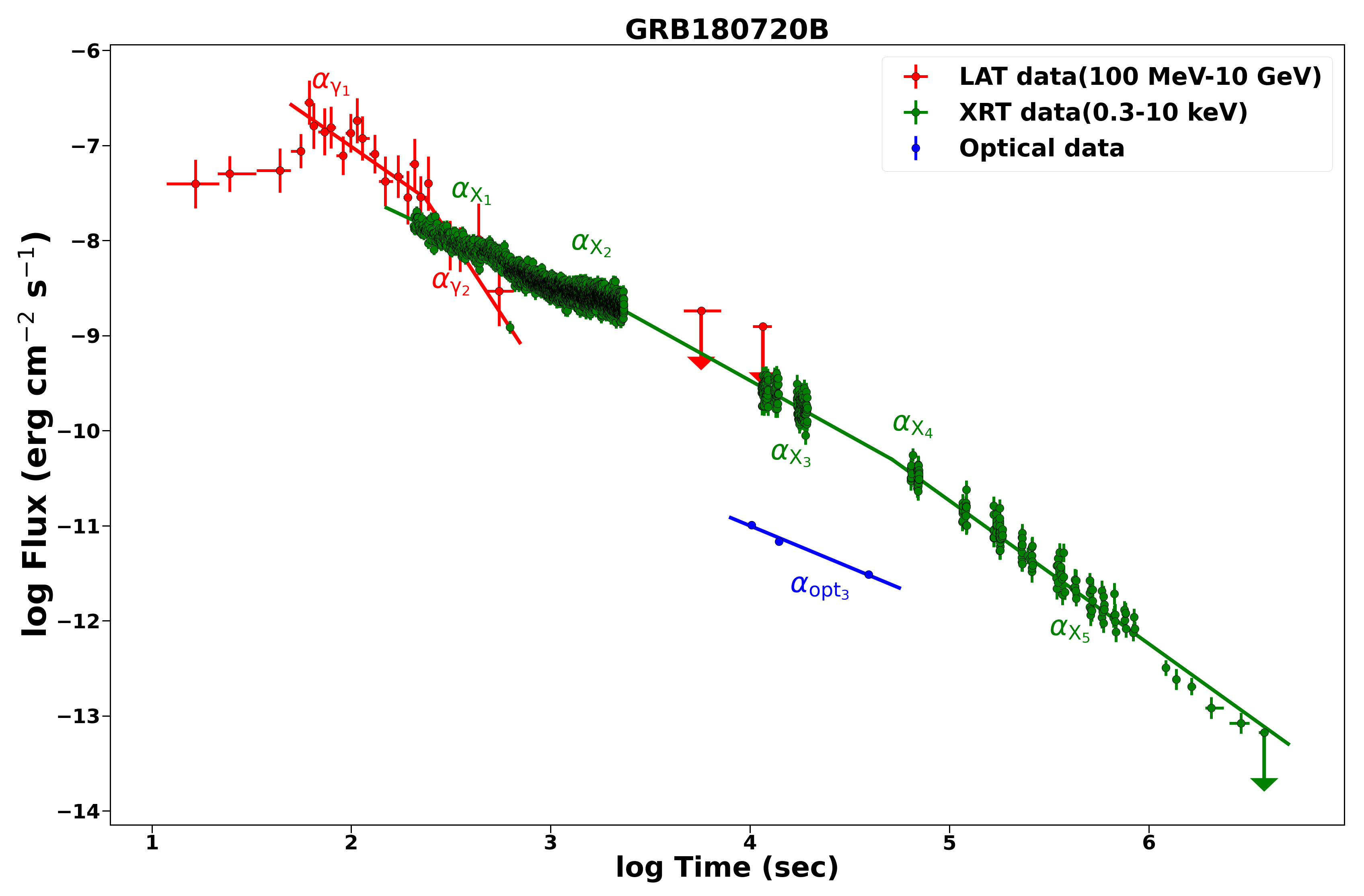}
    \end{minipage}
    \begin{minipage}[t]{0.32\textwidth}
        \centering
        \includegraphics[width=1\textwidth,height=0.18\textheight]{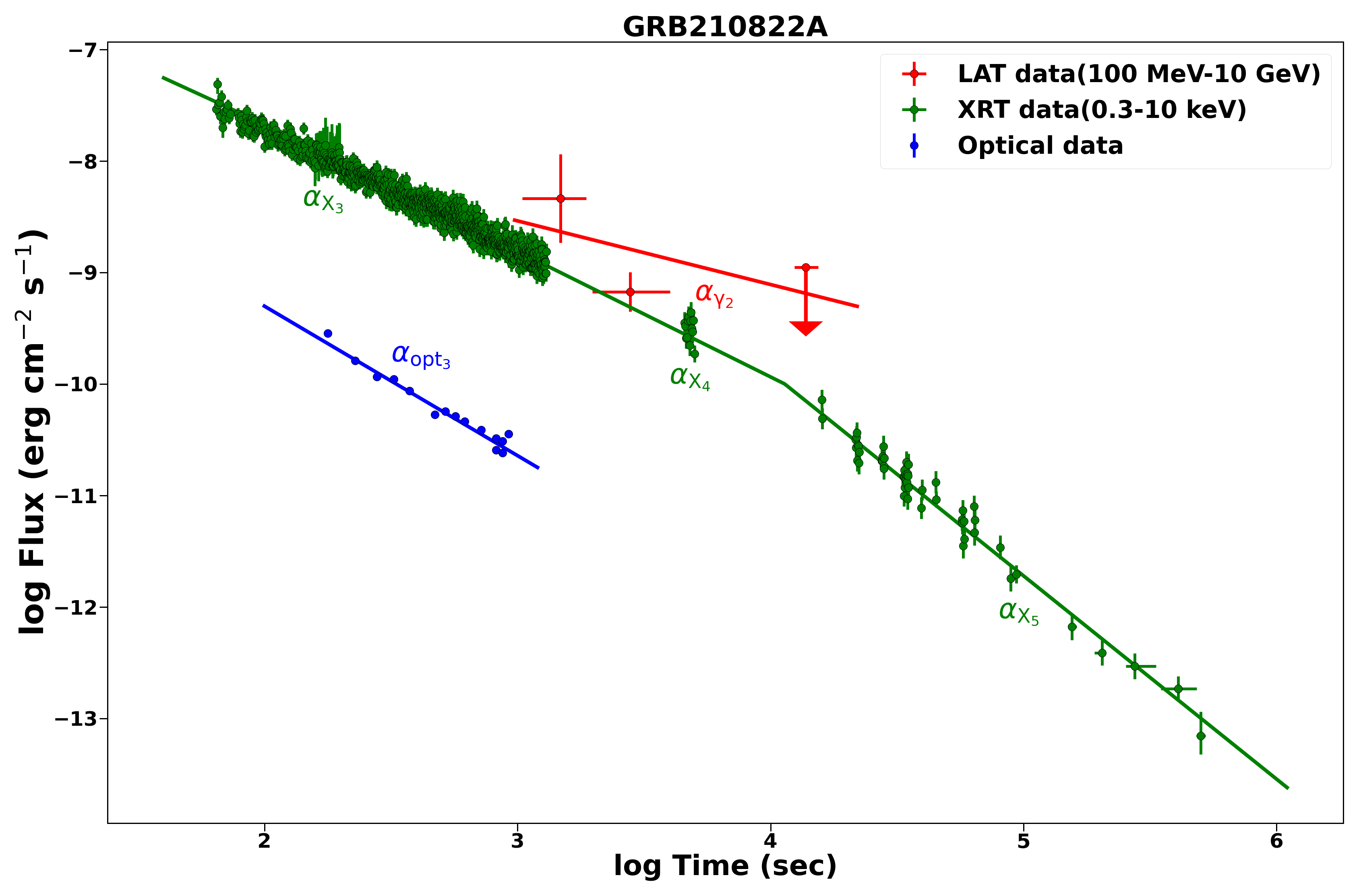}
    \end{minipage}

    \vspace{0.25cm} % Adjust space between rows

    \begin{minipage}[t]{0.32\textwidth}
        \centering
        \includegraphics[width=1\textwidth,height=0.18\textheight]{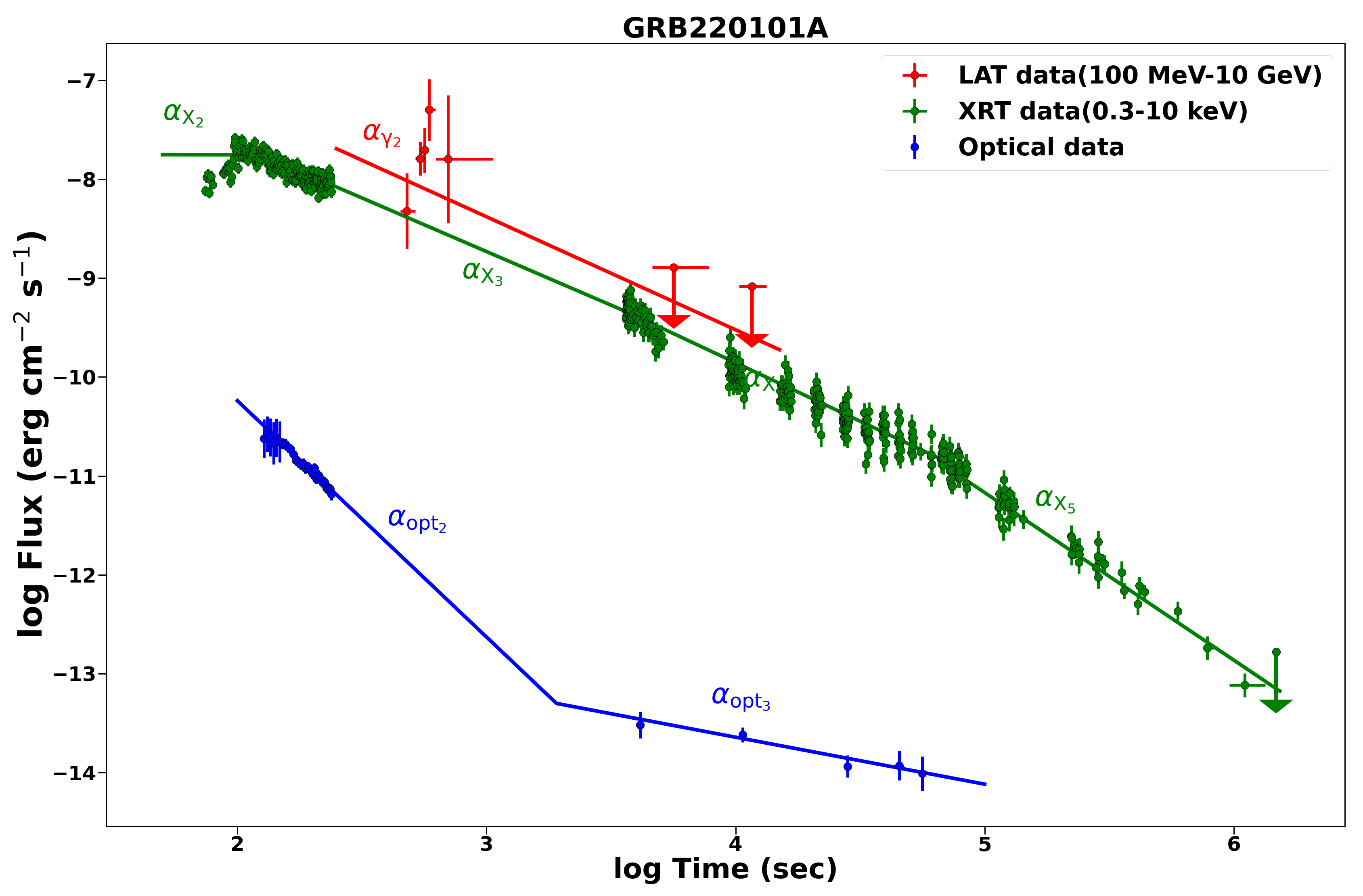}
    \end{minipage}
    \begin{minipage}[t]{0.32\textwidth}
        \centering
        \includegraphics[width=1\textwidth,height=0.18\textheight]{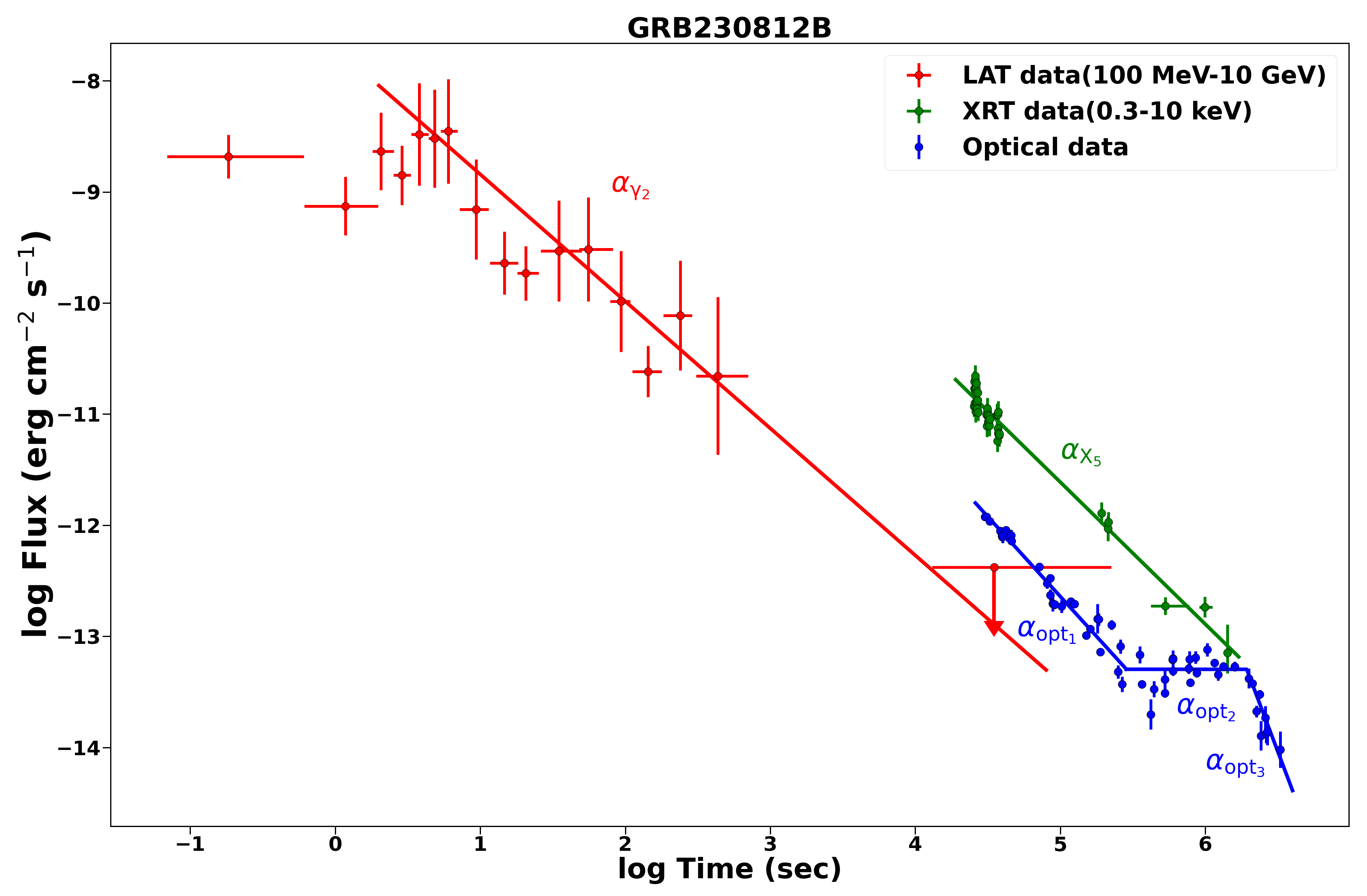}
    \end{minipage}

\caption{A comprehensive LC of 14 GRBs from our sample spanning across gamma-ray, X-ray, and optical wavelengths merged into a single plot. The corresponding fitting slopes ($\alpha$) respective to each wavelength are also indicated in the plot.}
\label{fig:LC_Plots_1}
\end{figure*}

\begin{figure*}
%\gridline{\fig{DistributionPlots/OPTICAL/alpha_1_distribution.png}{0.21\textwidth}{}
%\fig{DistributionPlots/OPTICAL/alpha_2_distribution.png}{0.21\textwidth}{}
%\fig{DistributionPlots/OPTICAL/alpha_3_distribution.png}{0.21\textwidth}{}
%\fig{DistributionPlots/OPTICAL/beta_distribution.png}{0.21\textwidth}{}
%}
%\gridline{\fig{DistributionPlots/LAT/alpha_1_distribution.png}{0.21\textwidth}{}
%\fig{DistributionPlots/LAT/alpha_2_distribution.png}{0.21\textwidth}{}
%\fig{DistributionPlots/LAT/beta_distribution.png}{0.21\textwidth}{}
%}
%\gridline{\fig{DistributionPlots/OPTICAL/alpha_1_distribution.png}{0.35\textwidth}{}
%\fig{DistributionPlots/LAT/alpha_1_distribution.png}{0.35\textwidth}{}
%}
%\gridline{\fig{DistributionPlots/OPTICAL/alpha_2_distribution.png}{0.35\textwidth}{}
%\fig{DistributionPlots/LAT/alpha_2_distribution.png}{0.35\textwidth}{}}    
%\gridline{\fig{DistributionPlots/OPTICAL/alpha_3_distribution.png}{0.35\textwidth}{}
%\fig{DistributionPlots/LAT/beta_distribution.png}{0.35\textwidth}{}
%}          
%\gridline{\fig{DistributionPlots/OPTICAL/beta_distribution.png}{0.35\textwidth}{}
%}
\gridline{\fig{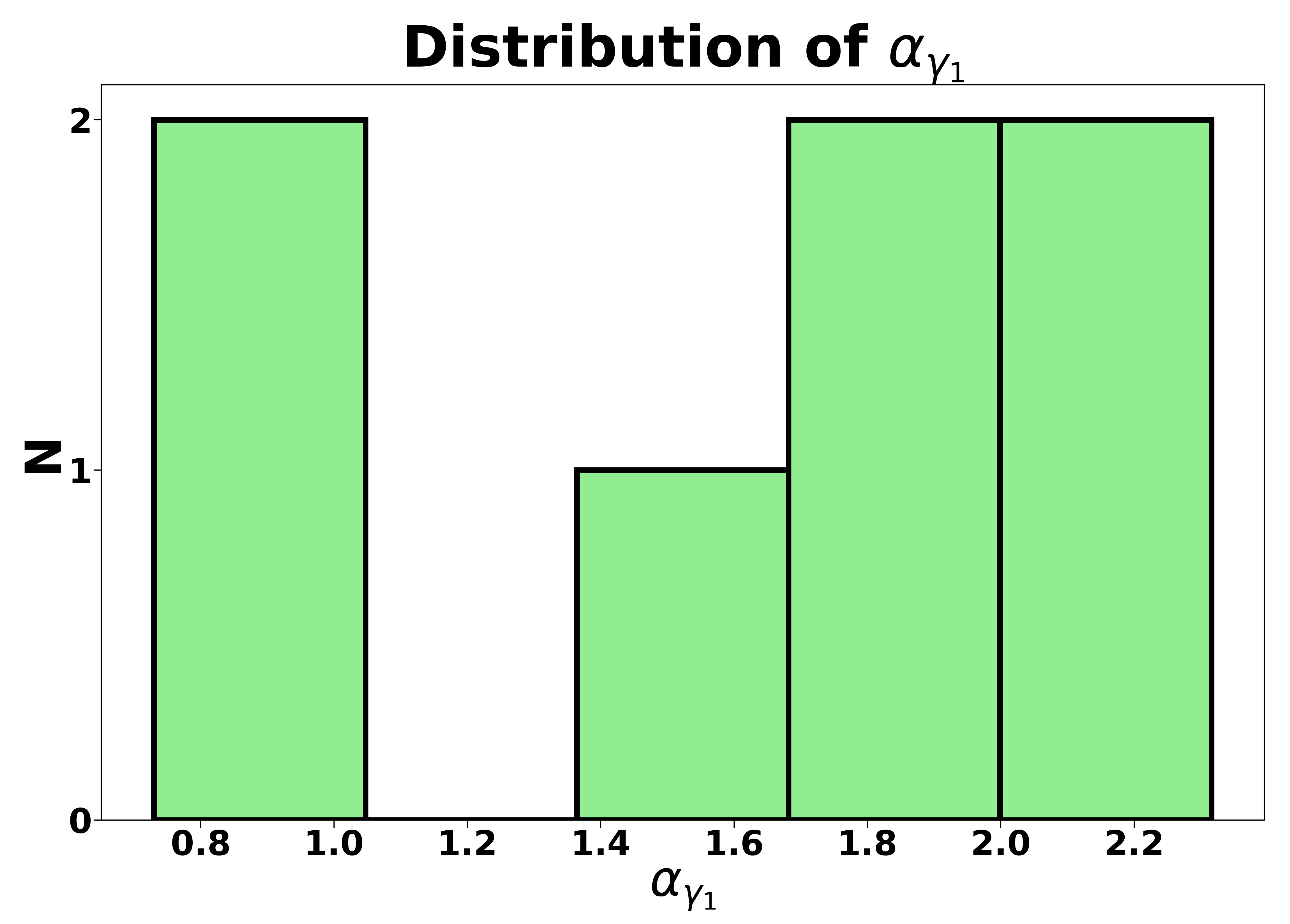}{0.27\textwidth}{}
\fig{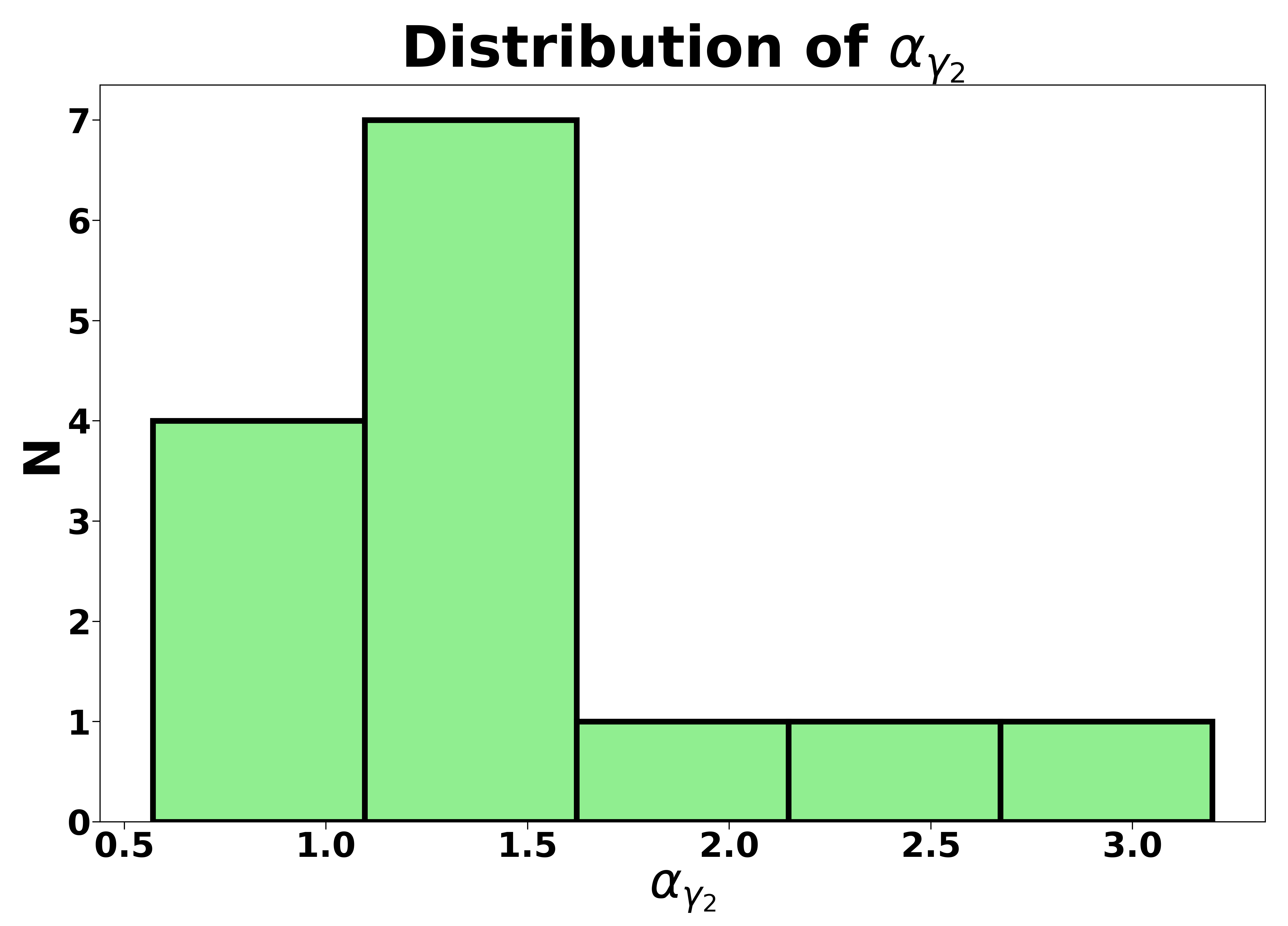}{0.27\textwidth}{}
\fig{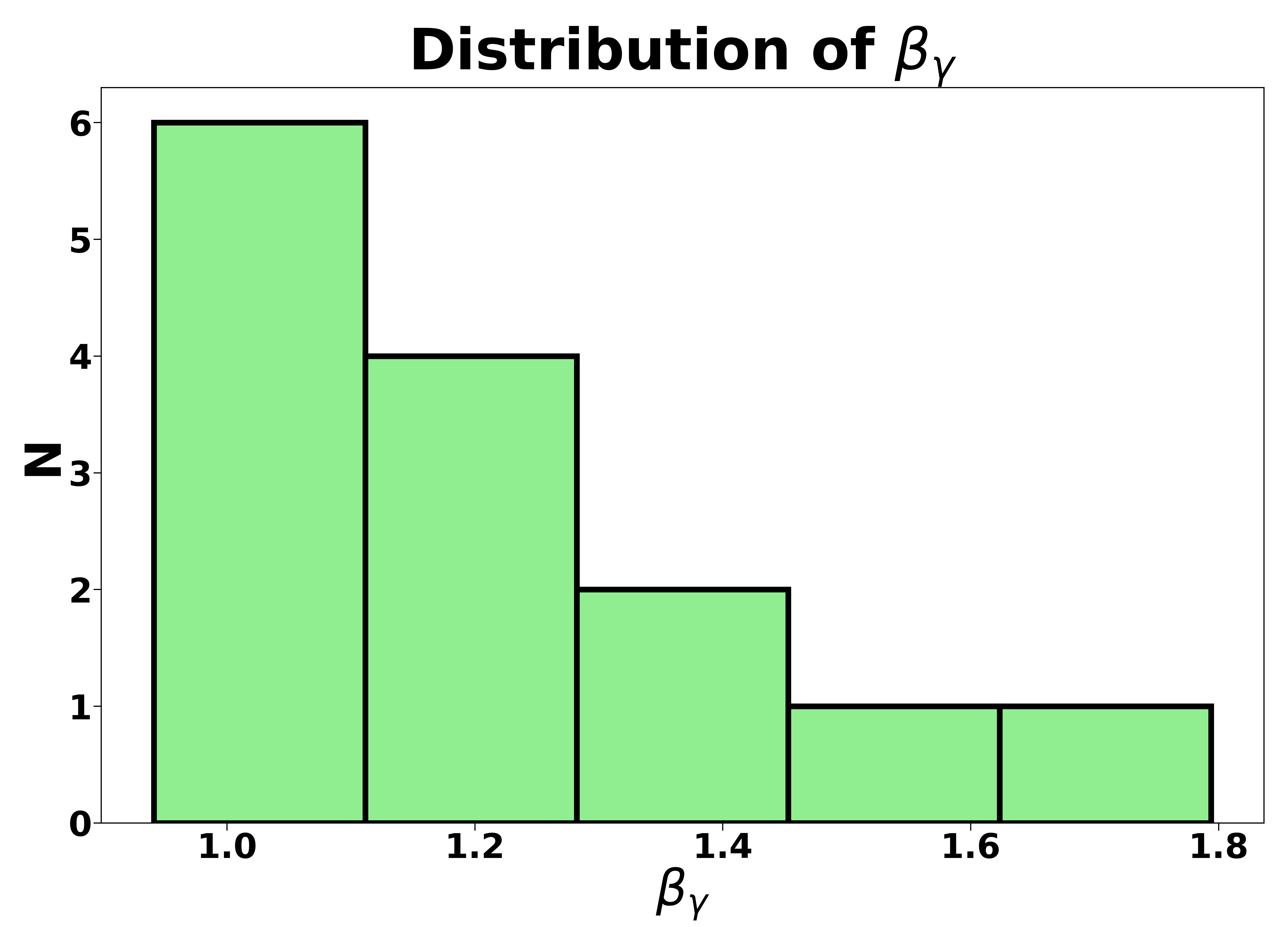}{0.27\textwidth}{}
}
\vspace{-25pt}
\gridline{\fig{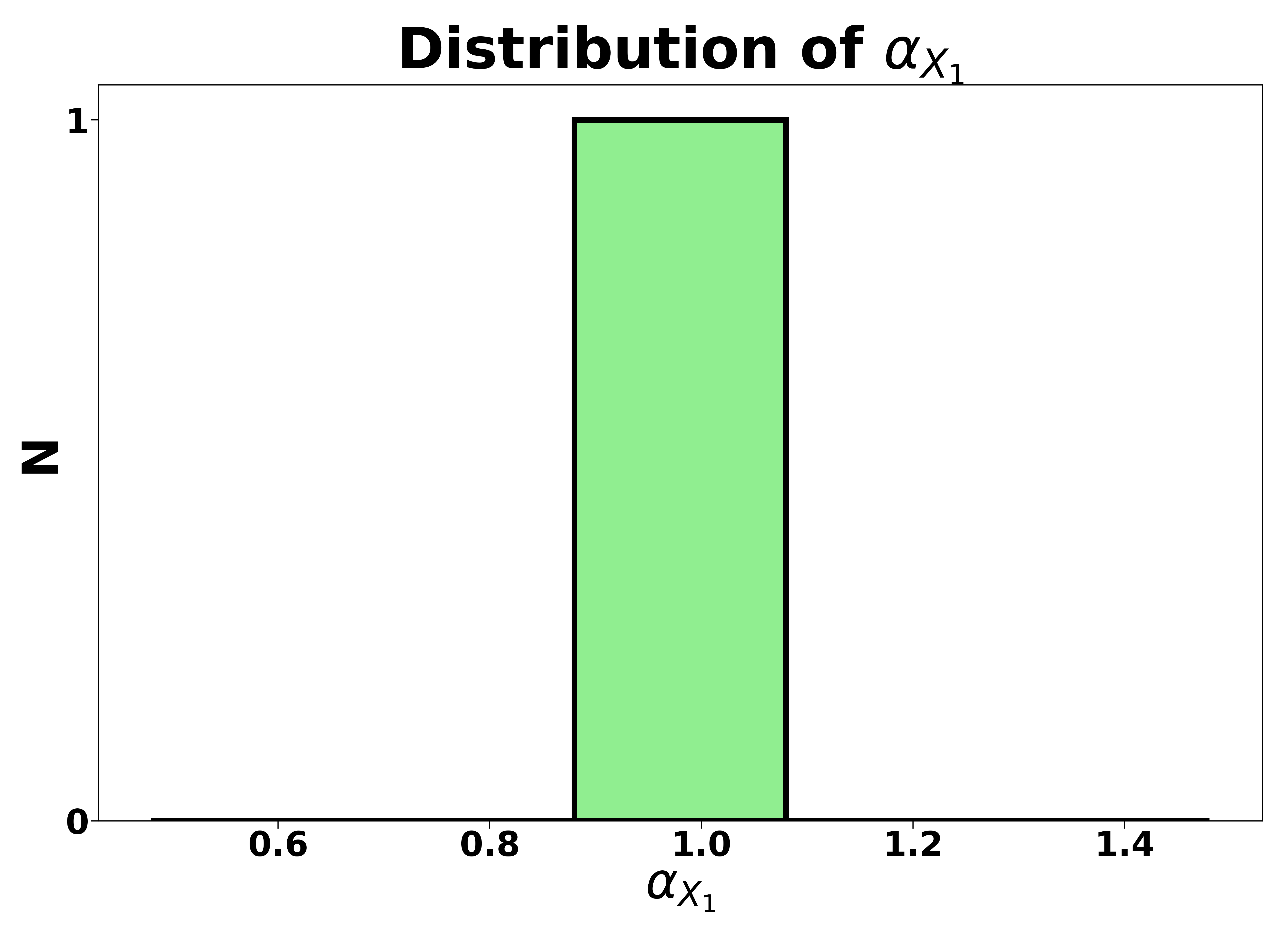}{0.27\textwidth}{}
\fig{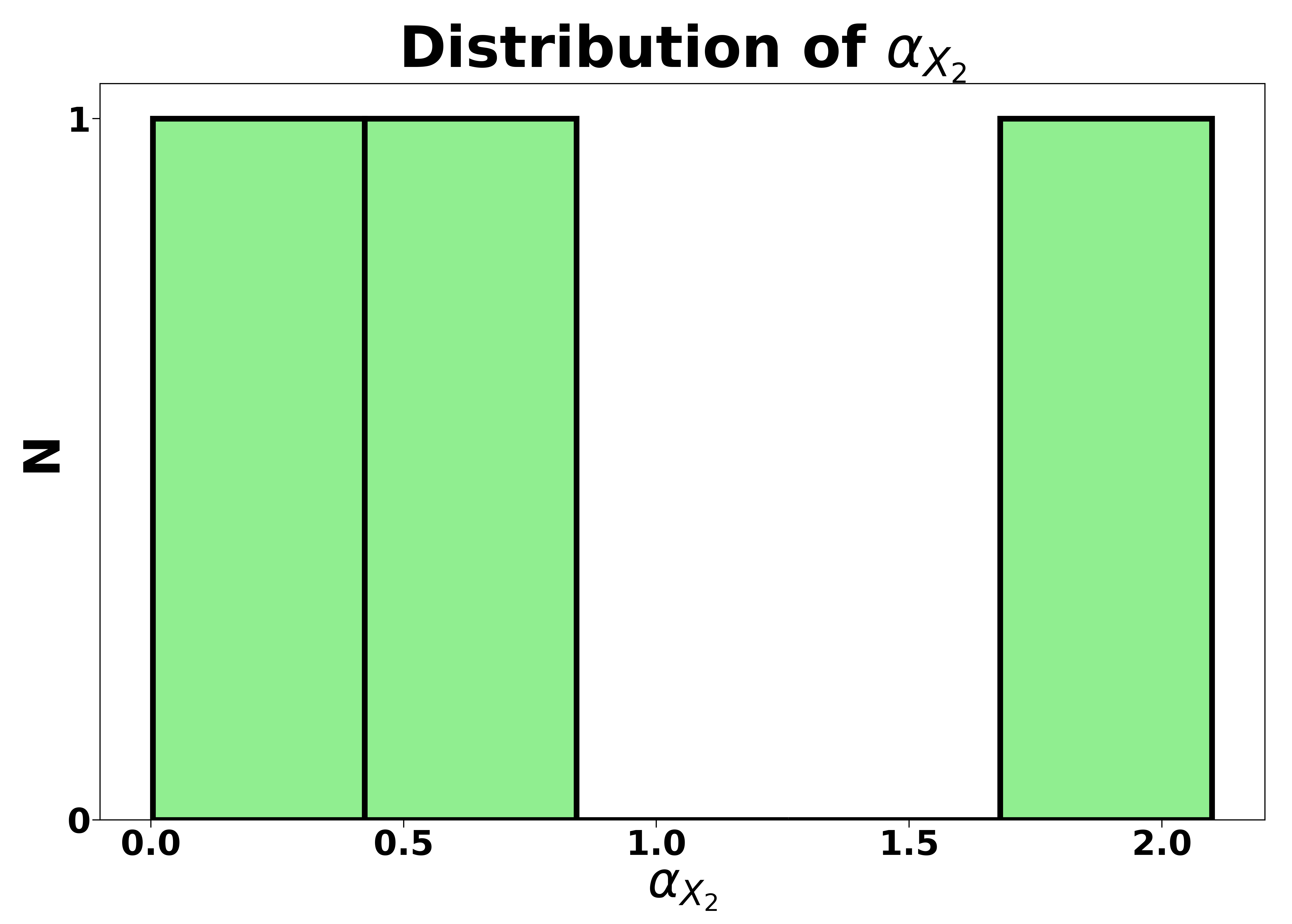}{0.27\textwidth}{}
\fig{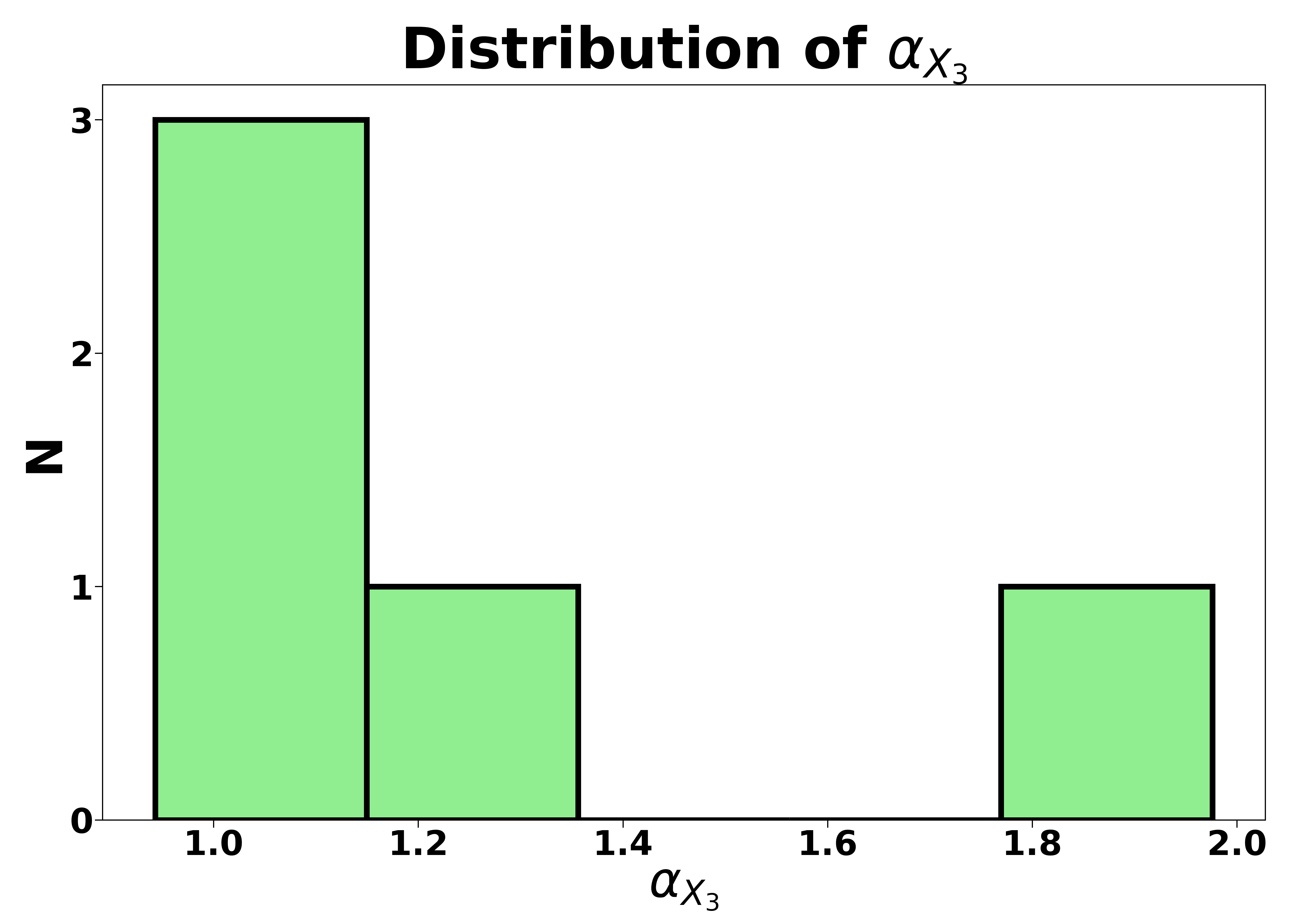}{0.27\textwidth}{}
}
\vspace{-25pt}
\gridline{\fig{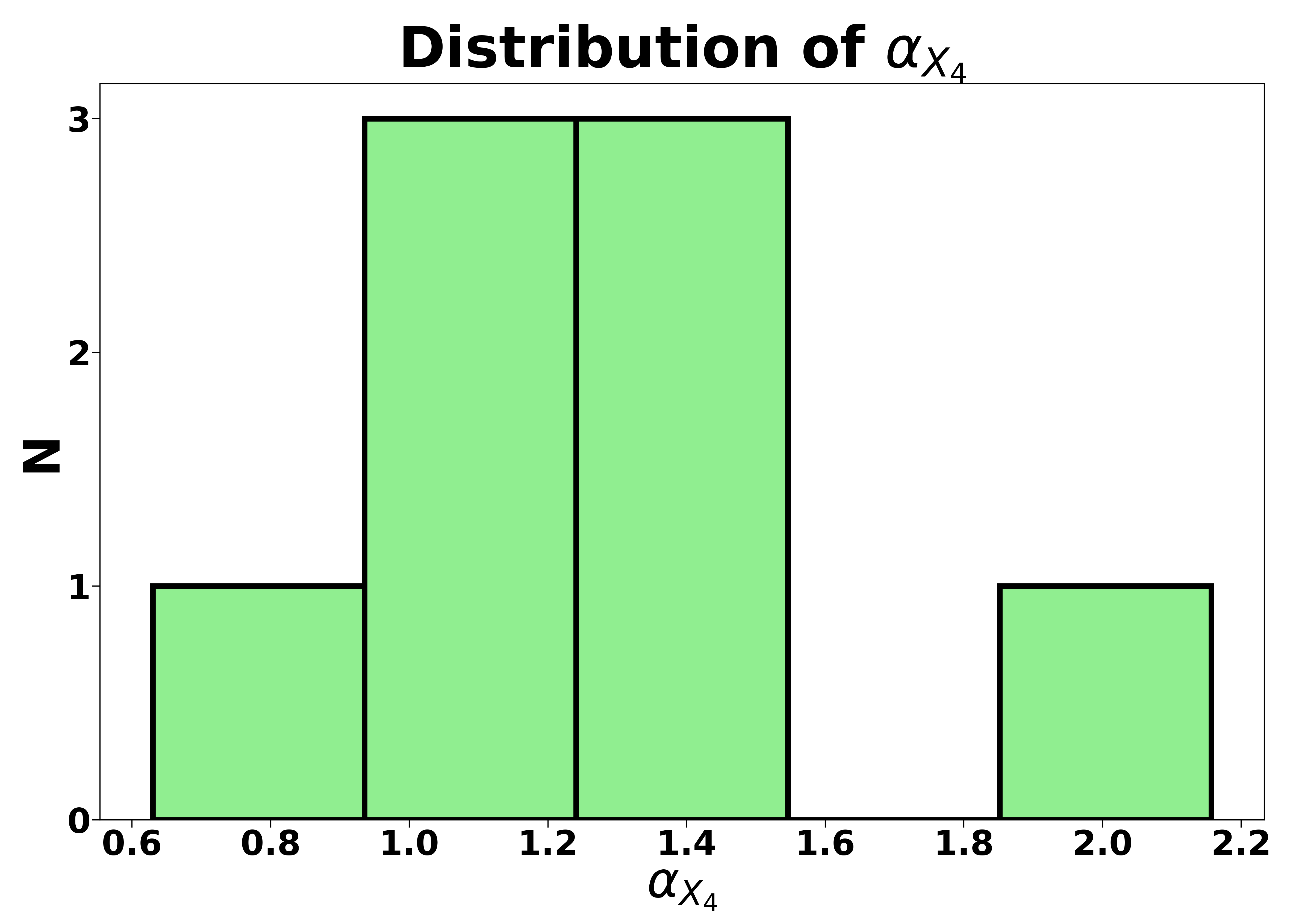}{0.27\textwidth}{}
\fig{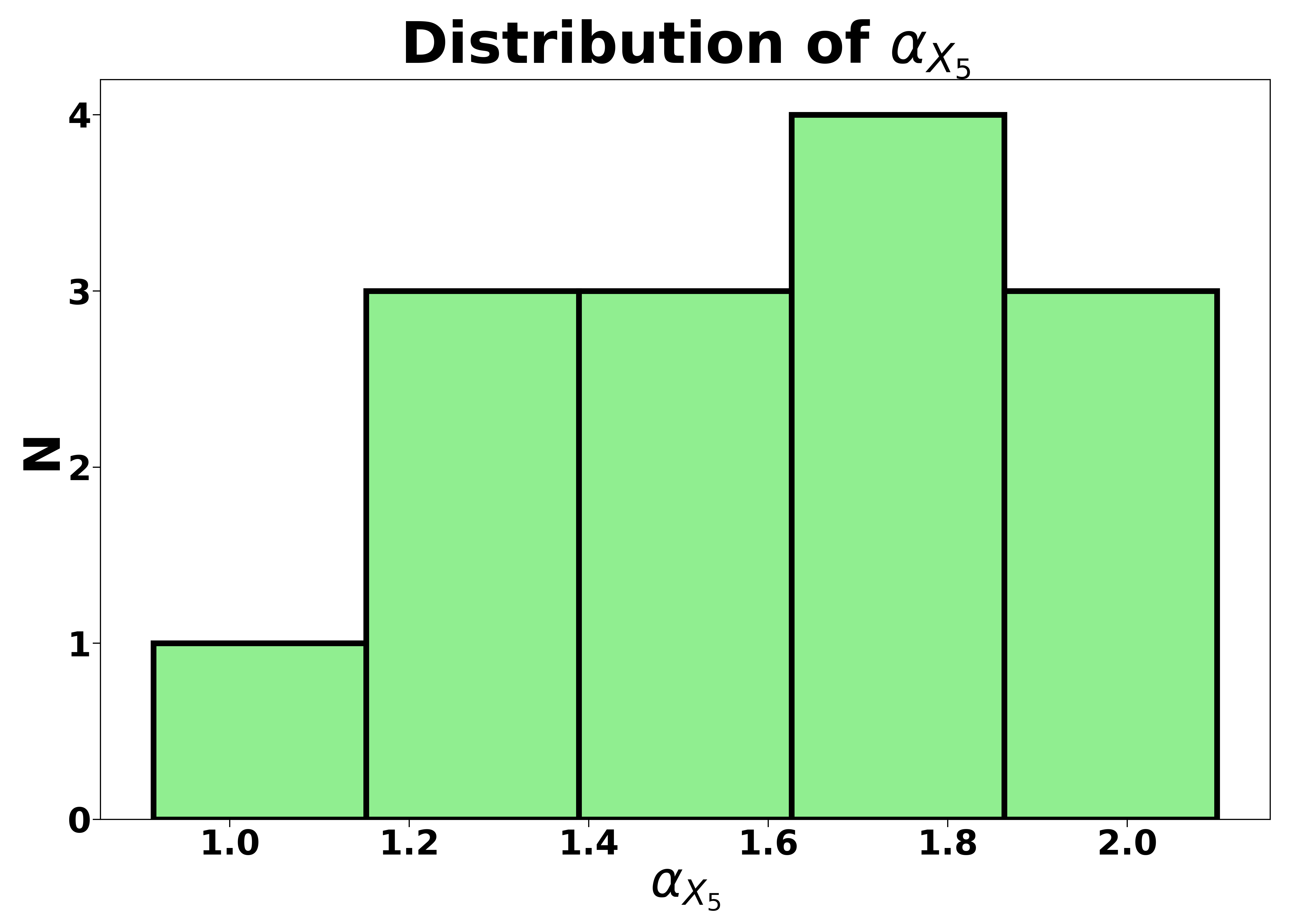}{0.27\textwidth}{}
\fig{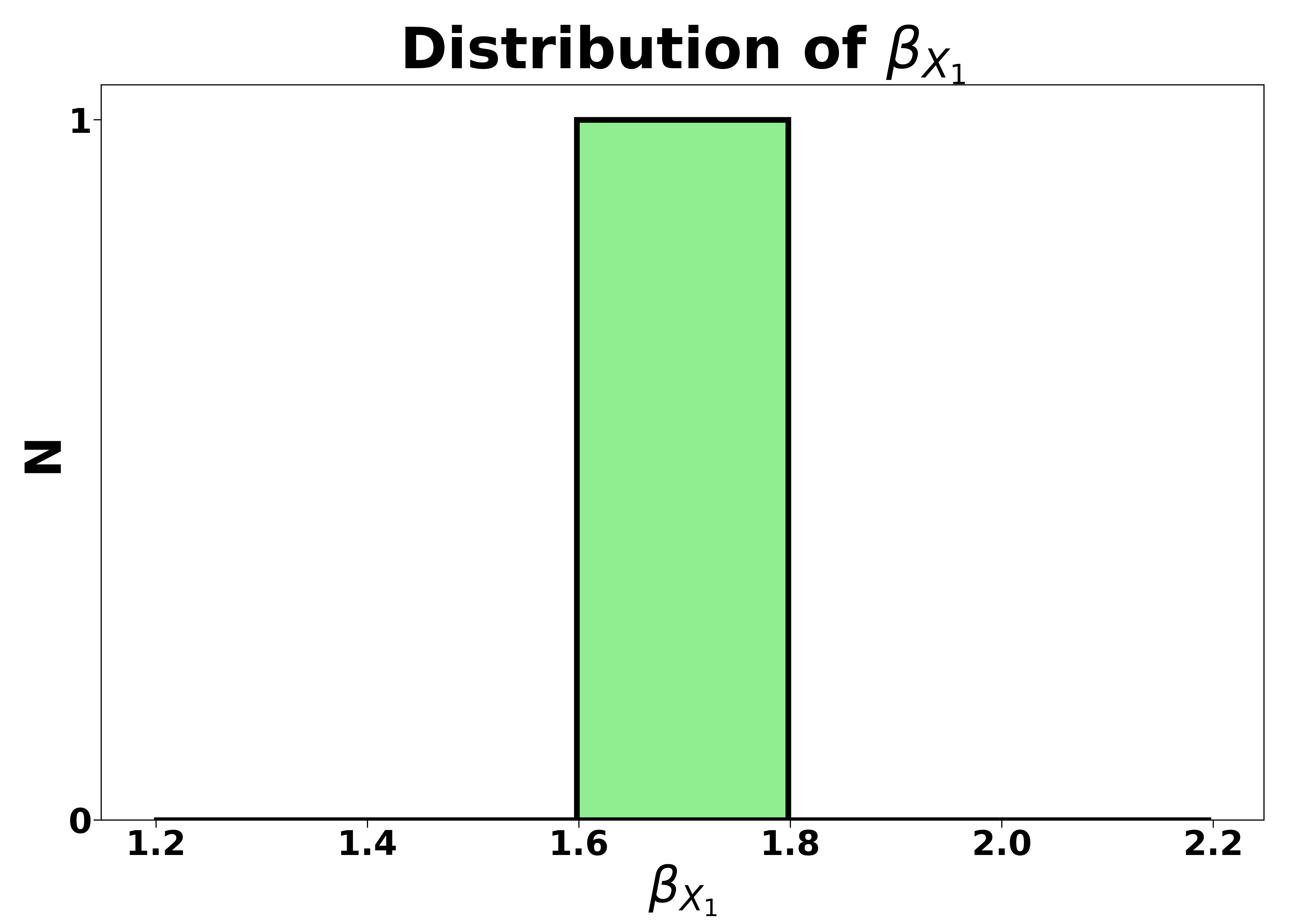}{0.27\textwidth}{}
}
\vspace{-25pt}
\gridline{\fig{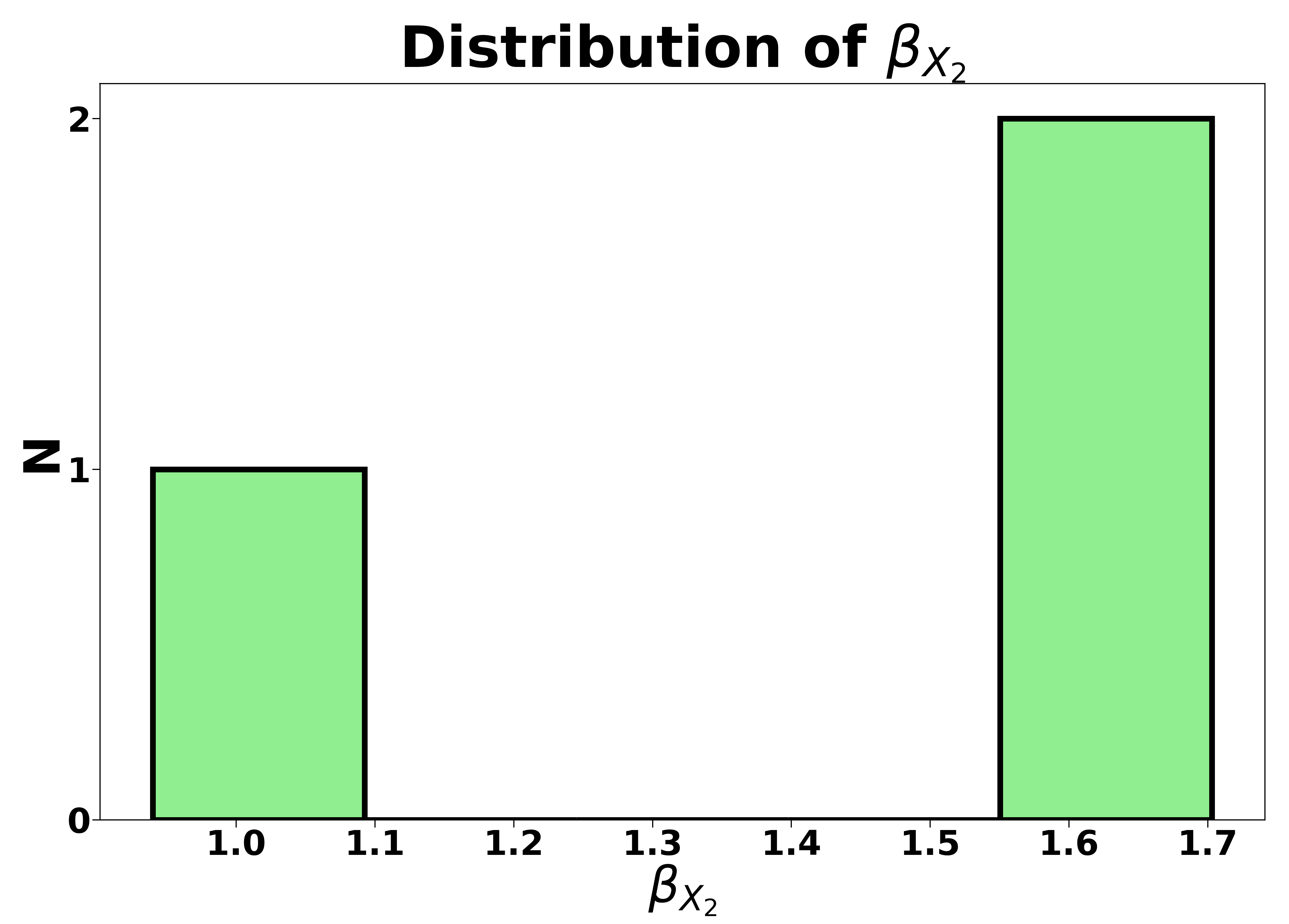}{0.27\textwidth}{}
\fig{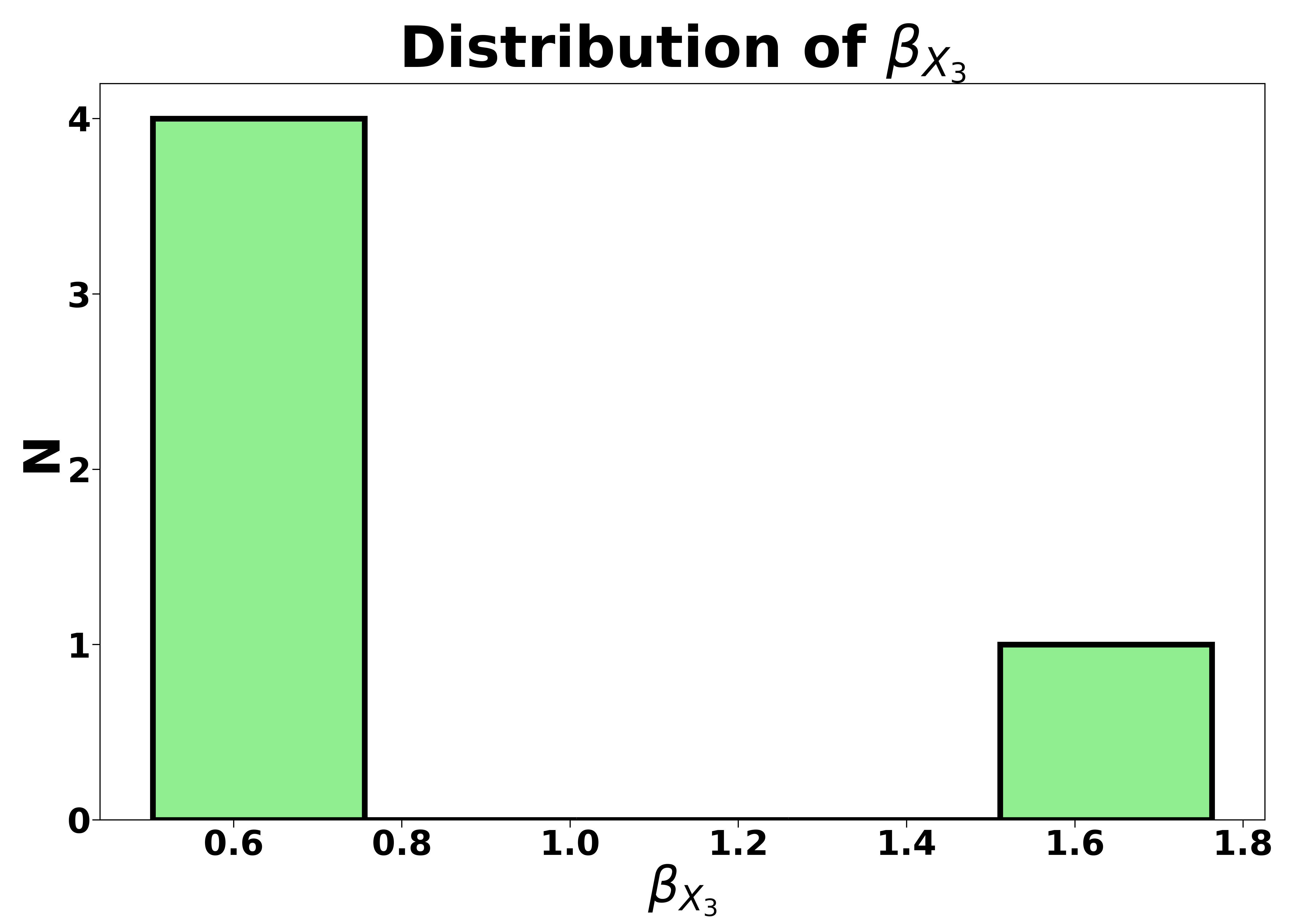}{0.27\textwidth}{}
\fig{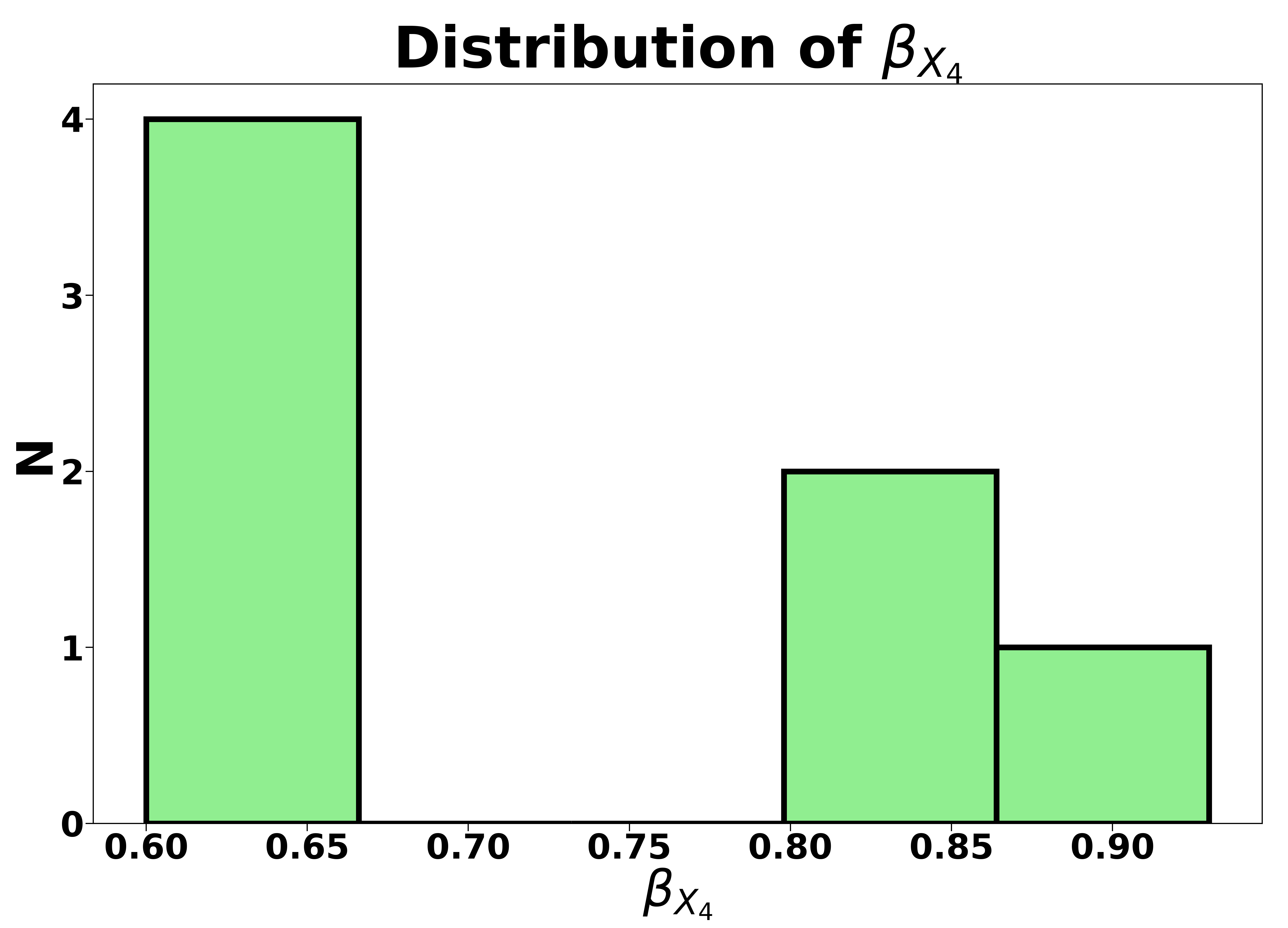}{0.27\textwidth}{}
}
\vspace{-25pt}
\gridline{\fig{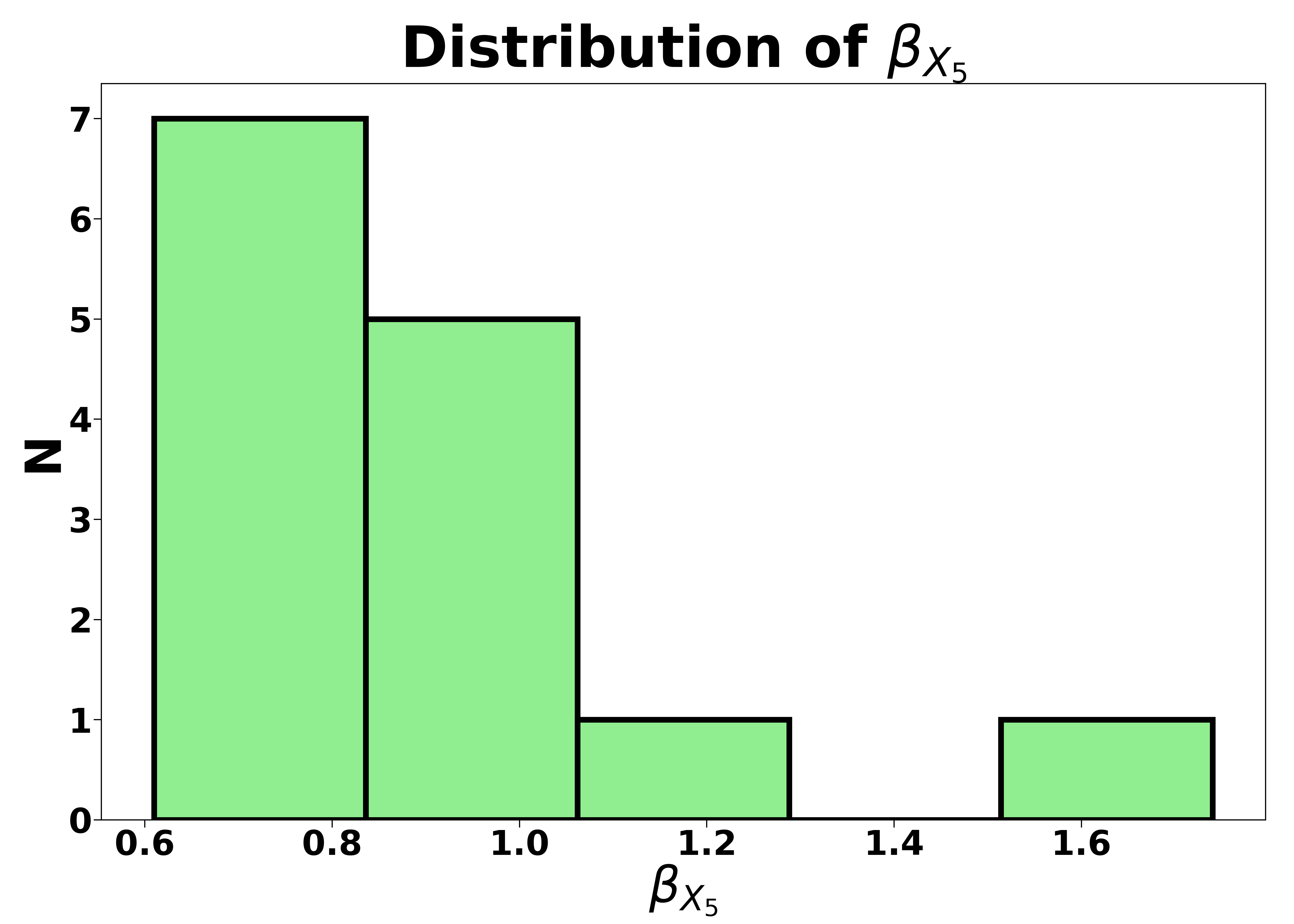}{0.27\textwidth}{}
\fig{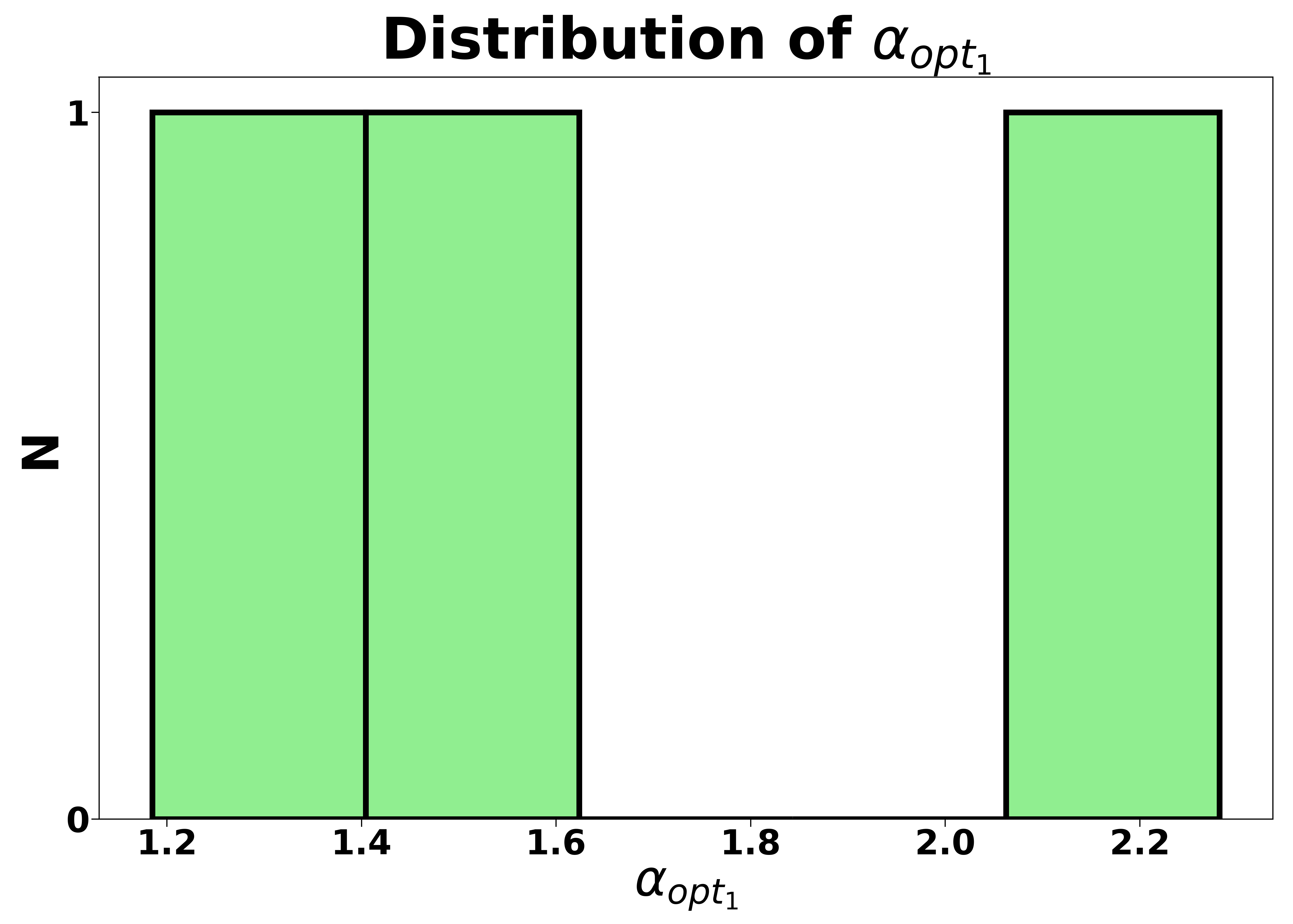}{0.27\textwidth}{}
\fig{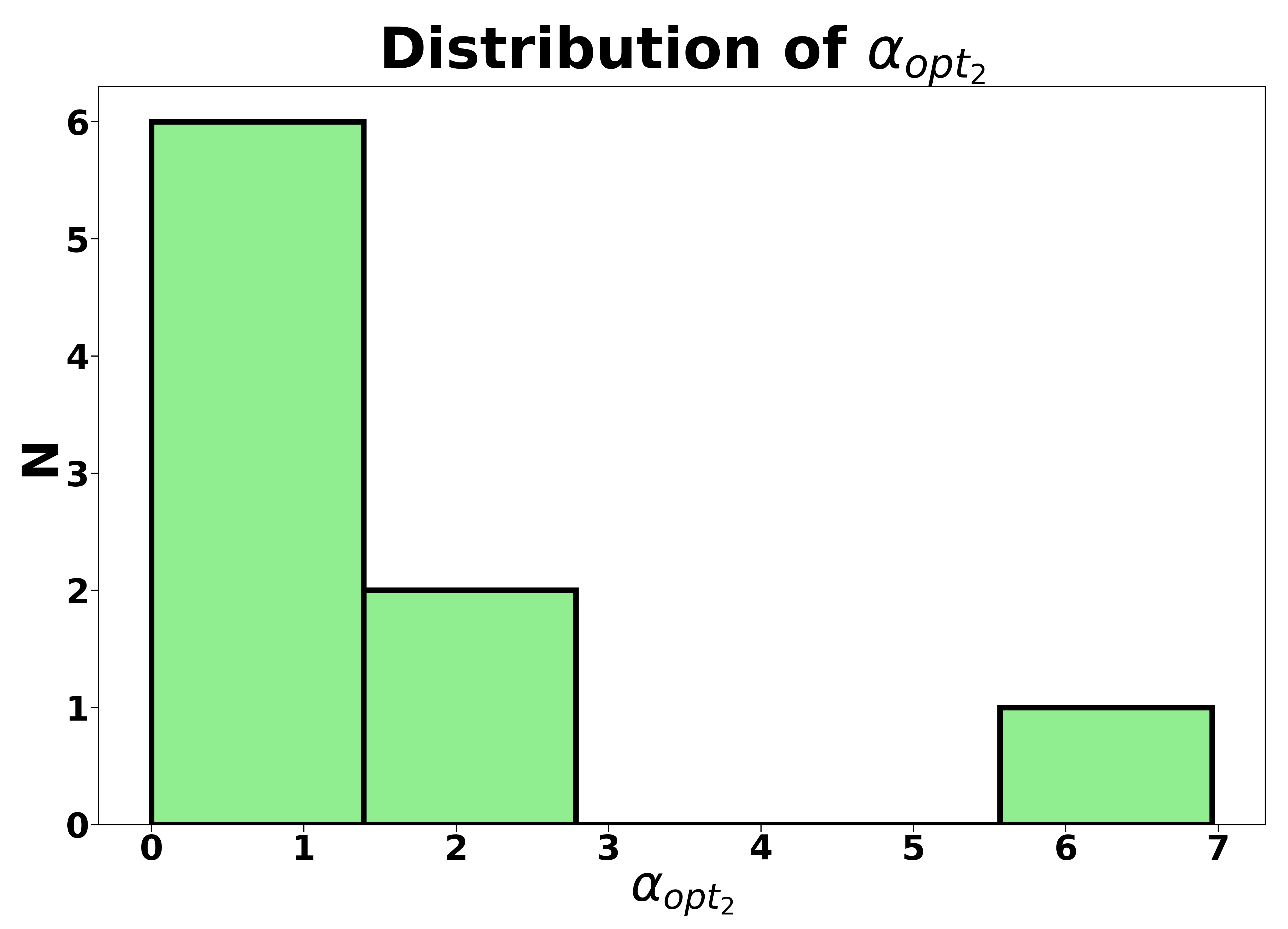}{0.27\textwidth}{}
}
\vspace{-25pt}
\gridline{\fig{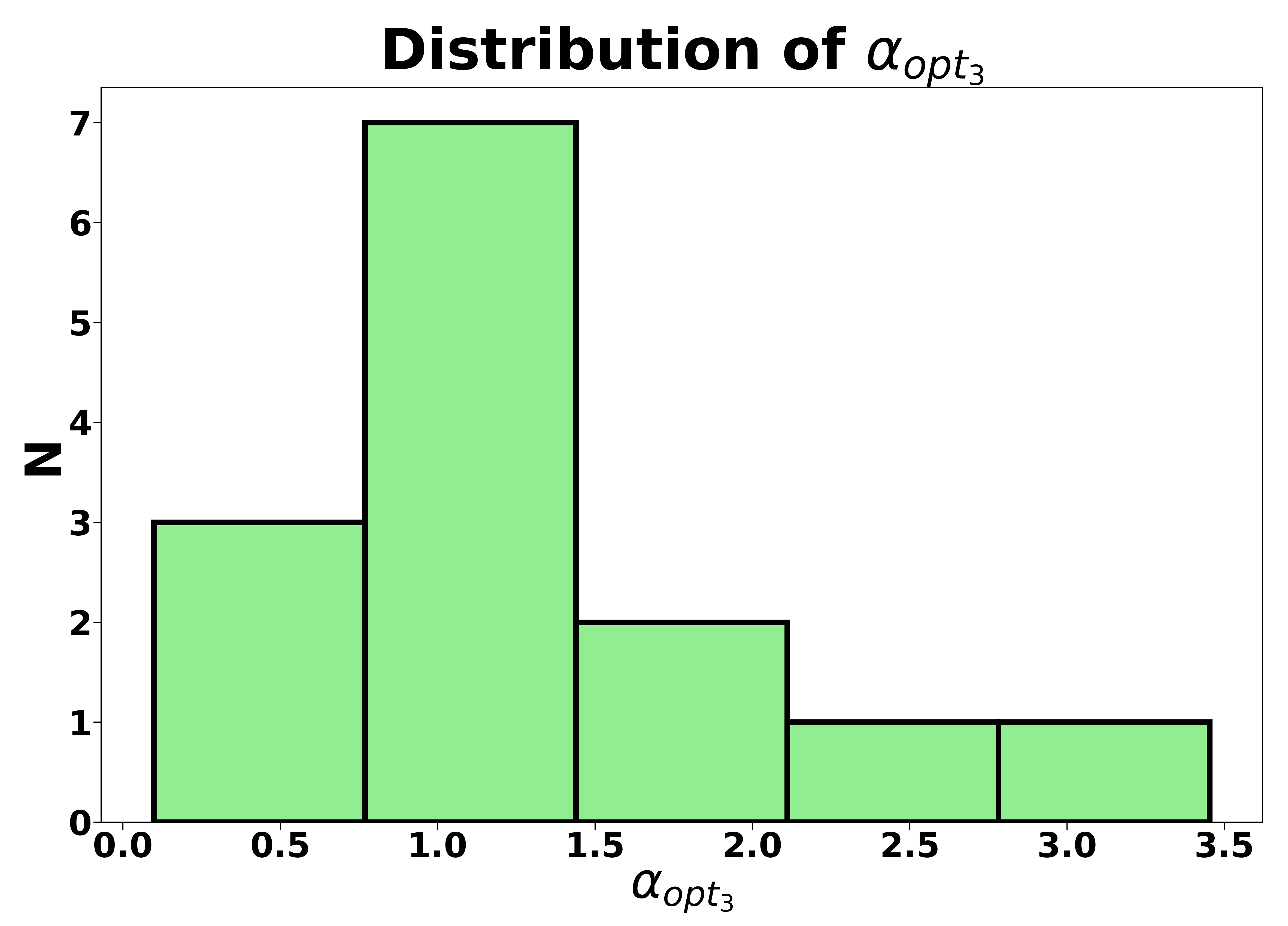}{0.28\textwidth}{}
\fig{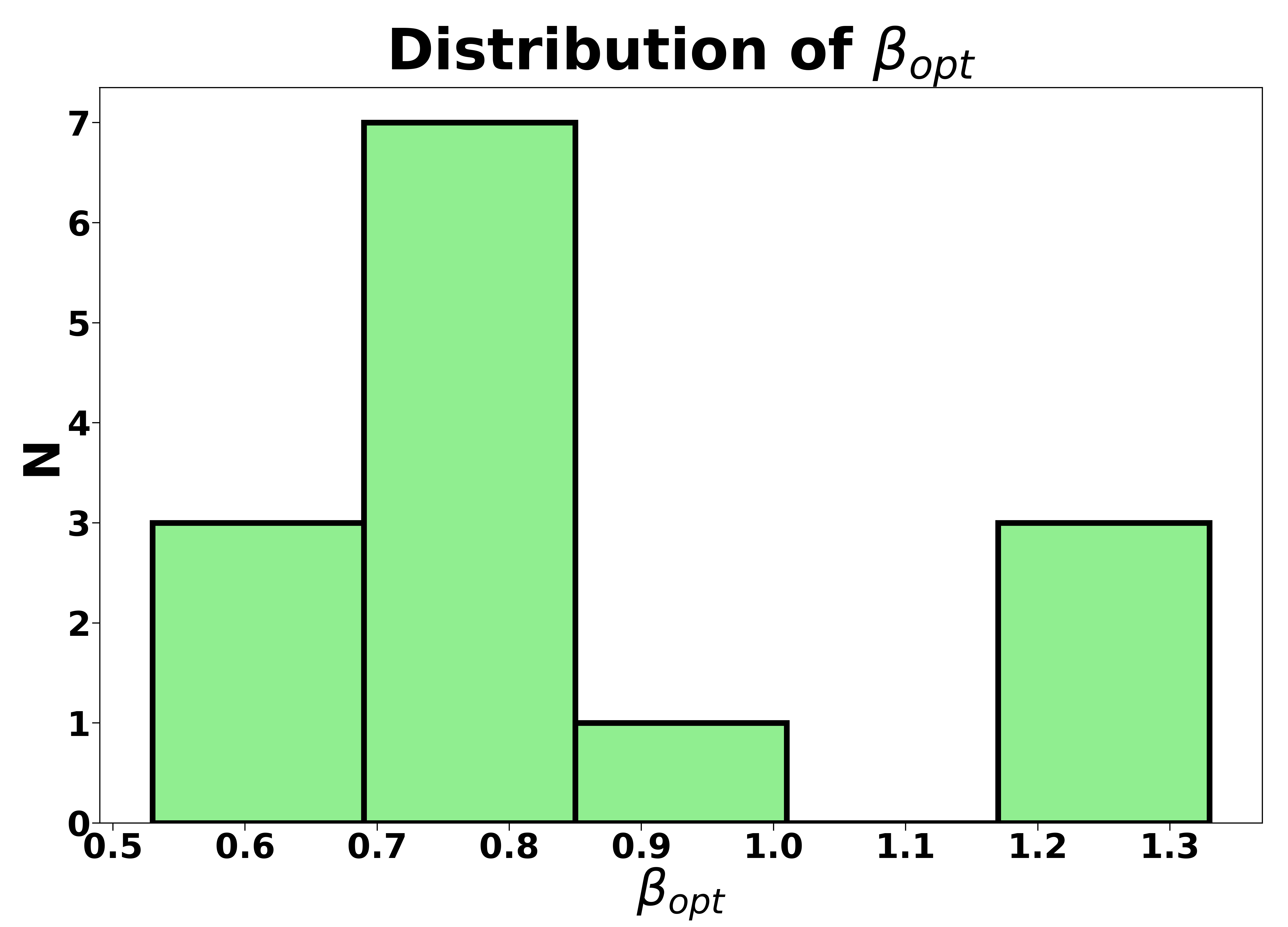}{0.28\textwidth}{}
}
\vspace{-25pt}
\caption{Distribution of $\alpha$, and  $\beta$ parameters across gamma-rays, Xrays, and the optical band for these 14 GRBs used in this study.}
\label{fig:distribution1}
\end{figure*}

%\textcolor{cyan}{As \cite{Dainotti2023Galax..11...25D} proposes, a comprehensive examination of the correlation between $\alpha$ and $\beta$ has been conducted. Various astrophysical environments, including the surrounding medium, electron population profile, and electron cooling regime, were taken into account for this purpose. Also, we have examined situations in which energy is injected and in which it is not. We have taken into account a constant ambient medium and density profiles, denoted by the power law $n \propto r^{k}$ with k values of 1.0, 1.5, 2.0, and 2.5.}
%%This approach aims to facilitate a comprehensive investigation into CRs.
%The inclusion of these density profiles accounts for the potential presence of a stratified medium alongside the conventional ISM and wind medium. 
%When considering the stratified medium, we uniformly explored the entire range of possible values of $k$, extending from $k = 0$ to $k = 2.9$. 
Prior studies conducted by \citet{2000ApJ...532..286K,2007ARA&A..45..177C,2012ApJ...746..122D,2012ApJ...751...57D,2013ApJ...776..120Y,hotokezaka2013progenitor,2020ApJ...895...94Y,Dainotti2023Galax..11...25D} have thoroughly discussed the presence of the stratified medium. 
In this study, we tested $k=$ 0, 1, 1.5, 2, and 2.5 for both energy injection and no energy injection scenarios considering three regimes: slow-cooling (SC, $\nu_{m} < \nu < \nu_{c}$), fast-cooling (FC, $\nu_{c} < \nu < \nu_{m}$), and slow/fast-cooling (SC/FC, $\nu > $ max\{$\nu_c,\nu_m$\}). 
We adopted the set of CRs for no energy injection scenario for $k = 0$ and 2 from \citet{Srinivasaragavan2020ApJ}, whereas the set of CRs that account for energy injection is adopted from \citet{2009ApJ...698...43R} and \citet{Gao2013}. 
The CRs are taken from \cite{Dainotti2023Galax..11...25D} for other density profiles. 
%\st{We examine each CR within specific \textit{p} value, namely $1 < p < 2$ or $p > 2$. 
%This $p$ range also dictates the $\beta-p$ relationship ($\beta = \beta(p)$). 
In our analysis, we analyze two sets of electron spectral distribution (\textit{p}) values for CRs without energy injection, one for $1 < p < 2$ and the other for $p > 2$. 
However, for CRs with energy injection, we exclusively focus on the \textit{p} value where  $p > 2$, following \citet{2009ApJ...698...43R}.
% It is important to note that these \textit{p} values dictate the $\beta - p$ relationship ($\beta = \beta$($p$)).} 
%\textcolor{orange}{For \textcolor{cyan}{the bursts detected by} \textit{Fermi}-LAT, we use $\alpha_{\rm{\gamma_2}}$ \st{for} \textcolor{cyan}{when} \st{both energy injection and without energy injection in cases of GRBs fitted with PL} \textcolor{cyan}{a GRB is fitted by a PL, with not distinction of energy injection and without energy injection}. In the case \st{of} \textcolor{cyan}{when a} GRBs \textcolor{cyan}{is} fitted with \textcolor{cyan}{a} BPL, \st{when examining without energy injection scenario, we employ $\alpha_{\rm{\gamma_2}}$}\textcolor{cyan}{we employ $\alpha_{\rm{\gamma_2}}$ when  we examine the no energy injection scenario}, while for energy injection scenario, we used $\alpha_{\rm{\gamma_1}}$.
%In the optical \textcolor{cyan}{band}, \st{for} \textcolor{cyan}{those} GRBs fitted with \textcolor{cyan}{a} PL \textcolor{cyan}{we employ} $\alpha_{\rm{opt_2}}$ \st{is utilized} for both energy injection and without energy injection \st{scenarios}\textcolor{cyan}{cases}. 
%For GRBs fitted with \st{models other}\textcolor{cyan}{any other model} than \textcolor{cyan}{a} PL, \st{in the case of with energy injection, we employ $\alpha_{\rm{opt_1}}$} \textcolor{cyan}{we employ $\alpha_{\rm{opt_1}}$ when energy injection is considered}, whereas for \textcolor{cyan}{the scenario} without energy injection, we utilize $\alpha_{\rm{opt_2}}$.}
For GRBs detected by the \textit{Fermi}-LAT, we use $\alpha_{\rm{\gamma_2}}$ when a GRB is fitted with a PL, regardless of whether it is in the context of energy injection or no energy injection scenarios. 
However, in the case when a GRB is fitted with a BPL, we employ $\alpha_{\rm{\gamma_2}}$ when we examine without energy injection scenario and $\alpha_{\rm{\gamma_1}}$ when energy injection is taken into account. The latter scenario indeed happens earlier in time in the segment of the LC. 
%We here show that LC segments in Figure \ref{fig:LC_Plots_1} (a), (b), (d), and (e). 
%In the case of bursts observed by \textit{Swift}-XRT, we use $\alpha_{\rm{X_4}}$ for GRBs fitted with a PL, both for scenarios with and without energy injection. 
For the X-ray analysis, we utilize different $\alpha_{\rm{X}}$, depending on the specific GRB and the scenario of energy injection. For GRBs fitted with PL, we employ $\alpha_{\rm{X_5}}$ in both scenarios: with and without energy injection. However, for GRBs 090510A and 170405A, we use $\alpha_{\rm{X_5}}$ when there is no energy injection, and $\alpha_{\rm{X_4}}$ when energy injection is considered. In the cases of GRBs 130427A and 220101A, $\alpha_{\rm{X_4}}$ is used for scenarios without energy injection, while $\alpha_{\rm{X_3}}$ and $\alpha_{\rm{X_2}}$ are applied when energy injection is considered, respectively. For GRB 180720B, $\alpha_{\rm{X_3}}$ and $\alpha_{\rm{X_2}}$ are used for no energy injection and energy injection cases, respectively. For GRB 210822A, $\alpha_{\rm{X_5}}$ is used for no energy injection scenario, as it does not exhibit energy injection in its LC. Similarly, GRB 160625B and 171010A show no signs of energy injection; hence, $\alpha_{\rm{X_4}}$ is used for their analysis in the absence of energy injection. 
For the optical analysis, we also employ different $\alpha_{\rm{opt}}$, depending on the specific GRB and the scenario of energy injection. In cases where GRB is fitted with PL, $\alpha_{\rm{opt_3}}$ is applied consistently across both scenarios, whether energy injection is present or not. However, GRB 090510A, 090902B, 130427A, 160625B, 171010A, and 220101A, $\alpha_{\rm{opt_2}}$ is used exclusively under the no energy injection scenario, as these GRBs do not show evidence of energy injection in their LC. For GRB 090926A, 120711A, and 230812B, $\alpha_{\rm{opt_3}}$ is used when there is no energy injection, whereas $\alpha_{\rm{opt_2}}$ is utilized when an energy injection scenario is taken into account.
We show the GRB LCs of our sample in all wavelengths in Figure \ref{fig:LC_Plots_1}, where the X-ray LCs of GRB 090510A, 090926A, 141028A, 170405A, and 230812B are rescaled for visualization purposes to their respective gamma-ray LCs; others remain unrescaled.
In Figure \ref{fig:distribution1}, we show the distributions of $\alpha_{\rm{\gamma_1}}$, $\alpha_{\rm{\gamma_2}}$, $\beta_{\rm{\gamma}}$, $\alpha_{\rm{X_1}}$, $\alpha_{\rm{X_2}}$, $\alpha_{\rm{X_3}}$, $\alpha_{\rm{X_4}}$, $\alpha_{\rm{X_5}}$, $\beta_{\rm{X_1}}$, $\beta_{\rm{X_2}}$, $\beta_{\rm{X_3}}$, $\beta_{\rm{X_4}}$, $\beta_{\rm{X_5}}$, $\alpha_{\rm{opt_1}}$, $\alpha_{\rm{opt_2}}$, $\alpha_{\rm{opt_3}}$, and $\beta_{\rm{opt}}$ parameters in our sample. 
%\textcolor{orange}{In Figure \ref{fig:distribution2}, we show the distributions of \textit{Fermi}-LAT $\log(T_{90})$ and $\log(E_{iso})$.}

In our analysis of CRs, we also consider the jet break scenario, using jet break equations derived from \cite{2022ApJ...940..189F}, which are quoted in Eq. \ref{Eq:jet-break-equation} :

%% LC PLOTS
%\begin{figure*}
%\gridline{\fig{LC Plots/GRB090328A.png}{0.27\textwidth}{}
%          \fig{LC Plots/GRB090510A.png}{0.27\textwidth}{}
%          \fig{LC Plots/GRB090902B.png}{0.27\textwidth}{}
%          }
%\gridline{\fig{LC Plots/GRB090926A.png}{0.27\textwidth}{}
%          \fig{LC Plots//GRB120711A.png}{0.27\textwidth}{}
%          \fig{LC Plots/GRB130427A.png}{0.27\textwidth}{}
%         }
%\gridline{\fig{LC Plots/GRB141028A.png}{0.27\textwidth}{}
%           \fig{LC Plots/GRB160625B.png}{0.27\textwidth}{}
%           \fig{LC Plots/GRB170405A.png}{0.27\textwidth}{}
%          }
%\gridline{\fig{LC Plots/GRB171010A.png}{0.27\textwidth}{}
%           \fig{LC Plots/GRB180720B.png}{0.27\textwidth}{}
%           \fig{LC Plots/GRB210822A.png}{0.27\textwidth}{}
%          }
%\gridline{\fig{LC Plots/GRB220101A.png}{0.27\textwidth}{}
%           \fig{LC Plots/GRB230812B.png}{0.27\textwidth}{}
%          }
%\caption{A comprehensive LC of 14 GRBs from our sample spanning across gamma-ray, X-ray, and optical wavelengths, merged into a single plot. The corresponding fitting slopes ($\alpha$) respective to each wavelength are also indicated in the plot.} \label{fig:LC_Plots_1}
%\end{figure*}

%% Distribution PLOTS

\begin{equation}
F_{\nu} \propto
\begin{cases}
t^{-1} \nu^{-\frac{1}{2}}, & \nu_{\rm c} < \nu < \nu_{\rm m} \\
t^{-p} \nu^{-\frac{(p-1)}{2}}, & \nu_{\rm m} < \nu < \nu_{\rm c} \\
t^{-p} \nu^{-\frac{p}{2}},   & {\rm{max}}\{\nu_{\rm m}, \nu_{\rm c}\} < \nu\, \\
\end{cases}
\label{Eq:jet-break-equation}
\end{equation}

 It is important to highlight that these jet break equations are independent of $k$. Since not all GRBs exhibit a jet break in their LCs, but for those that do, we apply the specific $\alpha$ parameters to account for the jet break in our CRs. Regarding X-rays, GRB 130427A, 160625B, 171010A, 180720B, and 220101A show evidence of a jet break whose slope index is $\alpha_{\rm{X_5}}$. Similarly, in the optical band, GRB 090510A, 090902B, 130427A, 160625B, 171010A, and 220101A exhibit a jet break whose slope index is $\alpha_{\rm{opt_3}}$.
\begin{figure*}
\gridline{%\fig{figures-UPDATED/LAT Without Energy Injection/k=0/SC.pdf}{0.18\textwidth}{(a)}
          %\fig{figures-UPDATED/LAT Without Energy Injection/k=0/FC.pdf}{0.18\textwidth}{(b)}
          \fig{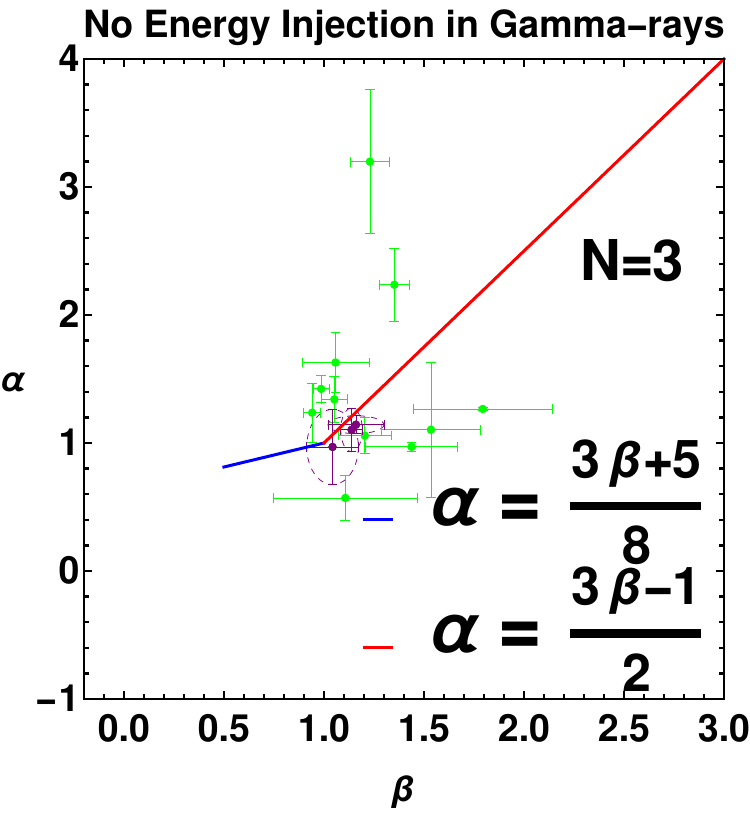}{0.23\textwidth}{(a)}
          \fig{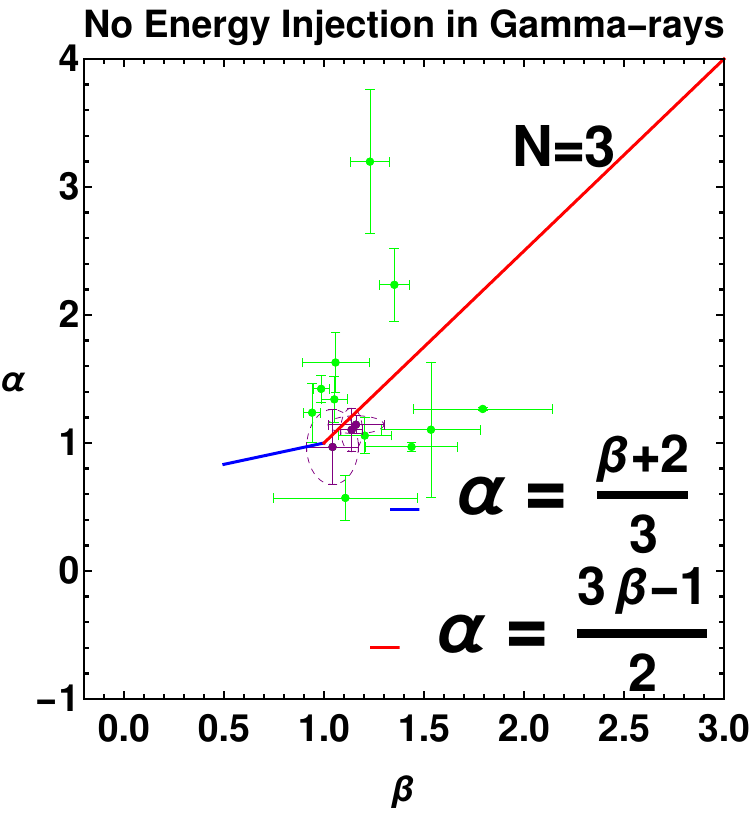}{0.23\textwidth}{(b)}
          \fig{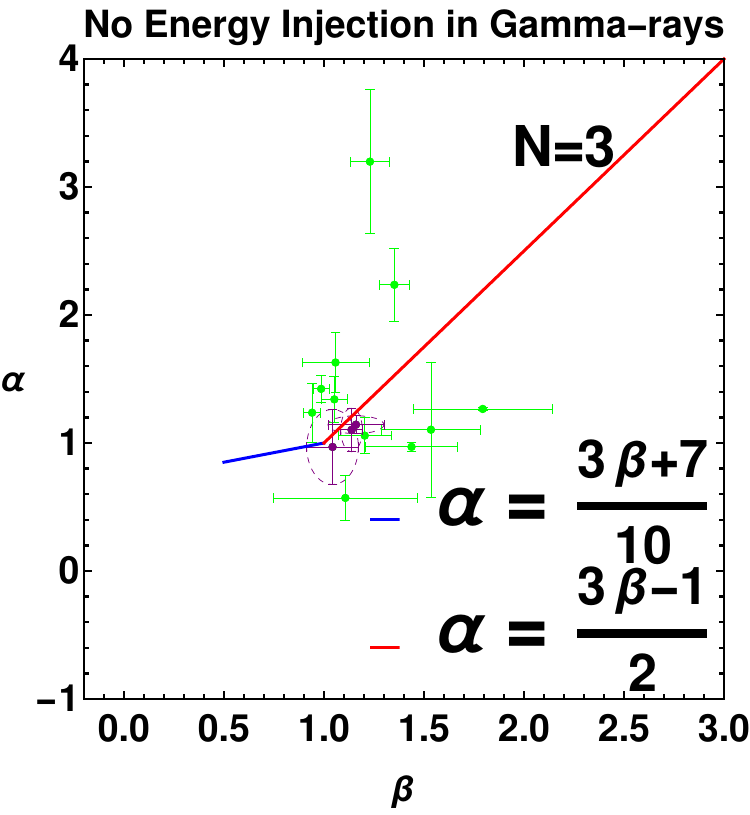}{0.23\textwidth}{(c)}
          }
%\gridline{%\fig{figures-UPDATED/LAT Without Energy Injection/k=1/SC.pdf}{0.18\textwidth}{(d)}
          %\fig{figures-UPDATED/LAT Without Energy Injection/k=1/FC.pdf}{0.18\textwidth}{(e)}
%          \fig{figures-UPDATED/LAT Without Energy Injection/k=1/SC FC.pdf}{0.18\textwidth}{(f)}
%          }
\vspace{-10pt}
\gridline{%\fig{figures-UPDATED/LAT Without Energy Injection/k=1.5/SC.pdf}{0.18\textwidth}{(g)}
          %\fig{figures-UPDATED/LAT Without Energy Injection/k=1.5/FC.pdf}{0.18\textwidth}{(h)}
          %fig{figures-UPDATED/LAT Without Energy Injection/k=1.5/SC FC.pdf}{0.18\textwidth}{(c)}
          \fig{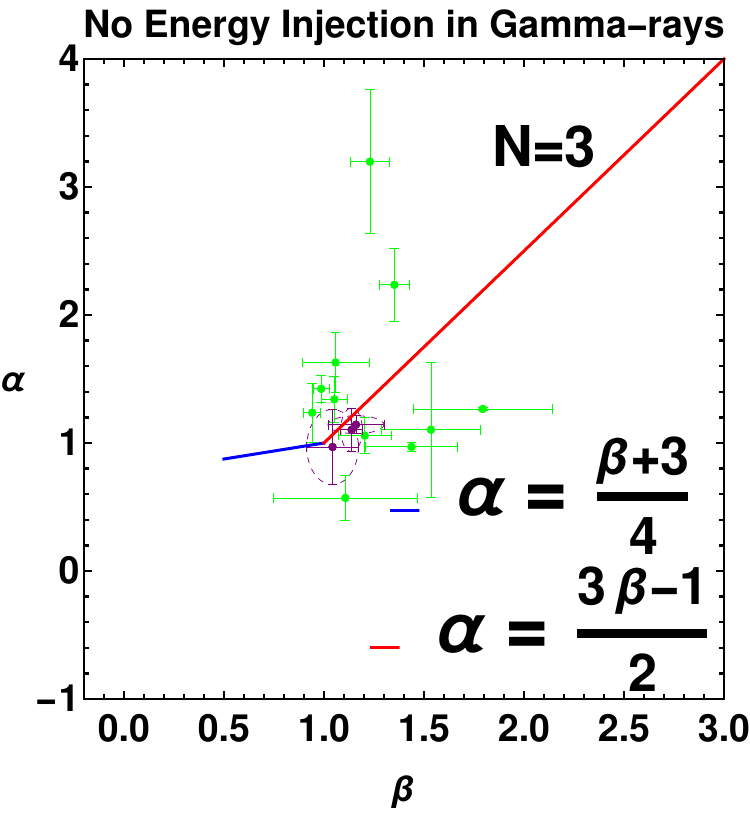}{0.23\textwidth}{(d)}
          \fig{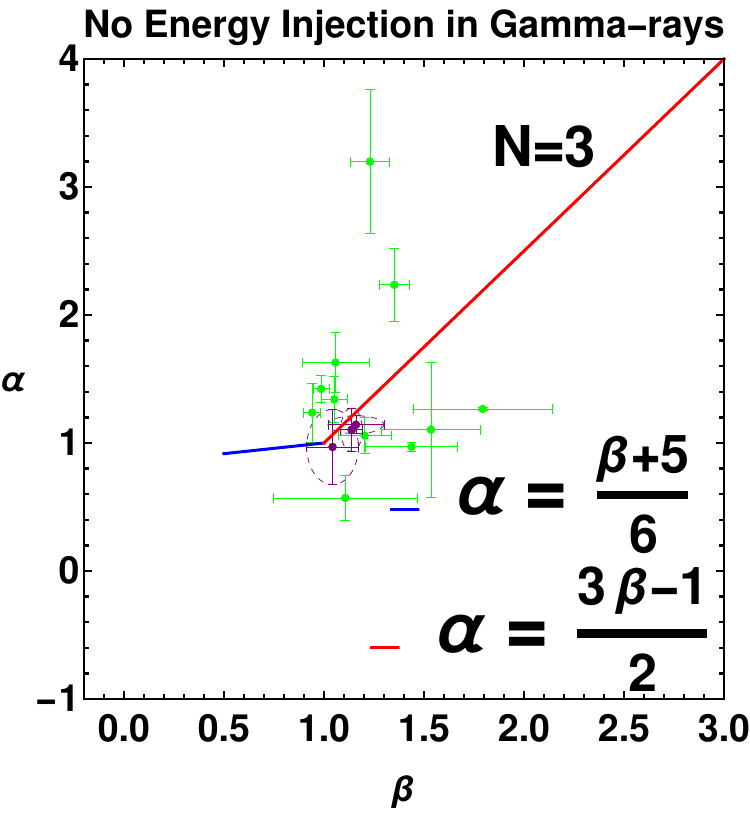}{0.23\textwidth}{(e)}
          }
%\gridline{%\fig{figures-UPDATED/LAT Without Energy Injection/k=2/SC.pdf}{0.18\textwidth}{(j)}
          %\fig{figures-UPDATED/LAT Without Energy Injection/k=2/FC.pdf}{0.18\textwidth}{(k)}
%          \fig{figures-UPDATED/LAT Without Energy Injection/k=2/SC FC.pdf}{0.18\textwidth}{(l)}
%          }

%\gridline{%\fig{figures-UPDATED/LAT Without Energy Injection/k=2.5/SC.pdf}{0.18\textwidth}{(m)}
          %\fig{figures-UPDATED/LAT Without Energy Injection/k=2.5/FC.pdf}{0.18\textwidth}{(n)}
 %         \fig{figures-UPDATED/LAT Without Energy Injection/k=2.5/SC FC.pdf}{0.18\textwidth}{(e)}
 %         }
          
% \gridline{\fig{KT_Eri.pdf}{0.3\textwidth}{(f)}}
\vspace{-10pt}
\gridline{%\fig{figures-UPDATED/LAT With Energy Injection/k=0/SC.pdf}{0.18\textwidth}{(a)}
          %\fig{figures-UPDATED/LAT With Energy Injection/k=0/FC.pdf}{0.18\textwidth}{(b)}
          \fig{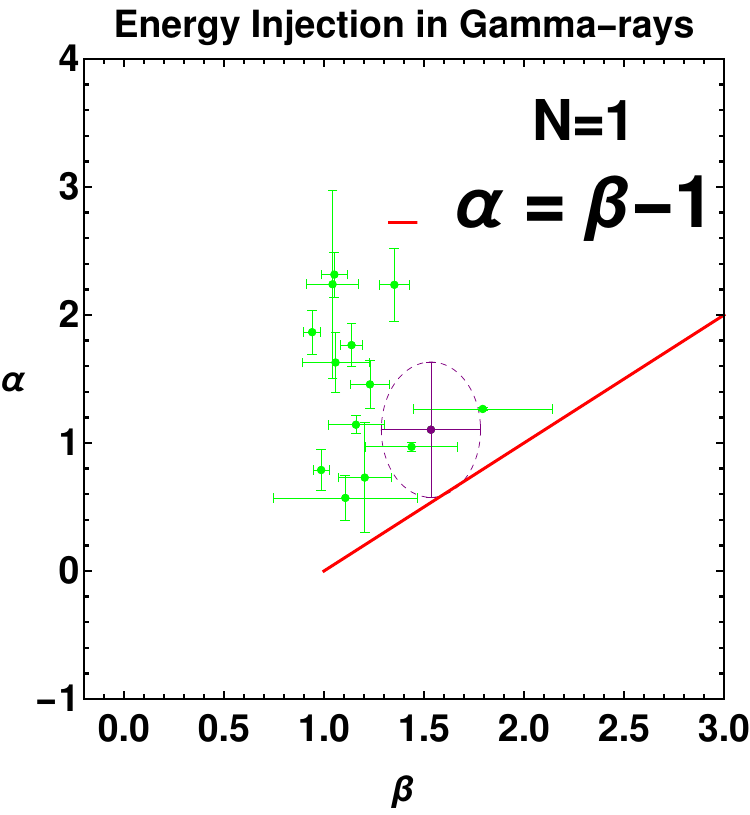}{0.23\textwidth}{(f)}
          \fig{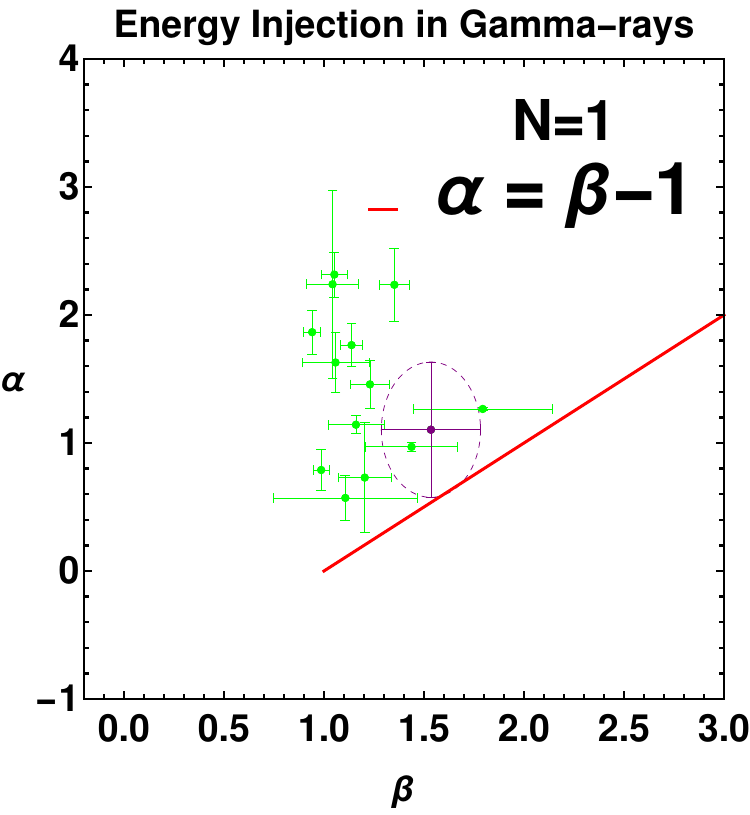}{0.23\textwidth}{(g)}
          \fig{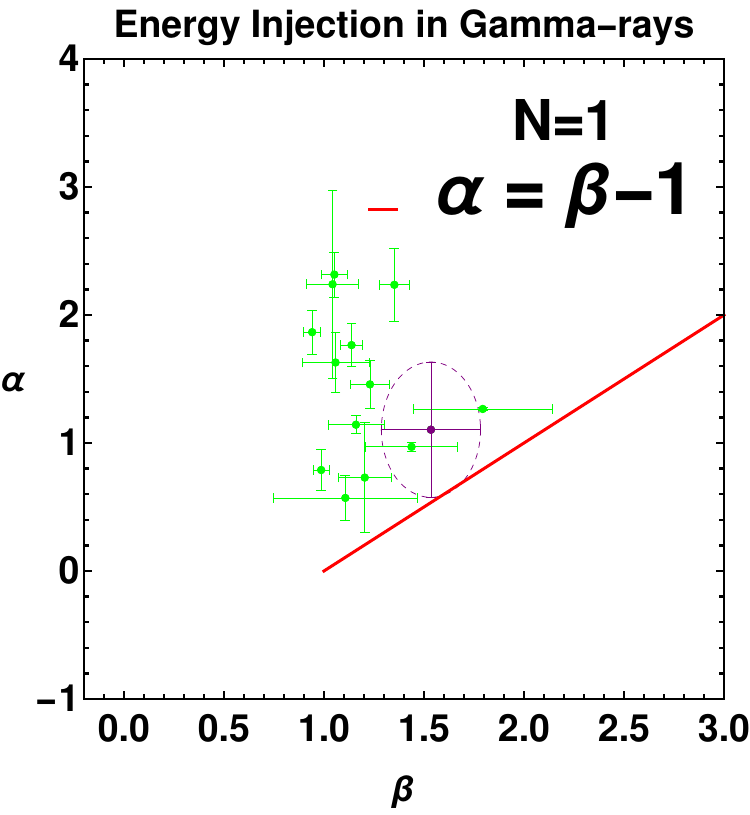}{0.23\textwidth}{(h)}
          }
\vspace{-10pt}
\gridline{%\fig{figures-UPDATED/LAT With Energy Injection/k=2/SC.pdf}{0.18\textwidth}{(j)}
          %\fig{figures-UPDATED/LAT With Energy Injection/k=2/FC.pdf}{0.18\textwidth}{(k)}
          \fig{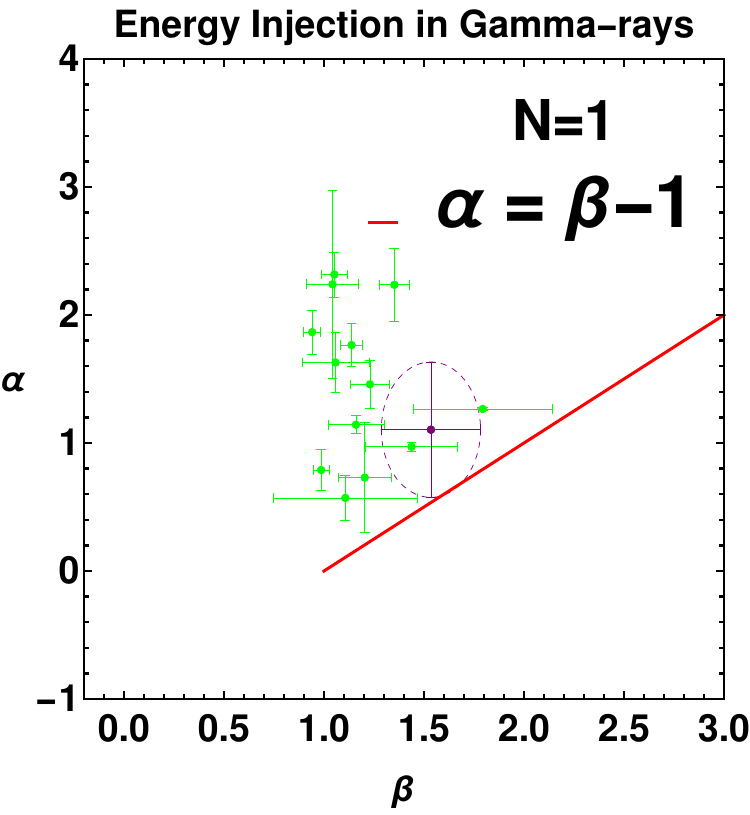}{0.23\textwidth}{(i)}
          \fig{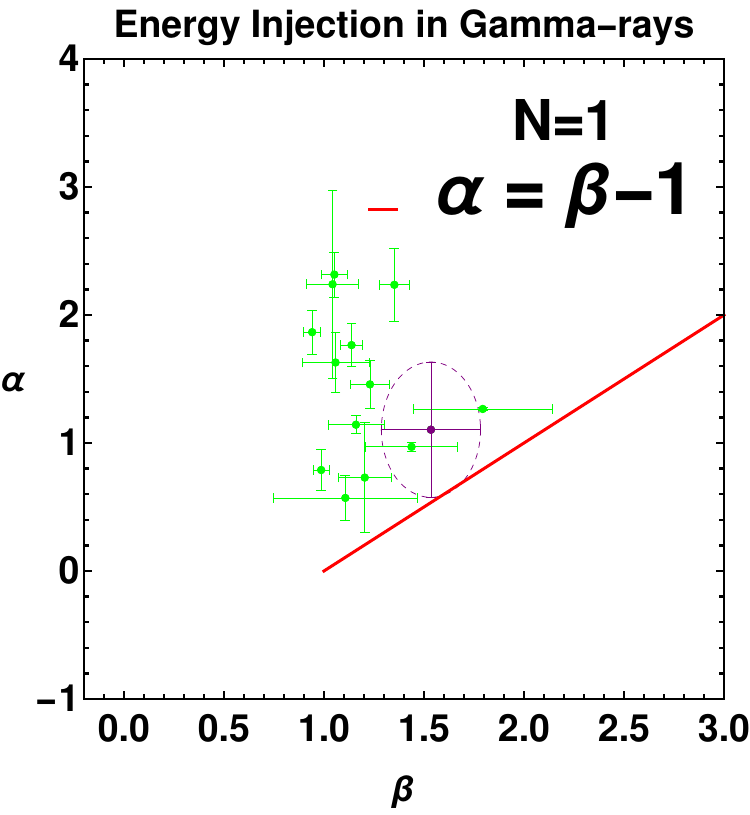}{0.23\textwidth}{(j)}
          }
\vspace{-10pt}
\caption{CRs corresponding to the synchrotron FS model for SC/FC regime for gamma-rays. Top two panels: no energy injection ($q=1$) for $k = 0$--$2.5$ (from (a) to (e), respectively). Bottom two panels: energy injection ($q=0$) for $k = 0$--$2.5$ (from (f) to (j), respectively). 
GRBs that satisfy the relations for gamma-ray parameters ($\alpha_{\rm{\gamma}}$ and $\beta_{\rm{\gamma}}$) are shown in purple; others are shown in green.}
\label{fig:LATnoinj}
\end{figure*}

\begin{table*}
\caption{Summary of results of the CRs obtained with gamma-ray parameters ($\alpha_{\rm{\gamma}}$ and $\beta_{\rm{\gamma}}$) without energy injection (\textit{q} = 1; upper panel) and with energy injection (\textit{q} = 0; lower panel), showing the number and occurrence rate of GRBs satisfying each relation between $\alpha$ and $\beta$ (where $\alpha=$ is omitted for brevity) out of a total of 14 GRBs. Since each occurrence rate is calculated independently from the total sample of 14 GRBs, there is no obligation for the rates to add up to 100\%.
}
\begin{center}
\begin{tabular}{lcccccccc}
\hline
\multicolumn{9}{c}{No Energy Injection (\textit{q} = 1) in Gamma-rays} \\
\hline
$n(r)$ & \text{Cooling} & $\nu$ \text{Range}    & $\beta(p)$ & \text{CR:}$1<p<2$      & \text{CR:}$p>2$ & \text{GRBs} & \text{Occurrence Rate} & \text{Figure}\\
%\hline
%$r^0$    & Slow & $\nu_m<\nu<\nu_c$ & $\frac{p-1}{2}$ & $\frac{6\beta+9}{16}$ & $\frac{3\beta}{2}$     & 4 & 25\% & (3a) \\
%$r^{-1}$ & Slow & $\nu_m<\nu<\nu_c$ & $\frac{p-1}{2}$ &  $\frac{4\beta+9}{12}$ & $\frac{9\beta+1}{6}$ & 1 & 6.3\% & (3d) \\ 
%$r^{-1.5}$ & Slow & $\nu_m<\nu<\nu_c$ & $\frac{p-1}{2}$ & $\frac{3\beta+9}{10}$ & $\frac{15\beta+3}{10}$ & 1 & 6.3\% & (3g) \\
%$r^{-2}$ & Slow & $\nu_m<\nu<\nu_c$ & $\frac{p-1}{2}$ & $\frac{2\beta+9}{8}$  & $\frac{3\beta+1}{2}$ & 1 & 6.3\% & (3j) \\
%$r^{-2.5}$ & Slow & $\nu_m<\nu<\nu_c$ & $\frac{p-1}{2}$ & $\frac{\beta + 9}{6}$ & $\frac{9\beta+5}{6}$ & 1 & 6.3\% & (3m) \\ 
%$r^0$ & Fast & $\nu_c<\nu<\nu_m$ & $\frac{1}{2}$ & $\frac{\beta}{2}$  & $\frac{\beta}{2}$ & 1 & 6.3\% & (3b) \\
%$r^{-1}$ & Fast & $\nu_c<\nu<\nu_m$ & $\frac{1}{2}$ & $\frac{\beta}{2}$ & $\frac{\beta}{2}$ & 1 & 6.3\% & (3e) \\
%$r^{-1.5}$ & Fast & $\nu_c<\nu<\nu_m$ & $\frac{1}{2}$ & $\frac{\beta}{2}$       & $\frac{\beta}{2}$ & 1 & 6.3\% & (3h) \\
%$r^{-2}$ & Fast & $\nu_c<\nu<\nu_m$ & $\frac{1}{2}$ & $\frac{\beta}{2}$ & $\frac{\beta}{2}$ & 1 & 6.3\%  & (3k) \\
%$r^{-2.5}$ & Fast & $\nu_c<\nu<\nu_m$ & $\frac{1}{2}$ & $\frac{\beta}{2}$    & $\frac{\beta}{2}$     & 1 & 6.3\% & (3n) \\
$r^0$  & Slow/Fast & $\nu > $ max\{$\nu_c,\nu_m$\} &$\frac{p}{2}$ & $\frac{3\beta+5}{8}$  & $\frac{3\beta-1}{2}$ & 3 & 21.4\% & (4a) \\
$r^{-1}$ & Slow/Fast & $\nu > $ max\{$\nu_c,\nu_m$\} & $\frac{p}{2}$ & $\frac{\beta+2}{3}$   & $\frac{3\beta-1}{2}$ & 3 & 21.4\%  & (4b) \\
$r^{-1.5}$ & Slow/Fast & $\nu > $ max\{$\nu_c,\nu_m$\} & $\frac{p}{2}$ & $\frac{3\beta+7}{10}$   & $\frac{3\beta-1}{2}$ & 3 & 21.4\%  & (4c) \\
$r^{-2}$ & Slow/Fast & $\nu > $ max\{$\nu_c,\nu_m$\} & $\frac{p}{2}$ & $\frac{\beta+3}{4}$   & $\frac{3\beta-1}{2}$ & 3 & 21.4\% & (4d) \\
$r^{-2.5}$ & Slow/Fast & $\nu > $ max\{$\nu_c,\nu_m$\} & $\frac{p}{2}$& $\frac{\beta + 5}{6}$ & $\frac{3\beta-1}{2}$ & 3 & 21.4\% & (4e) \\
 \hline
 \hline
\multicolumn{8}{c}{Energy Injection ($q=0$) in Gamma-rays} \\
\hline
 $n(r)$ & Cooling & $\nu$ \text{ Range}  & $\beta(p)$ & \text{CR:}$p>2$ & \text{GRBs} & \text{Occurrence Rate} & \text{Figure} \\
 \hline
 %$r^0$ & Slow   & $\nu_m<\nu<\nu_c$ & $\frac{p-1}{2}$ & $\beta-1$ & 1 & 6.3\%& \text{(4a)} \\
 %$r^{-1}$ & Slow   & $\nu_m<\nu<\nu_c$ & $\frac{p-1}{2}$ & $\beta-\frac{2}{3}$ & 4 & 25\%& \text{(4d)} \\
  %$r^{-1.5}$ & Slow   & $\nu_m<\nu<\nu_c$ & $\frac{p-1}{2}$ & $\beta-\frac{2}{5}$ & 4 & 25\%& \text{(4g)} \\
 %$r^{-2}$ & Slow    & $\nu_m<\nu<\nu_c$ & $\frac{p-1}{2}$& $\beta$ & 1 & 6.3\% & \text{(4j)}\\
 %$r^{-2.5}$ & Slow    & $\nu_m<\nu<\nu_c$ & $\frac{p-1}{2}$& $\frac{\beta+2}{3}$ & 7 & 43.8\% & \text{(4m)}\\
%$r^0$ & Fast &$ \nu_c<\nu<\nu_m$ & $\frac{1}{2}$ &$ -\beta$ & 0&
%  0\% & \text{(4b)}\\
%$r^{-1}$ & Fast &$ \nu_c<\nu<\nu_m$ & $\frac{1}{2}$ &$ -\beta$ & 0&
 % 0\% & \text{(4e)}\\
%$r^{-1.5}$ & Fast &$ \nu_c<\nu<\nu_m$ & $\frac{1}{2}$ &$ -\beta$ & 0&
 % 0\% & \text{(4h)}\\
%$r^{-2}$ & Fast    &$ \nu_c<\nu<\nu_m$ & $\frac{1}{2}$ &$ -\beta$ & 0 &
%  0\% & \text{(4k)}\\
%$r^{-2.5}$ & Fast    &$ \nu_c<\nu<\nu_m$ & $\frac{1}{2}$ &$ -\beta$ & 0 &
%  0\% & \text{(4n)}\\
$r^0$ & Slow/Fast  & $\nu > $ max\{$\nu_c,\nu_m$\} & $\frac{p}{2}$ & $\beta-1$ & 1 & 7.14\% & \text{(4f)}\\
$r^{-1}$ & Slow/Fast  & $\nu > $ max\{$\nu_c,\nu_m$\} & $\frac{p}{2}$ & $\beta-1$ & 1 & 7.14\% & \text{(4g)}\\
$r^{-1.5}$ & Slow/Fast  & $\nu > $ max\{$\nu_c,\nu_m$\} & $\frac{p}{2}$ & $\beta-1$ & 1 & 7.14\% & \text{(4h)}\\
$r^{-2}$ & Slow/Fast  & $\nu > $ max\{$\nu_c,\nu_m$\} & $\frac{p}{2}$ & $\beta-1$ & 1 & 7.14\% & \text{(4i)}\\
$r^{-2.5}$ & Slow/Fast  & $\nu > $ max\{$\nu_c,\nu_m$\} & $\frac{p}{2}$ & $\beta-1$ & 1 & 7.14\% & \text{(4j)}\\
\hline
\end{tabular}
\end{center}
\label{table:LAT crSummary}
\end{table*}

%% SWIFT XRT RESULTS TABLE
\begin{table*}
\caption{Summary of results of the CRs obtained with X-ray parameters ($\alpha_{\rm{X}}$ and $\beta_{\rm{X}}$) without energy injection (\textit{q} = 1; upper panel), with energy injection (\textit{q} = 0; middle panel), and jet break (bottom panel), showing the number and occurrence rate of GRBs satisfying each relation (where $\alpha=$ is omitted for brevity) out of a total of 14 GRBs. Since each occurrence rate is calculated independently from the total sample of 14 GRBs, there is no obligation for the rates to add up to 100\%.
}
\begin{center}
%\resizebox{1.0\textwidth}{!}{
\scalebox{0.75}{
\begin{tabular}{lcccccccc}
\hline
\multicolumn{9}{c}{No Energy Injection in X-rays (\textit{q} = 1)} \\
\hline
$n(r)$ & \text{Cooling} & $\nu$ \text{Range}    & $\beta(p)$ & \text{CR:}$1<p<2$      & \text{CR:}$p>2$ & \text{GRBs} & \text{Occurrence Rate} & \text{Figure}\\
%\hline
$r^0$    & Slow & $\nu_m<\nu<\nu_c$ & $\frac{p-1}{2}$ & $\frac{6\beta+9}{16}$ & $\frac{3\beta}{2}$     & 9 & 64.3\% & (5a) \\
 
$r^{-1}$ & Slow & $\nu_m<\nu<\nu_c$ & $\frac{p-1}{2}$ &  $\frac{4\beta+9}{12}$ & $\frac{9\beta+1}{6}$ & 8 & 57.1\% & (5d) \\
 
$r^{-1.5}$ & Slow & $\nu_m<\nu<\nu_c$ & $\frac{p-1}{2}$ & $\frac{3\beta+9}{10}$ & $\frac{15\beta+3}{10}$ & 8 & 57.1\% & (5g) \\

$r^{-2}$ & Slow & $\nu_m<\nu<\nu_c$ & $\frac{p-1}{2}$ & $\frac{2\beta+9}{8}$  & $\frac{3\beta+1}{2}$ & 5 & 35.7\% & (5j) \\
 
$r^{-2.5}$ & Slow & $\nu_m<\nu<\nu_c$ & $\frac{p-1}{2}$ & $\frac{\beta + 9}{6}$ & $\frac{9\beta+5}{6}$ & 1 & 7.14\% & (5m) \\

$r^0$ & Fast & $\nu_c<\nu<\nu_m$ & $\frac{1}{2}$ & $\frac{\beta}{2}$  & $\frac{\beta}{2}$ & 0 & 0\% & (5b) \\
 
$r^{-1}$ & Fast & $\nu_c<\nu<\nu_m$ & $\frac{1}{2}$ & $\frac{\beta}{2}$ & $\frac{\beta}{2}$ & 0 & 0\% & (5e) \\
 
$r^{-1.5}$ & Fast & $\nu_c<\nu<\nu_m$ & $\frac{1}{2}$ & $\frac{\beta}{2}$       & $\frac{\beta}{2}$      & 0 & 0\% & (5h) \\
 
$r^{-2}$ & Fast & $\nu_c<\nu<\nu_m$ & $\frac{1}{2}$ & $\frac{\beta}{2}$ & $\frac{\beta}{2}$ & 0 & 0\%  & (5k) \\
 
$r^{-2.5}$ & Fast & $\nu_c<\nu<\nu_m$ & $\frac{1}{2}$ & $\frac{\beta}{2}$    & $\frac{\beta}{2}$     & 0 & 0\% & (5n) \\

$r^0$  & Slow/Fast & $\nu > $ max\{$\nu_c,\nu_m$\} &$\frac{p}{2}$ & $\frac{3\beta+5}{8}$  & $\frac{3\beta-1}{2}$ & 3 & 21.4\% & (5c) \\
 
$r^{-1}$ & Slow/Fast & $\nu > $ max\{$\nu_c,\nu_m$\} & $\frac{p}{2}$ & $\frac{\beta+2}{3}$   & $\frac{3\beta-1}{2}$ & 3 & 21.4\%  & (5f) \\
 
$r^{-1.5}$ & Slow/Fast & $\nu > $ max\{$\nu_c,\nu_m$\} & $\frac{p}{2}$ & $\frac{3\beta+7}{10}$   & $\frac{3\beta-1}{2}$ & 3 & 21.4\%  & (5i) \\
 
$r^{-2}$ & Slow/Fast & $\nu > $ max\{$\nu_c,\nu_m$\} & $\frac{p}{2}$ & $\frac{\beta+3}{4}$   & $\frac{3\beta-1}{2}$ & 3 & 21.4\% & (5l) \\
 
$r^{-2.5}$ & Slow/Fast & $\nu > $ max\{$\nu_c,\nu_m$\} & $\frac{p}{2}$& $\frac{\beta + 5}{6}$ & $\frac{3\beta-1}{2}$ & 3 & 21.4\% & (5o) \\
 
 \hline
 \hline

\multicolumn{8}{c}{Energy Injection in X-rays ($q=0$)} \\
\hline
 $n(r)$ & Cooling & $\nu$ \text{ Range}  & $\beta(p)$ & \text{CR:}$p>2$ & \text{GRBs} & \text{Occurrence Rate} & \text{Figure} \\
 \hline

 $r^0$ & Slow   & $\nu_m<\nu<\nu_c$ & $\frac{p-1}{2}$ & $\beta-1$ & 0 & 0\%& \text{(6a)} \\

 $r^{-1}$ & Slow   & $\nu_m<\nu<\nu_c$ & $\frac{p-1}{2}$ & $\beta-\frac{2}{3}$ & 0 & 0\%& \text{(6d)} \\

$r^{-1.5}$ & Slow   & $\nu_m<\nu<\nu_c$ & $\frac{p-1}{2}$ & $\beta-\frac{2}{5}$ & 1 & 7.14\%& \text{(6g)} \\

 $r^{-2}$ & Slow    & $\nu_m<\nu<\nu_c$ & $\frac{p-1}{2}$& $\beta$ & 3 & 21.4\% & \text{(6j)}\\

 $r^{-2.5}$ & Slow    & $\nu_m<\nu<\nu_c$ & $\frac{p-1}{2}$& $\frac{\beta+2}{3}$ & 3 & 21.4\% & \text{(6m)}\\

$r^0$ & Fast &$ \nu_c<\nu<\nu_m$ & $\frac{1}{2}$ &$ -\beta$ & 0& 0\% & \text{(6b)}\\

$r^{-1}$ & Fast &$ \nu_c<\nu<\nu_m$ & $\frac{1}{2}$ &$ -\beta$ & 0& 0\% & \text{(6e)}\\

$r^{-1.5}$ & Fast &$ \nu_c<\nu<\nu_m$ & $\frac{1}{2}$ &$ -\beta$ & 0& 0\% & \text{(6h)}\\

$r^{-2}$ & Fast    &$ \nu_c<\nu<\nu_m$ & $\frac{1}{2}$ &$ -\beta$ & 0 & 0\% & \text{(6k)}\\

$r^{-2.5}$ & Fast    &$ \nu_c<\nu<\nu_m$ & $\frac{1}{2}$ &$ -\beta$ & 0 & 0\% & \text{(6n)}\\

$r^0$ & Slow/Fast  & $\nu > $ max\{$\nu_c,\nu_m$\} & $\frac{p}{2}$ & $\beta-1$ & 0 & 0\% & \text{(6c)}\\

$r^{-1}$ & Slow/Fast  & $\nu > $ max\{$\nu_c,\nu_m$\} & $\frac{p}{2}$ & $\beta-1$ & 0 & 0\% & \text{(6f)}\\

$r^{-1.5}$ & Slow/Fast  & $\nu > $ max\{$\nu_c,\nu_m$\} & $\frac{p}{2}$ & $\beta-1$ & 0 & 0\% & \text{(6i)}\\

$r^{-2}$ & Slow/Fast  & $\nu > $ max\{$\nu_c,\nu_m$\} & $\frac{p}{2}$ & $\beta-1$ & 0 & 0\% & \text{(6l)}\\

$r^{-2.5}$ & Slow/Fast  & $\nu > $ max\{$\nu_c,\nu_m$\} & $\frac{p}{2}$ & $\beta-1$ & 0 & 0\% & \text{(6o)}\\

\hline
\hline

\multicolumn{7}{c}{Jet Break in X-rays} \\
\hline
 Cooling & $\nu$ \text{ Range}  & $\beta(p)$ & \text{CR:}$1<p<2$ & \text{GRBs} & \text{Occurrence Rate} & \text{Figure} \\
 \hline

 Slow   & $\nu_m<\nu<\nu_c$ & $\frac{p-1}{2}$ & $2\beta+1$ & 0 & 0\%& \text{(9a)} \\

Fast &$ \nu_c<\nu<\nu_m$ & $1$ & $\frac{1}{2}$ & 0& 0\% & \text{(9b)}\\

Slow/Fast  & $\nu > $ max\{$\nu_c,\nu_m$\} & $p$ & $\frac{\beta}{2}$ & 2 & 14.3\% & \text{(9c)}\\

\hline
\end{tabular}
}
\end{center}
\label{table:XRT crSummary}
\end{table*}

\begin{figure*}
\gridline{\fig{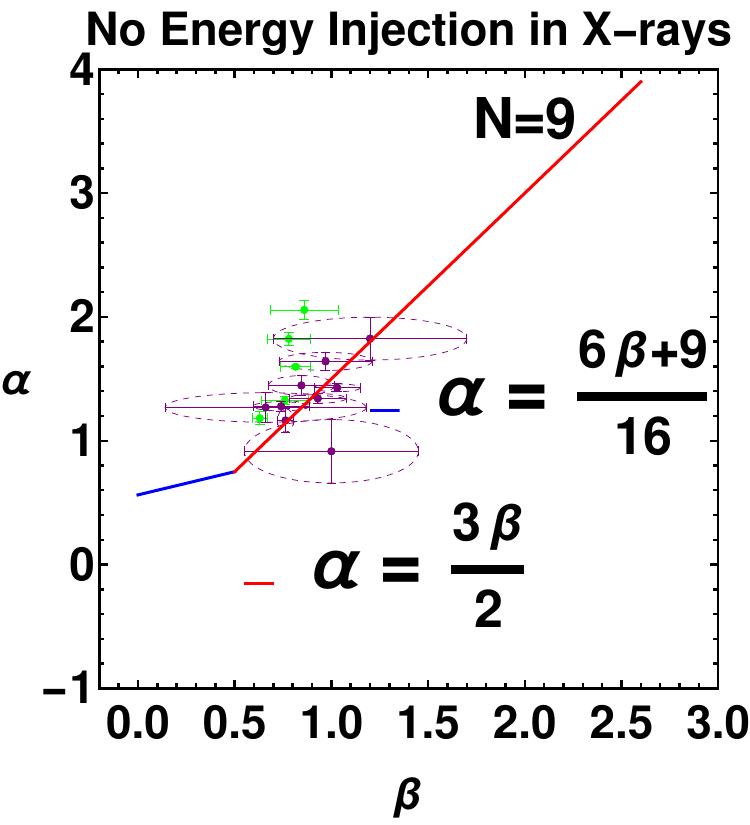}{0.18\textwidth}{(a)}
          \fig{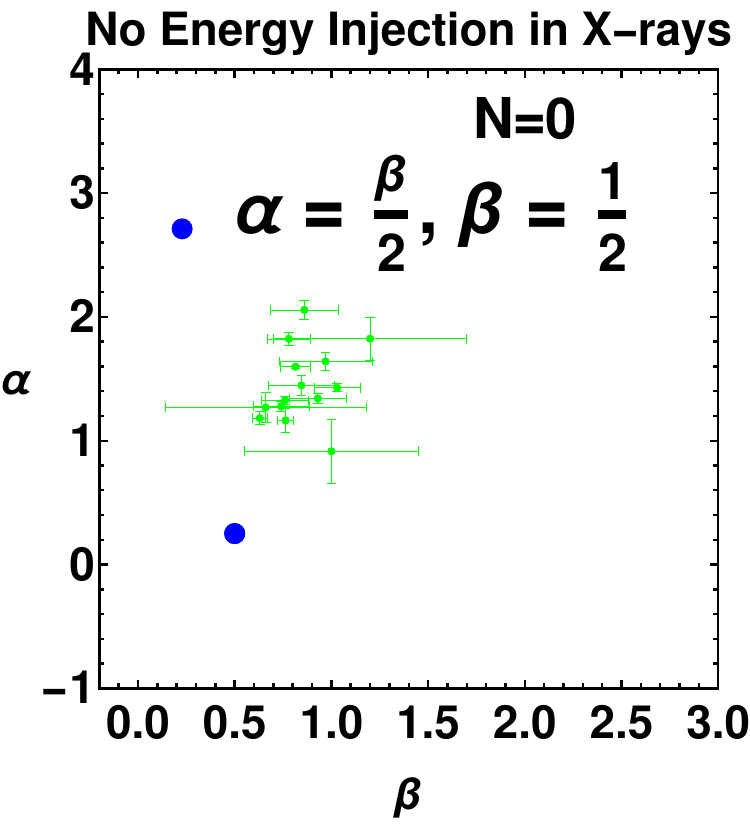}{0.18\textwidth}{(b)}
          \fig{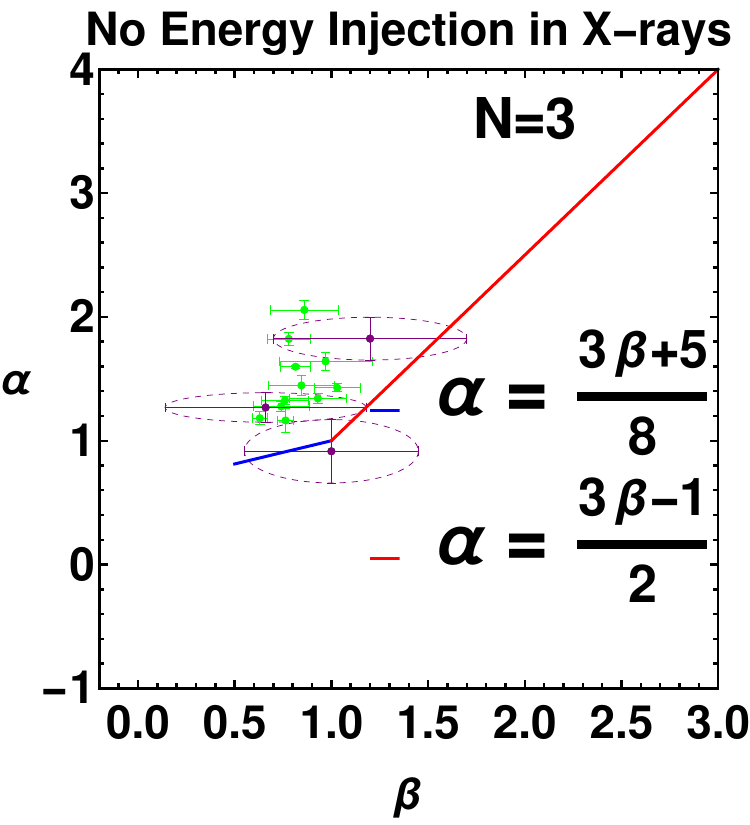}{0.18\textwidth}{(c)}
          }
\vspace{-8.7pt}
\gridline{\fig{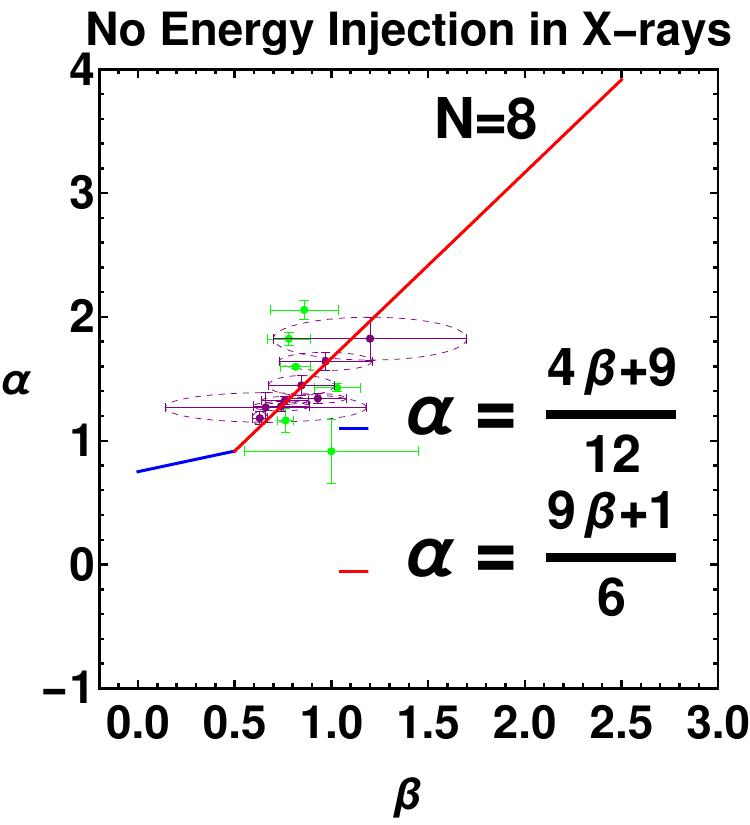}{0.18\textwidth}{(d)}
          \fig{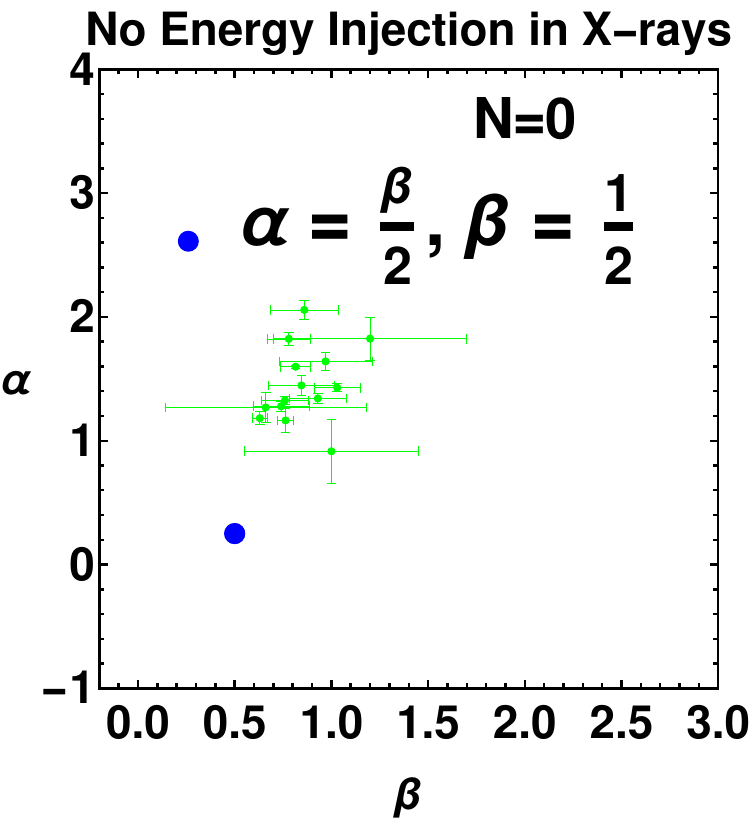}{0.18\textwidth}{(e)}
          \fig{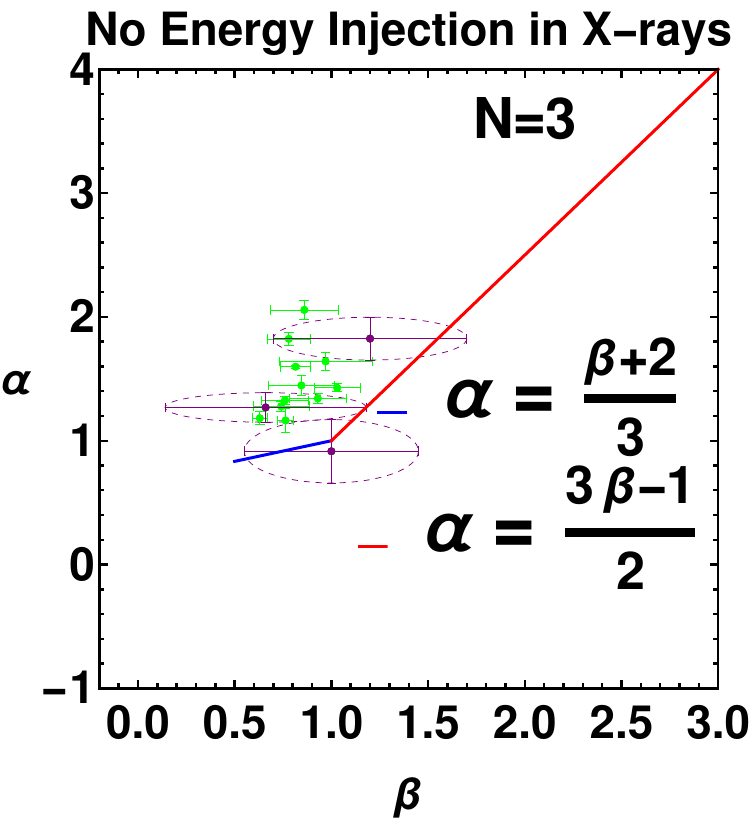}{0.18\textwidth}{(f)}
          }
\vspace{-8.7pt}
\gridline{\fig{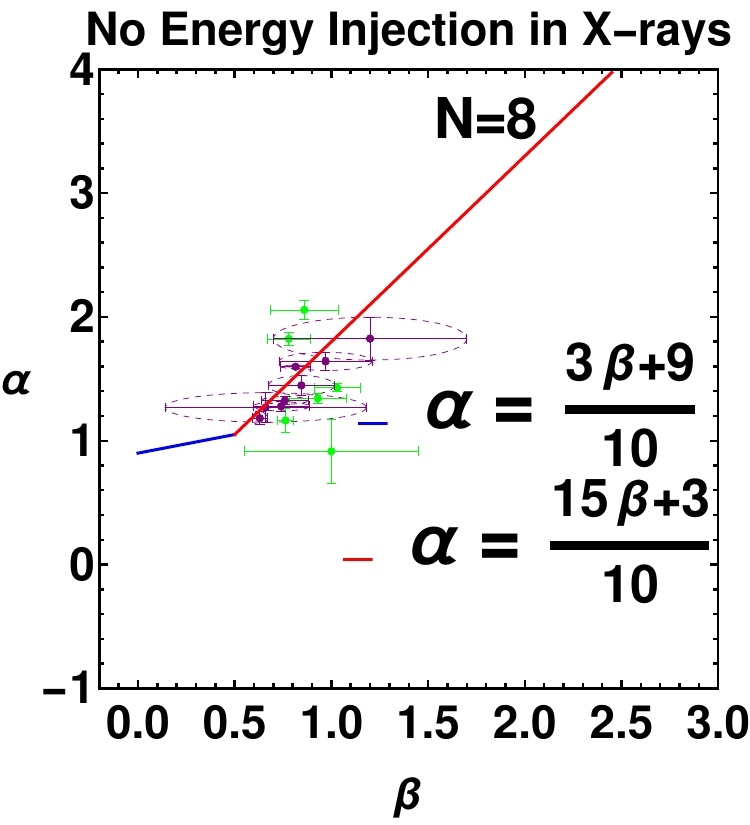}{0.18\textwidth}{(g)}
          \fig{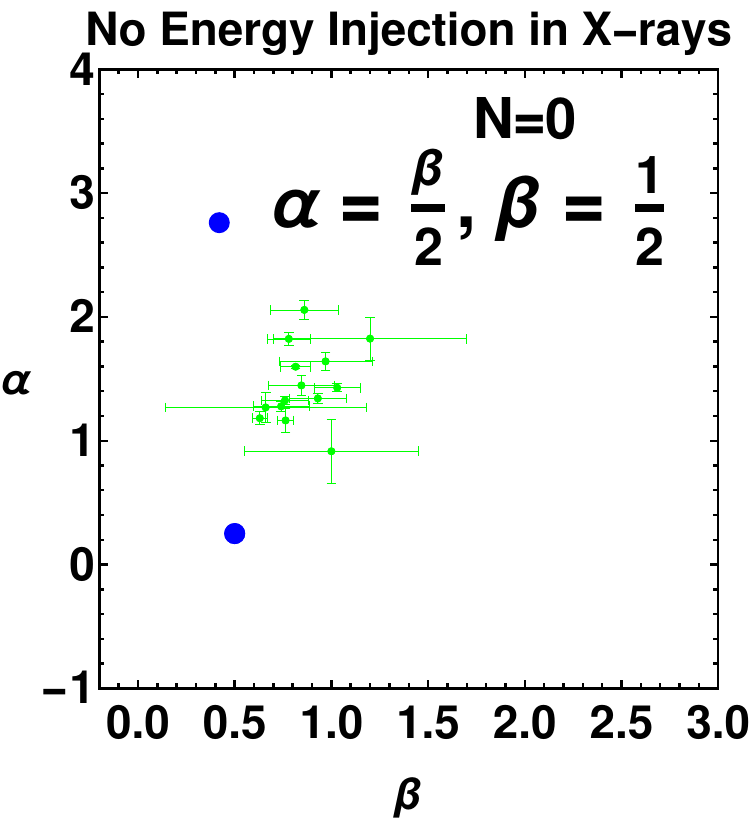}{0.18\textwidth}{(h)}
          \fig{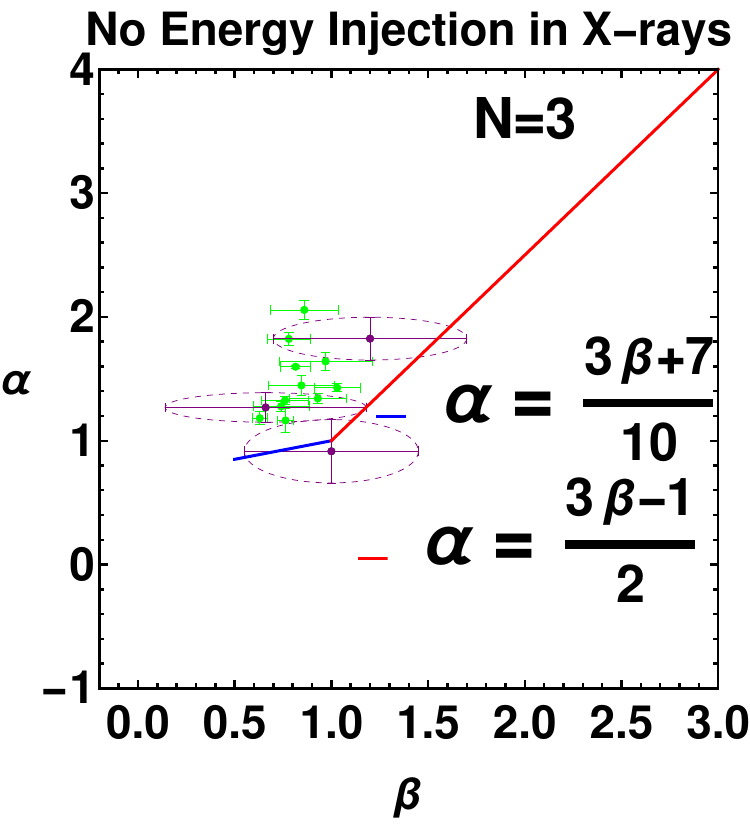}{0.18\textwidth}{(i)}
          }
\vspace{-8.7pt}
\gridline{\fig{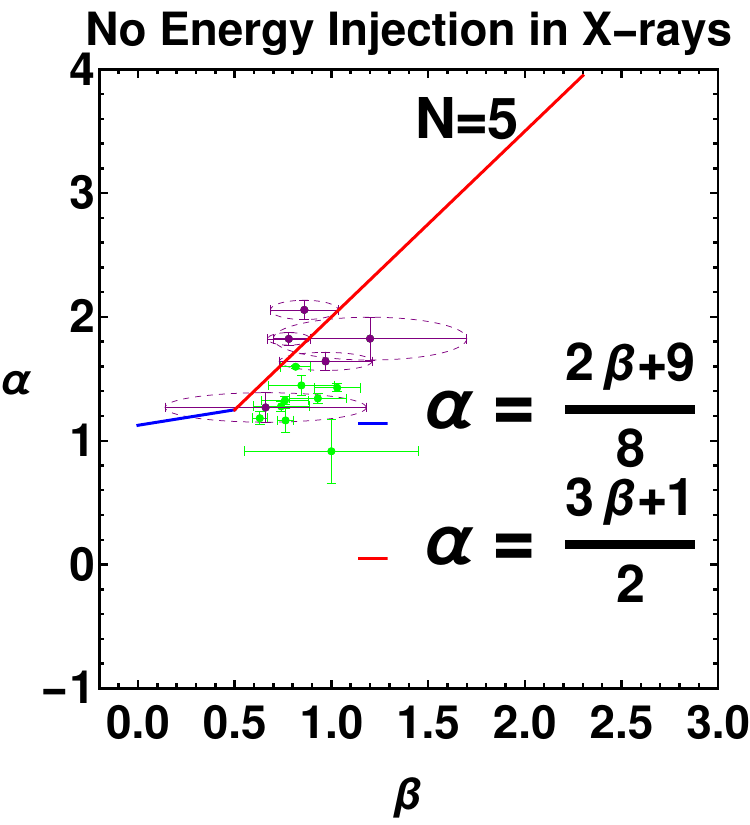}{0.18\textwidth}{(j)}
          \fig{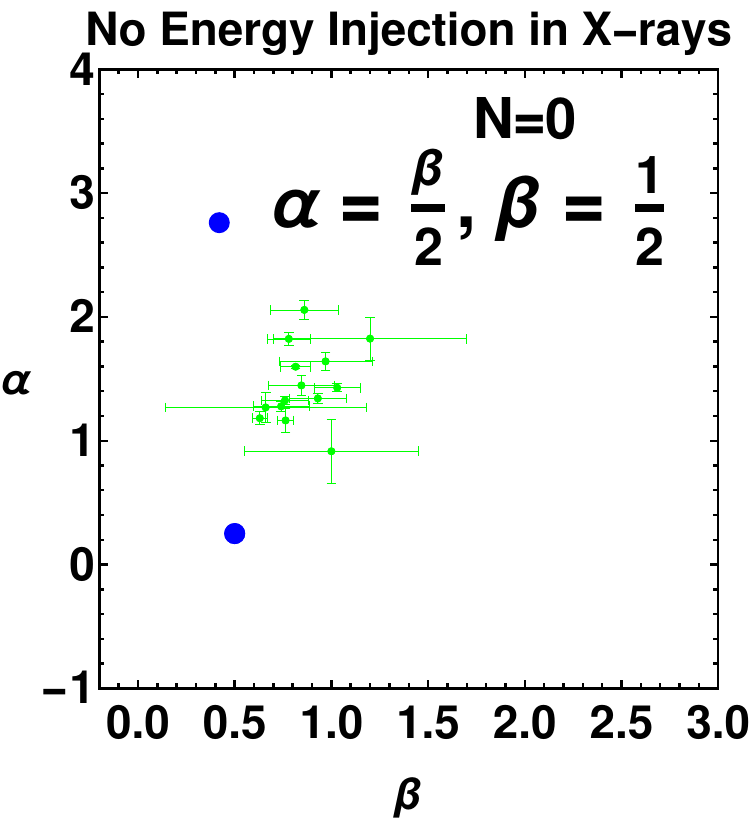}{0.18\textwidth}{(k)}
          \fig{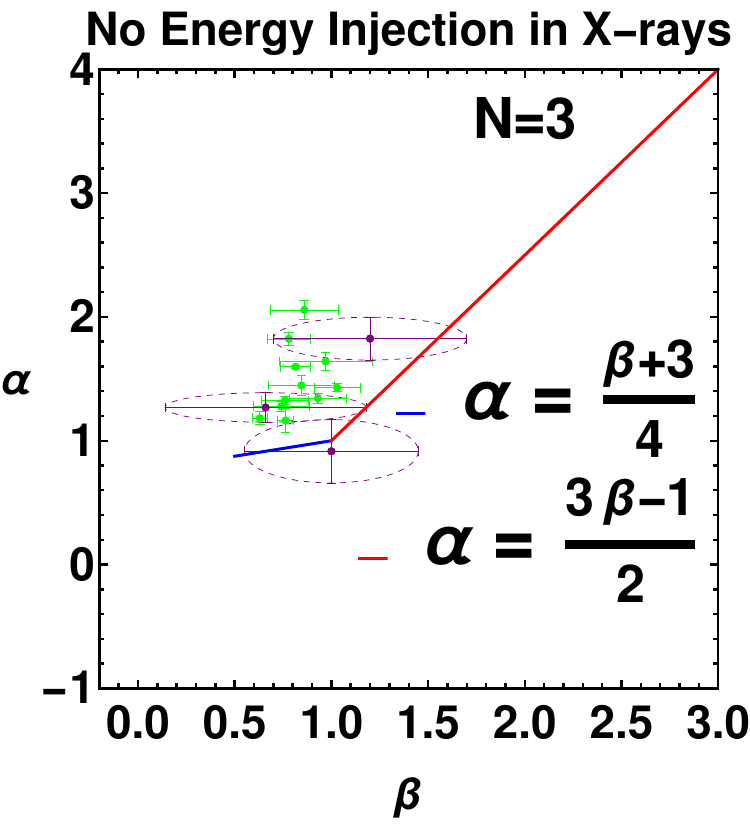}{0.18\textwidth}{(l)}
          }
\vspace{-8.7pt}
\gridline{\fig{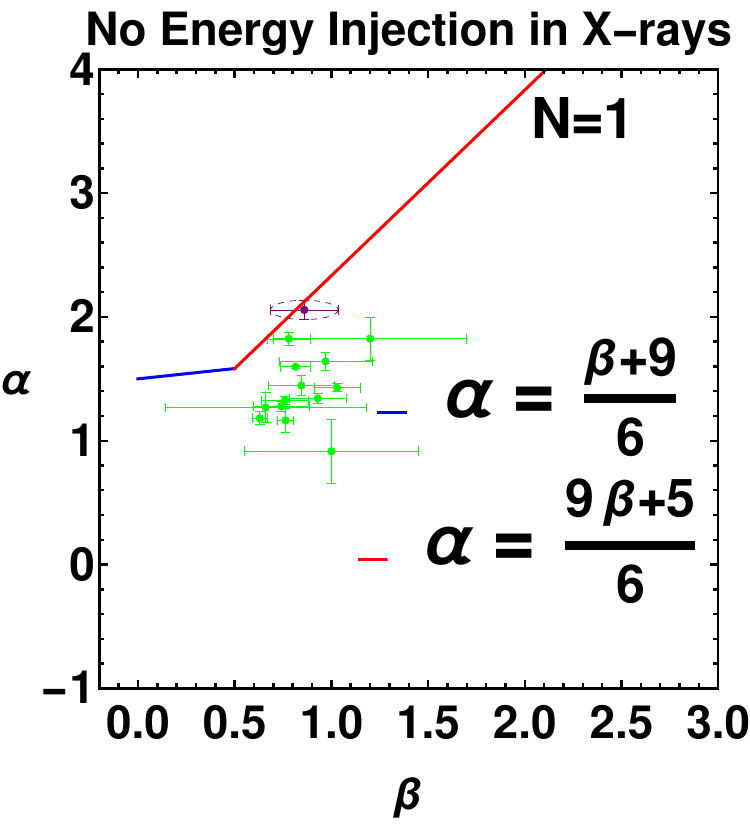}{0.18\textwidth}{(m)}
          \fig{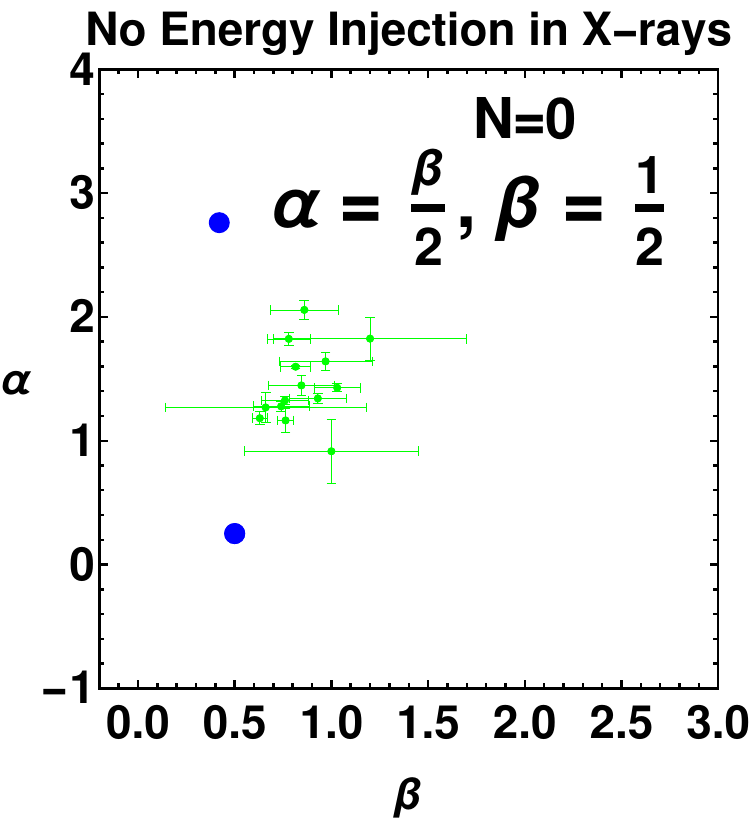}{0.18\textwidth}{(n)}
          \fig{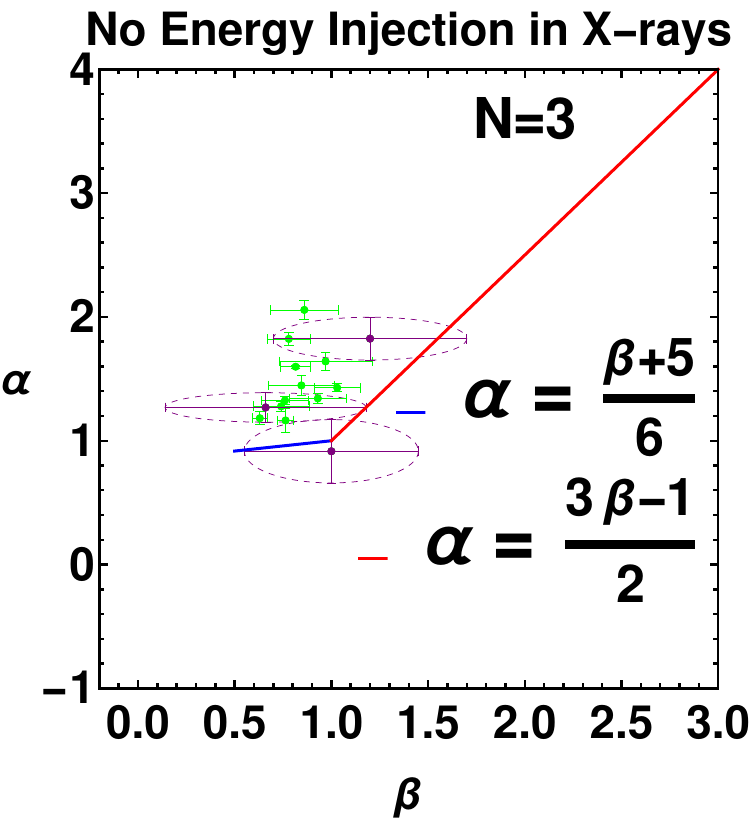}{0.18\textwidth}{(o)}
         }
\vspace{-6.21614pt}
\caption{CRs in X-rays corresponding to the synchrotron FS model for $k = 0$--$2.5$ (from top to bottom) with no energy injection ($q=1$). Columns denote SC, FC, and SC/FC regimes from left to right. GRBs that satisfy the relations for X-ray parameters ($\alpha_{\rm{X}}$ and $\beta_{\rm{X}}$) are shown in purple; others are shown in green.}
\label{fig:XRTnoinj}
%\vspace{-16.64456pt}
\end{figure*}

%% SWIFT XRT RESULTS FIGURES EI
\begin{figure*}
\gridline{\fig{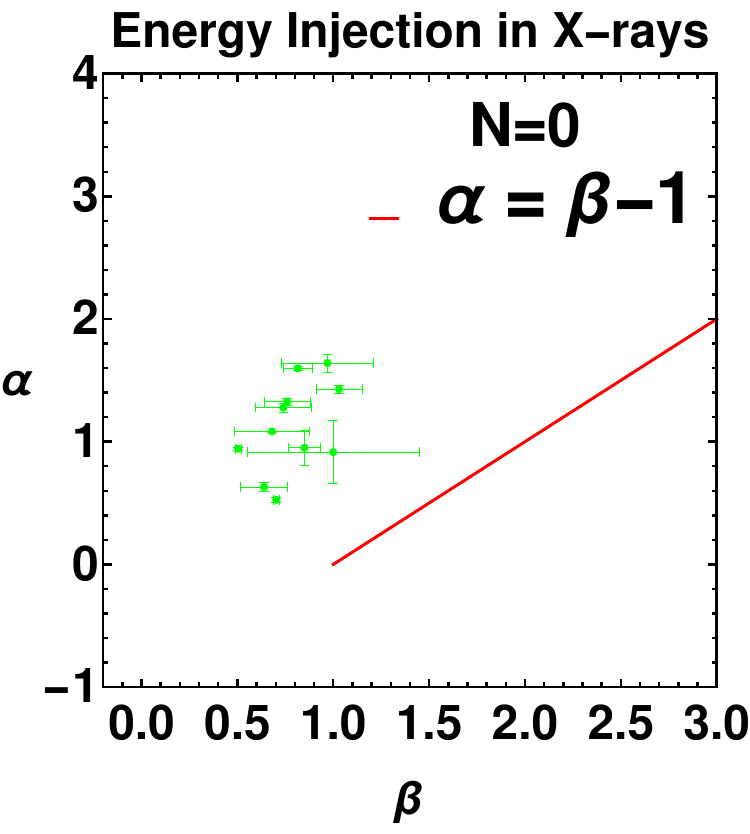}{0.18\textwidth}{(a)}
          \fig{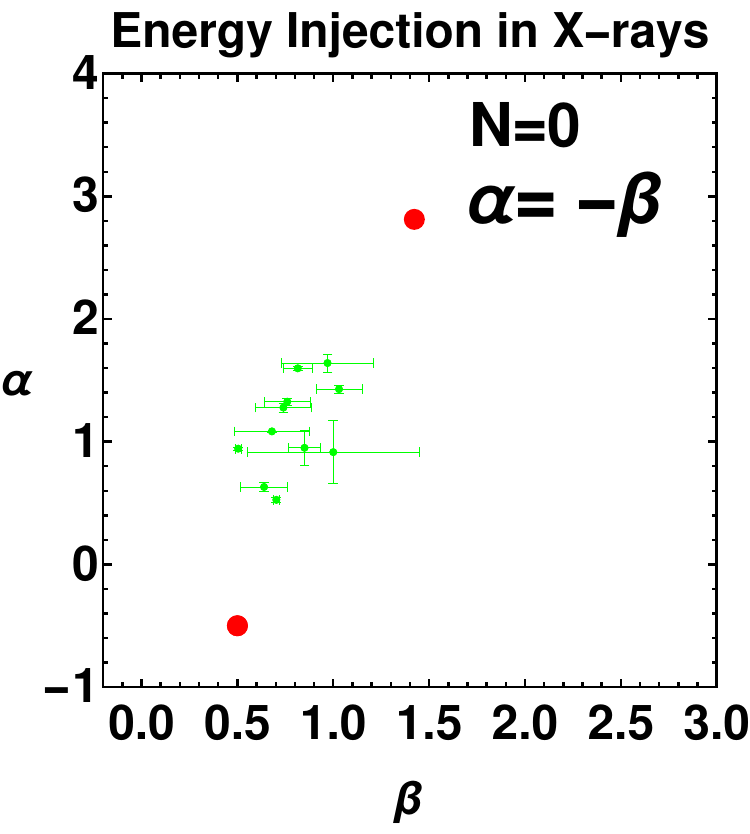}{0.18\textwidth}{(b)}
          \fig{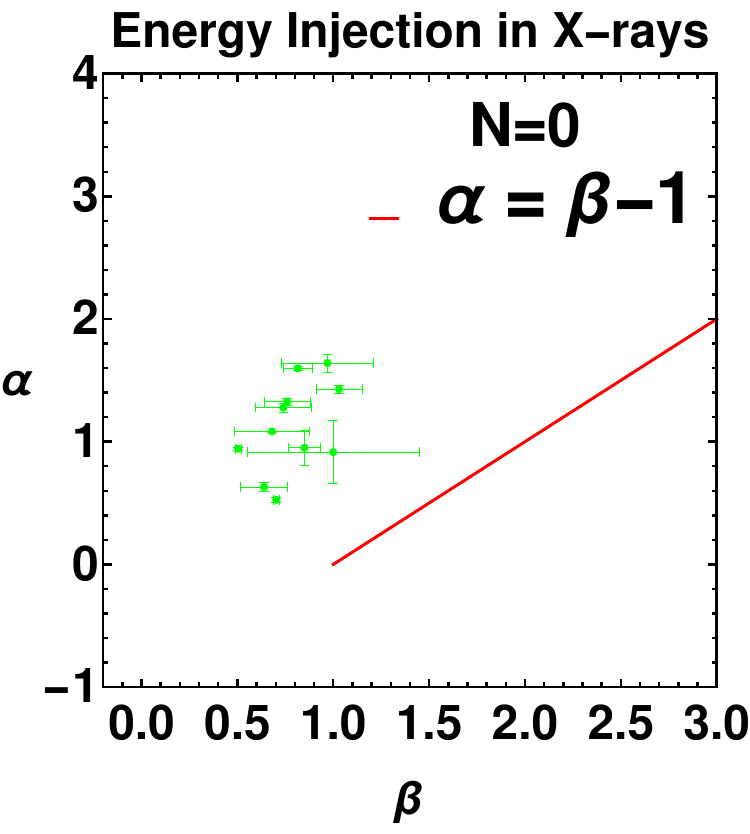}{0.18\textwidth}{(c)}
          }
\vspace{-8.7pt}
\gridline{\fig{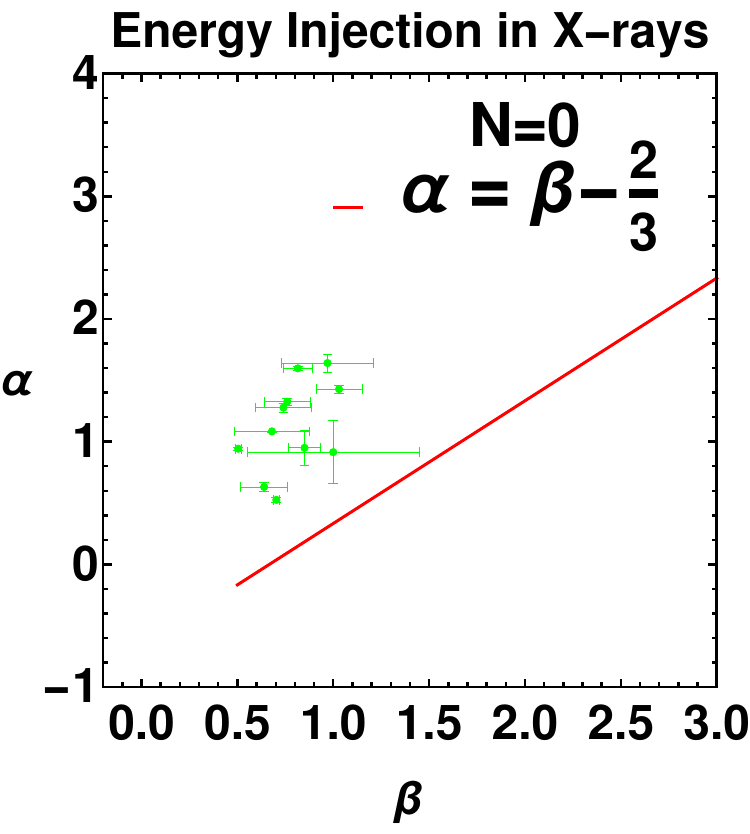}{0.18\textwidth}{(d)}
          \fig{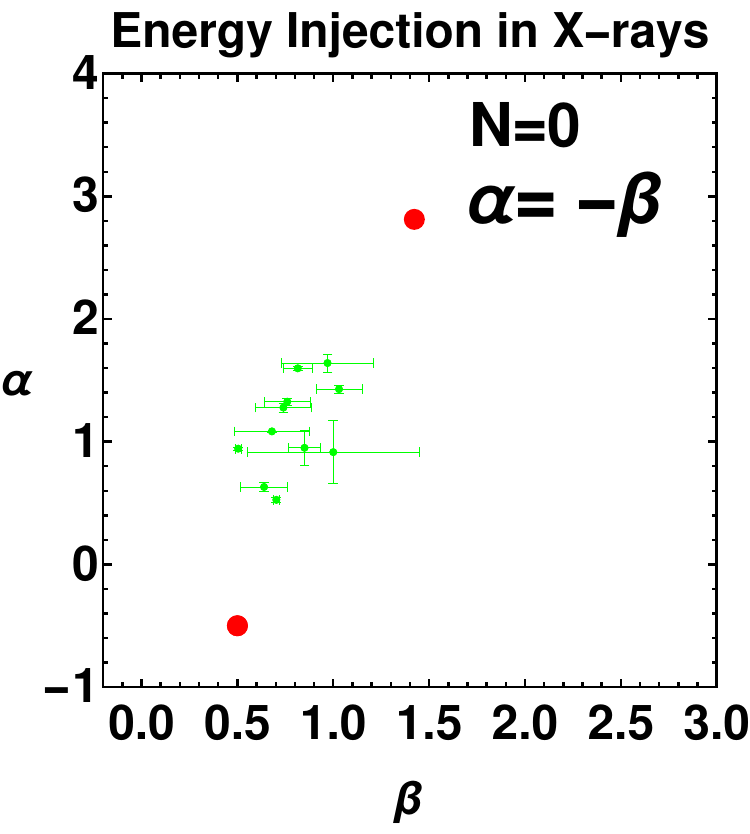}{0.18\textwidth}{(e)}
          \fig{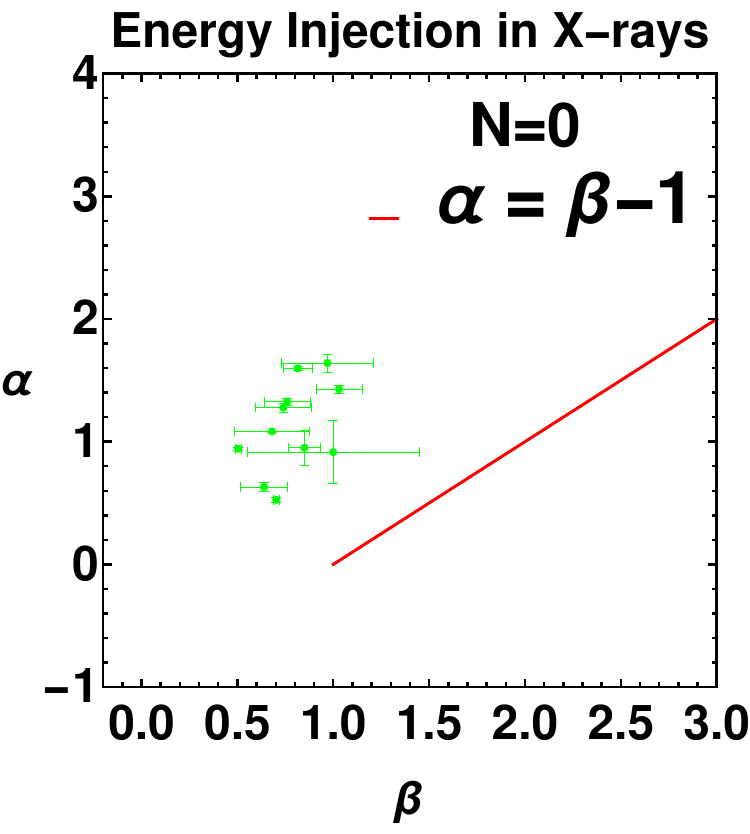}{0.18\textwidth}{(f)}
          }
\vspace{-8.7pt}
\gridline{\fig{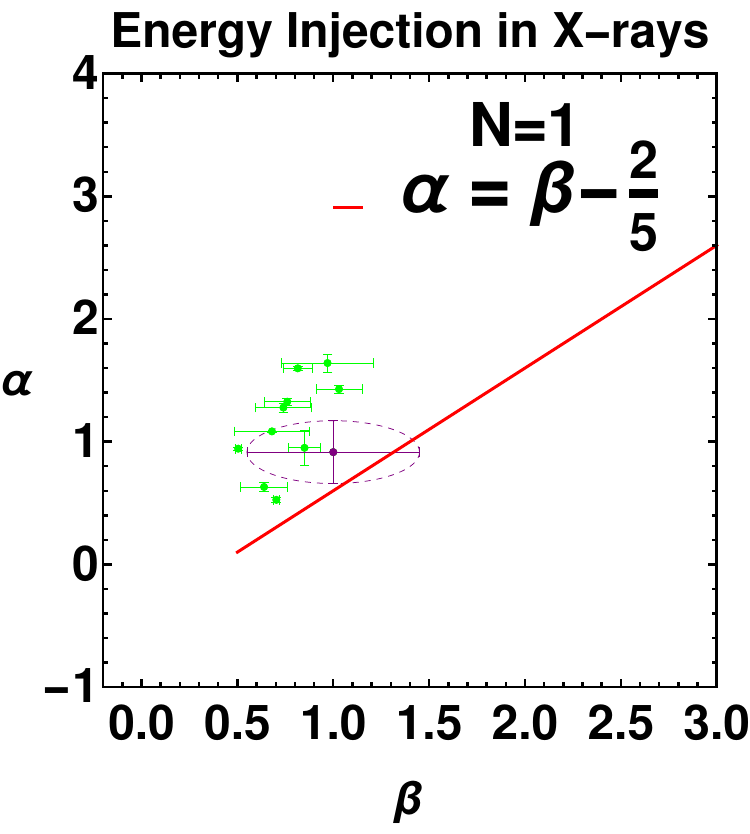}{0.18\textwidth}{(g)}
          \fig{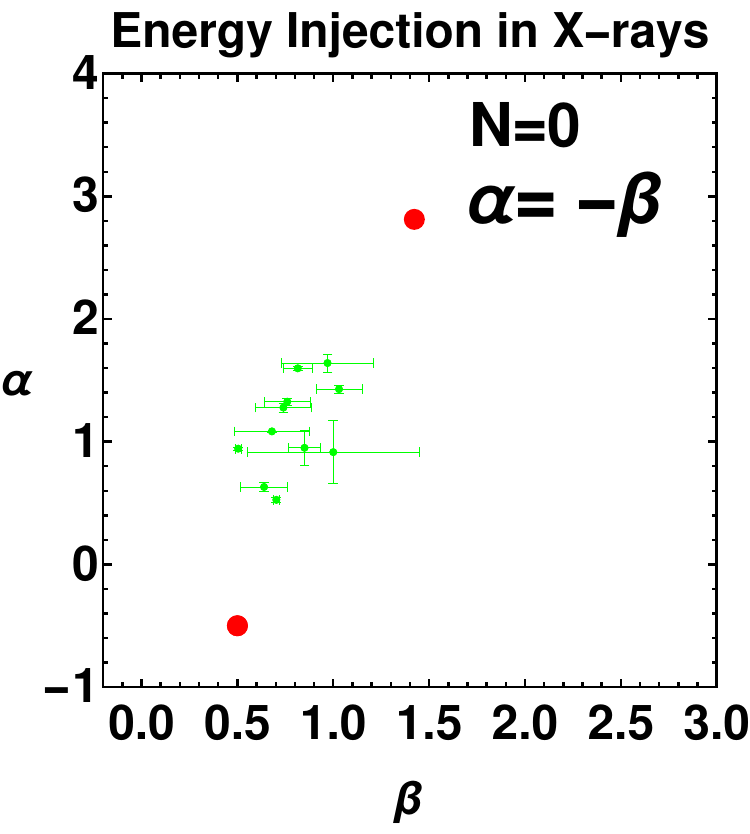}{0.18\textwidth}{(h)}
          \fig{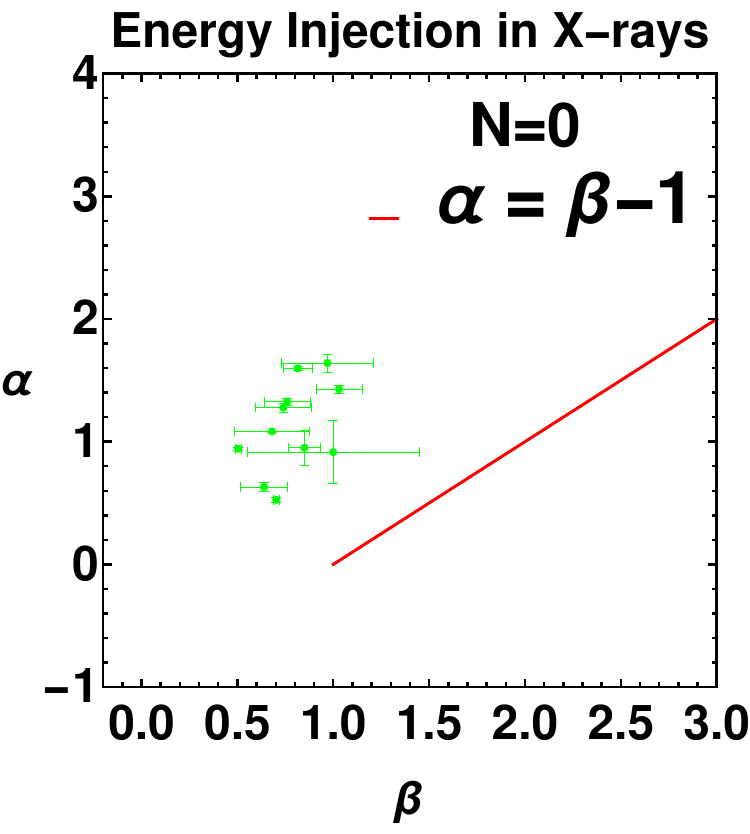}{0.18\textwidth}{(i)}
          }
\vspace{-8.7pt}
\gridline{\fig{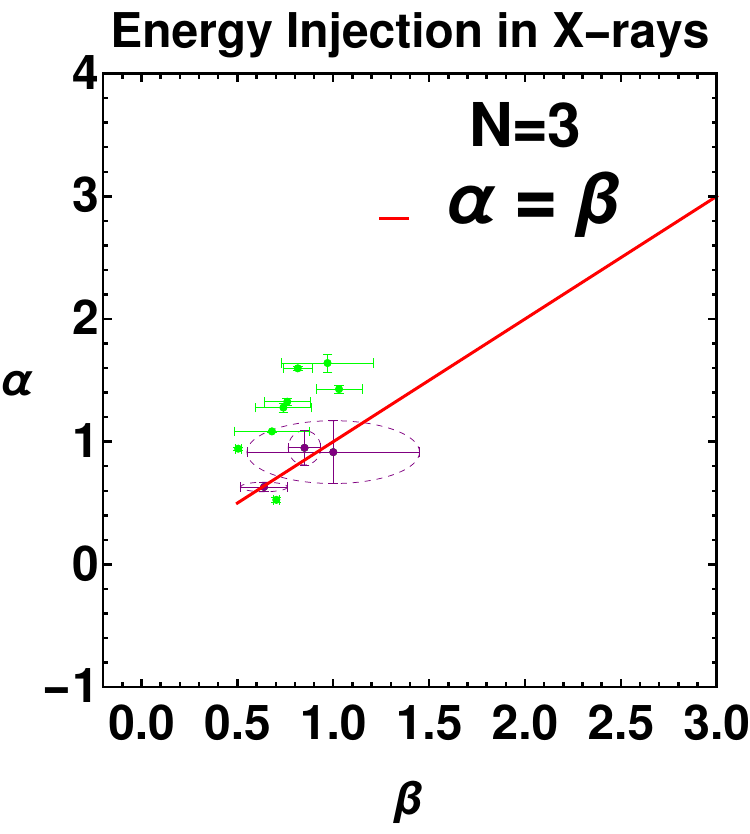}{0.18\textwidth}{(j)}
          \fig{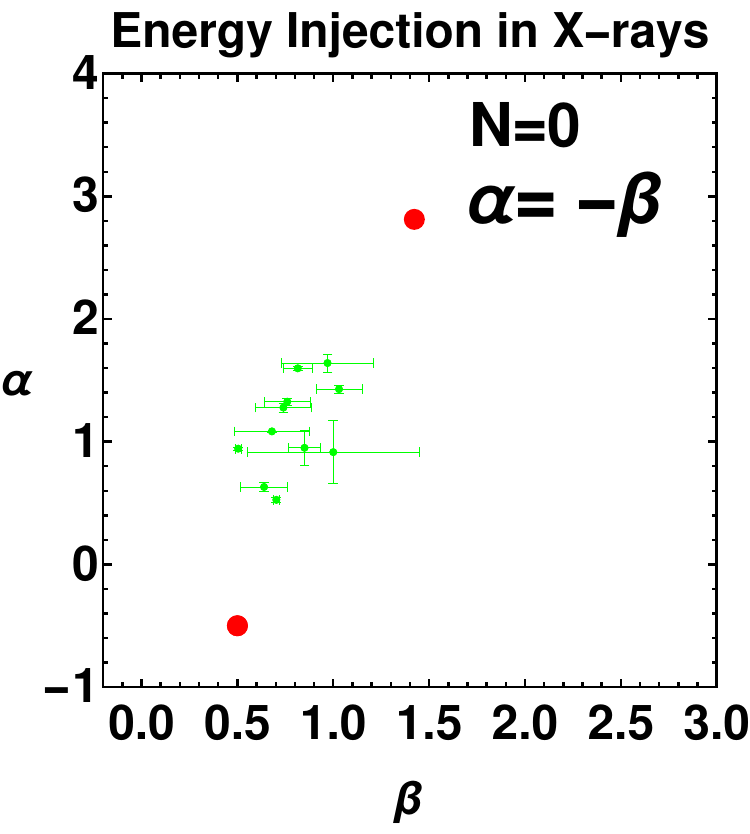}{0.18\textwidth}{(k)}
          \fig{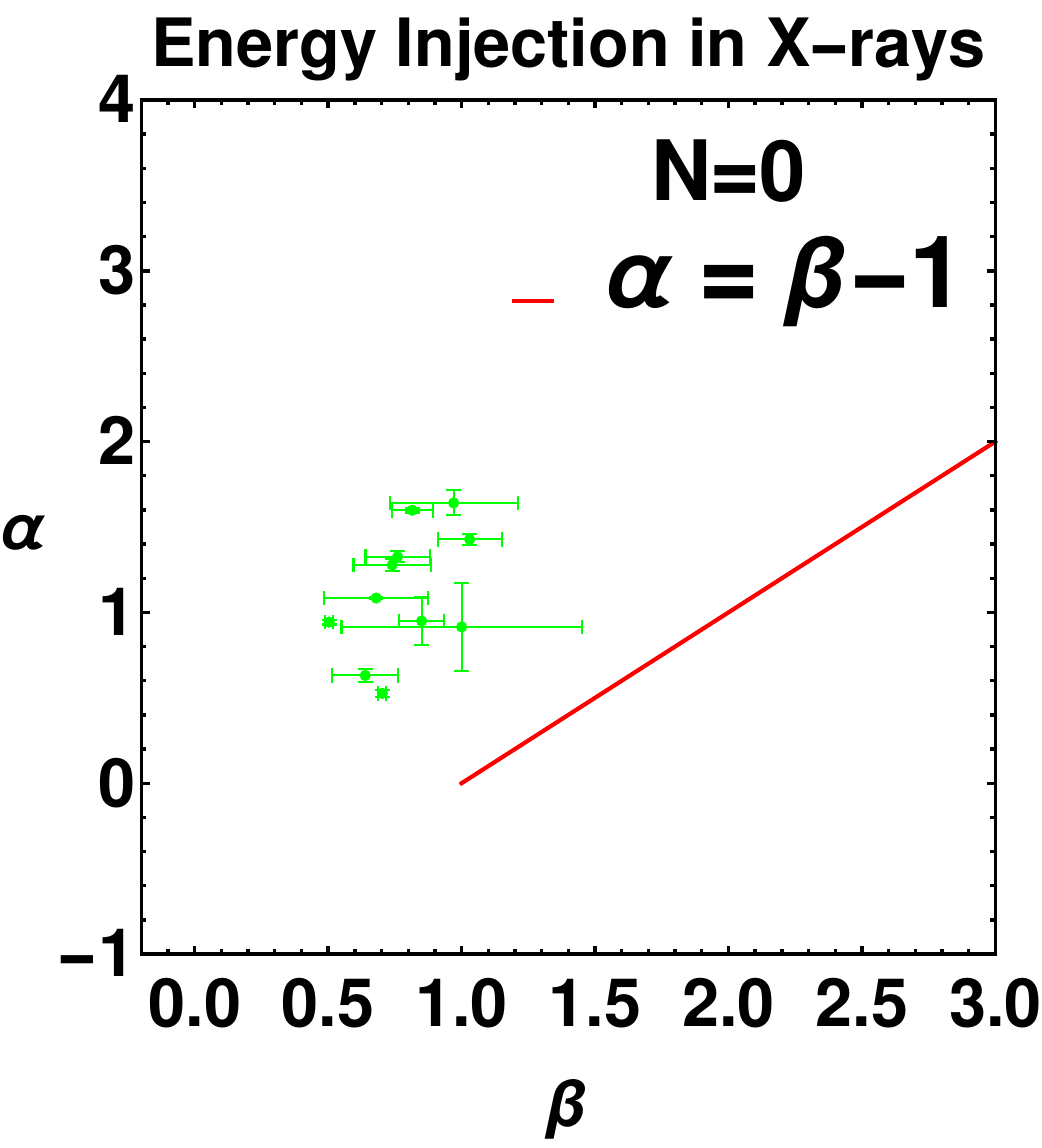}{0.18\textwidth}{(l)}
          }
\vspace{-8.7pt}
\gridline{\fig{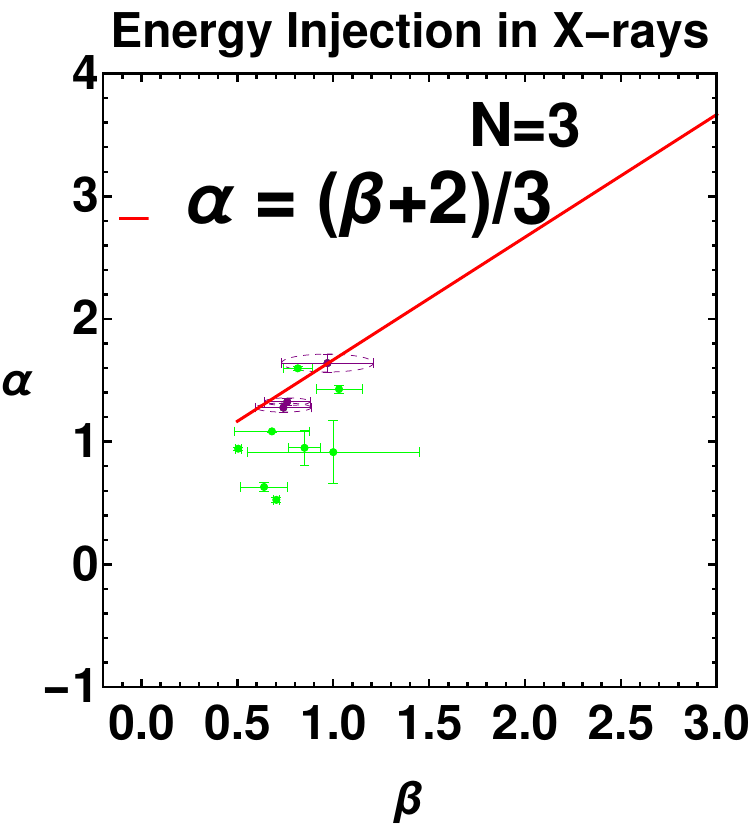}{0.18\textwidth}{(m)}
          \fig{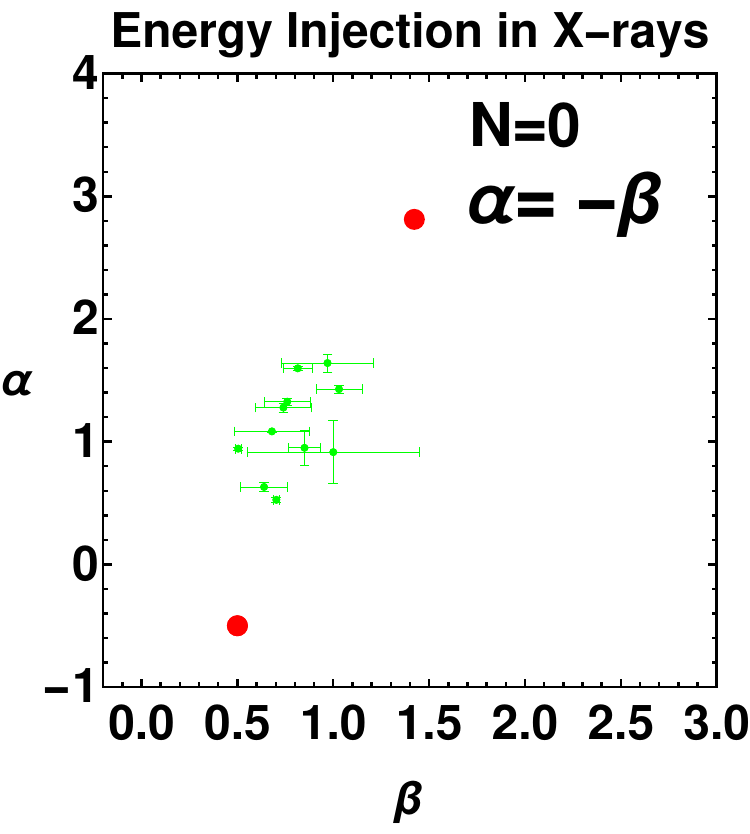}{0.18\textwidth}{(n)}
          \fig{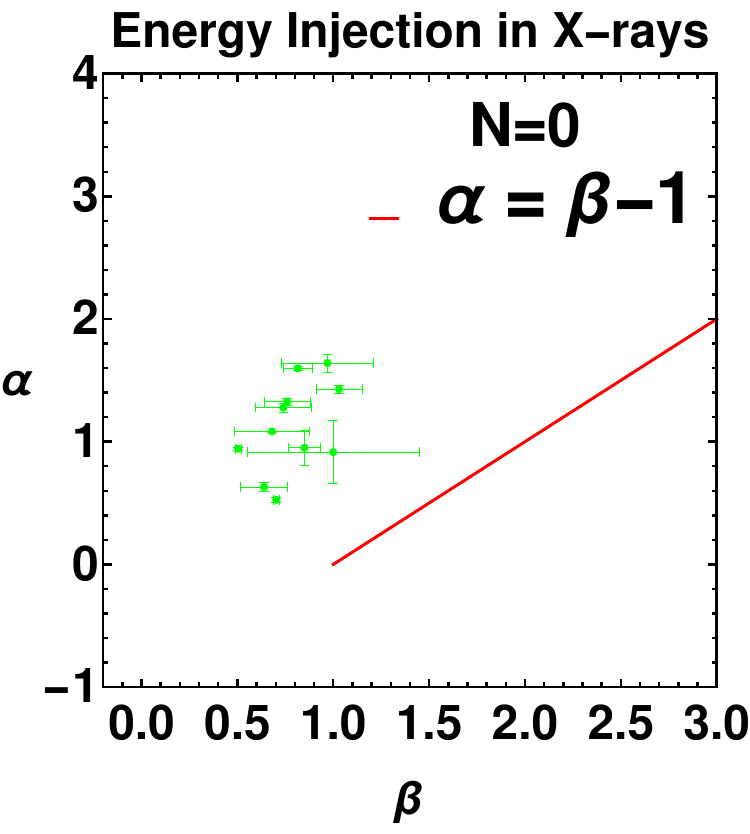}{0.18\textwidth}{(o)}
          }
\vspace{-6.21614pt}
\caption{CRs in X-rays corresponding to the synchrotron FS model for $k = 0$--$2.5$ (from top to bottom) with energy injection ($q=0$) and $q=0$ denoting instantaneous energy injection. Columns denote SC, FC, and SC/FC regimes from left to right. GRBs that satisfy the relations for X-ray parameters ($\alpha_{\rm{X}}$ and $\beta_{\rm{X}}$) are shown in purple; others are shown in green.}
\label{fig:XRTinj}
%\vspace{-20.65765pt}
\end{figure*}

\begin{table*}
\caption{%Summary of results of the CRs obtained with 
CRs determined by optical parameters ($\alpha_{\rm{opt}}$ and $\beta_{\rm{opt}}$) without energy injection (\textit{q} = 1; upper panel) and with energy injection (\textit{q} = 0; middle panel), and jet break (bottom panel), showing the number and occurrence rate of GRBs satisfying each relation (where $\alpha=$ is omitted for brevity) out of a total of 14 GRBs. Since each occurrence rate is calculated independently from the total sample of 14 GRBs, there is no obligation for the rates to add up to 100\%.
}
\begin{center}
%\resizebox{1.0\textwidth}{!}{
\scalebox{0.75}{
\begin{tabular}{lccccccccc}
%\begin{tabular}{p{1.0cm} p{1.5cm} p{3cm} p{1cm} p{2cm} p{1.5cm} p{1cm} p{2.3cm} p{1cm}}
%\begin{tabularx}{1.0\textwidth}{lXccccccccc}
\hline
\multicolumn{9}{c}{No Energy Injection (\textit{q} = 1) in Optical} \\
\hline
$n(r)$ & \text{Cooling} & $\nu$ \text{Range}    & $\beta(p)$ & \text{CR:}$1<p<2$ & \text{CR:}$p>2$ & \text{GRBs} & \text{Occurrence Rate} & \text{Figure}\\
\hline
$r^0$    & Slow & $\nu_m<\nu<\nu_c$ & $\frac{p-1}{2}$ & $\frac{6\beta+9}{16}$ & $\frac{3\beta}{2}$     & 0 & 0\% & (7a) \\
 
$r^{-1}$ & Slow & $\nu_m<\nu<\nu_c$ & $\frac{p-1}{2}$ &  $\frac{4\beta+9}{12}$ & $\frac{9\beta+1}{6}$ & 0 & 0\% & (7d) \\
 
$r^{-1.5}$ & Slow & $\nu_m<\nu<\nu_c$ & $\frac{p-1}{2}$ & $\frac{3\beta+9}{10}$ & $\frac{15\beta+3}{10}$ & 0 & 0\% & (7g) \\

$r^{-2}$ & Slow & $\nu_m<\nu<\nu_c$ & $\frac{p-1}{2}$ & $\frac{2\beta+9}{8}$  & $\frac{3\beta+1}{2}$ & 0 & 0\% & (7j) \\
 
$r^{-2.5}$ & Slow & $\nu_m<\nu<\nu_c$ & $\frac{p-1}{2}$ & $\frac{\beta + 9}{6}$ & $\frac{9\beta+5}{6}$ & 0 & 0\% & (7m) \\

$r^0$ & Fast & $\nu_c<\nu<\nu_m$ & $\frac{1}{2}$ & $\frac{\beta}{2}$  & $\frac{\beta}{2}$ & 0 & 0\% & (7b) \\
 
$r^{-1}$ & Fast & $\nu_c<\nu<\nu_m$ & $\frac{1}{2}$ & $\frac{\beta}{2}$ & $\frac{\beta}{2}$ & 0 & 0\% & (7e) \\
 
$r^{-1.5}$ & Fast & $\nu_c<\nu<\nu_m$ & $\frac{1}{2}$ & $\frac{\beta}{2}$       & $\frac{\beta}{2}$  & 0 & 0\% & (7h) \\
 
$r^{-2}$ & Fast & $\nu_c<\nu<\nu_m$ & $\frac{1}{2}$ & $\frac{\beta}{2}$ & $\frac{\beta}{2}$ & 0 & 0\%  & (7k) \\
 
$r^{-2.5}$ & Fast & $\nu_c<\nu<\nu_m$ & $\frac{1}{2}$ & $\frac{\beta}{2}$    & $\frac{\beta}{2}$     & 0 & 0\% & (7n) \\

$r^0$  & Slow/Fast & $\nu > $ max\{$\nu_c,\nu_m$\} &$\frac{p}{2}$ & $\frac{3\beta+5}{8}$  & $\frac{3\beta-1}{2}$ & 2 & 14.3\% & (7c) \\
 
$r^{-1}$ & Slow/Fast & $\nu > $ max\{$\nu_c,\nu_m$\} & $\frac{p}{2}$ & $\frac{\beta+2}{3}$   & $\frac{3\beta-1}{2}$ & 2 & 14.3\%  & (7f) \\
 
$r^{-1.5}$ & Slow/Fast & $\nu > $ max\{$\nu_c,\nu_m$\} & $\frac{p}{2}$ & $\frac{3\beta+7}{10}$   & $\frac{3\beta-1}{2}$ & 2 & 14.3\%  & (7i) \\
 
$r^{-2}$ & Slow/Fast & $\nu > $ max\{$\nu_c,\nu_m$\} & $\frac{p}{2}$ & $\frac{\beta+3}{4}$   & $\frac{3\beta-1}{2}$ & 2 & 14.3\% & (7l) \\
 
$r^{-2.5}$ & Slow/Fast & $\nu > $ max\{$\nu_c,\nu_m$\} & $\frac{p}{2}$& $\frac{\beta + 5}{6}$ & $\frac{3\beta-1}{2}$ & 3 & 21.4\% & (7o) \\
 
 \hline
 \hline

\multicolumn{8}{c}{Energy Injection ($q=0$) in Optical} \\
\hline
 $n(r)$ & Cooling & $\nu$ \text{ Range}  & $\beta(p)$ & \text{CR:}$p>2$ & \text{GRBs} & \text{Occurrence Rate} & \text{Figure} \\
 \hline

 $r^0$ & Slow   & $\nu_m<\nu<\nu_c$ & $\frac{p-1}{2}$ & $\beta-1$ & 0 & 0\%& \text{(8a)} \\

 $r^{-1}$ & Slow   & $\nu_m<\nu<\nu_c$ & $\frac{p-1}{2}$ & $\beta-\frac{2}{3}$ & 1 & 7.14\%& \text{(8d)} \\

  $r^{-1.5}$ & Slow   & $\nu_m<\nu<\nu_c$ & $\frac{p-1}{2}$ & $\beta$ & 0 & 0\%& \text{(8g)} \\

 $r^{-2}$ & Slow    & $\nu_m<\nu<\nu_c$ & $\frac{p-1}{2}$ & $\beta-\frac{2}{5}$ & 1 & 7.14\% & \text{(8j)}\\

 $r^{-2.5}$ & Slow  & $\nu_m<\nu<\nu_c$ & $\frac{p-1}{2}$ & $\frac{\beta+2}{3}$ & 1 & 7.14\% & \text{(8m)}\\

$r^0$ & Fast &$ \nu_c<\nu<\nu_m$ & $\frac{1}{2}$ & $-\beta$ & 0&
  0\% & \text{(8b)}\\

$r^{-1}$ & Fast &$ \nu_c<\nu<\nu_m$ & $\frac{1}{2}$ & $-\beta$ & 0 & 0\% & \text{(8e)}\\

$r^{-1.5}$ & Fast &$ \nu_c<\nu<\nu_m$ & $\frac{1}{2}$ & $-\beta$ & 0 & 0\% & \text{(8h)}\\

$r^{-2}$ & Fast    &$ \nu_c<\nu<\nu_m$ & $\frac{1}{2}$ & $-\beta$ & 0 & 0\% & \text{(8k)}\\

$r^{-2.5}$ & Fast    &$ \nu_c<\nu<\nu_m$ & $\frac{1}{2}$ & $-\beta$ & 0 & 0\% & \text{(8n)}\\

$r^0$ & Slow/Fast  & $\nu > $ max\{$\nu_c,\nu_m$\} & $\frac{p}{2}$ & $\beta-1$ & 0 & 0\% & \text{(8c)}\\

$r^{-1}$ & Slow/Fast  & $\nu > $ max\{$\nu_c,\nu_m$\} & $\frac{p}{2}$ & $\beta-1$ & 0 & 0\% & \text{(8f)}\\

$r^{-1.5}$ & Slow/Fast  & $\nu > $ max\{$\nu_c,\nu_m$\} & $\frac{p}{2}$ & $\beta-1$ & 0 & 0\% & \text{(8i)}\\

$r^{-2}$ & Slow/Fast  & $\nu > $ max\{$\nu_c,\nu_m$\} & $\frac{p}{2}$ & $\beta-1$ & 0 & 0\% & \text{(8l)}\\

$r^{-2.5}$ & Slow/Fast  & $\nu > $ max\{$\nu_c,\nu_m$\} & $\frac{p}{2}$ & $\beta-1$ & 0 & 0\% & \text{(8o)}\\

\hline
\hline

\multicolumn{7}{c}{Jet Break in Optical} \\
\hline
 Cooling & $\nu$ \text{ Range}  & $\beta(p)$ & \text{CR:}$1<p<2$ & \text{GRBs} & \text{Occurrence Rate} & \text{Figure} \\
 \hline

 Slow   & $\nu_m<\nu<\nu_c$ & $\frac{p-1}{2}$ & $2\beta+1$ & 0 & 0\%& \text{(9d)} \\

Fast &$ \nu_c<\nu<\nu_m$ & $1$ & $\frac{1}{2}$ & 0& 0\% & \text{(9e)}\\

Slow/Fast  & $\nu > $ max\{$\nu_c,\nu_m$\} & $p$ & $\frac{\beta}{2}$ & 0 & 0\% & \text{(9f)}\\

\hline
\end{tabular}
}
%\end{tabularx}
\end{center}
\label{table: OPT crSummary}
\end{table*}

\begin{figure*}
\gridline{\fig{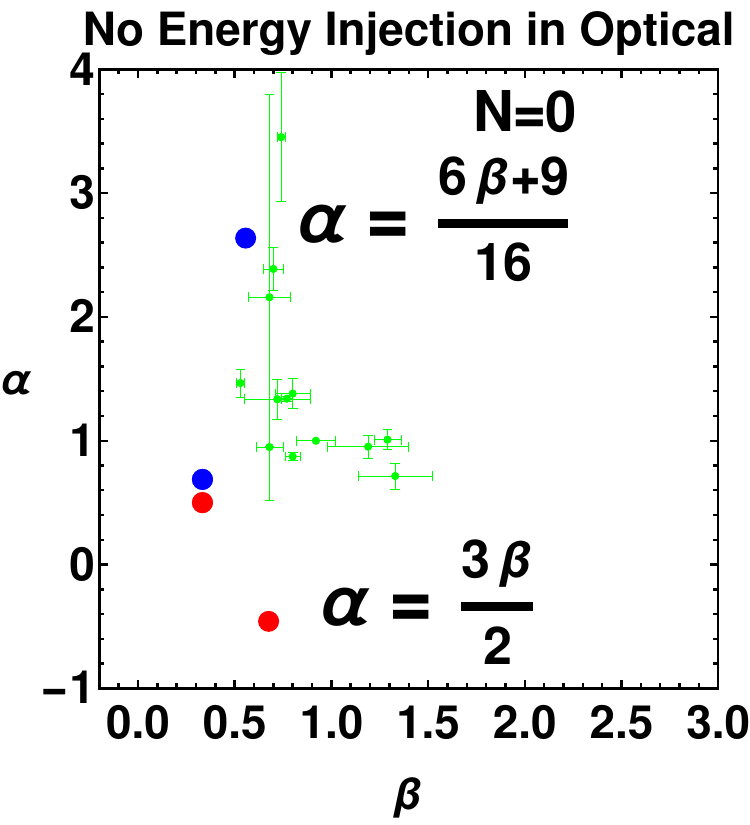}{0.18\textwidth}{(a)}
          \fig{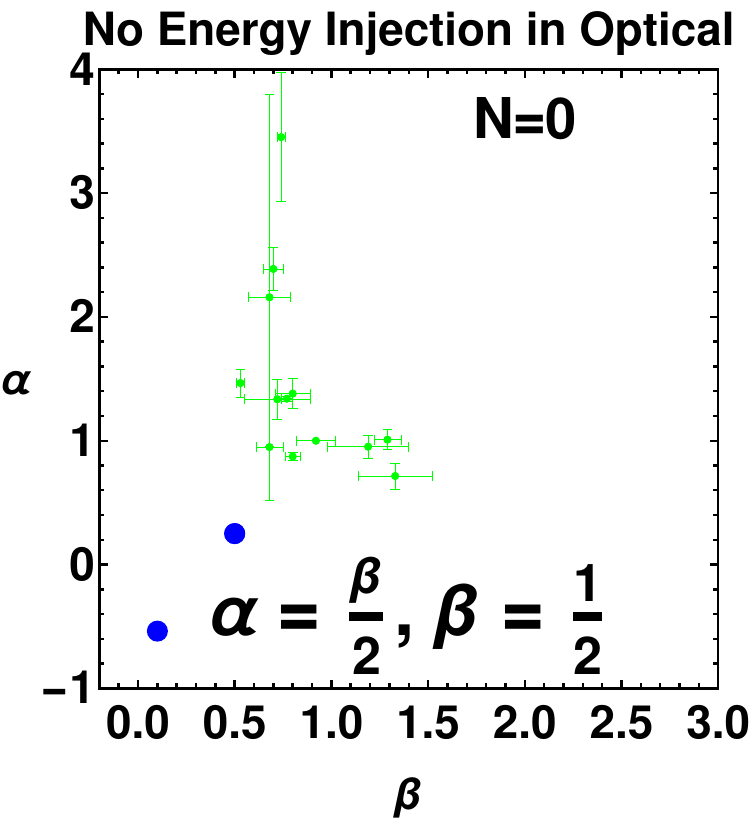}{0.18\textwidth}{(b)}
          \fig{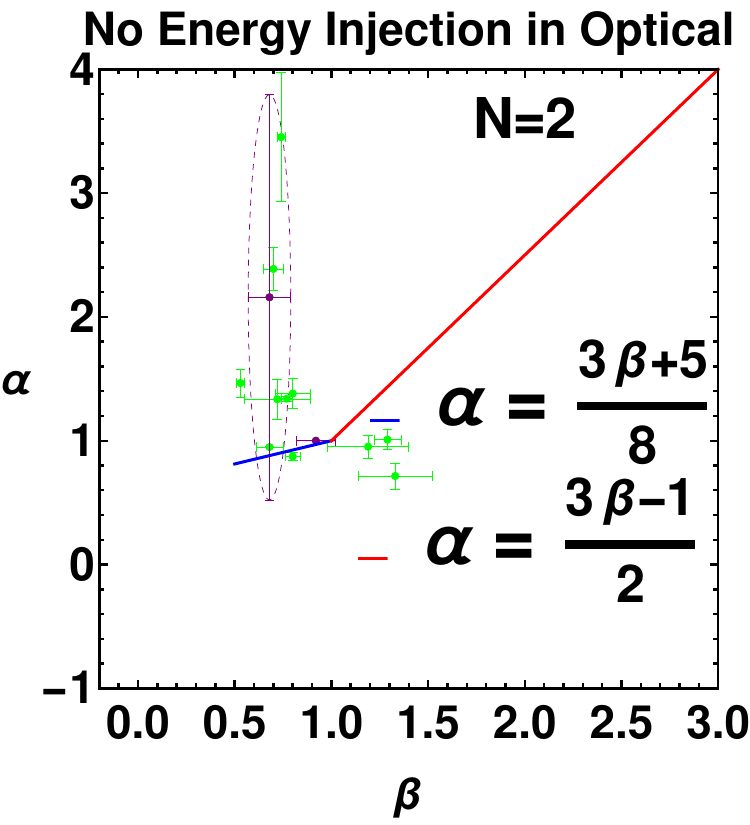}{0.18\textwidth}{(c)}
          }
\vspace{-8.7pt}
\gridline{\fig{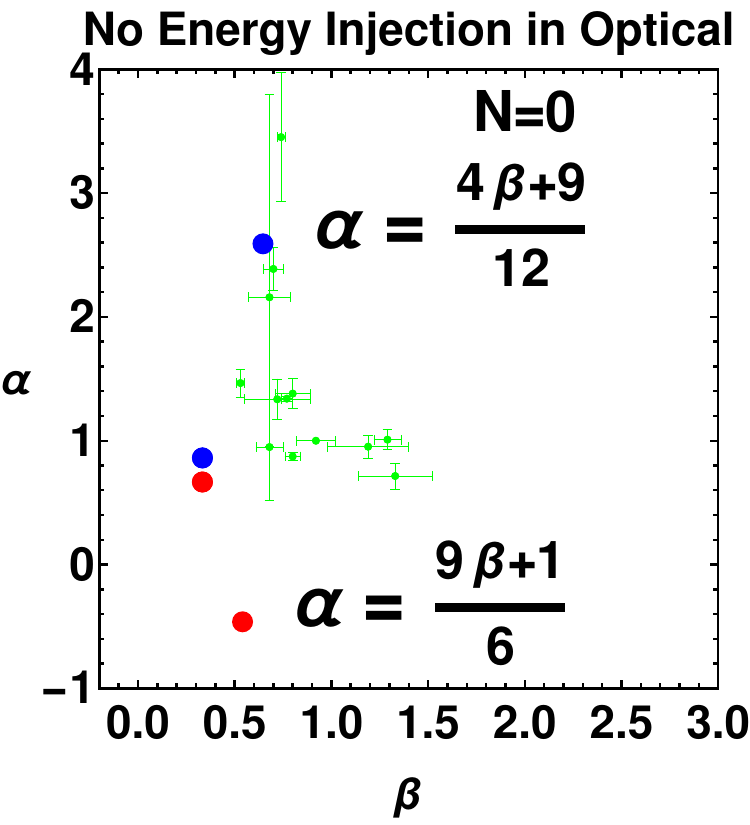}{0.18\textwidth}{(d)}
          \fig{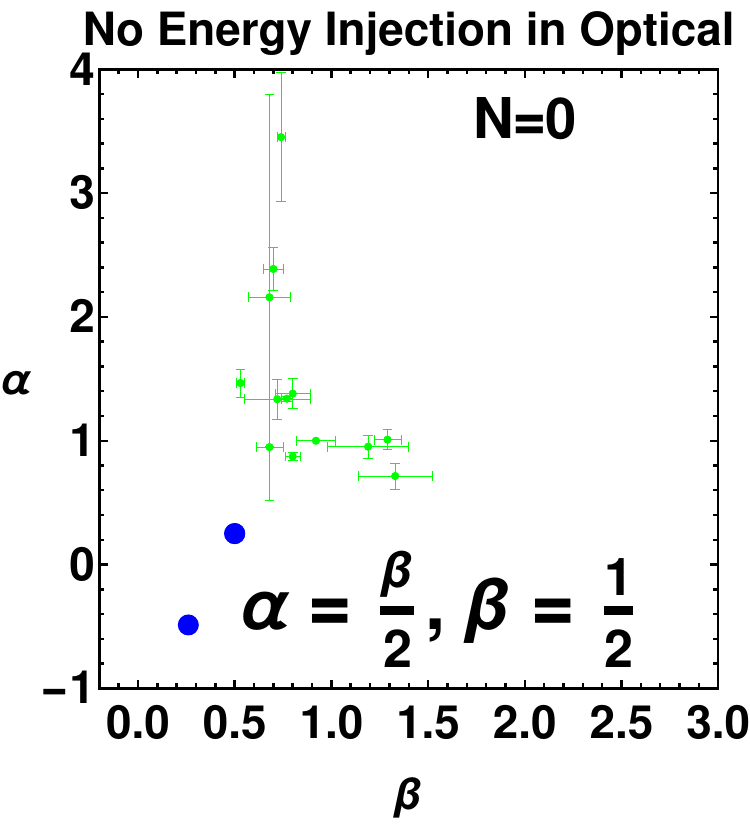}{0.18\textwidth}{(e)}
          \fig{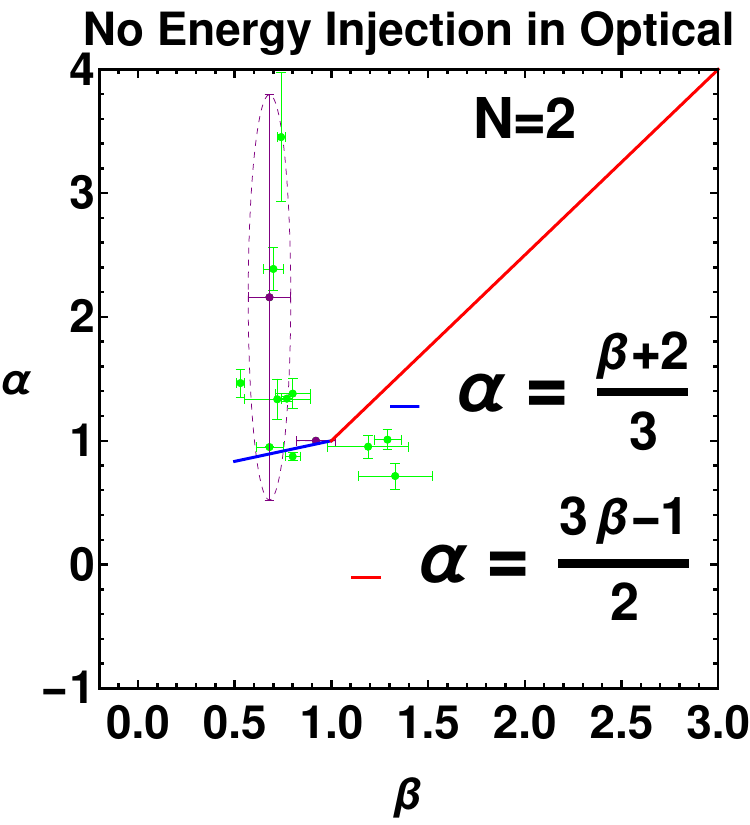}{0.18\textwidth}{(f)}
          }
\vspace{-8.7pt}
\gridline{\fig{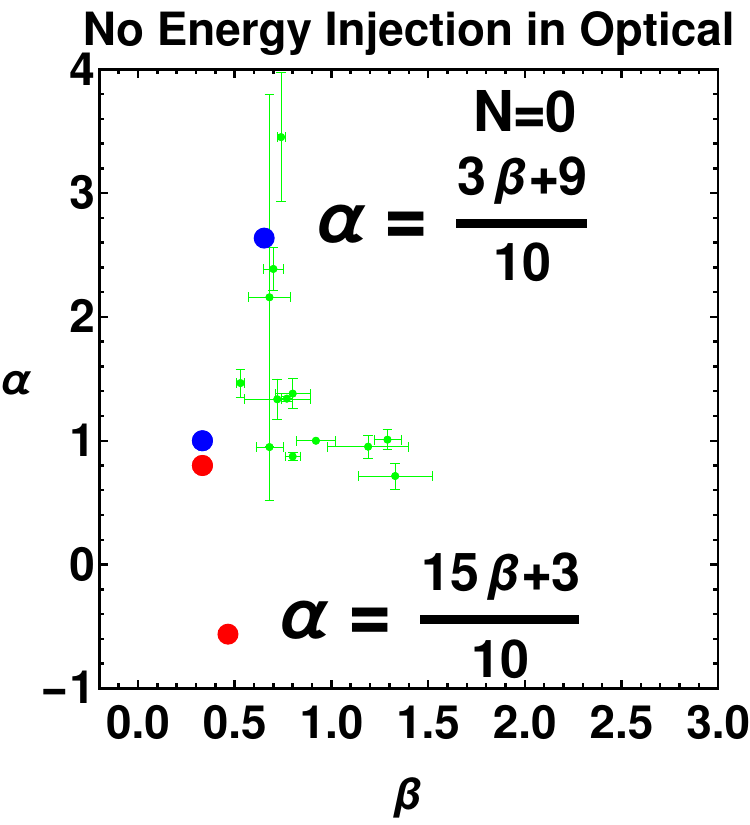}{0.18\textwidth}{(g)}
          \fig{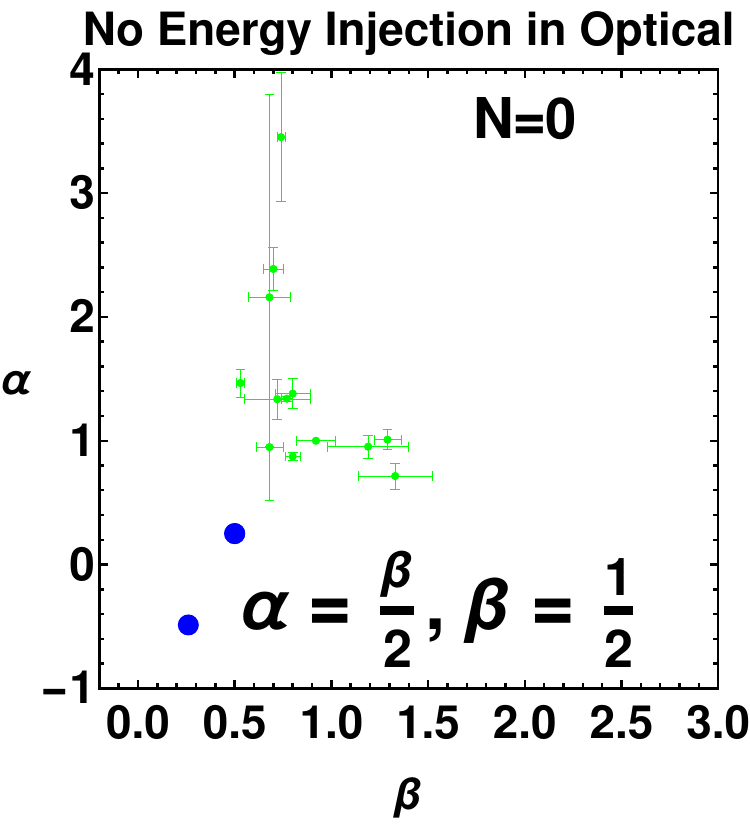}{0.18\textwidth}{(h)}
          \fig{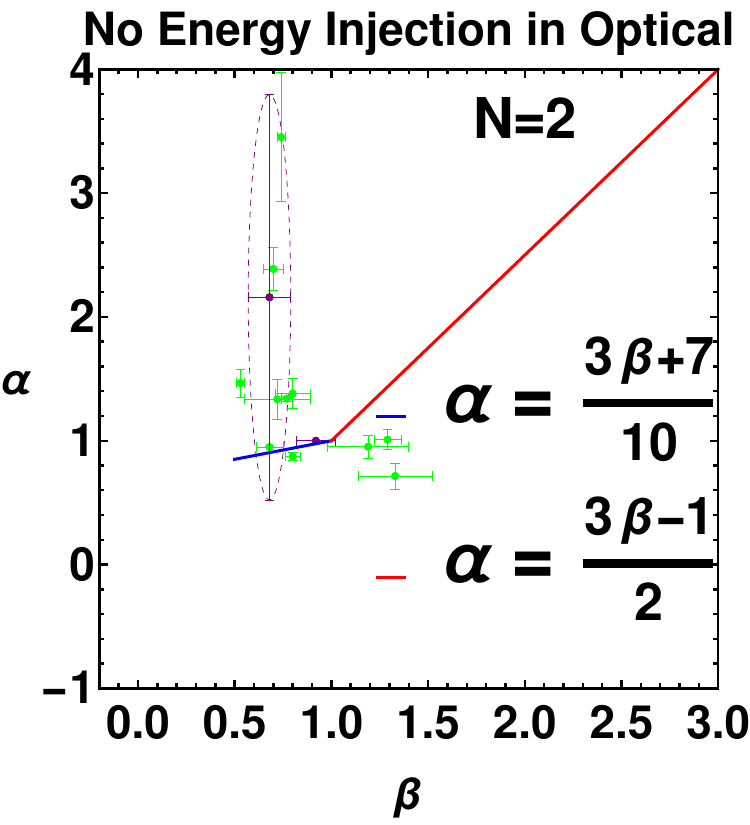}{0.18\textwidth}{(i)}
          }
\vspace{-8.7pt}
\gridline{\fig{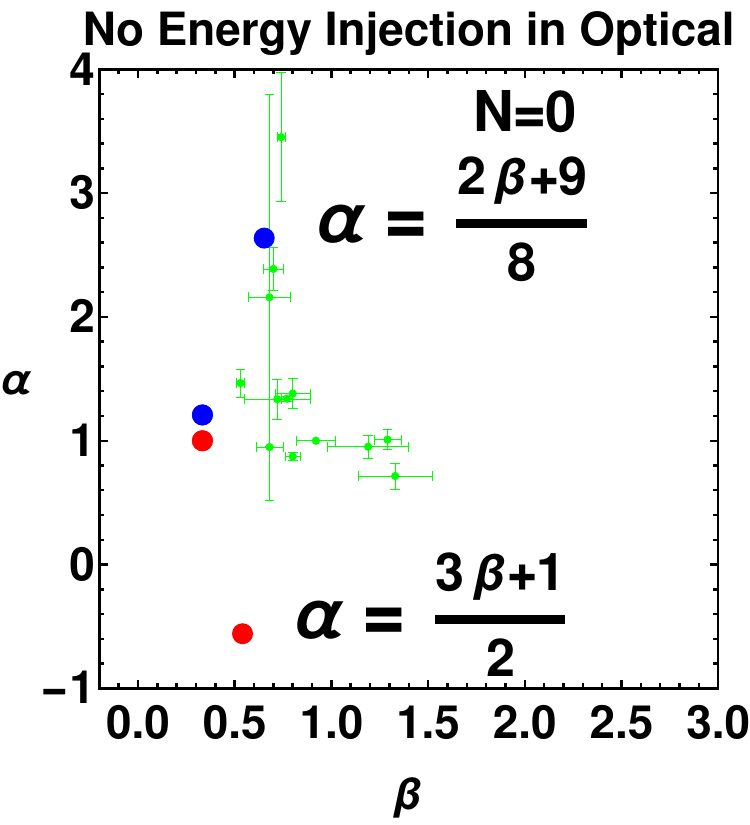}{0.18\textwidth}{(j)}
          \fig{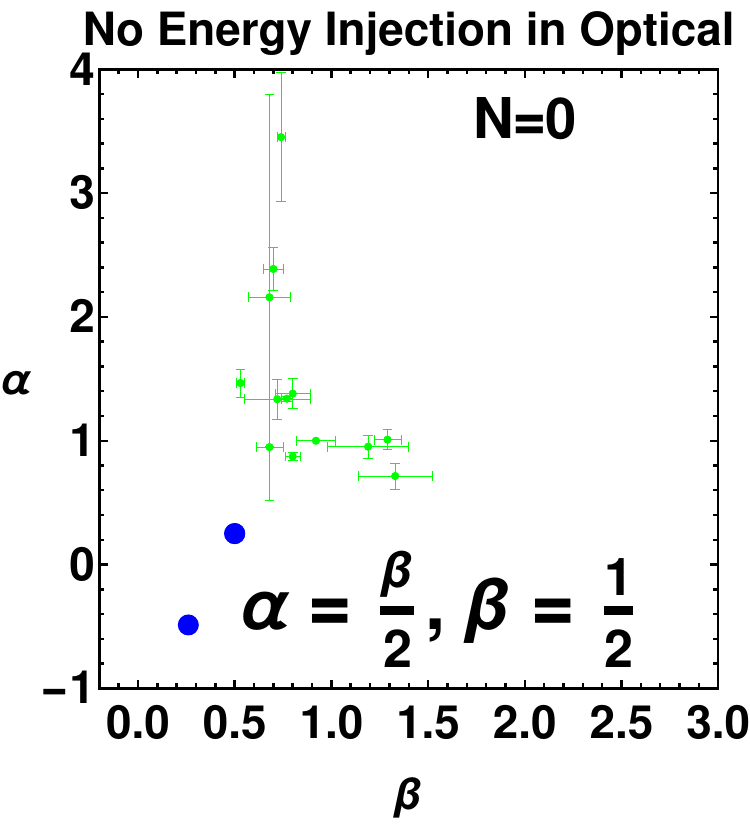}{0.18\textwidth}{(k)}
          \fig{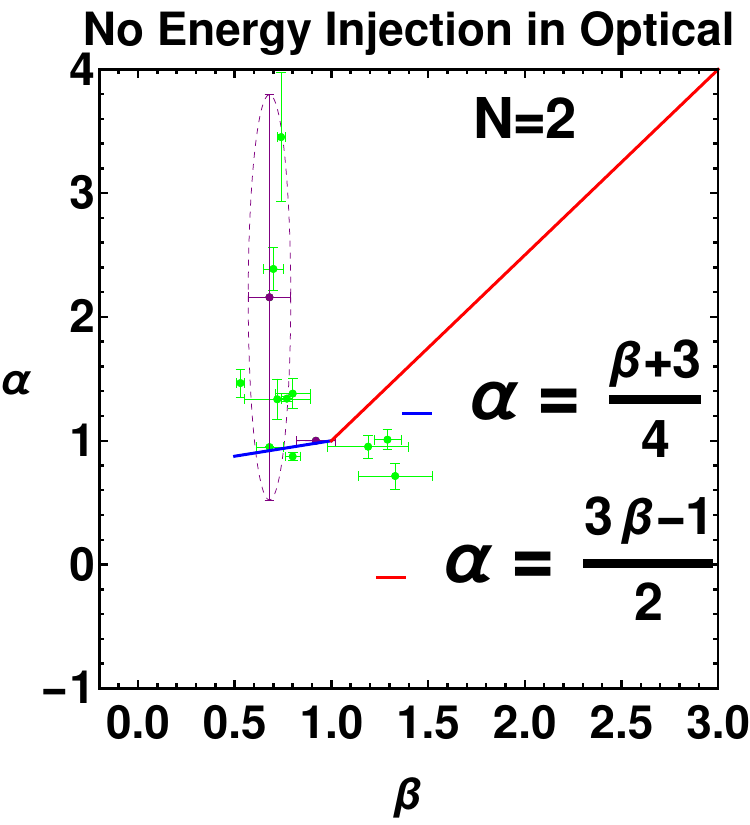}{0.18\textwidth}{(l)}
          }
\vspace{-8.7pt}
\gridline{\fig{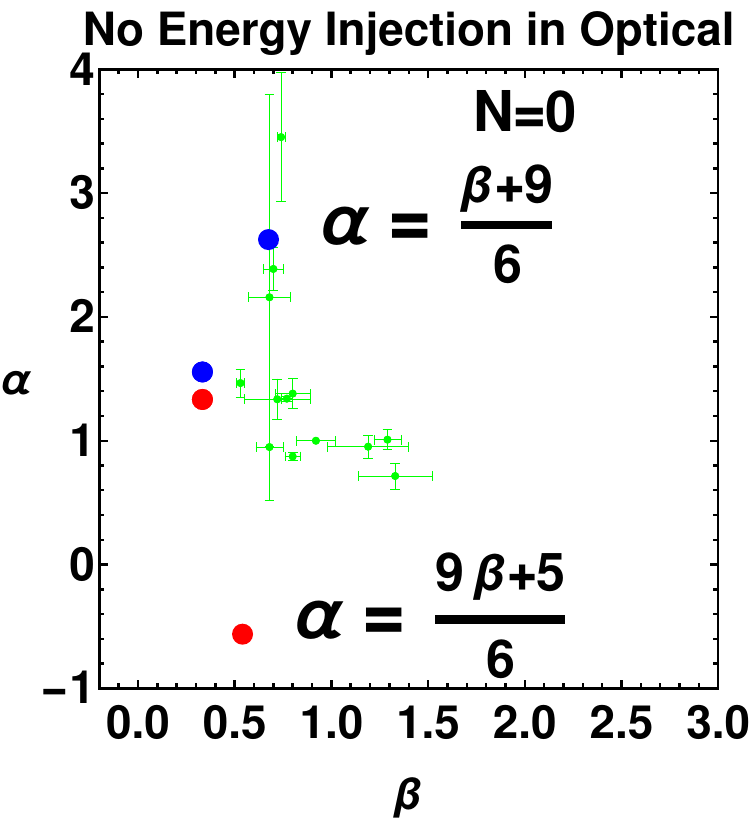}{0.18\textwidth}{(m)}
          \fig{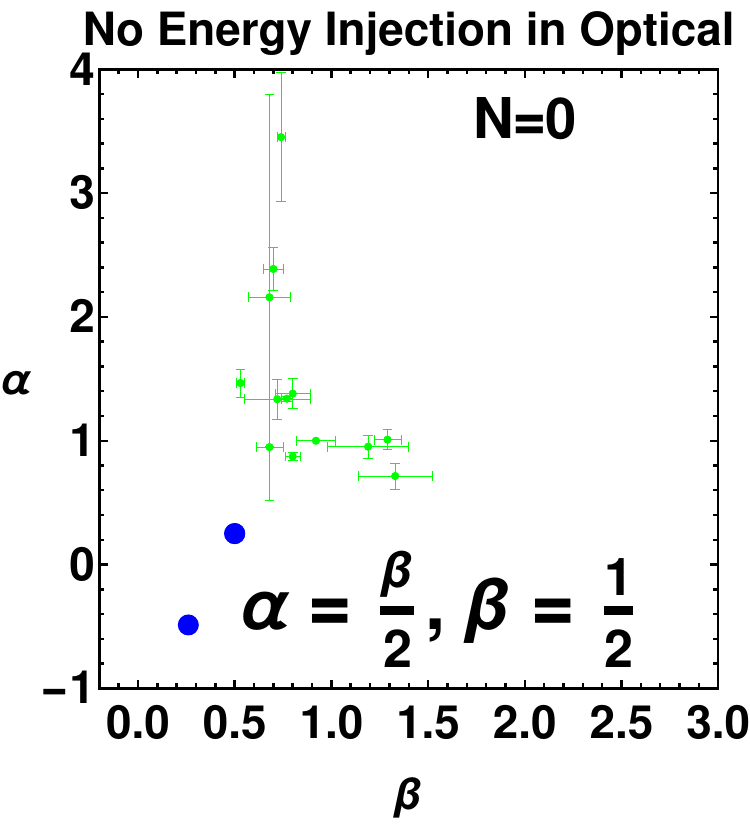}{0.18\textwidth}{(n)}
          \fig{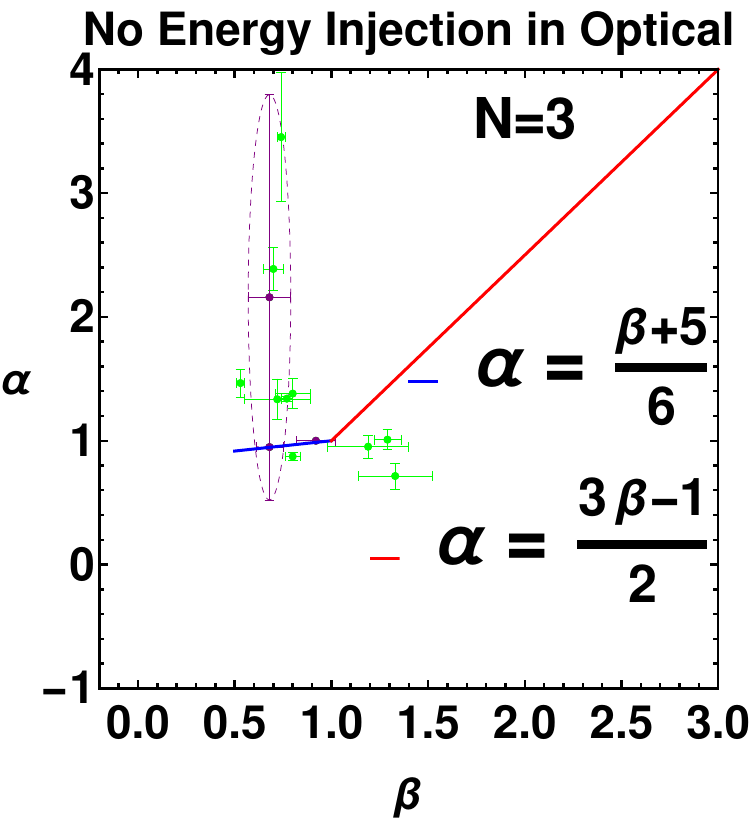}{0.18\textwidth}{(o)}
          }
\vspace{-6.21614pt}
\caption{CRs in optical corresponding to the synchrotron FS model for $k = 0$--$2.5$ (from top to bottom) with no energy injection ($q=1$). Columns denote SC, FC, and SC/FC regimes from left to right. GRBs that satisfy the relations for optical parameters ($\alpha_{\rm{opt}}$ and $\beta_{\rm{opt}}$) are shown in purple; others are shown in green.}
\label{fig:OPTnoinj}
%\vspace{-16.64456pt}
\end{figure*}

%% OPTICAL RESULTS FIGURES EI
\begin{figure*}
\gridline{\fig{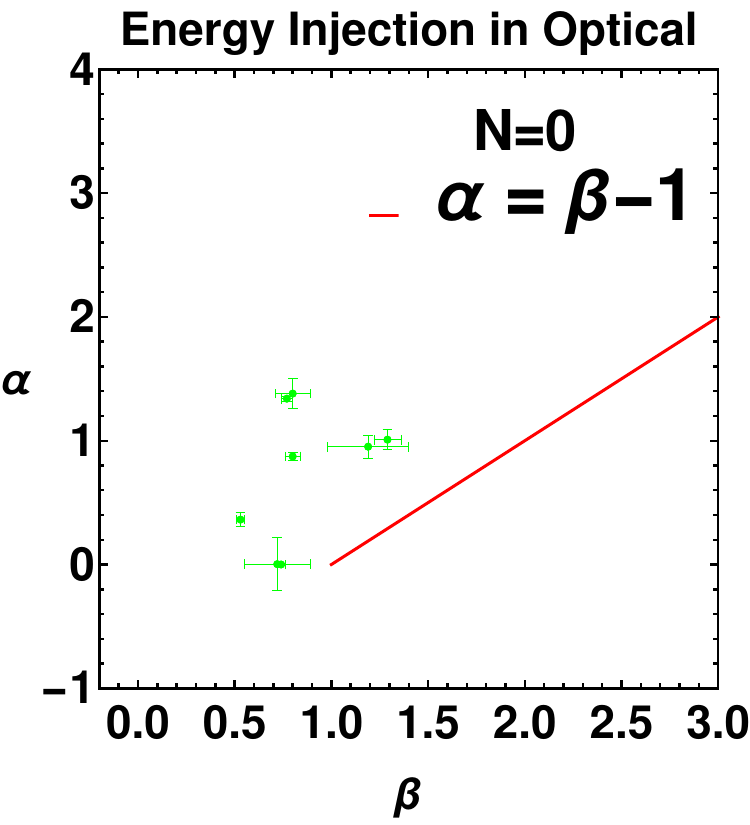}{0.18\textwidth}{(a)}
          \fig{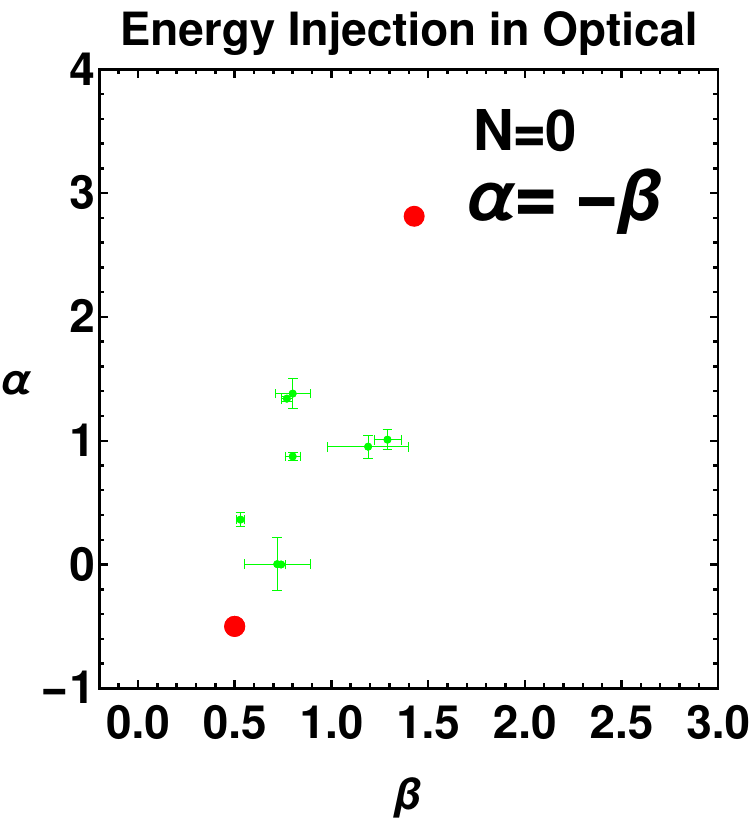}{0.18\textwidth}{(b)}
          \fig{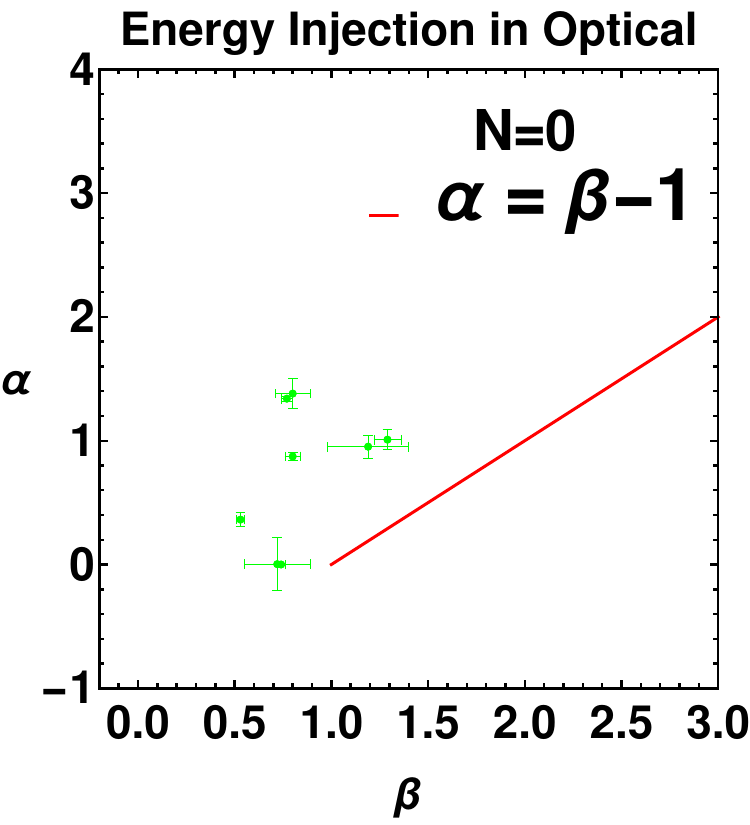}{0.18\textwidth}{(c)}
          }
\vspace{-8.7pt}
\gridline{\fig{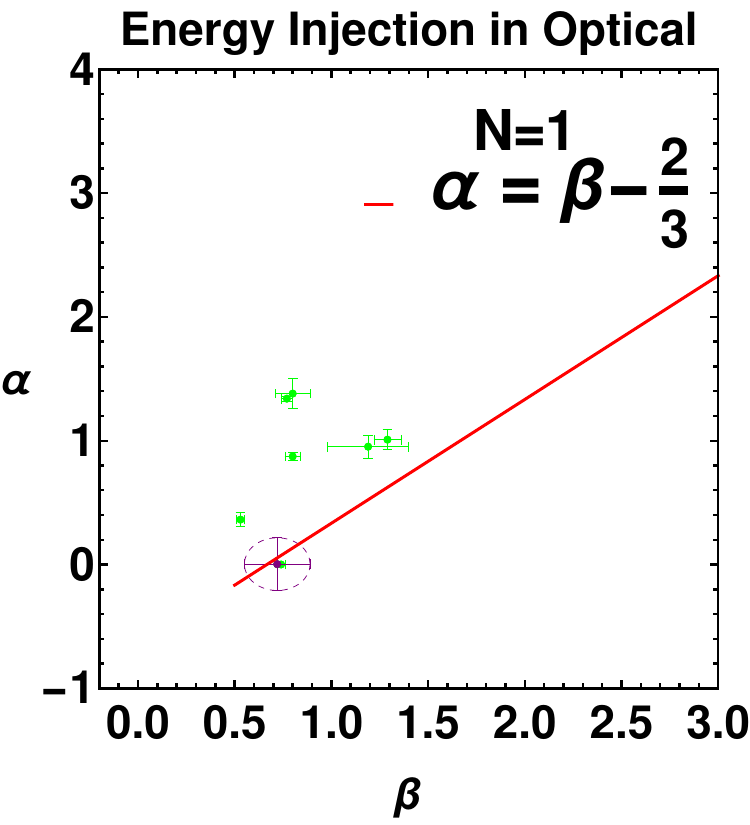}{0.18\textwidth}{(d)}
          \fig{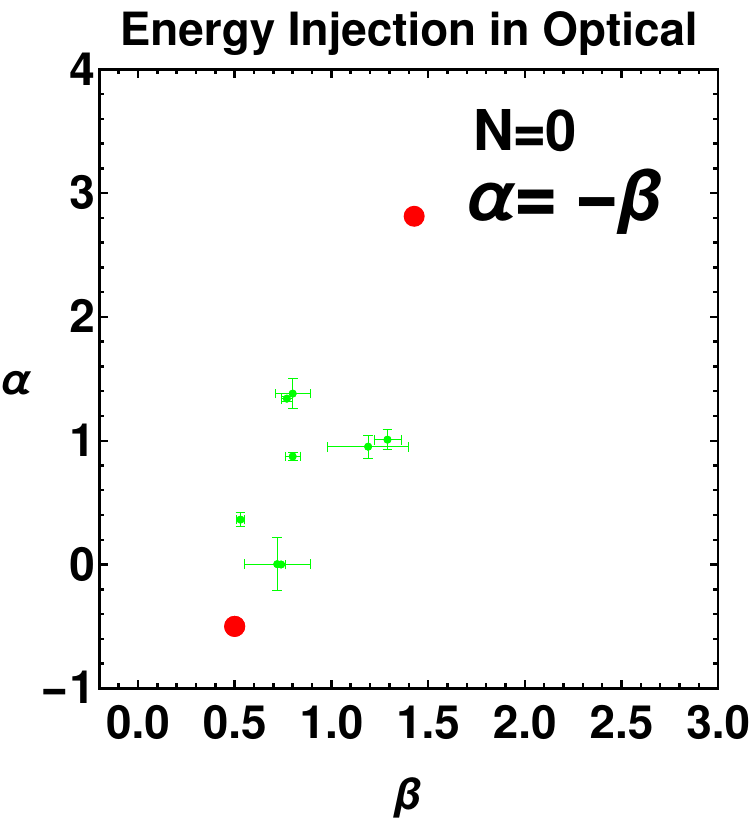}{0.18\textwidth}{(e)}
          \fig{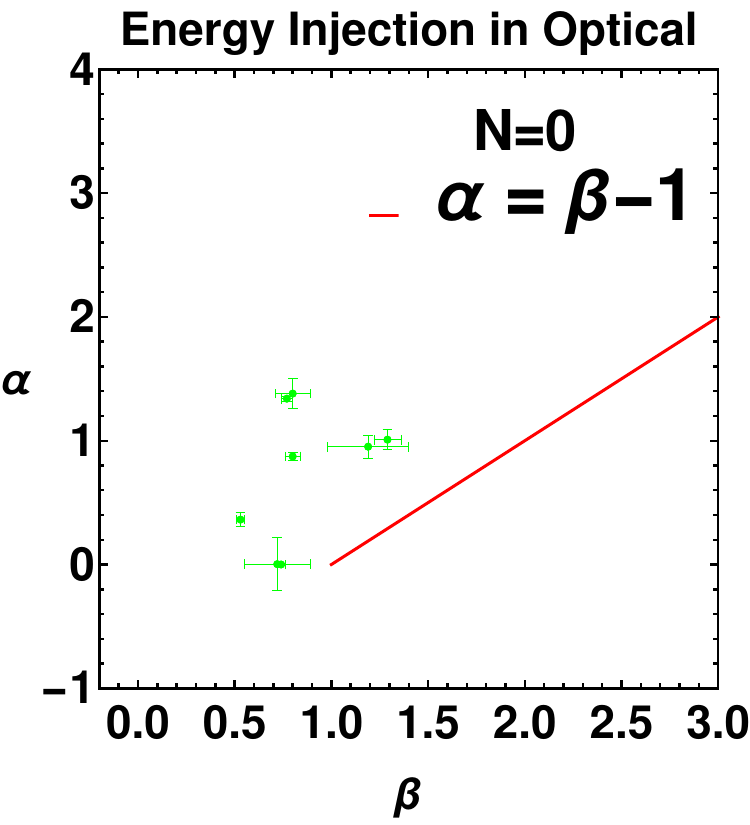}{0.18\textwidth}{(f)}
          }
\vspace{-8.7pt}
\gridline{\fig{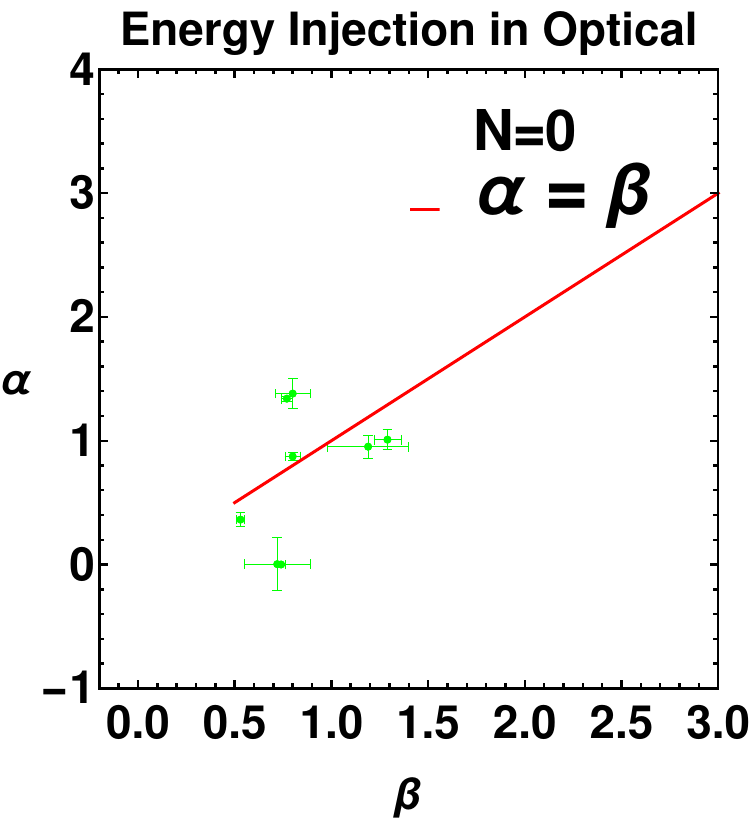}{0.18\textwidth}{(g)}
          \fig{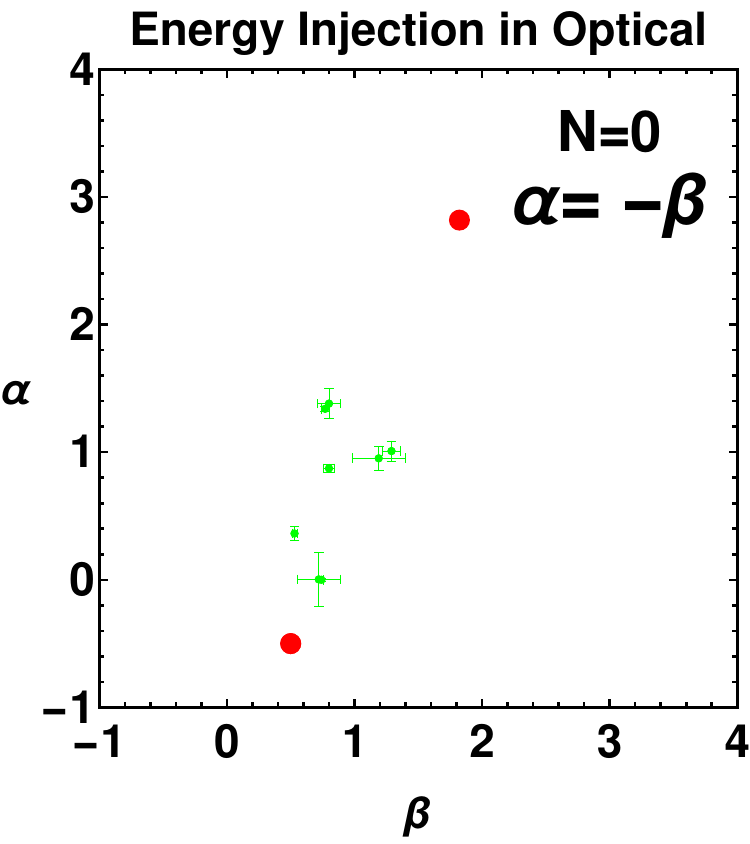}{0.18\textwidth}{(h)}
          \fig{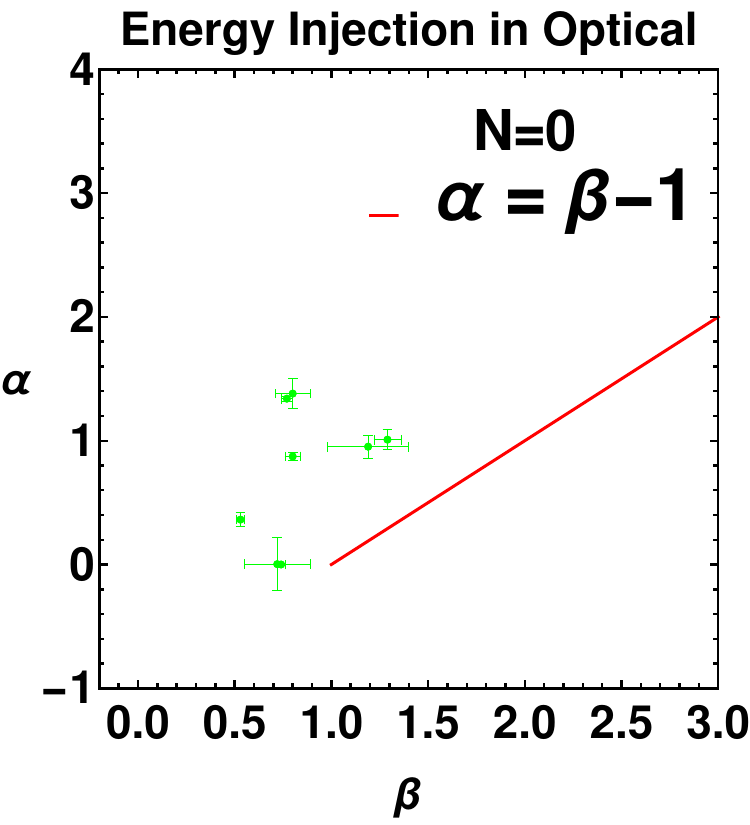}{0.18\textwidth}{(i)}
          }
\vspace{-8.7pt}
\gridline{\fig{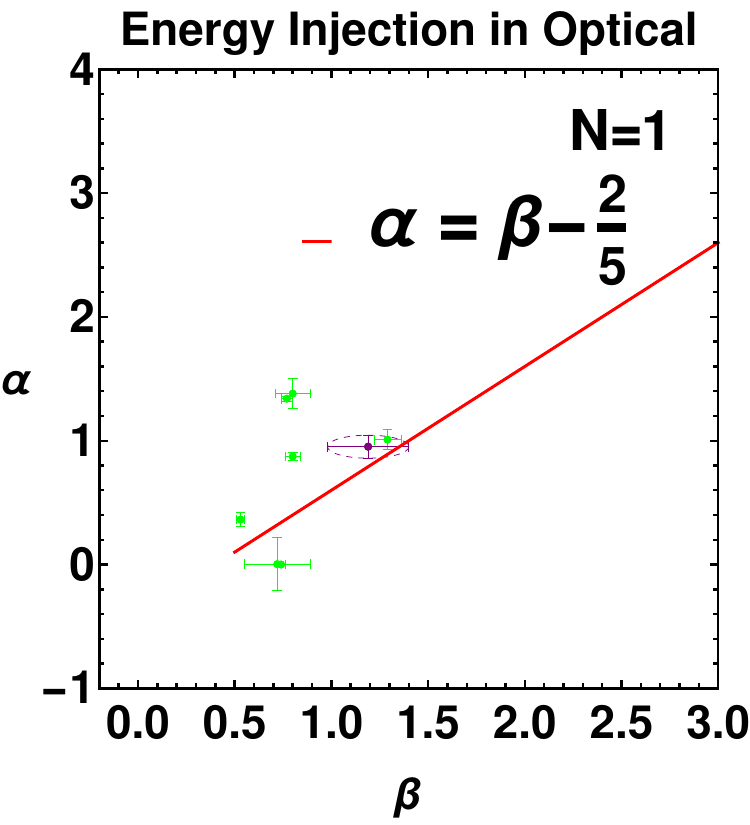}{0.18\textwidth}{(j)}
          \fig{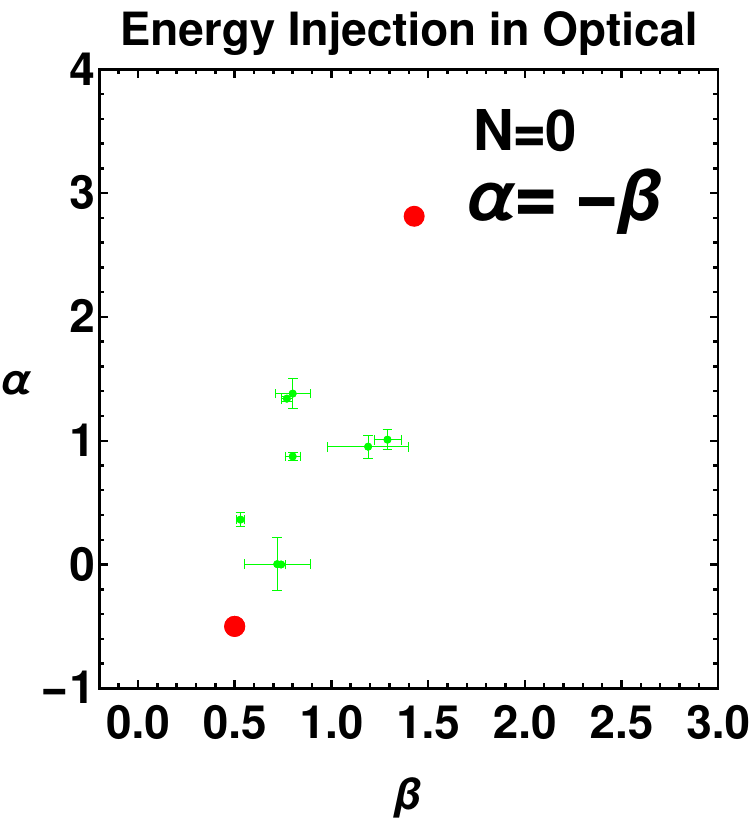}{0.18\textwidth}{(k)}
          \fig{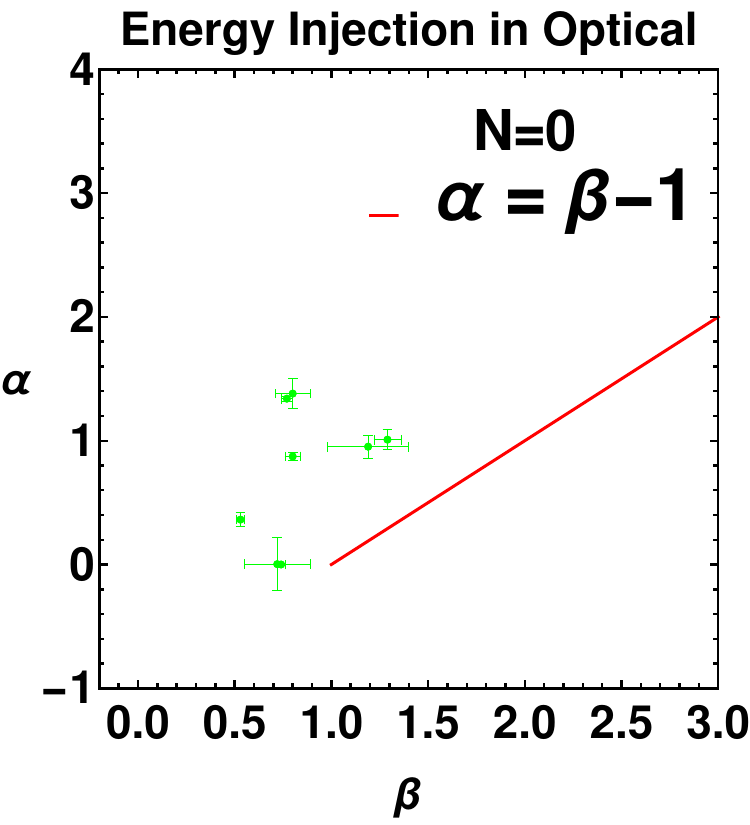}{0.18\textwidth}{(l)}
          }
\vspace{-8.7pt}
\gridline{\fig{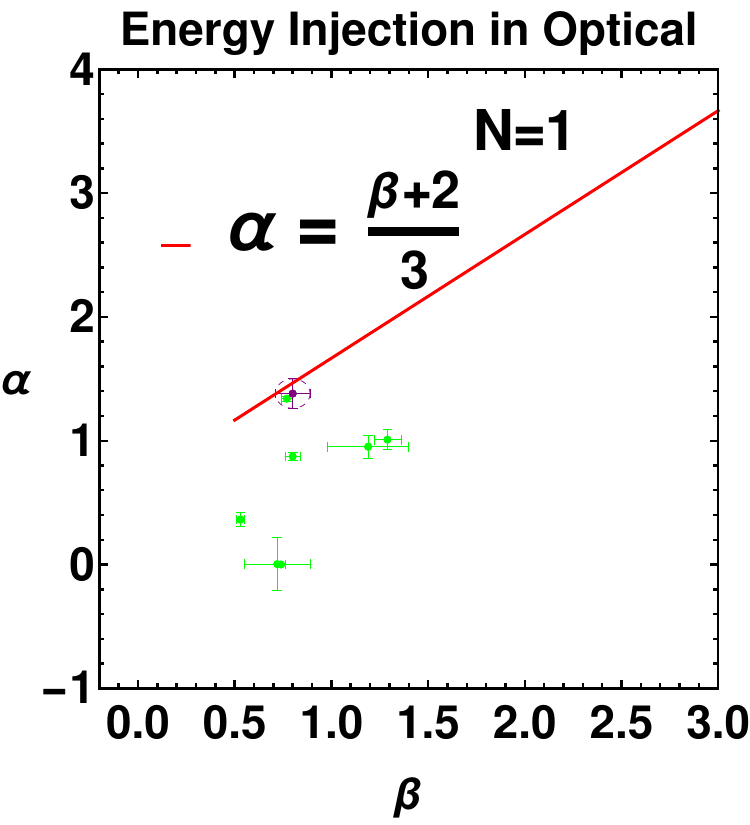}{0.18\textwidth}{(m)}
          \fig{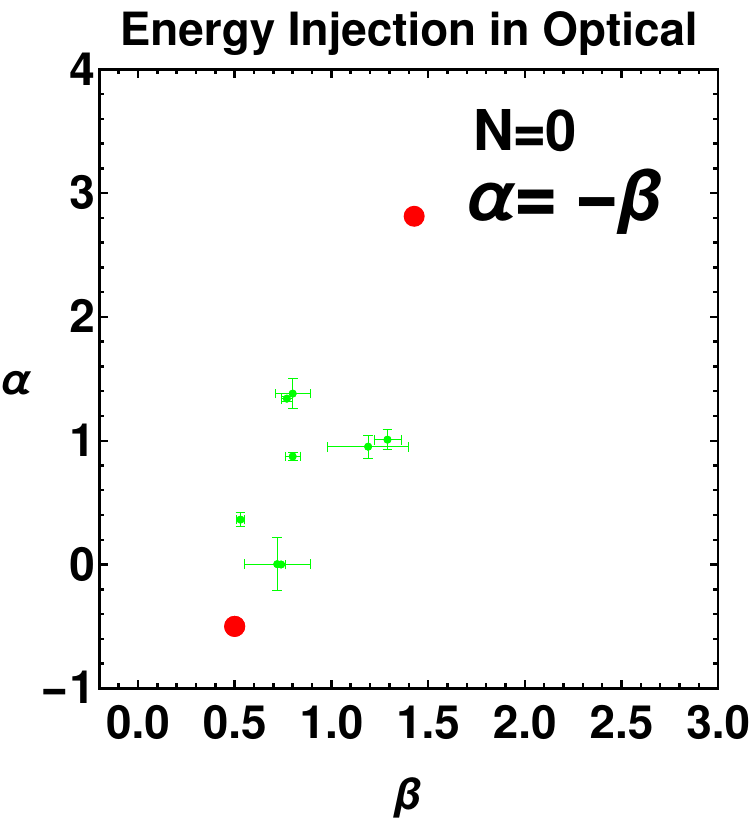}{0.18\textwidth}{(n)}
          \fig{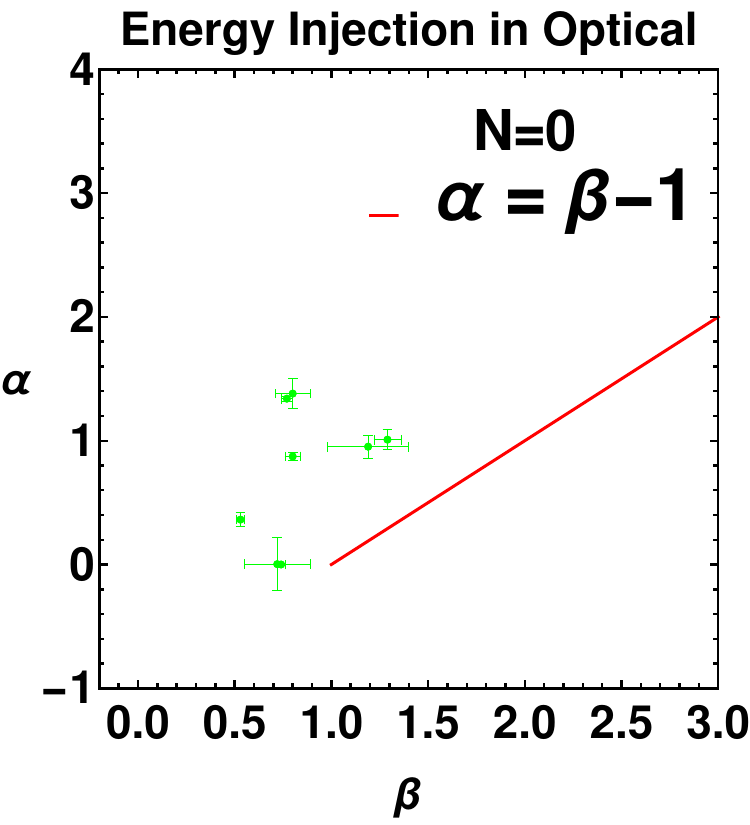}{0.18\textwidth}{(o)}
          }
\vspace{-6.21614pt}
\caption{CRs in optical corresponding to the synchrotron FS model for $k = 0$--$2.5$ (from top to bottom) with energy injection ($q=0$) and $q=0$ denoting instantaneous energy injection. Columns denote SC, FC, and SC/FC regimes from left to right. GRBs that satisfy the relations for optical parameters ($\alpha_{\rm{opt}}$ and $\beta_{\rm{opt}}$) are shown in purple; others are shown in green.}
\label{fig:OPTinj}
%\vspace{-28.17075pt}
\end{figure*}

%% OPTICAL RESULTS FIGURES JB
\begin{figure*}
\gridline{\fig{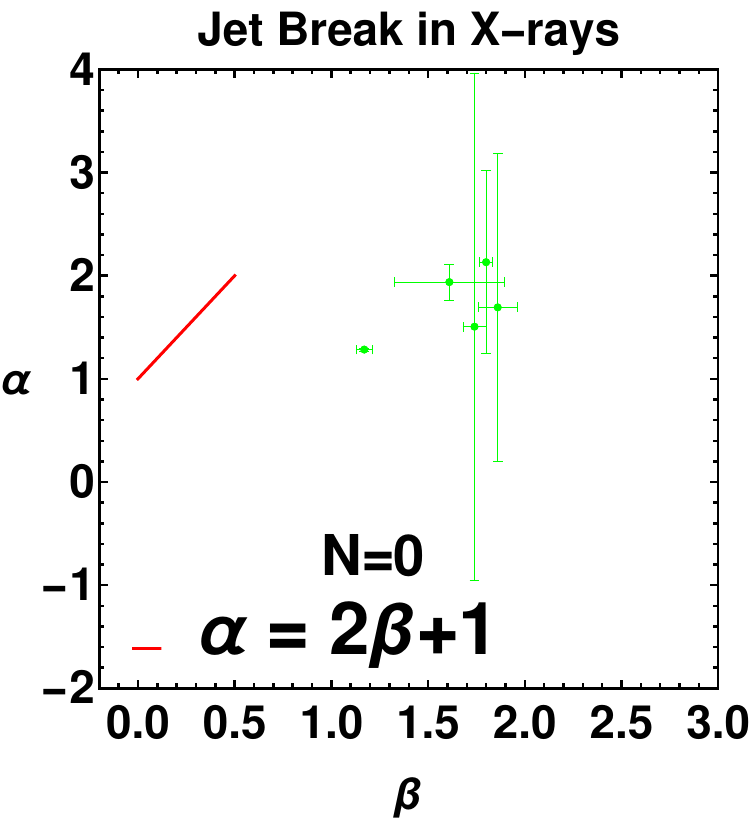}{0.18\textwidth}{(a)}
          \fig{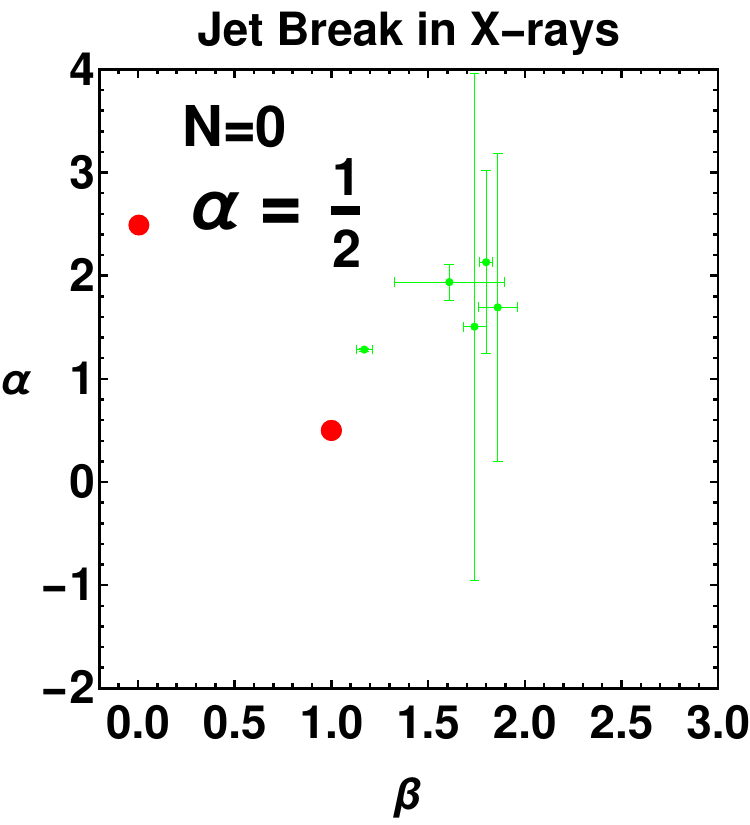}{0.18\textwidth}{(b)}
          \fig{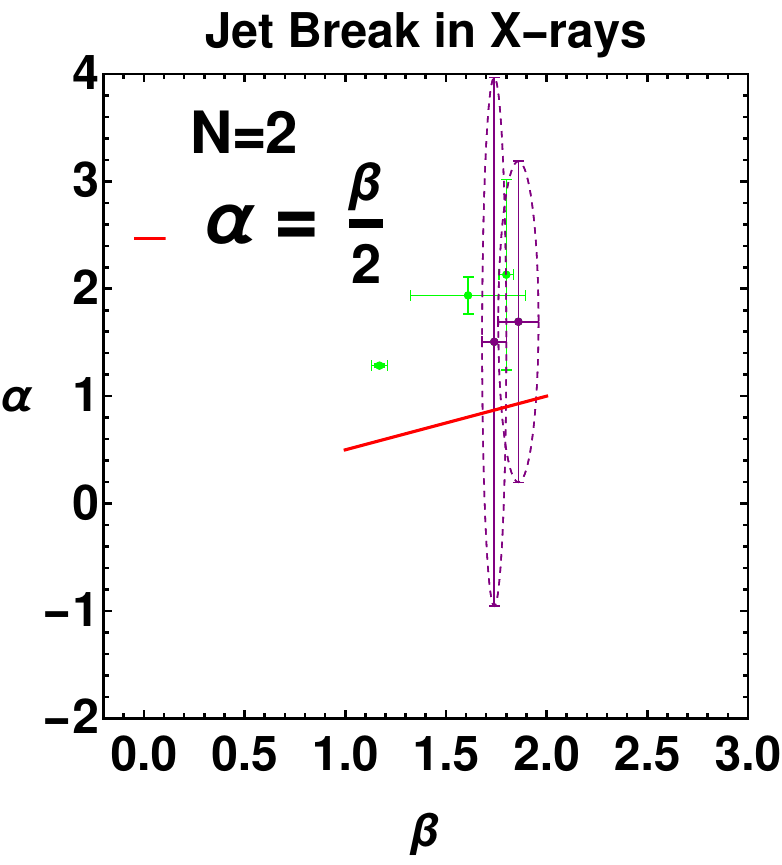}{0.18\textwidth}{(c)}
          }
\gridline{\fig{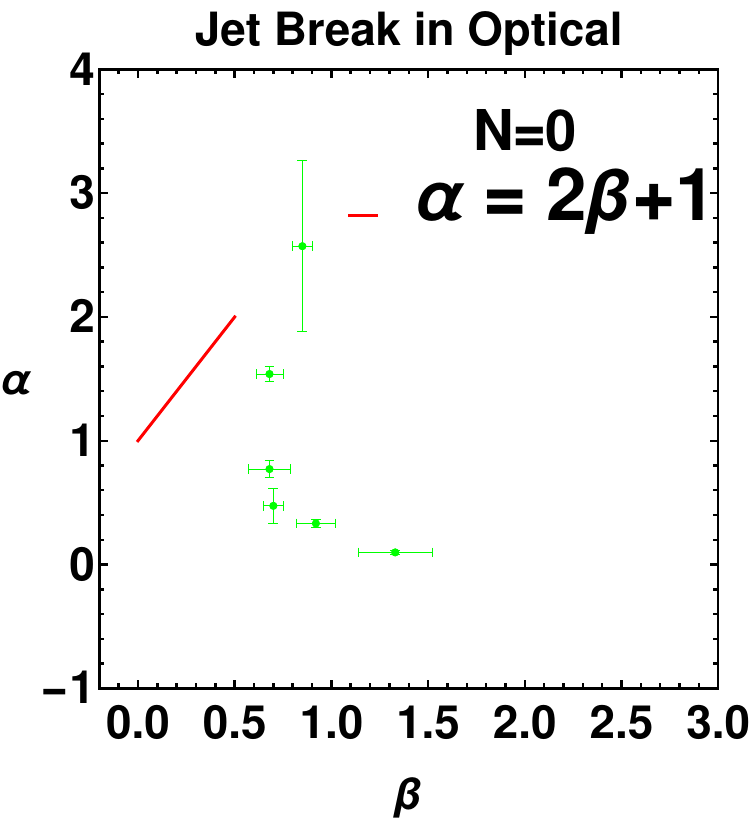}{0.18\textwidth}{(d)}
          \fig{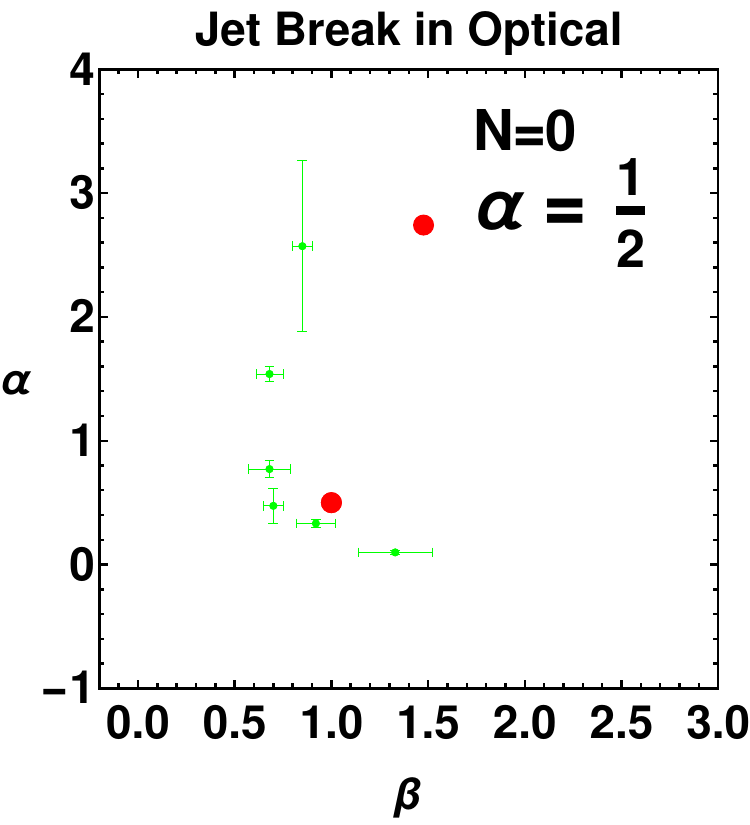}{0.18\textwidth}{(e)}
          \fig{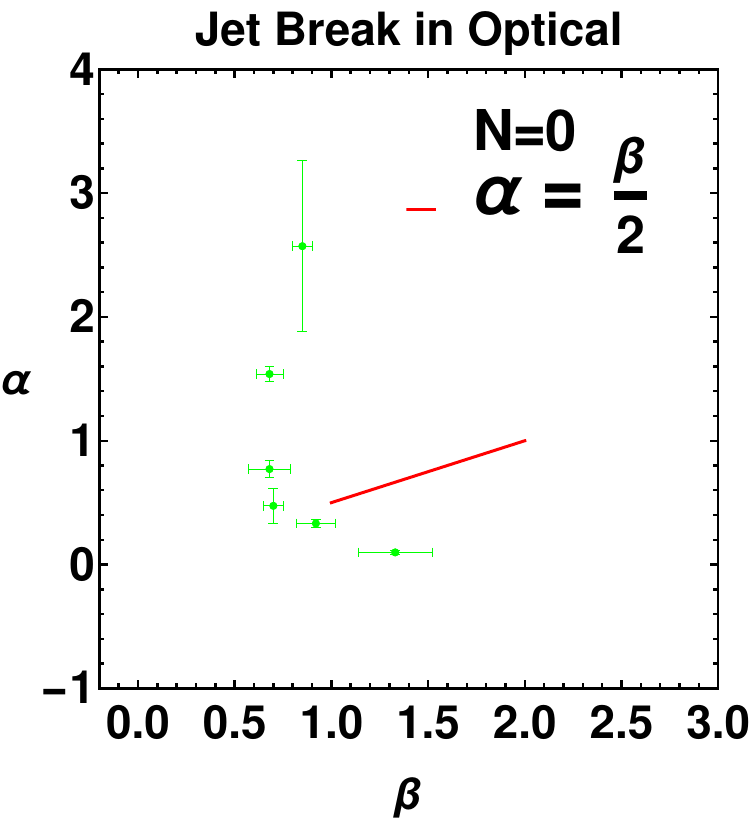}{0.18\textwidth}{(f)}
          }
 \caption{CRs corresponding to the synchrotron FS model with jet-break for X-rays (top row) and optical (bottom row). Columns denote SC, FC, and SC/FC regimes from left to right. GRBs that satisfy the relations for X-ray parameters ($\alpha_{\rm{X}}$ and $\beta_{\rm{X}}$) and optical parameters ($\alpha_{\rm{opt}}$ and $\beta_{\rm{opt}}$) are shown in purple; others are shown in green.}
\label{fig:OPT-JB}
\vspace{-16.64456pt}
\end{figure*}

\section{Results}\label{sec:results}
Here, we present the results of the CRs for each wavelength in separate subsections. Figures \ref{fig:LATnoinj} to \ref{fig:OPT-JB} display the sample of 14 GRBs represented in green. GRBs that satisfy the CRs within errors are depicted in purple.
We note that since the uncertainties are correlated, these must be represented with ellipses instead of a rectangular shape. The fulfillment or not of the CRs is computed mathematically between the regions of the ellipses and the lines representing the CRs. For better visualization, the points showing the CRs are large.
The N symbol in the picture shows the number of GRBs fulfilling that particular CR for each sub-panel. For X-ray and optical, we offer consistently in Figure \ref{fig:XRTnoinj}, \ref{fig:XRTinj}, \ref{fig:OPTnoinj}, and \ref{fig:OPTinj} the SC regime (first column), FC regime (second column), SC/FC regime (third column) for the following density media: $k=0$ (first row), $k=1$ (second row), $k=1.5$ (third row), $k=2$ (fourth row), and $k=2.5$ (fifth row). For gamma-ray, we exclusively show the SC/FC ($\nu > $ max\{$\nu_c,\nu_m$\}) regime for all $k$ values (0, 1, 1.5, 2, and 2.5), depicted in Figures \ref{fig:LATnoinj}. 
Figure \ref{fig:OPT-JB}, we show the jet break CRs for X-ray and optical with density medium SC regime (first column), FC regime (second column), SC/FC regime (third column).
The CRs are graphically represented in all wavelengths as lines or points depending on the values of  $\alpha$ and $\beta$. 
When $p > 2$, CRs are denoted by red lines or points for both with or without energy injection cases. 
For CRs without energy injection where $1 < p < 2$, they are represented by blue a line or point. 
CRs with a jet break where $1 < p < 2$, they are represented by a red line or point.
%Notably, uncertainties within 1 $\sigma$ for GRBs satisfying CRs are displayed as ellipses rather than rectangles because we assume independence between $\alpha$ and $\beta$ parameters.

\subsection{Gamma-ray Closure Relations}
We summarize the CR results for the gamma-rays in Table \ref{table:LAT crSummary}. We here decide to discuss and present only the SC/FC ($\nu > $ max\{$\nu_c,\nu_m$\}) regime since the other scenarios carry small information and are less likely to happen in the high energy gamma-rays.
Under the no energy injection scenario, in the SC/FC ($\nu > $ max\{$\nu_c,\nu_m$\}) regime, 3 GRBs (21.4\%) satisfy the CRs for all values of $k$. This means that there is degeneracy among the several possible media.  GRBs are: 090926A, 171010A, 230812B.
However, in this scenario, only the case of GRB 220101A (7.14\%) satisfies the CRs in the energy injection for all the $k$ values.

\subsection{X-Ray Closure Relations}
\label{CRs in X-rays}
In the case of X-rays, Table \ref{table:XRT crSummary} summarizes the results for the CRs.
In the no energy injection scenario, the most preferred regime is the SC regime where 9 GRBs (64.3\%) in $k=0$ satisfy the CRs followed by $k=1$ and 1.5 with 8 GRBs (57.1\%), 5 GRBs (35.7\%) for $k=2$, and 1 GRB (7.14\%) for $k=2.5$ conform to the CRs. %\textcolor{orange}{where only the SC/FC regime holds significant information and is the most likely scenario}, 
In this case, the ISM is the most preferred environment.

The second most preferred regime is the SC/FC ($\nu > $ max\{$\nu_c,\nu_m$\}) regime where 3 GRBs (21.4\%) satisfy the CRs for all values of $k$.
The least preferred regime is the FC regime ($\nu_c<\nu<\nu_m$) for all $k$ values, with none (0\%) of the GRBs satisfying the CRs.
Similarly, in the energy injection scenario, the SC regime is favored the most, with a stratified medium of $k=2.5$ and wind medium ($k=2$) with 3 GRBs (21.4\%) satisfying the CRs, and only 1 GRB (7.14\%) fulfilling the CR for $k=1.5$. 
In this case, the most preferred environment is the stratified medium with  $k=2.5$ along with the wind medium.
The least preferred regime is the FC and SC/FC regime, where no (0\%) GRB satisfies the CRs for all values of $k$. 

In the jet break scenario, the preferred regime is the SC/FC with 2 GRBs (14.3\%) conforming to the CRs. In this case, both SC and FC regimes are the least favored regimes, with no (0\%) GRBs fulfilling the CR.

\subsection{Optical Closure Relations}
\label{CRs in optical}
For optical wavelengths, Table \ref{table: OPT crSummary} summarizes the results for CRs.
In the no energy injection scenario, the most preferred regime is the SC/FC regime ($\nu > $ max\{$\nu_c,\nu_m$\}), with 3 GRBs (21.4\%) satisfying the CRs for the  $k$ = 2.5 followed by the constant medium $k$ = 0 and $k = 1, 1.5,$ and 2 with only 2 GRB (14.3\%). In this case, the most preferred environment is the stratified medium with $k=2.5$.
%The second most favored regime is the SC regime for $k=2$, with only 1 GRB (7.14\%)  satisfying the CRs.  
The least preferred regimes are the SC and FC regimes for all $k$ values, with none (0\%) of the GRBs satisfying the CRs.
However, in the scenario with energy injection, we observed the most preferred is the SC regime for $k=1, 2$ and 2.5 with only 1 GRB (7.14\%) satisfying the CRs. 
For the same regime $k=0$ and, none (0\%) of the GRBs conforms to the CRs.
%The second most preferred regime is the SC/FC regime, where 1 GRB (6.67\%) satisfies the CRs for all $k$ values.
%For all other cases, no GRB satisfies the CRs for either value of $k$.
The least preferred are the FC and SC/FC regimes, where none (0\%) of the GRBs satisfy the CRs for all values of $k$. 
Also, in this case, the most preferred environment is the stratified medium with $k=1$ and 1.5, along with the wind medium.
In the jet break scenario, no (0\%) GRB fulfils the CR for any regime.

\begin{longrotatetable}
\begin{deluxetable*}{lcccccccccccccccccl}
%\begin{deluxetable*}{lllrrrrrrll}
%\tablenum{2}
\tablecaption{Table summarizes the fulfillment of CRs in multiple wavelengths for the no energy injection scenario (top panel) and energy injection scenario (bottom panel). Column 1 provides the GRB names, while Column 2 indicates the corresponding wavelengths for which a GRB fulfills a given CR - $\gamma$ for gamma-rays, X for X-rays, and O for optical. Columns 3 to 7 indicate the fulfillment status of the given CR in the environment of the SC, FC, and SC/FC regimes for $k$ = 0 (ISM), $k=2$ (stellar wind environment), $k=1, 1.5, 2.5$ (stratified medium) for each GRB. Column 8 shows the preferred medium for each GRB. \label{table:comparisonWEI}}
\tablewidth{700pt}
\tabletypesize{\scriptsize}
\tablehead{
\colhead{GRB Name} & \colhead{Wavelengths} & \multicolumn{3}{c}{$k$ = 0} & \multicolumn{3}{c}{$k$ = 1} & \multicolumn{3}{c}{$k$ = 1.5} & \multicolumn{3}{c}{$k$ = 2} & \multicolumn{3}{c}{$k$ = 2.5} &\multicolumn{1}{c}{preferred medium} \\
\hline
& & SC & FC & SC/FC  & SC & FC & SC/FC  & SC & FC & SC/FC  & SC & FC & SC/FC  & SC & FC & SC/FC 
} 
%\hline
%\decimalcolnumbers
\startdata
\hline
\multicolumn{18}{c}{No Energy Injection ($q=1$)} \\
\hline
%080916C & X, O & X & - & X & X & - & X & X & - & X & O & - & X & O & - & X & all \\
%080916C & $\gamma$, X, O & X & $\gamma$ & X & \textcolor{blue}{X} & $\gamma$ & X & X & $\gamma$ & X & O & $\gamma$ & X & O & $\gamma$ & X\\
090328A & X & X & - & - & X & - & - & X & - & - & X & - & - & - & - & - & $k=0, 1, 1.5, 2$  \\ 
090510A & X & - & - & - & - & - & - & - & - &  - & X & - & - & X & - & - & $k=2, 2.5$ \\
090902B & X, O & - & - & O & X & - & O & X & - & O & - & - & O & - & - & O & all \\
%090902B & $\gamma$, X, O & $\gamma$ & - & - & X & - & - & \textcolor{orange}{X} & - & - & - & - & - & - & - & - \\
090926A & $\gamma$, X & X & - & $\gamma$ & - & - & $\gamma$ & - & - & $\gamma$ & - & - & $\gamma$ & - & - & $\gamma$ & all\\
%100414A & X, O & X & - & O & X & - & O & X & - & O & X & - & O & X & - & O & all \\
%110731A & $\gamma$, X, O & X & - & $\gamma$ & X & - & $\gamma$ & - & - & $\gamma$, O & - & - & $\gamma$, O & - & - & $\gamma$, O & all \\
120711A & X & - & - & - & - & - & - & X & - & - & - & - & - & - & - & - & $k=1.5$ \\
%120711A & $\gamma$, X, O & $\gamma$ & - & - & $\gamma$ & - & - & $\gamma$, \textcolor{orange}{X} & - & - & - & - & - & - & - & - \\
130427A & X, O & X & - & O & X & - & O & X & - & O & - & - & O & - & - & O & all \\
141028A & X & X & - & X & - & - & X & - & - & X & - & - & X & - & - & X & all \\
%160509A & $\gamma$ & - & - & $\gamma$ & - & - & $\gamma$ & - & - & $\gamma$ & - & - & $\gamma$ & - & - & $\gamma$ & all \\
160625B & X & X & - & X & X & - & X & X & - & X & X & - & X & - & - & X & all  \\
170405A & X, O & X & - & X & X & - & X & X & - & X & X & - & X & - & - & X, O & all \\ 
 171010A & $\gamma$, X & X & - & $\gamma$ & X & - & $\gamma$ & - & - & $\gamma$ & - & - & $\gamma$ & - & - & $\gamma$ & all \\
180720B & X & X & - & - & - & - & - & - & - & - & - & - & - & - & - & - & $k=0$ \\
210822A & X & - & - & - & - & - & - & - & - & - & X & - & - & - & - & - & $k=2$ \\
220101A & X & - & - & - & X & - & - & X & - & - & - & - & - & - & - & - & $k=1, 1.5$ \\
% 221009A & $\gamma$, X & - & - & $\gamma$ & - & - & $\gamma$ & - & - & $\gamma$ & X & - & $\gamma$ & - & - & $\gamma$ & all \\
230812B & $\gamma$, X & X & - & $\gamma$ & X & - & $\gamma$ & X & - & $\gamma$ & - & - & $\gamma$ & - & - & $\gamma$ & all \\
\hline
\hline
\multicolumn{18}{c}{Energy Injection ($q=0$)} \\
\hline
%080916C & - & - & - & - & - & - & - & - & - & - & - & - & - & - & - & - & none \\
090328A & X, O & - & - & - & - & - & - & - & - & - & O & - & - & X & - & - & $k=2, 2.5$ \\ 
090510A & X & - & - & - & - & - & - & - & - & - & X & - & - & - & - & - &  $k = 2$\\
090902B & X & - & - & - & - & - & - & - & - & - & - & - & - & X & - & - & $k=2.5$\\
090926A & O & - & - & - & O & - & - & - & - & - & - & - & - & - & - & - & $k=1$ \\
%100414A & X, O & - & - & - & - & - & - & O & - & - & - & - & - & X, O & - & -  & $k=1.5, 2.5$ \\
%110731A & - & - & - & - & - & - & - & - & - & - & - & - & - & - & - & - & none \\
120711A & - & - & - & - & - & - & - & - & - & - & - & - & - & - & - & - & none \\
130427A & - & - & - & - & - & - & - & - & - & - & - & - & - & - & - & - &  none \\
141028A & X & - & - & - & - & - & - & X & - & - & X & - & - & - & - & - & $k=1.5, 2$ \\
%160509A & X, O & - & - & - & - & - & - & X, O & - & - & - & - & - & - & - & - & $k=1.5$\\
160625B & - & - & - & - & - & - & - & - & - & - & - & - & - & - & - & - & none \\
170405A & X, O & - & - & - & - & - & - & - & - & - & X & - & - & O & - & - & $k=2, 2.5$\\
171010A & - & - & - & - & - & - & - & - & - & - & - & - & - & - & - & - & none \\
180720B & - & - & - & - & - & - & - & - & - & - & - & - & - & - & - & - & none \\
210822A & - & - & - & - & - & - & - & - & - & - & - & - & - & - & - & - & none \\
220101A & $\gamma$ & - & - & $\gamma$ & - & - & $\gamma$ & - & - & $\gamma$ & - & - & $\gamma$ & - & - & $\gamma$ & all \\
% 221009A & - & - & - & - & - & - & - & - & - & - & - & - & - & - & - & - & none \\
230812B & X & - & - &- & - & - &- & - & - &- & - & - & - & X & - & - & $k=2$
\enddata
\end{deluxetable*}
%\end{landscape}
%\end{sideways}
\end{longrotatetable}

\begin{table*}
\centering
\caption{Comparison of the preferred environment of 7 GRBs common between our study and the analysis of \cite{gompertz2018environments}.
}
%\begin{center}
\begin{tabular}{lccccl}
\hline
\hline
\multicolumn{4}{c}{Comparison of preferred environment with \cite{gompertz2018environments}} \\
\hline
GRB & \cite{gompertz2018environments} & Our Study & Our Study \\
Name &  & No Energy Injection ($q=1$) & Energy Injection ($q=0$) \\
\hline
\hline

090328A & $k=0$ & $k=0, 1, 1.5, 2$ & $k=2, 2.5$ \\
090902B & $k=0$ & $k=0, 1, 1.5, 2, 2.5$ & $k=2.5$ \\
090926A & $k=2$ & $k=0, 1, 1.5, 2, 2.5$ & $k=1$ \\
120711A & $k=2$ & $k=1.5$ & none \\
130427A & $k=2$ & $k=0, 1, 1.5, 2, 2.5$ & none \\
141028A & Unknown & $k=0, 1, 1.5, 2, 2.5$ & $k=1.5, 2$ \\
160625B & $k=0$ & $k=0, 1, 1.5, 2, 2.5$ & none \\

\hline
\hline
\end{tabular}
%\end{center}
\label{table:comparison}
\end{table*}

\section{Discussion} \label{sec:discussion}
\begin{figure}
    \centering
    \includegraphics[scale=0.68]%,\height=0.60]
    {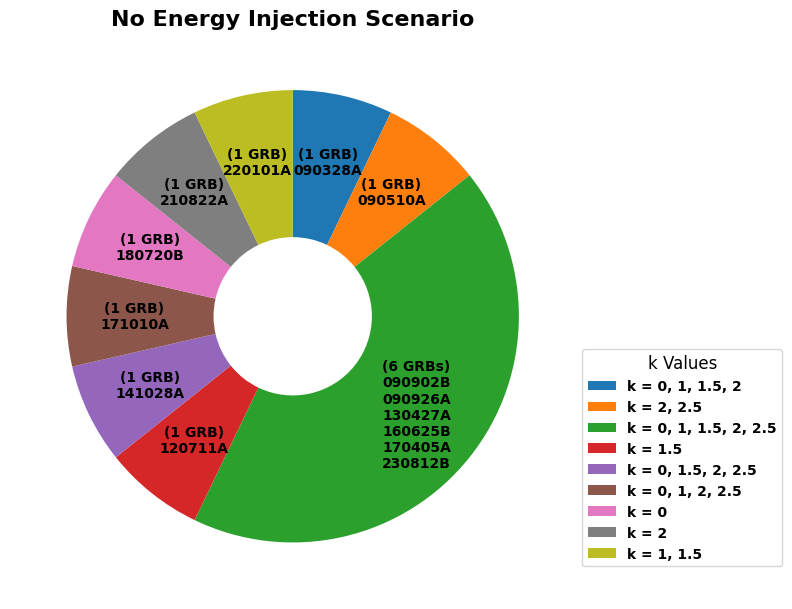}
    % \centering
     \includegraphics[scale=0.68]%,\height=0.60]
    {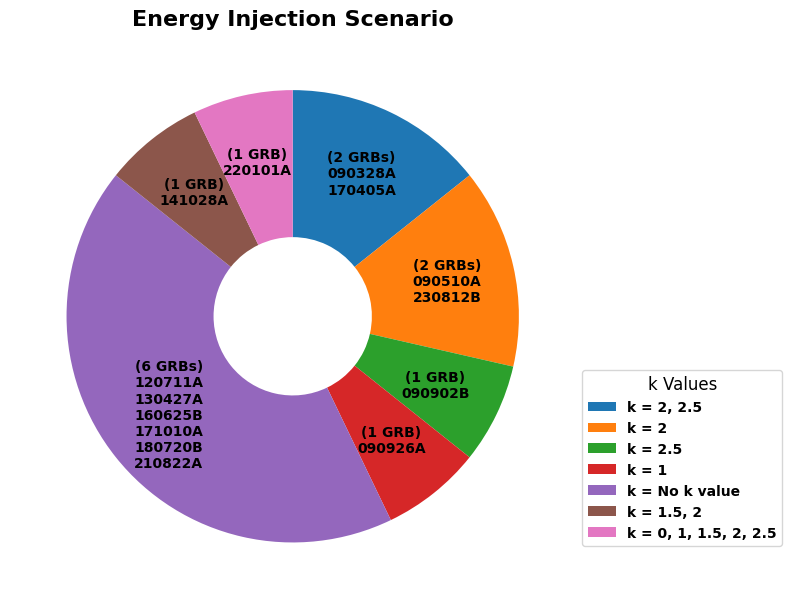}
       \caption{Pie chart representing the environment determination for each GRBs. The color coding indicates different $k$ values or different combinations of those values, as reported in the legend on the right. The upper and lower panels show the cases of no energy and energy injection, respectively.}
       \label{piechart}
    
\end{figure}

In Table \ref{table:comparisonWEI}, we present a detailed comparative analysis of each GRB for both no-energy (top panel) and energy injection scenarios (bottom panel), respectively. 
The individualized comparative analysis provides insights into the energy mechanism and the ambient environment for each GRB across these wavelengths.
The table is comprehensive, and thus, we omit to discuss the singular cases, but we instead provide pie charts (Figure \ref{piechart}) for the analysis in relation to the density environments.

\begin{figure}
   \centering
   \includegraphics[scale=0.7]%,\height=0.60]
  {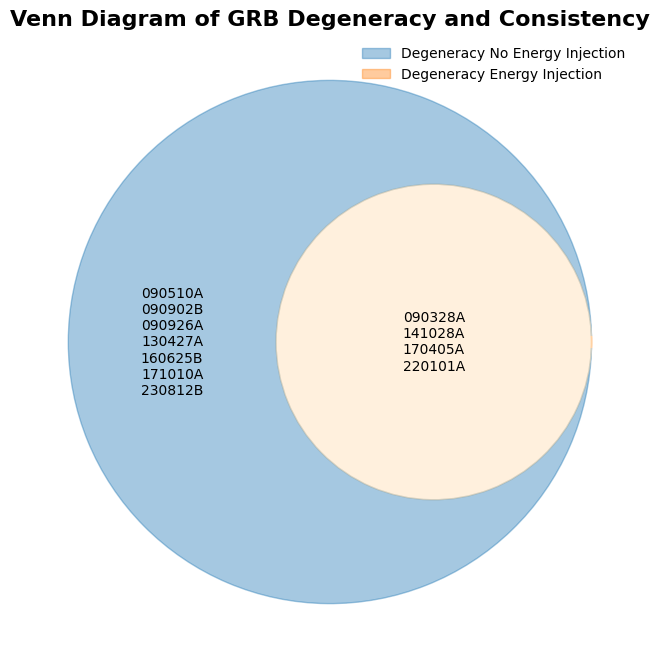}
\caption{The Venn diagram represents the degeneracy compared to the k values among the 14 GRBs in both with and without energy injection scenarios. The color coding represents the two regimes: energy injection in orange and no-energy injection in blue.}
     \label{fig:vienndiag}
\end{figure}

\subsection{The no-energy injection and energy injection scenario}
To simplify the visualization of the energy and no-energy injection cases, we have shown a Venn diagram (see Figure \ref{fig:vienndiag}) illustrating which GRBs undergo degeneracies. 
In the absence of energy injection (see the top panel of Table \ref{table:comparisonWEI}), some GRBs exhibit a preference for a specific value, while some show a preference for more than one value and and some of them that have no values for the energy injection scenario.

Differently from the no-energy injection scenario in which all GRBs satisfy at least one CR in at least one energy band, for the case of energy injection, there are GRBs for which no CR is fulfilled in any of the bands. These are GRBs 120711A, 130427A, 160625B, 171010A, 180720B, and 210822A. On average, it becomes evident that CRs without energy injection is preferred over energy injection. 

We have also considered the jet break, and we notice that the only scenario in which the CRs are satisfied is in X-rays, with 2 cases only for the SC/FC regime.

%Furthermore, \cite{2019ApJ...883..134T} studied a sample of 59 \textit{Fermi}-LAT GRBs and found that only 48 GRBs satisfied at least one CR; the rest did not satisfy any of the CR,which strongly suggests that the energy injection scenario is not necessarily needed to explain the GRB afterglows same as we found in our study. Following this, \cite{2019ApJ...878...52A} claimed that the GRBs (GRB 090902B, GRB 100116A, GRB 100511A, GRB 110625A, GRB 120526A, GRB 120624B, GRB 120709A, GRB 130518A, GRB 140206B, GRB 160325A, and GRB 171120A) which did not satisfy any CRs in the study done by \cite{2019ApJ...883..134T}, exhibit a slower temporal decay with $\alpha < 1$, suggesting the need for continuous energy injection from the central engine to explain these extended emission features. 

When considering different cases, our analysis reveals a predominant preference considering all wavelengths for the SC/FC ($\nu > $ max\{$\nu_c,\nu_m$\}) in the no-energy injection scenario with a total of 36 cases and the second most preferred regime with 31 cases is the SC ($\nu_m<\nu<\nu_c$) regime.  In the energy injection scenario, the SC ($\nu_m<\nu<\nu_c$) regime is most preferred, with 10 cases total.
In the energy injection scenario, out of 10 cases, only one is driven by a constant-density medium. In the no-energy injection scenario, 8 out of 36 fulfill the CRs with a constant density medium.
For the optical data only, in the no energy injection scenario, 11 cases are fulfilled for the SC/FC  with a preference of k=2.5; in the energy injection scenario, the preferred scenario is the one with the stratified medium for $k = 1, 2.5$, and it also equivalently prefers a wind medium.
In the optical data analysis, an identical outcome for the $k=2$ and ISM in the optical was found in the investigation conducted by \citet{2022ApJ...940..169D}.

\subsection{ Discussion regarding the environment}
One necessary takeaway from this study is that we could pinpoint a definite environment for some GRBs.  Two pie charts, in Figure \ref{piechart}, show the GRB names along with their corresponding k values color-coded for both the no-energy injection (upper panel) and energy-injection (lower panel) scenarios. In the no-energy injection regime, GRB 120711A has $k=1.5$, GRB 180720B has $k=0$, and GRB 210822A has $k=2$, while other cases are degenerate in the density profile. However, if we explore the energy injection scenario, the following GRBs have definite medium: GRB 090510A has $k=2$ (but also 2.5 in the no-energy injection scenario), GRB 090902B has $k=2.5$, GRB 090926A has $k=1$, and GRB 230812B has $k=2$. 
The result of GRB 090510A agrees with \cite{2011MNRAS.414.1379P}, who model the multi-wavelength observations with a stratified medium.
We here stress that we have provided in Eq. \ref{Eq:bulklorentz} the relation between the Lorentz $\Gamma$ factor and the kinetic energy, the density profile, and redshift. Thus, one can infer the radii at which the model is a valid approximation. For example, if we consider GRB 130427's time of the last break $t_{break_{3}}= 80909$s and a Lorentz factor of 455 as stated in Ackerman et al. (2013), we will obtain radii of $7.36 \times 10^{20}$ cm. This would be the radii at which there will be a transition towards the interstellar environment about this GRB, and the radii from the progenitor will be marked as the model of the stratified medium that can still be applicable. This is roughly the order of magnitude for the radii of the other GRBs.

\subsection{Comparison with other studies}
The fact that the no-energy injection scenario is favored is an exciting conclusion drawn from our study since many works invoked energy injection scenario to interpret the plateau phase observed in GRB afterglows, especially for \textit{Swift}-XRT LCs, but also for a high-energy component observed in the \textit{Fermi}-LAT LCs. Indeed, \cite{Zhang2006, Nousek2006, 2007ApJ...666.1002Z} explored the \textit{Swift}-XRT LCs and proposed energy injection as a potential scenario to account for the plateau phase or shallow decay phase in the GRB afterglows \citep{Zhang2006,Nousek2006}. However, \cite{GRB110731A:Fermi} examined GRB 110731A using multi-wavelength observations and indicated that the afterglows of some \textit{Fermi}-LAT GRBs are dominated by the bright FS emission rather than prolonged energy injection episode. Furthermore, as discussed in Section \ref{CRs in gamma-rays} \cite{2019ApJ...883..134T} found that 48/59 GRBs observed by Fermi-LAT satisfied at least one CR, thus showing that the standard fireball model is able to explain most of the observations. \cite{2019ApJ...883..134T} show that the preferred density is the constant density medium, while in our case, both the constant and stellar wind are almost equally probable for the no-energy injection scenario. However, the difference between \cite{2019ApJ...883..134T}, and our study is that in our study, we use a deterministic approach with stratified stellar wind, with several values of $k$, while in \cite{2019ApJ...883..134T}, a probabilistic approach is taken and only a constant density medium and a wind medium have been investigated. 
Indeed, \citet{2021ApJS..255...13D} reinforces our results again, as they also found that most of their \textit{Fermi}-LAT GRBs adhere to the CRs for the no-energy injection scenario. These studies and our study strongly suggest that the no-injection scenario is sufficient to explain the dynamics of a significant portion of GRB afterglows in the high-energy regime. This conclusion poses challenges to the previously accepted paradigm of invoking the energy injection scenario. It highlights the reliability of the standard fireball model in explaining GRB afterglows without invoking continuous energy injection for many LAT-detected bursts.

When comparing the seven GRBs (090328A, 090902B, 090926A, 120711A, 130427A, 141028A, and 160625B) common to both our study and the analysis conducted by \cite{gompertz2018environments}, we observed some discrepancies in the preferred environments, as outlined in Table \ref{table:comparison}. 
There are two notable cases that have defined medium in our study: 120711A with $k=1.5$ in the no-energy injection scenario, GRB 090926A with $k=1$ in the energy-injection scenario, while \cite{gompertz2018environments} both have a wind medium. However, one has to note that the case of GRB 120711 is more controversial since, in the energy injection scenario, none of the $k$ values can be determined. On the other hand, GRB 090926A in the no-energy injection scenario has degeneracy since all values of k are possible and, therefore, also the wind medium. Another notable case, undetermined by \cite{gompertz2018environments}, is GRB 141028A, where we find that $k=1.5$ and 2 are viable for the energy injection scenario, and additionally $k=0$ and 2.5 are possible for the no-energy injection case. Given the degeneracies within the same GRBs, it is often challenging to draw a definite conclusion on the environments.
The key difference between \cite{gompertz2018environments} and our study is that they estimated the best-fit value of the electron distribution power-law index, $p$, by calculating a weighted mean of the $p$ values, including the errorbars which were determined using both the $\alpha$ and $\beta$ indices from X-ray and optical data. They analyzed five different scenarios based on the relative positions of the following frequencies: cooling, $\nu_{c}$, the X-ray, $\nu_{x}$, and optical in the R-band, $\nu_{R}$. These scenarios included: $\nu_{c} < \nu_{R} < \nu_{x}$, $\nu_{R} < \nu_{c} < \nu_{x}$, $\nu_{R} < \nu_{x} < \nu_{c}$ in an ISM environment, $\nu_{R} < \nu_{c} < \nu_{x}$ and $\nu_{R} < \nu_{x} < \nu_{c}$ in a wind environment. 
Instead of computing the $p$ values, we directly fit the CRs using the observed $\alpha$ and $\beta$ values from gamma-ray, X-ray, and optical data for both energy and no-energy injection. 
However, in our case, we also have a detailed method for computing the error bars by defining ellipses regions (see Section \ref{sec:results}).
A key difference is that we also explored the stratified medium ($k=1, 1.5, 2.5$) along with the wind ($k=2$) and ISM ($k=0$), while they focused on $k=0$ and 2. Another difference in the preferred environments between our study and their analysis resides in the  LC fitting. \cite{gompertz2018environments} relied on the automated light curve fits available in the \textit{Swift}-XRT catalog for X-ray data. For the optical data, they primarily used values from existing literature when available and, in cases where they were not, fitted the light curves using data from GCN circulars, applying either a PL or BPL model. In contrast, we fit both X-ray and optical LCs following the GRBLC package \citep{Dainotti2024MNRAS.533.4023D} with either a PL or BPL, or their combinations, as appropriate. Additionally, we performed a homogeneous analysis of the optical data (see Section \ref{sec:methodology}).

As a matter of fact, given that our optical LCs have been processed separately in a uniform way by adding all the available data in the literature, including GCN and private communication, has guaranteed a more uniform treatment of the sample, also in view of the color evolution analysis. Continuing on the comparison between the optical and X-ray data analyzed together, it is worth mentioning the treatment by \cite{Wang2015ApJS..219....9W}, as we have mentioned in Section \ref{CRs in X-rays}, where 85 GRBs with simultaneous observations in optical and X-rays have been discussed. The treatment of \cite{Wang2015ApJS..219....9W} is analytical in nature, similar to our approach. Also, the general conclusion about the reliability of the standard forward shock model is similar since at least half of the GRBs in the sample can be explained within the simple afterglow FS model, but there are cases, segments in the LCs in which the CRs are not fulfilled and 3/85 cases in which none of the CR is satisfied. The conclusion of \cite{Wang2015ApJS..219....9W} is that one can indeed recover more cases following the standard model if one also includes the long-lasting reverse shock, structured jets, and stratified circumburst medium density profile with values of $k$ different from 0 and 2. Indeed, this is similar to what we recovered in our study since the number of cases following the CRs is characterized by a stratified medium that includes $k=1$, 1.5, and 2.5.

In the X-ray data, in this current analysis without energy injection, the SC ($\nu_m<\nu<\nu_c$) regime with constant-density medium (with $k= 0$  there are 9 cases of fulfillment) emerges as the most favored, followed by the stratified medium for $k=1$ and 1.5 with 8 cases. Also, \citet{Srinivasaragavan2020ApJ} under the case with no-energy injection showed a disagreement between CRs at high and low energies, suggesting that the emission mechanism may differ at high energies for those without energy injection. 
It is worth noting that both in the no-energy and energy injection scenario, none of the GRBs in our sample demonstrates the preference for the FC ($\nu_c<\nu<\nu_m$) regime for any value of $k$ in either X-rays or the optical band. 
In scenarios involving energy injection, our investigation highlights that the SC ($\nu_m<\nu<\nu_c$) regime across X-ray and optical wavelengths (7 GRBs: GRB 090328A, 090510A, 090902B, 090926A, 141028A, 170405A, and 230812B) is the most preferred vs. the other regimes for example, the SC/FC has no case fulfilled. 
Here, we also note that in another study conducted by \citet{2021PASJ...73..970D} in X-rays, the SC ($\nu_m<\nu<\nu_c$) regime was found to be the preferred regime within the context of energy injection.
FC ($\nu_c<\nu<\nu_m$) regime remains the least favored regime, with none of the examined GRBs satisfying it across X-rays and optical wavelength for any $k$ value.

\section{Summary and Conclusions}
We tested the external FS model through a set of CRs with a sample of  14 GRBs observed contemporaneously in gamma-rays, X-rays, and optical bands by various space-based and ground-based telescopes. 
We observe that most of our sample satisfies at least one CR, indicating that the external FS model can explain numerous characteristics observed in GRBs.
If a GRB satisfies multiple CRs, it indicates the existence of equally plausible scenarios, indicating a more complex nature of the GRB mechanism.
It is crucial to acknowledge that, in numerous instances, there is a degeneracy of scenarios. However, our results remain significant since they enable us to exclude certain scenarios. The inclusion of multiple wavelengths provides an opportunity to break these degeneracies.
If we wish to reconduct this study to the constant or wind-medium, we can assess that the number of GRBs with $k=0$ is 10, while the number of GRBs with $k=2$ is 11 in the no-energy injection scenario. In the energy injection scenario, we have one case only for $k=0$ and 6 cases with $k=2$. \cite{Gompertz2018} have discussed the equal split and how this relates to the progenitor. Almost equality may arise from two options: a diverse progenitor or the same progenitor but with a different environment relating to a different evolutionary stage. One could imagine that a star with a weak stellar wind in its final stage will influence the environment less than a star characterized by a strong wind type. Simplistically, one can assume that more massive stars emit more energy and radiate a strong stellar wind. However, this hypothesis, which was tested via a Kolmogorov-Smirnov Test, does not yield a significant difference between the cumulative distribution of the GRBs characterized by the wind and the constant medium. The interesting perspective discussed in \cite{Gompertz2018} is that there is room for fitting a medium with $k=1$ with some GRBs. Indeed, this is what we did by expanding the stratified medium to different values of $k$. Hence, it is still worth enlarging the sample and investigating this issue more in the future.

Another take away of this study is that the degeneracy we found is challenging to break, and the stratified medium scenario can occur when the emission is unstable. Still, another possibility to look at this scenario and to attempt to break this degeneracy is to take one step back and consider only the two possibilities of wind and constant medium but enlarge the cases with different values of the $q$ parameter (e.g., -0.5, -0.3, 0.3, and 0.5). This analysis will be the object of a forthcoming paper.

\section{Acknowledgment} 
\label{sec:Acknowledgment}
We thank the work of Rishav Bhattacharjee for his help in developing the initial code for showing the multi-wavelength LCs. We are particularly grateful to Prof. Bing Zhang for the very useful comments about the conclusion on the stratified medium vs. the constant medium and the problem of the degeneration of the parameters. E. Bissaldi is grateful to NAOJ to have hosted her at the Comos Lodge in June 2023, where the collaboration on this paper started. E. Bissaldi is also grateful to the Project Title: H2020-MSCA-RISE-2016 NEWS, Project ID Number: 734303 for supporting her visit to NAOJ.

\bibliography{sample631}{}
\bibliographystyle{aasjournal}



\end{document}